# Logical Relations for Monadic Types[†]

JEAN GOUBAULT-LARRECQ[1 ‡]

SŁAWOMIR LASOTA[2 §]

DAVID NOWAK[3 ¶]

[1] *LSV, ENS Cachan, CNRS, INRIA*

[2] *Institute of Informatics, Warsaw University*

[3] *Research Center for Information Security, AIST*

Logical relations and their generalizations are a fundamental tool in proving properties of lambda-calculi, e.g., yielding sound principles for observational equivalence. We propose a natural notion of logical relations able to deal with the monadic types of Moggi's computational lambda-calculus. The treatment is categorical, and is based on notions of subsconing, mono factorization systems, and monad morphisms. Our approach has a number of interesting applications, including cases for lambda-calculi with non-determinism (where being in logical relation means being bisimilar), dynamic name creation, and probabilistic systems.

## Contents

† A preliminary version of this paper was presented at the 11th Annual Conference of the European Association for Computer Science Logic (CSL'02), Edinburgh, Scotland, 22–25 September 2002 (Goubault-Larrecq et al., 2002).
‡ Work partially supported by the RNTL project EVA, and the ACI jeunes chercheurs "Sécurité informatique, protocoles cryptographiques et détection d'intrusions".
§ Work partially supported by the Polish KBN grant 7 T11C 002 21, and the IST-2005-16004 Integrated Project SENSORIA: Software Engineering for Service-Oriented Overlay Computers.
¶ Work partially supported by the ACI jeunes chercheurs "Sécurité informatique, protocoles cryptographiques et détection d'intrusions".

## 1. Introduction

1.1. *Motivation and context.*

Logical relations and their generalizations (Mitchell, 1996) are a fundamental tool in proving properties of lambda-calculi, e.g., characterizing lambda-definability (Plotkin, 1980; Jung and Tiuryn, 1993; Alimohamed, 1995; Fiore and Simpson, 1999), proving equational completeness (Statman, 1985; Mitchell, 1996), studying parametric polymorphism (Reynolds, 1983; Ma and Reynolds, 1992; Lazić and Nowak, 2000) notably. On the other hand, Moggi's computational lambda-calculus (Moggi, 1991) has proved useful to define various notions of computations on top of the lambda-calculus: side-effects, input-output, continuations, non-determinism (Wadler, 1992), probabilistic computation (Ramsey and Pfeffer, 2002) in particular.

What should then be a natural notion of logical relation for Moggi's computational lambda-calculus? Although there is no unique answer to this question, we propose one that is satisfying in practice. We shall demonstrate the relevance of our approach by illustrating our construction on monads for non-determinism, dynamic name creation, and probabilistic computation.

Moggi's insight is based on categorical semantics: while categorical models of the $\lambda$-calculus are cartesian closed categories (CCCs), the computational lambda-calculus requires CCCs with a strong monad $(\boldsymbol{T}, \boldsymbol{\eta}, \boldsymbol{\mu}, \boldsymbol{t})$. The *monadic types* of the computational lambda-calculus are given by the syntax:

$$\tau ::= b|\tau \to \tau|\tau \times \tau|\boldsymbol{T}(\tau)$$

where $b$ ranges over a set $B$ of so-called *base types*, and $\boldsymbol{T}(\tau)$ is meant to denote the type of *computations* of type $\tau$. Compared to the lambda-calculus, Moggi's calculus has an additional **val** operation, of type $\tau \to \boldsymbol{T}(\tau)$, and an additional **let** $x = u$ **in** $v$ construct, of type $\boldsymbol{T}(\tau')$ provided $u$ has type $\boldsymbol{T}(\tau)$ and $v$ has type $\boldsymbol{T}(\tau')$ under the assumption $x : \tau$. Every computational lambda-term has a unique interpretation as a morphism in a CCC with a strong monad. In fact the category $\boldsymbol{Comp}$ whose objects are types and whose morphisms are terms up to $\beta\eta$-conversion is the free CCC-with-a-strong-monad over the set $B$.

Accordingly, our study will rest on categorical principles. While there is a flurry of generalizations of logical relations (Kripke logical relations (Mitchell, 1996), lax logical relations (Plotkin et al., 2000), pre-logical relations (Honsell and Sannella, 2002), etc.), we use *subscones* (Mitchell and Scedrov, 1993) as a unifying framework for defining logical relations. Recall that subscones over $\boldsymbol{Set}$ allow one to define logical relations, and subscones over the presheaf category $\boldsymbol{Set}^{\mathcal{I}}$ lead to $\mathcal{I}$-indexed Kripke logical relations (Mitchell and Scedrov, 1993). Technically, the development in (Mitchell and Scedrov, 1993) is based on unique *lifting* of the CCC structure to the subscone. Our whole endeavor then reduces to finding appropriate liftings of monads on categories $\boldsymbol{C}$ to the subscone category $(\mathbb{C} \mathrel{\restriction\!\!\downarrow} |_|)$ (see Section 4).

The important property of logical relations is the so-called Basic Lemma (Mitchell, 1996): meanings of a lambda-term in different models w.r.t. related environments are related. This is immediate for subscones, and stems from the fact that $\boldsymbol{Comp}$ is the free CCC-with-a-strong-monad on $B$ (a trivial adaptation of Proposition 5.2 in (Mitchell and Scedrov, 1993)). In particular, that any two closed terms that are in logical relation are observationally equivalent is immediate.

1.2. *Outline.*

We return to preliminaries in Section 2. We then define liftings of monads to *scones* in Section 3; this is simpler than for subscones, and of independent interest. The construction is based on the use of monad morphisms. We then lift monads to *sub*scones in Section 4, using a mono factorization system. The important case where the target category $\boldsymbol{C}$ is a product of two categories is investigated in Section 5: this is where *binary* logical relations arise, allowing us to compare two models. We terminate our lifting construction by lifting the monoidal structure and monad strength in Sections 6 and 7, respectively. Section 8 establishes a result by which adjunctions give rise to monad morphisms. In Section 9, we return to the basics of subscone theory. While the standard construction of the CCC structure over the subscone requires a functor $|_|$ that commutes with finite products, we show that the use of mono factorization systems, as in Section 4, allows us to relax this requirement to $|_|$ being only *monoidal.* While we do not make any use here of this observation, monoidal functors are more natural from a categorical point of view than product preserving ones, and we feel it should be interesting in future applications (we have some already, but we refrain from including them in this paper).

It remains to test the relevance of our construction (Section 10): the logical relations thus defined characterize bisimulations when $\boldsymbol{T}$ is the non-determinism monad (as suggested in (Lazić and Nowak, 2000)), a generalization of Larsen and Skou's probabilistic bisimulations (Larsen and Skou, 1991) when $\boldsymbol{T}$ is a measure monad (Giry, 1981; Jones, 1990), and a notion close to Pitts and Stark's logical relations for observational equivalence of programs that create names dynamically (Pitts and Stark, 1993; Stark, 1998). We comment on related work in Section 11 and conclude in Section 12.

## 2. Preliminaries

Fix two categories $\boldsymbol{C}$ and $\mathbb{C}$ and a functor $|_| : \boldsymbol{C} \to \mathbb{C}$. Consider the comma category $(\mathbb{C} \downarrow |_|)$, whose objects are tuples $\langle S, f, A\rangle$, with $f : S \to |A|$ in $\mathbb{C}$ and whose morphisms are pairs $\langle g, h\rangle : \langle S, f, A\rangle \to \langle S', f', A'\rangle$, $g : S \to S'$ in $\mathbb{C}$ and $h : A \to A'$ in $\boldsymbol{C}$, such that the diagram on the right commutes in $\mathbb{C}$.

$$\begin{array}{ccc} S & \xrightarrow{f} & |A| \\ {\scriptstyle g}\downarrow & & \downarrow{\scriptstyle |h|} \\ S' & \xrightarrow[f']{} & |A'| \end{array} \qquad (1)$$

This category is the *scone of* $\boldsymbol{C}$ *over* $\mathbb{C}$ via $|_|$, $(\mathbb{C} \downarrow |_|)$. (We extend here terminology of (Mitchell and Scedrov, 1993), where the name *scone* was reserved to the case $\mathbb{C} = \boldsymbol{Set}$, $|_| = \boldsymbol{C}(1, _)$ only.) The projection functor $U : (\mathbb{C} \downarrow |_|) \to \boldsymbol{C}$ maps $\langle S, f, A\rangle$ to $A$ and the morphism $\langle g, h\rangle$ to $h$.

In the sequel we shall be especially interested in the case where $\mathbb{C} = \boldsymbol{Set}$, and $|_| = \boldsymbol{C}(1, _)$ is the *global section* functor, where 1 is terminal in $\boldsymbol{C}$. Another interesting situation arises when $\boldsymbol{C} = \mathbb{C}\times\mathbb{C}$ and $|(A, B)| = A\times B$, assuming that $\mathbb{C}$ has finite products. Objects of the scone then represent binary relations between objects in $\mathbb{C}$. In this case, given two functors $|_|_1 : \boldsymbol{C}_1 \to \mathbb{C}$ and $|_|_2 : \boldsymbol{C}_2 \to \mathbb{C}$, we may define $|_| : \boldsymbol{C} \to \mathbb{C}$, for $\boldsymbol{C} = \boldsymbol{C}_1\times\boldsymbol{C}_2$, by $|(A_1, A_2)| = |A_1|_1\times|A_2|_2$.

Further assume we are given a monad $(\boldsymbol{T}, \boldsymbol{\eta}, \boldsymbol{\mu})$ on $\boldsymbol{C}$. When $\boldsymbol{C} = \boldsymbol{C}_1\times\boldsymbol{C}_2$, the monad $\boldsymbol{T}$ on $\boldsymbol{C}$ will be usually defined pointwise, by two monads $\boldsymbol{T}_1$ and $\boldsymbol{T}_2$ on $\boldsymbol{C}_1$ and $\boldsymbol{C}_2$, respectively: $\boldsymbol{T}(A_1, A_2) = (\boldsymbol{T}_1(A_1), \boldsymbol{T}_2(A_2))$.

## 3. Lifting of a Monad to a Scone

By *lifting of a monad* $(\boldsymbol{T}, \boldsymbol{\eta}, \boldsymbol{\mu})$ to the scone $(\mathbb{C} \downarrow |_|)$ of $\boldsymbol{C}$ over $\mathbb{C}$ we mean a monad $(\widetilde{T}, \widetilde{\eta}, \widetilde{\mu})$ on $(\mathbb{C} \downarrow |_|)$ such that the diagram

$$\begin{array}{ccc} (\mathbb{C} \downarrow |_|) & \xrightarrow{\widetilde{T}} & (\mathbb{C} \downarrow |_|) \\ {\scriptstyle U}\downarrow & & \downarrow{\scriptstyle U} \\ \boldsymbol{C} & \xrightarrow[\boldsymbol{T}]{} & \boldsymbol{C} \end{array} \tag{2}$$

commutes, i.e. $U \circ \widetilde{T} = \boldsymbol{T} \circ U$ and moreover

$$U\widetilde{\eta} = \boldsymbol{\eta}_U \quad \text{and} \quad U\widetilde{\mu} = \boldsymbol{\mu}_U. \tag{3}$$

By $U\widetilde{\eta}$ and $\boldsymbol{\eta}_U$ we mean the two possible compositions of a natural transformation with $U$, similarly $U\widetilde{\mu}$ and $\boldsymbol{\mu}_U$. The equations (3) amount to the requirement that the two diagrams on the right commute, for all objects $X$ in $(\mathbb{C} \downarrow |_|)$:

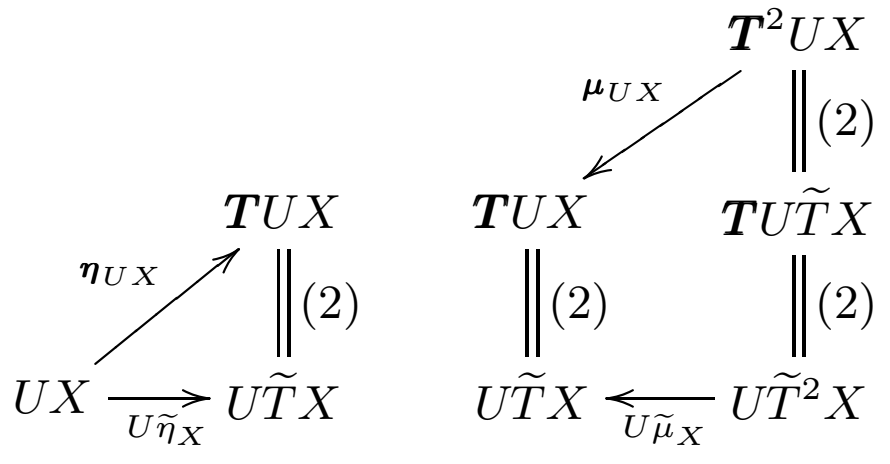

In other words, the functor $U$ together with the identity natural transformation is a morphism of monads from $\widetilde{T}$ to $\boldsymbol{T}$. (We recall monad morphisms shortly.) Note that the equations (3) determine the $\boldsymbol{C}$-components of $\widetilde{\eta}$ and $\widetilde{\mu}$ unambiguously. Moreover, diagram (2) determines the $\boldsymbol{C}$-component of the action of $\widetilde{T}$ on objects and morphisms, i.e. $\langle S, f, A\rangle$ is necessarily mapped to $\langle\widetilde{S}, \widetilde{f}, \boldsymbol{T}A\rangle$, for some $\widetilde{S}, \widetilde{f}$ and a morphism $\langle g, h\rangle$ is necessarily mapped to $\langle\widetilde{g}, \boldsymbol{T}h\rangle$, for some $\widetilde{g}$.

Our notion of lifting could be stated more generally, for an arbitrary pair of categories, a functor from the first one to the second one and a monad on the second category. In fact, in the next section we consider a lifting of $\boldsymbol{T}$ to another category, namely a suitable subcategory of $(\mathbb{C} \downarrow |_|)$.

To be able to give an appropriate lifting we assume another monad $(\mathbb{T}, \eta, \mu)$ on $\mathbb{C}$ such that the two monads $\boldsymbol{T}$ and $\mathbb{T}$ are related by a *monad morphism* from $\boldsymbol{T}$ to $\mathbb{T}$, i.e. a natural transformation

$$\sigma : \mathbb{T}|_| \;\Rightarrow\; |\boldsymbol{T}|$$

making the following two diagrams commute, for each object $A$ in $\boldsymbol{C}$:

$$\begin{array}{ccccc}
 & & & & \mathbb{T}^2|A| \\
 & & & \swarrow^{\mathbb{\mu}_{|A|}} & \downarrow{\scriptstyle \mathbb{T}\sigma_A} \\
 & \mathbb{T}|A| & \mathbb{T}|A| & & \mathbb{T}|\boldsymbol{T}A| \\
\nearrow^{\mathbb{\eta}_{|A|}} & \downarrow{\scriptstyle \sigma_A} & \downarrow{\scriptstyle \sigma_A} & & \downarrow{\scriptstyle \sigma_{\boldsymbol{T}A}} \\
|A| \xrightarrow[|\boldsymbol{\eta}_A|]{} & |\boldsymbol{T}A| & |\boldsymbol{T}A| & \xleftarrow[|\boldsymbol{\mu}_A|]{} & |\boldsymbol{T}^2A|
\end{array} \tag{4}$$

To be formal, a monad morphism is a *pair* $(|_|, \sigma)$ satisfying the equations above. We shall however continue to say that $\sigma$ is a monad morphism, when $|_|$ is understood.

Having $\sigma$, we define $\widetilde{T}$ on objects by

$$\langle S, f, A\rangle \qquad \longmapsto \qquad \langle \mathbb{T}S, \sigma_A \circ \mathbb{T}f, \boldsymbol{T}A\rangle$$

exploiting that if $S \xrightarrow{f} |A|$ then $\mathbb{T}S \xrightarrow{\mathbb{T}f} \mathbb{T}|A| \xrightarrow{\sigma_A} |\boldsymbol{T}A|$ is a morphism. On morphisms we define $\widetilde{T}$ by

$$\langle g, h\rangle \qquad \longmapsto \qquad \langle \mathbb{T}g, \boldsymbol{T}h\rangle$$

since from $\begin{array}{ccc} S & \xrightarrow{f} & |A| \\ {\scriptstyle g}\downarrow & & \downarrow{\scriptstyle |h|} \\ S' & \xrightarrow[f']{} & |A'| \end{array}$ we deduce that $\begin{array}{ccccc} \mathbb{T}S & \xrightarrow{\mathbb{T}f} & \mathbb{T}|A| & \xrightarrow{\sigma_A} & |\boldsymbol{T}A| \\ {\scriptstyle \mathbb{T}g}\downarrow & & & & \downarrow{\scriptstyle |\boldsymbol{T}h|} \\ \mathbb{T}S' & \xrightarrow[\mathbb{T}f']{} & \mathbb{T}|A'| & \xrightarrow[\sigma_{A'}]{} & |\boldsymbol{T}A'| \end{array}$ commutes, by naturality of $\sigma$. Moreover, we put

$$\widetilde{\eta}_{\langle S,f,A\rangle} = \langle \mathbb{\eta}_S, \boldsymbol{\eta}_A\rangle \quad\text{and}\quad \widetilde{\mu}_{\langle S,f,A\rangle} = \langle \mathbb{\mu}_S, \boldsymbol{\mu}_A\rangle.$$

Checking that this defines a monad is straightforward. First, to check that unit and multiplication are well defined it is sufficient to merge the commuting diagrams (4) and complete them with naturality squares for $\mathbb{\eta}$ and $\mathbb{\mu}$ as shown on the right.

$$\begin{array}{ccccc}
S & \xrightarrow{\mathbb{\eta}_S} & \mathbb{T}S & \xleftarrow{\mathbb{\mu}_S} & \mathbb{T}^2S \\
 & & & & \downarrow{\scriptstyle \mathbb{T}^2 f} \\
{\scriptstyle f}\downarrow & & \downarrow{\scriptstyle \mathbb{T}f} & & \mathbb{T}^2|A| \\
 & & & \swarrow^{\mathbb{\mu}_{|A|}} & \downarrow{\scriptstyle \mathbb{T}\sigma_A} \\
 & & \mathbb{T}|A| & & \mathbb{T}|\boldsymbol{T}A| \\
 & \nearrow^{\mathbb{\eta}_{|A|}} & \downarrow{\scriptstyle \sigma_A} & & \downarrow{\scriptstyle \sigma_{\boldsymbol{T}A}} \\
|A| & \xrightarrow[|\boldsymbol{\eta}_A|]{} & |\boldsymbol{T}A| & \xleftarrow[|\boldsymbol{\mu}_A|]{} & |\boldsymbol{T}^2A|
\end{array}$$

Unit $\widetilde{\eta}$ and multiplication $\widetilde{\mu}$ are natural since they are defined pointwise and $\boldsymbol{\eta}$, $\boldsymbol{\mu}$, $\mathbb{\eta}$ and $\mathbb{\mu}$ are. Verifying monad laws is immediate, by the same argument.

The monad morphism (4) can be equivalently given by a lifting of $|_|$ to the categories of algebras of the monads, i.e. by a functor $\widetilde{|_|} : \boldsymbol{C_T} \to \mathbb{C}_{\mathbb{T}}$ making the diagram commute:

$$\begin{array}{ccc} \boldsymbol{C_T} & \xrightarrow{\widetilde{|_|}} & \mathbb{C}_{\mathbb{T}} \\ {\scriptstyle U_{\boldsymbol{T}}}\big\downarrow & & \big\downarrow{\scriptstyle U_{\mathbb{T}}} \\ \boldsymbol{C} & \xrightarrow[|_|]{} & \mathbb{C} \end{array}$$

where $\boldsymbol{C_T}$ and $\mathbb{C}_{\mathbb{T}}$ denote categories of algebras of the monads and $U_{\boldsymbol{T}}$ and $U_{\mathbb{T}}$ denote the obvious forgetful functors. In fact, for fixed monads $\boldsymbol{T}$ and $\mathbb{T}$ and a fixed functor $|_|$, there is a one-to-one correspondence between the monad morphisms $\sigma$ and liftings of $|_|$ to algebras. The proof of this fact can be found in (Appelgate, 1965) or in (Johnstone, 1975).

## 4. Lifting of a Monad to a Subscone

Following, and slightly extending (Mitchell and Scedrov, 1993), we may call the *subscone of* $\boldsymbol{C}$ *over* $\mathbb{C}$ the full subcategory $(\mathbb{C} \hookdownarrow |_|)$ of $(\mathbb{C} \downarrow |_|)$ consisting of all objects $\langle S, f, A\rangle$ with $f$ a mono, written $S \overset{f}{\hookrightarrow} |A|$. (We shall actually define the subscone slightly differently below.)

When $\mathbb{C} = \boldsymbol{Set}$ and $|A|$ is given by $\boldsymbol{C}(1, A)$, each object $S \hookrightarrow |A|$ in the subscone represents a subset of global elements of $A$. In the binary case, i.e. when $\boldsymbol{C} = \boldsymbol{C}_1 \times \boldsymbol{C}_2$ and $|(A_1, A_2)| = \boldsymbol{C}_1(1_1, A_1) \times \boldsymbol{C}_2(1_2, A_2)$, $S \hookrightarrow |(A_1, A_2)|$ corresponds to a binary relation on global elements of $A_1$ and $A_2$—when $A_1$ and $A_2$ are the respective denotations of type $\tau$ in two given models, this will be the logical relation at type $\tau$.

For technical reasons, we require that $\mathbb{C}$ has a *mono factorization system.* This is essentially an epi-mono factorization (Adamek et al., 1990), except we relax part of the definition: we keep the mono part but do not require the epis in the sequel. Alternately, this is a factorization system where one of the classes of morphisms is required to consist of monos only.

Formally, a mono factorization system is given by two distinguished subclasses of morphisms in $\mathbb{C}$, the so-called *pseudoepis* $\twoheadrightarrow$ and the so-called *relevant monos* $\hookrightarrow$. The latter must be monos, while the former are not required to be epis. Both classes must be closed under composition with isomorphisms.

Each morphism $f$ in $\mathbb{C}$ must factor as $f = m \circ e$ for some pseudoepi $e$ and some relevant mono $m$; and each commuting square (5) has a diagonal making both triangles commute as in (6). We call this diagonal morphism the *diagonal fill-in.* Note that the diagonal fill-in is necessarily unique and that whenever the lower-right triangle commutes, the upper-left triangle does too. Furthermore, the latter property guarantees that the factorization $f = m \circ e$ is unique up to iso.

$$\begin{array}{ccc} \cdot & \twoheadrightarrow & \cdot \\ \big\downarrow & & \big\downarrow \\ \cdot & \hookrightarrow & \cdot \end{array} \qquad (5)$$

$$\begin{array}{ccc} \cdot & \twoheadrightarrow & \cdot \\ \big\downarrow & \swarrow & \big\downarrow \\ \cdot & \hookrightarrow & \cdot \end{array} \qquad (6)$$

In particular, we do not require neither pseudoepis nor relevant monos to be closed under composition, which holds true for an epi-mono factorization system, see e.g. (Adamek et al., 1990,

Chapter 14). But it is easy to deduce from the diagonal fill-in property that a composition of two pseudoepis is pseudoepi indeed, and similarly a composition of two relevant monos is a relevant mono, see e.g. (Barr, 1998). It is also proved there that both classes contain all isomorphisms.

In fact, the factorization of $f$ as $m \circ e$ determines uniquely a so-called *relevant subobject* of the codomain, defined as follows. Two relevant monos in $\mathbb{C}$ with the same codomain, $S_1 \overset{f_1}{\hookrightarrow} S$ and $S_2 \overset{f_2}{\hookrightarrow} S$ are called equivalent if and only if there exist $g_1$ and $g_2$ making the two triangles commute:

$$\begin{array}{ccc} S_1 & \underset{g_1}{\overset{g_2}{\rightleftarrows}} & S_2 \\ & {\scriptstyle f_1}\searrow \qquad \swarrow{\scriptstyle f_2} & \\ & S & \end{array}$$

i.e., $f_1 \circ g_1 = f_2$ and $f_2 \circ g_2 = f_1$. A *relevant subobject* of $S$ is an equivalence class of relevant monos with codomain $S$. Equivalently, we could take as objects of the subscone all relevant subobjects of $|A|$, for all objects $A$ in $\boldsymbol{C}$. We prefer to keep the simpler presentation, despite the fact that this implies that some constructions in the sequel are only determined up to isomorphism, e.g., (7) below.

We come back to the definition of the subscone:

**Definition 4.1.** Given two categories $\boldsymbol{C}$, $\mathbb{C}$, a functor $|_| : \boldsymbol{C} \to \mathbb{C}$, and a mono factorization system on $\mathbb{C}$, the *subscone of* $\boldsymbol{C}$ *over* $\mathbb{C}$ is the full subcategory $(\mathbb{C} \mathbin{\rotatebox[origin=c]{-90}{$\hookrightarrow$}} |_|)$ of $(\mathbb{C} \downarrow |_|)$ consisting of all objects $\langle S, f, A \rangle$ with $f$ a relevant mono $S \overset{f}{\hookrightarrow} |A|$ .

It may seem that the notation $(\mathbb{C} \mathbin{\rotatebox[origin=c]{-90}{$\hookrightarrow$}} |_|)$ is too vague, as it does not mention $\boldsymbol{C}$ or the mono factorization system explicitly. It will be clear that making all parameters explicit would make the notation extremely heavy.

Additionally, we shall assume that $\mathbb{T}e$ is pseudoepi for every morphism $e$ in a subclass of all pseudoepis called *relevant pseudoepis*, which we shall define shortly. This will be used in Diagram (11) below. In most applications, it will suffice to check that $\mathbb{T}$ preserves pseudoepis.

Note the following simple and important fact:

**Fact 4.2.** The first component $g$ of a morphism $\langle g, h \rangle$ (recall that $\begin{array}{ccc} S & \hookrightarrow & |A| \\ {\scriptstyle g}\downarrow & & \downarrow{\scriptstyle |h|} \\ S' & \hookrightarrow & |A'| \end{array}$ commutes) in a subscone is uniquely determined by the second component $h$.

This is because the bottom arrow is now mono.

Let us define a lifting of the monad to the subscone by analogy with (2) and (3) for the scone. In the binary case mentioned at the beginning of this section, this corresponds to a lifting of a monad to the category of binary relations (as objects) and relation preserving functions (as morphisms).

4.1. $\widetilde{T}$ *on objects.*

The lifting $\widetilde{T}$ on objects is given by the mono part of the mono factorization of the lifting of the previous section: $\langle S, f, A\rangle$ is taken to $\langle \widetilde{S}, m, \boldsymbol{T}A\rangle$ given by the diagram on the right.

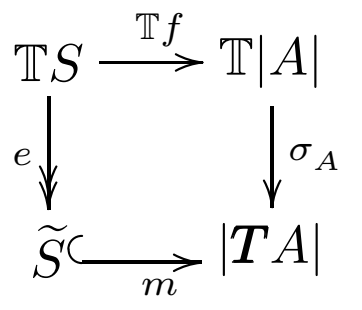

(7)

We call pseudoepis $e$ arising in this way $\mathbb{T}, \sigma$*-relevant pseudoepis.* That is, a $\mathbb{T}, \sigma$-relevant pseudoepi is the pseudoepi part of a factorization of a morphism of the form $\sigma_A \circ \mathbb{T}f$, where $f$ is a relevant mono. For short, we shall call them *relevant pseudoepis* when $\mathbb{T}$ and $\sigma$ are clear from context.

Clearly $\widetilde{T}$ is defined only up to iso. Formally, the construction would be unambiguous if we worked with subobjects of $|\boldsymbol{T}A|$, which are determined uniquely.

4.2. $\widetilde{T}$ *on morphisms.*

Given a morphism $\langle g, h\rangle$, the diagram on the right commutes. Then the action of $\widetilde{T}$ on $\langle g, h\rangle$ will be obtained from the unique diagonal guaranteed by (6). We construct diagram (9) below from two copies of (7).

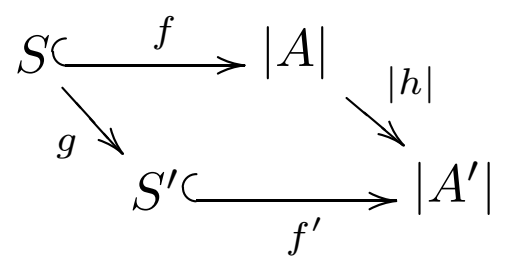

(8)

All four given faces of the cube commute. Both front and back faces commute by definition of $\widetilde{T}$ on objects: they are copies of diagram (7). The right-hand face is a naturality square of $\sigma$; the top face is by application of $\mathbb{T}$ to diagram (8), hence commutes by definition of morphisms in the subscone.

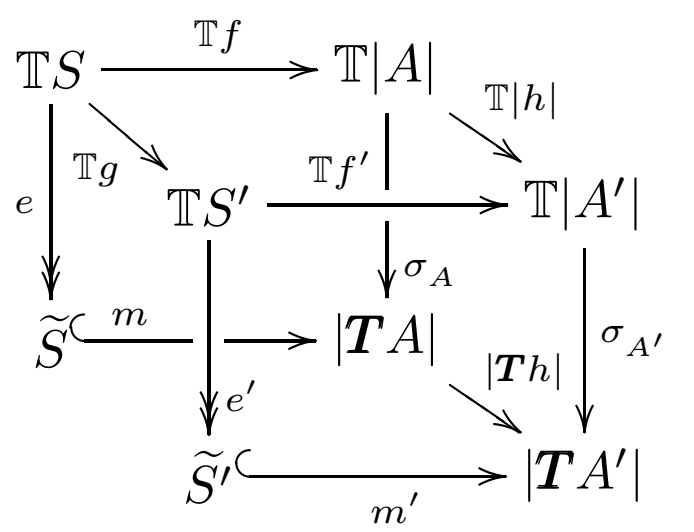

(9)

Now, an instance of diagram (5) can be found in (9) by two walks from $\mathbb{T}S$ to $|\boldsymbol{T}A'|$: one starts with the pseudoepi $\mathbb{T}S \xrightarrow{e} \widetilde{S}$, the other ends with the relevant mono $\widetilde{S}' \xrightarrow{m'} |\boldsymbol{T}A'|$. Since all faces commute, there is an arrow $\widetilde{S} \dashrightarrow \widetilde{S}'$ as in diagram (6), making the two newly created faces of the cube commute. This arrow is unique by Fact 4.2. Now $\widetilde{T}\langle g, h\rangle$ is given by the bottom face. Functoriality follows immediately from uniqueness of the diagonal arrow in (6).

4.3. *Unit* $\widetilde{\eta}$.

The ($\mathbb{C}$-component of the) unit $\widetilde{\eta}_{\langle S,f,A\rangle}$ is defined by post-composing $\eta_S$ with the pseudoepi part of the mono factorization in (7). This is well-defined since everything in sight in the diagram on the right commutes. Indeed, the right triangle is the monad morphism diagram (4) (left), the upper square is the naturality of $\eta$ while the lower one is a copy of (7).

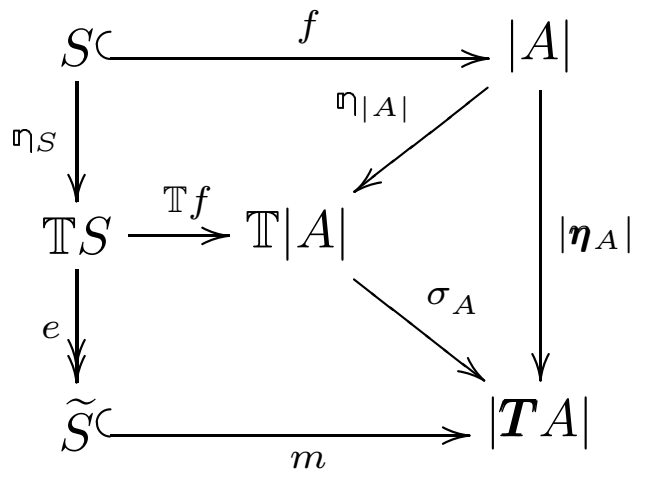

(10)

4.4. *Multiplication* $\widetilde{\mu}$.

The ($\mathbb{C}$-component of the) multiplication $\widetilde{\mu}_{\langle S,f,A\rangle}$ will be induced by a diagram similar to (9) (below).

Again, all the faces not having the dashed arrow or the required dotted arrow as edge commute. The front face and the lower half of the back face are instances of (7), defining $\widetilde{T}\langle S,f,A\rangle$ and $\widetilde{T}^2\langle S,f,A\rangle$, respectively. The upper half of the back face is by application of $\mathbb{T}$ to the front face. The right-hand face is the other monad morphism diagram (4) (right), which we had not used yet, while the upper one is a naturality square for $\mathbb{\mu}$.

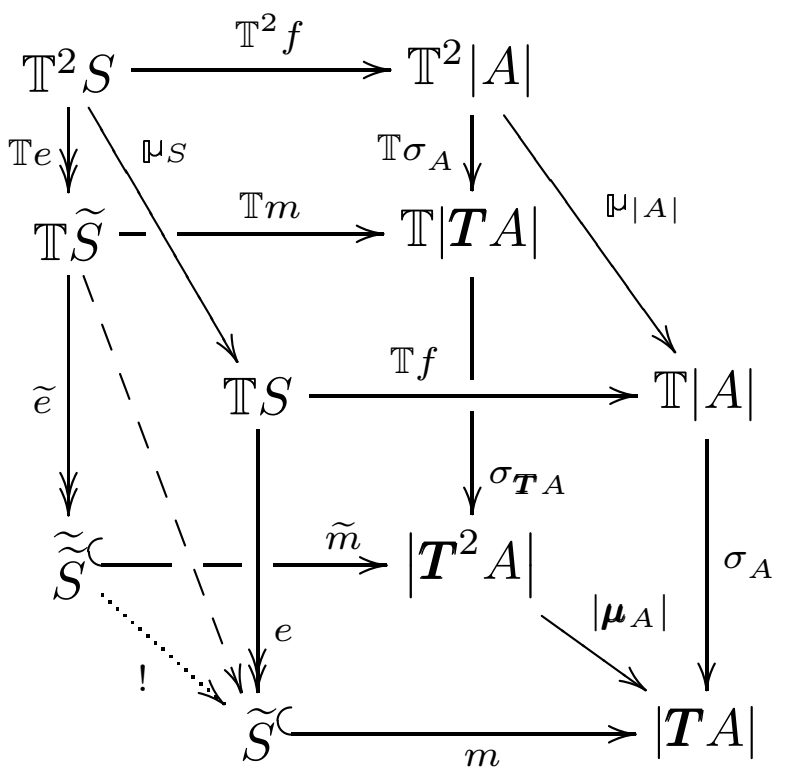

(11)

Note that $\mathbb{T}e$ is a pseudoepi, since $e$ is a relevant pseudoepi by construction, and $\mathbb{T}$ maps relevant pseudoepis to pseudoepis. The composition $\widetilde{e} \circ \mathbb{T}e$ is necessarily a pseudoepi as well (Barr, 1998). We may use this result, or use a diagonal (6) twice. Here, and in some other cases later, we prefer to do so.

First, similarly as in diagram (9) we find an instance of diagram (5) by two walks from $\mathbb{T}^2 S$ to $|\boldsymbol{T}A|$, one starting with $\mathbb{T}e$ and the other ending with $m$. Hence, the unique dashed arrow exists and makes the two triangles commute. One of them, involving the pseudoepi $\mathbb{T}e$, is the upper part of the left-hand side. The other one, namely that involving the relevant mono $m$, allows us to apply (5) again, since the following two walks from $\mathbb{T}\widetilde{S}$ to $|\boldsymbol{T}A|$ commute: one starting with the pseudoepi $\widetilde{e}$ and the other consisting of the dashed arrow followed by $m$. Hence, the unique dotted arrow exists and makes the bottom face as well as the triangle in the left-hand face commute. The multiplication $\widetilde{\mu}_{\langle S,f,A\rangle}$ is then defined by the bottom face of the cube.

Verification of the monad laws is a formality due to the following:

**Fact 4.3.** Given two parallel arrows in $(\boldsymbol{C} \downarrow |_|)$, say $\langle g_1, h_1\rangle$ and $\langle g_2, h_2\rangle$, they are equal whenever the second components $h_1$ and $h_2$ are.

The proof is immediate by Fact 4.2. Using this fact, and knowing that second components of $\widetilde{\eta}$ and $\widetilde{\mu}$ satisfy the monad laws (as they are unit and multiplication of $\boldsymbol{T}$, respectively), we deduce immediately that $\widetilde{\eta}$ and $\widetilde{\mu}$ satisfy the monad laws too. Similarly one proves naturality of $\widetilde{\eta}$ and $\widetilde{\mu}$. We shall use this argument extremely often in the sequel.

It is useful to summarize the ingredients we have used here. To lift a monad $(\boldsymbol{T}, \boldsymbol{\eta}, \boldsymbol{\mu})$ on $\boldsymbol{C}$ to $(\mathbb{C} \downarrow |_|)$, we need:

**(i)** a category $\mathbb{C}$ and a functor $|_| : \boldsymbol{C} \to \mathbb{C}$,
**(ii)** a monad $(\mathbb{T}, \eta, \mu)$ on $\mathbb{C}$,
related to $(\boldsymbol{T}, \boldsymbol{\eta}, \boldsymbol{\mu})$ by a monad morphism $(|_|, \sigma)$ from $\boldsymbol{T}$ to $\mathbb{T}$,
**(iii)** a mono factorization system on $\mathbb{C}$,
**(iv)** $\mathbb{T}$ maps relevant pseudoepis to pseudoepis.

Recall that to lift the CCC structure of $\boldsymbol{C}$ to the subscone, we additionally require $\mathbb{C}$ to be a CCC with pullbacks, and $|_|$ to preserve finite products (Mitchell and Scedrov, 1993). Description of the construction can be found e.g., in (Goubault-Larrecq and Goubault, 2003), Section 5.4. We shall see in Section 9 that the existence of a mono factorization system on $\mathbb{C}$ allows us to relax the requirements on $|_|$ somewhat.

## 5. Lifting of a Monad to Relations

Recall that we would like to lift monads to categories of binary relations as objects. Hence, assume in this section that $\boldsymbol{C}$ is a product category, $\boldsymbol{C} = \boldsymbol{C}_1 \times \boldsymbol{C}_2$ and that both $\boldsymbol{C}_1$ and $\boldsymbol{C}_2$ are equipped with monads $\boldsymbol{T}_1$ and $\boldsymbol{T}_2$, and functors $|_|_1 : \boldsymbol{C}_1 \to \mathbb{C}$ and $|_|_2 : \boldsymbol{C}_2 \to \mathbb{C}$. A monad $\boldsymbol{T}$ on $\boldsymbol{C}$ can be defined pairwise: $\boldsymbol{T}\langle A_1, A_2\rangle = \langle \boldsymbol{T}_1 A_1, \boldsymbol{T}_2 A_2\rangle$ and similarly we define $|_| : \boldsymbol{C} \to \mathbb{C}$ by $|(A_1, A_2)| = |A_1|_1 \times |A_2|_2$.

To this aim we assume binary products in $\mathbb{C}$, i.e., for each pair of objects $A_1$, $A_2$ of $\mathbb{C}$, an object $A_1 \times A_2$ in $\mathbb{C}$, together with two morphisms $\pi_1 : A_1 \times A_2 \to A_1$ and $\pi_2 : A_1 \times A_2 \to A_2$ satisfying the requirement that for every morphisms $f_1$ and $f_2$, there is a unique morphism $h$ making the whole diagram commute. We write $\langle f_1, f_2\rangle$ for $h$.

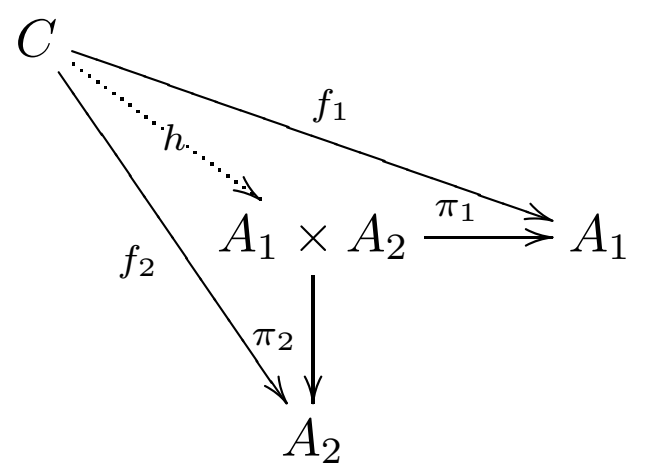

In the same vein monad morphisms from $\boldsymbol{T}_1$ to $\mathbb{T}$ and from $\boldsymbol{T}_2$ to $\mathbb{T}$ induce a monad morphism from $\boldsymbol{T}$ to $\mathbb{T}$. Indeed, given any two monad morphisms

$$\sigma^1 : \mathbb{T}|_|_1 \;\Rightarrow\; |\boldsymbol{T}_1|_1 \quad \text{and} \quad \sigma^2 : \mathbb{T}|_|_2 \;\Rightarrow\; |\boldsymbol{T}_2|_2,$$

we can define $\sigma_{(A_1,A_2)} : \mathbb{T}(|A_1|_1 \times |A_2|_2) \to |\boldsymbol{T}_1 A_1|_1 \times |\boldsymbol{T}_2 A_2|_2$ by

$$\sigma_{(A_1,A_2)} = \langle \sigma^1_{A_1} \circ \mathbb{T}\pi_1, \sigma^2_{A_2} \circ \mathbb{T}\pi_2\rangle, \tag{12}$$

where $\pi_1$ and $\pi_2$ denote the projections from $|A_1|_1 \times |A_2|_2$.

The situation gets much simpler when $\mathbb{C} = \boldsymbol{Set}$, $|_|_1 = \boldsymbol{C}_1(1_1, _)$ and $|_|_2 = \boldsymbol{C}_2(1_2, _)$, where we assume that $\boldsymbol{C}_1$ and $\boldsymbol{C}_2$ have terminal objects, $1_1$ and $1_2$ respectively. Each object $S \hookrightarrow |(A_1, A_2)|$ in the subscone defines a binary relation (again noted $S$) on global elements of $A_1$ and $A_2$. Obviously $\boldsymbol{Set}$ satisfies all requirements from previous sections, with surjections as pseudoepis and injections as relevant monos.

For a moment imagine that $\boldsymbol{T}_1$ and $\boldsymbol{T}_2$ are *strong* monads and that we are able to lift strong monads to subscones – this will be tackled in detail in the following Sections 6 and 7. Given two CCCs $\boldsymbol{C}_1$ and $\boldsymbol{C}_2$ with respective strong monads $\boldsymbol{T}_1$ and $\boldsymbol{T}_2$, the fact that ***Comp*** is the free CCC with strong monad on the set $B$ of base types means that there are two representations of CCCs-with-strong-monads, $[\![_]\!]_1$ and $[\![_]\!]_2$, from ***Comp*** to $\boldsymbol{C}_1$ and $\boldsymbol{C}_2$ respectively: they are the natural meaning functions for monadic types and computational $\lambda$-terms.

Our construction of a lifting together with standard constructions on subscones (Mitchell and Scedrov, 1993) yield another representation of CCCs-with-strong-monads $[\![_]\!]$ from ***Comp*** to $(\boldsymbol{Set} \downarrow |_|)$. That $[\![_]\!]$ is a lifting means that $U \circ [\![_]\!] = \langle [\![_]\!]_1, [\![_]\!]_2 \rangle$, i.e., the diagram on the right commutes. When $\boldsymbol{C}_1$ and $\boldsymbol{C}_2$ are concrete categories, this means that

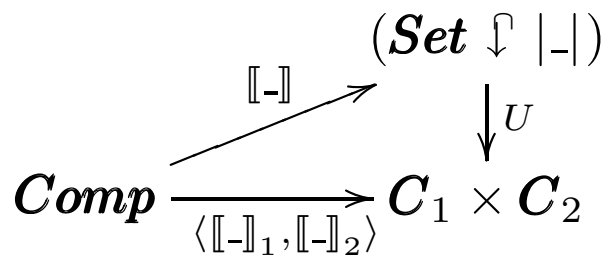

$$\forall a_1 \in [\![\Gamma]\!]_1, a_2 \in [\![\Gamma]\!]_2 . (a_1, a_2) \in [\![\Gamma]\!] \Rightarrow ([\![t]\!]_1 (a_1), [\![t]\!]_2 (a_2)) \in [\![\tau]\!] \tag{13}$$

for all terms $t$ of type $\tau$ in the context $\Gamma = x_1 : \tau_1, \ldots, x_n : \tau_n$; representations of $\Gamma$ are taken to be products of the representations of $\tau_1, \ldots, \tau_n$; $[\![\tau]\!]$ is a relation between $[\![\tau]\!]_1$ and $[\![\tau]\!]_2$, defined by induction on types $\tau$ (the case where $\tau$ is a base type is arbitrary):

$$\begin{aligned} (f_1, f_2) \in [\![\tau \to \tau']\!] &\iff \forall (a_1, a_2) \in [\![\tau]\!] . (f_1(a_1), f_2(a_2)) \in [\![\tau']\!] \\ ((a_1, a_1'), (a_2, a_2')) \in [\![\tau \times \tau']\!] &\iff (a_1, a_2) \in [\![\tau]\!] \wedge (a_1', a_2') \in [\![\tau']\!] \\ (B_1, B_2) \in [\![\boldsymbol{T}\tau]\!] &\iff (B_1, B_2) \in \widetilde{T}[\![\tau]\!] \end{aligned}$$

These equations (except possibly the last one) are the standard definition of a *logical relation.* (13) is the already cited *Basic Lemma.*

Further simplification is gained when $\boldsymbol{C}_1 = \boldsymbol{C}_2 = \boldsymbol{Set}$, the three monads $\boldsymbol{T}_1$, $\boldsymbol{T}_2$ and $\mathbb{T}$ are identical and both $|_|_1$ and $|_|_2$ are identity functors. The monad morphism $\sigma$ reduces to distributivity of the monad $\mathbb{T}$ over binary product, and (12) rewrites to

$$\sigma_{(A_1, A_2)} = \langle \mathbb{T}\pi_1, \mathbb{T}\pi_2 \rangle : \mathbb{T}(A_1 \times A_2) \to \mathbb{T}A_1 \times \mathbb{T}A_2$$

where by $\mathbb{T}$ we denote a given single monad on ***Set***. This is a particularly important special case, so we study it in more detail.

Every binary relation $S \subseteq A_1 \times A_2$ has a representation $S \xhookrightarrow{\langle \pi^S{}_1, \pi^S{}_2 \rangle} A_1 \times A_2$ where the arrow is the inclusion induced by two projections $\pi^S{}_1 : S \to A_1$ and $\pi^S{}_2 : S \to A_2$. In fact, the full subcategory of the subscone consisting exclusively of inclusions instead of all injections is equivalent to the whole subscone, so without loss of generality we consider only inclusions in the rest of this section.

Recall the action of a lifted monad $\widetilde{T}$ on a relation $S \xhookrightarrow{\langle \pi^S{}_1, \pi^S{}_2 \rangle} A_1 \times A_2$ :

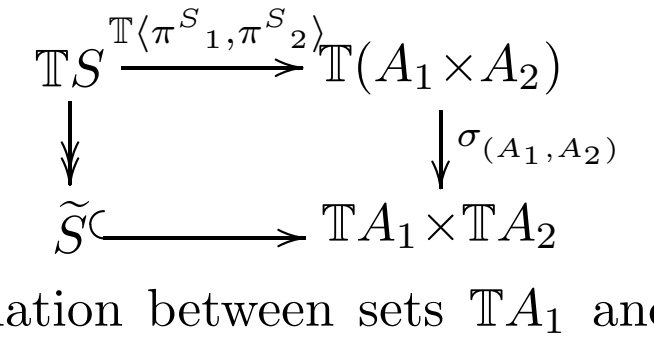

The functor $\widetilde{T}$ maps a relation $S$ to the relation between sets $\mathbb{T}A_1$ and $\mathbb{T}A_2$ defined as the direct image of the function $\langle \mathbb{T}\pi^S{}_1, \mathbb{T}\pi^S{}_2 \rangle : TS \to \mathbb{T}A_1 \times \mathbb{T}A_2$, since the middle (dashed) triangle in the following diagram commutes by the universal property of product

(together with all other triangles):

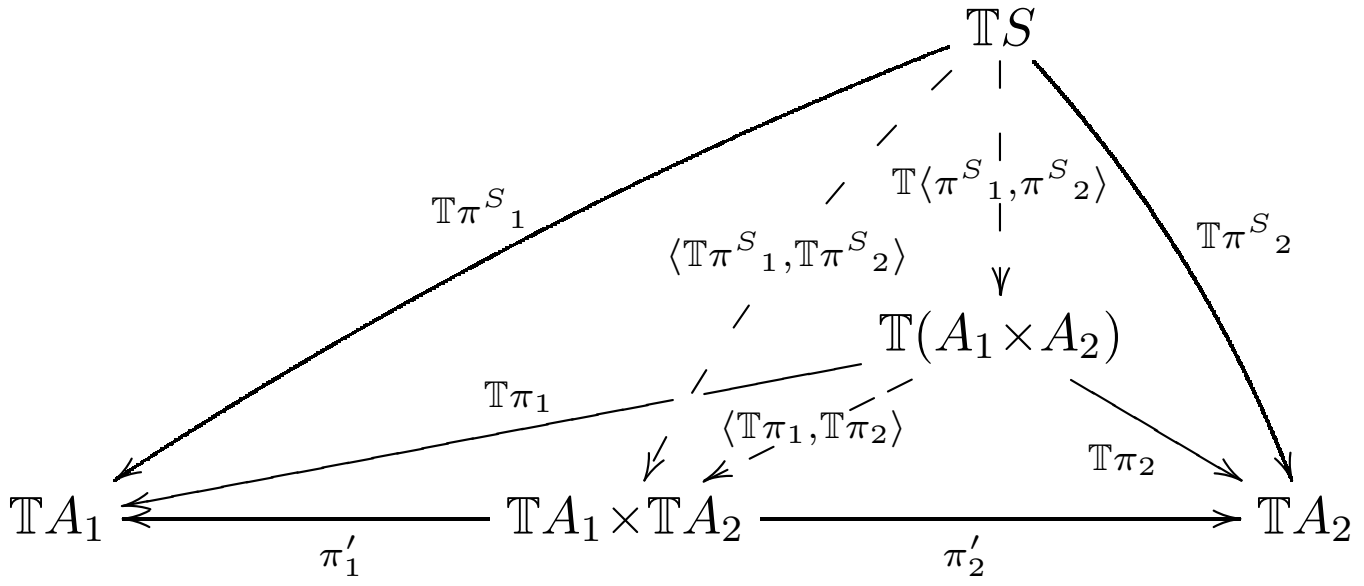

where by $\pi'_1$ and $\pi'_2$ we denote two projections from $\mathbb{T}A_1 \times \mathbb{T}A_2$.

It is instructive to look at some concrete example before going to more technical Sections 6 and 7. Further examples are presented in more detail in Section 10. Consider $\mathbb{T}A = \mathbb{P}_{\mathrm{fin}}(A)$, the finite powerset monad on ***Set***. If we assume for simplicity that ${\pi^S}_1$ and ${\pi^S}_2$ are simply inclusions, then the function $\mathbb{T}{\pi^S}_1$ takes a finite relation $R \subset S$ to its domain, i.e. $\{x | \exists y \cdot (x, y) \in R\}$, and $\mathbb{T}{\pi^S}_2$ takes $R$ to its codomain, i.e., $\{y | \exists x \cdot (x, y) \in R\}$. Hence, the image of the function $\langle \mathbb{T}{\pi^S}_1, \mathbb{T}{\pi^S}_2 \rangle$ is a relation $\widetilde{S}$ between finite subsets of $A_1$ and $A_2$ that contains precisely domain-codomain pairs of finite relations $R \subseteq S$. Hence $(B_1, B_2) \in \widetilde{S}$ iff

$$\forall b_1 \in B_1 . \exists b_2 \in B_2 . (b_1, b_2) \in S \qquad \wedge \qquad \forall b_2 \in B_2 . \exists b_1 \in B_1 . (b_1, b_2) \in S.$$

(Pitts, 1996) also considers lifting of certain constructions on objects to corresponding constructions on relations over these objects. The concept of lifting is similar to ours, although technically different. The notion of "relation" used in the paper is given by a so called *relational structure* on a category, and is fairly abstract. On the other hand, Pitts restricts to categories of domains, usually defined as least solutions of recursive domain equations. The interest of the author is mainly in questions related to domain theory. One of main results gives conditions for existence of lifting of the solution of an equation in the following sense: given a domain defined by $D = \Phi(D)$, is there a relation $\Delta$ on $D$ such that $\Delta = \Phi(\Delta)$? Strong monads and their liftings are not considered in the work of Pitts. Although, for certain types of relational structures the approach of Pitts seems to yield similar liftings as ours. E.g., in the case of binary relations, the function space is lifted precisely as logical relations are usually defined on a function type, and hence exactly like in the case of lifting to subscones.

## 6. Lifting Monoidal Structures

### 6.1. *Monoidal Categories*

In this section, and the following ones, we assume that each of the categories $\boldsymbol{C}$ and $\mathbb{C}$ is equipped with a monoidal structure. In other words, we assume that $(\boldsymbol{C}, \otimes, \boldsymbol{I}, \boldsymbol{\alpha}, \boldsymbol{\ell}, \boldsymbol{r})$ and $(\mathbb{C}, \otimes, \mathbb{I}, \mathbb{a}, \mathbb{l}, \mathbb{r})$ are *monoidal categories* (Mac Lane, 1971). This will allow us to extend our lifting of a monad to one of a strong monad in the following Section 7.

This means that $\mathbb{I}$ is an object of $\mathbb{C}$, $\otimes$ is a functor from $\mathbb{C} \times \mathbb{C}$ to $\mathbb{C}$, and $\mathbb{a}_{A,B,C}$ :

$(A \otimes B) \otimes C \to A \otimes (B \otimes C)$, $\mathbb{l}_A : \mathbb{I} \otimes A \to A$, $\mathbb{r}_A : A \otimes \mathbb{I} \to A$ are natural isomorphisms called the associativity, the left unit and the right unit laws respectively, making the following squares commute.

$$\begin{array}{ccccc} ((A \otimes B) \otimes C) \otimes D & \xrightarrow{\mathbb{\alpha}_{A\otimes B,C,D}} & (A \otimes B) \otimes (C \otimes D) & \xrightarrow{\mathbb{\alpha}_{A,B,C\otimes D}} & A \otimes (B \otimes (C \otimes D)) \\ \downarrow{\scriptstyle \mathbb{\alpha}_{A,B,C}\otimes \mathrm{id}_D} & & & & \uparrow{\scriptstyle \mathrm{id}_A \otimes \mathbb{\alpha}_{B,C,D}} \\ (A \otimes (B \otimes C)) \otimes D & & \xrightarrow{\mathbb{\alpha}_{A,B\otimes C,D}} & & A \otimes ((B \otimes C) \otimes D) \end{array} \tag{14}$$

$$\begin{array}{ccc} (A \otimes \mathbb{I}) \otimes B & \xrightarrow{\mathbb{\alpha}_{A,\mathbb{I},B}} & A \otimes (\mathbb{I} \otimes B) \\ & {\scriptstyle \mathbb{r}_A \otimes \mathrm{id}_B} \searrow \quad \swarrow {\scriptstyle \mathrm{id}_A \otimes \mathbb{l}_B} & \\ & A \otimes B & \end{array} \tag{15}$$

And similarly for $\boldsymbol{I}$, $\boldsymbol{\otimes}$, $\boldsymbol{\alpha}$, $\boldsymbol{\ell}$, $\boldsymbol{r}$.

We prefer to work in a slightly more general setting compared to (Mitchell and Scedrov, 1993), where cartesian structure was assumed. In Section 9 we show how this added generality can be exploited for a fragment of linear lambda calculus. Typically, our categories will have finite products, then $\mathbb{I}$ will be a terminal object, $\otimes$ will be binary product, and $\mathbb{\alpha}$, $\mathbb{l}$ and $\mathbb{r}$ will be the obvious isomorphisms.

We also assume that $|_|$ is a *monoidal functor* (Eilenberg and Kelly, 1966). That is, there is a *mediating pair* $(\mathbb{\theta}, \mathbb{i})$, composed of a natural transformation $\mathbb{\theta}_{A,B} : |A| \otimes |B| \to |A \boldsymbol{\otimes} B|$ and a morphism $\mathbb{i} : \mathbb{I} \to |\boldsymbol{I}|$ satisfying the following *coherence conditions*:

$$\begin{array}{ccc} (|A| \otimes |B|) \otimes |C| & \xrightarrow{\mathbb{\alpha}_{|A|,|B|,|C|}} & |A| \otimes (|B| \otimes |C|) \\ \downarrow{\scriptstyle \mathbb{\theta}_{A,B}\otimes \mathrm{id}_{|C|}} & & \downarrow{\scriptstyle \mathrm{id}_{|A|}\otimes \mathbb{\theta}_{B,C}} \\ |A \boldsymbol{\otimes} B| \otimes |C| & & |A| \otimes |B \boldsymbol{\otimes} C| \\ \downarrow{\scriptstyle \mathbb{\theta}_{A\boldsymbol{\otimes} B,C}} & & \downarrow{\scriptstyle \mathbb{\theta}_{A,B\boldsymbol{\otimes} C}} \\ |(A \boldsymbol{\otimes} B) \boldsymbol{\otimes} C| & \xrightarrow{|\boldsymbol{\alpha}_{A,B,C}|} & |A \boldsymbol{\otimes} (B \boldsymbol{\otimes} C)| \end{array} \tag{16}$$

$$\begin{array}{ccc} \mathbb{I} \otimes |A| & & \\ \downarrow{\scriptstyle \mathbb{i}\otimes \mathrm{id}_{|A|}} & \searrow{\scriptstyle \mathbb{l}_{|A|}} & \\ |\boldsymbol{I}| \otimes |A| & & |A| \\ \downarrow{\scriptstyle \mathbb{\theta}_{\boldsymbol{I},A}} & \nearrow{\scriptstyle |\boldsymbol{\ell}_A|} & \\ |\boldsymbol{I} \boldsymbol{\otimes} A| & & \end{array} \tag{17} \qquad \begin{array}{ccc} |A| \otimes \mathbb{I} & & \\ \downarrow{\scriptstyle \mathrm{id}_{|A|}\otimes \mathbb{i}} & \searrow{\scriptstyle \mathbb{r}_{|A|}} & \\ |A| \otimes |\boldsymbol{I}| & & |A| \\ \downarrow{\scriptstyle \mathbb{\theta}_{A,\boldsymbol{I}}} & \nearrow{\scriptstyle |\boldsymbol{r}_A|} & \\ |A \boldsymbol{\otimes} \boldsymbol{I}| & & \end{array} \tag{18}$$

Finally, we assume that pseudoepis and relevant monos form a so-called *monoidal mono factorization system*, i.e., for every two pseudoepis $e_1$, $e_2$, then $e_1 \otimes e_2$ is again a pseudoepi. This name stems from (Ambler, 1991), Definition 5.2.1, p.91.

We define below a lifting of the monoidal structure to the subscone: we show that the subscone is a monoidal category $((\mathbb{C} \downarrow |_|), \widetilde{\otimes}, \widetilde{I}, \widetilde{\alpha}, \widetilde{\ell}, \widetilde{r})$ in such a way that $U(\widetilde{A}\widetilde{\otimes}\widetilde{B}) = U\widetilde{A} \otimes U\widetilde{B}$, $U\widetilde{I} = \boldsymbol{I}$, $U\widetilde{\alpha}_{\widetilde{A},\widetilde{B},\widetilde{C}} = \boldsymbol{\alpha}_{U\widetilde{A},U\widetilde{B},U\widetilde{C}}$, $U\widetilde{\ell}_{\widetilde{A}} = \boldsymbol{\ell}_{U\widetilde{A}}$, $U\widetilde{r}_{\widetilde{A}} = \boldsymbol{r}_{U\widetilde{A}}$. Lifting to scones is omited here but can be easily extracted from diagrams below by dropping all factorizations.

6.2. *Unit element* $\widetilde{I}$.

Let $\widetilde{I}$ be the triple $\langle \overline{I}, \bar{\mathbb{i}}, \boldsymbol{I} \rangle$ as built from the diagram on the right, obtained from a factorization of $\mathbb{i}$.

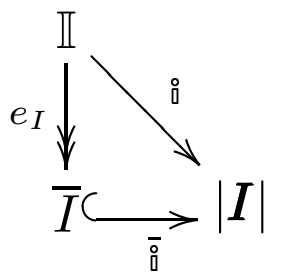

(19)

6.3. *Tensor product* $\widetilde{\otimes}$.

We build the tensor product $\langle S_1, m_1, A_1 \rangle \widetilde{\otimes} \langle S_2, m_2, A_2 \rangle$ in the obvious way: compose $m_1 \otimes m_2$ with the mediating natural transformation $\mathbb{\theta}$, and factorize.

The tensor product is then given by $\langle S_{12}, m_{12}, A_1 \boldsymbol{\otimes} A_2 \rangle$ on the right. This is similar to the construction of $\widetilde{T}$.

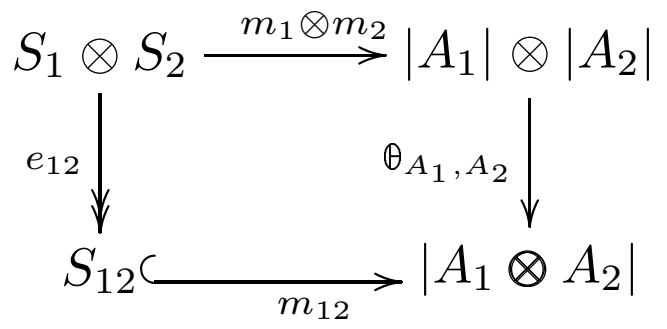

(20)

6.4. *Associativity* $\widetilde{\alpha}$.

This is more involved, but basically similar to the construction of the multiplication of the monad in the subscone. In the diagram below, $e_{12} \otimes \mathrm{id}_{S_3}$ is pseudoepi because both $e_{12}$ (given by Diagram 20) and $\mathrm{id}_{S_3}$ are pseudoepis, and because our mono factorization system is monoidal. The two front faces and the two back faces are derived from the definition of $\widetilde{\otimes}$, the top face is a naturality square for $\mathbb{a}$, the right face is the coherence condition (16). The dashed arrow, and then the dotted arrow $\widehat{\alpha}$, are by the diagonal fill-in property of our factorization system. The desired associativity morphism is then the pair $(\widehat{\alpha}, \boldsymbol{\alpha}_{A_1,A_2,A_3})$ (bottom face).

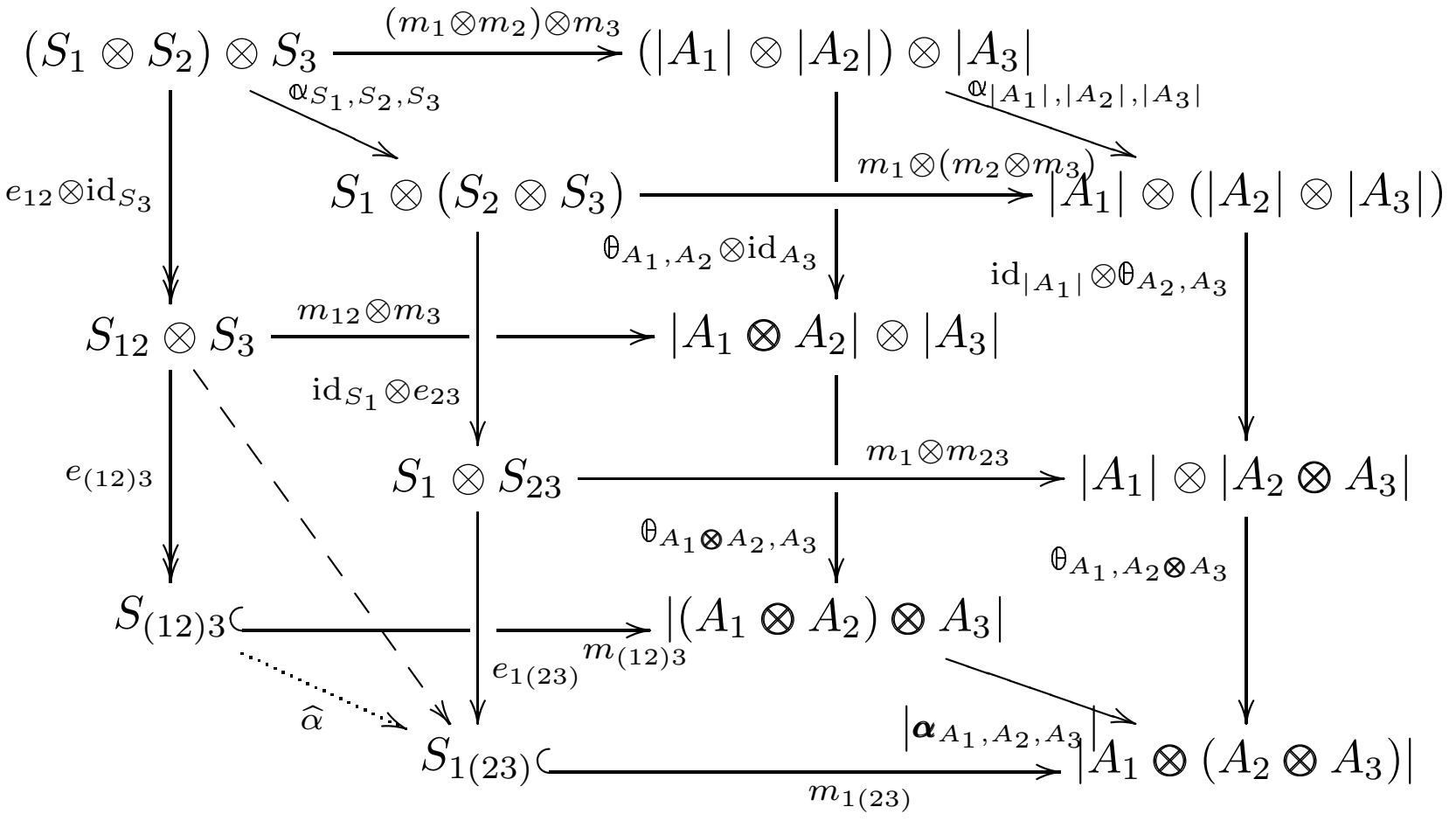

(21)

The inverse is given by a very similar diagram (below).

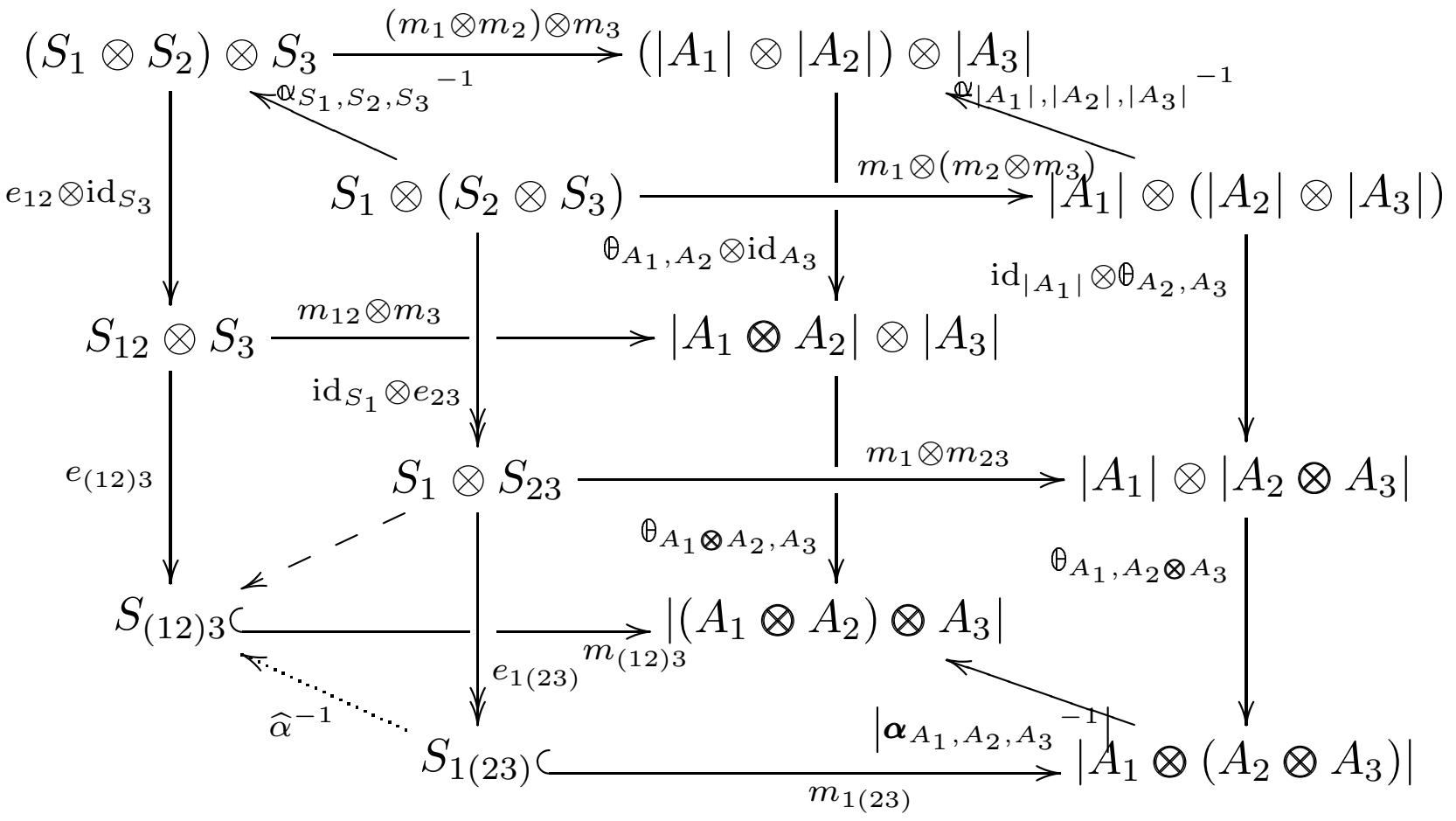

(22)

6.5. *Left unit* $\widetilde{\ell}$.

Let $\langle S, m, A \rangle$ be any object of the subscone. We build the diagram below. The left triangle in the upper back face is the definition of $\widetilde{I}$, the rest of this face corresponds to two ways of writing $\mathfrak{i} \otimes m$, the lower back face is the definition of $\widetilde{I} \widetilde{\otimes} \langle S, m, A \rangle$. The upper, slanted face is a naturality square for $\mathbb{l}$.

Finally, the rightmost triangle is the coherence condition (17). As usual, we first derive the dashed, then the dotted arrow $\widehat{\mathbb{l}}$ by diagonal fill-ins. The desired left unit is $(\widehat{\mathbb{l}}, \boldsymbol{\ell}_A)$.

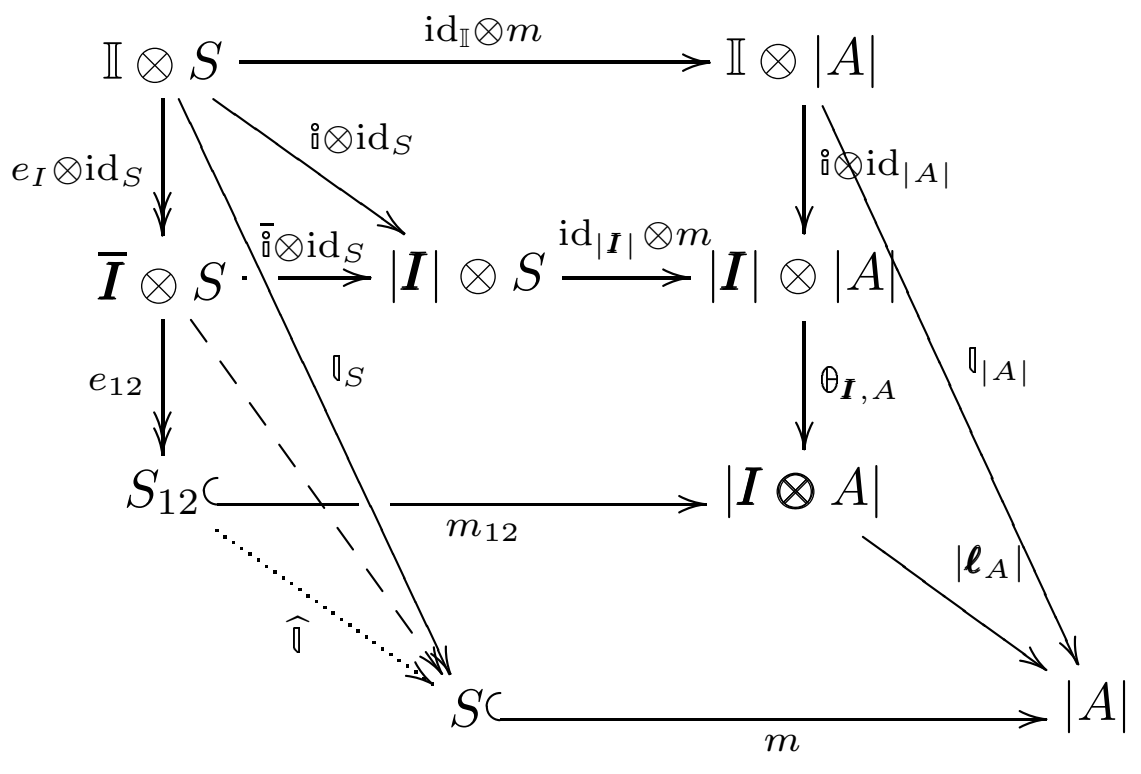

(23)

The inverse to $\widehat{\mathbb{l}}$ is also given by a diagonal fill-in. Start from $S$, then go to $S$ (again) by the identity morphism—this is a pseudoepi—, then follow $m$, $\left|\boldsymbol{\ell}_A{}^{-1}\right|$ to $|\boldsymbol{I} \otimes A|$; or start from $S$, climb along $\mathbb{l}_S^{-1}$, then follow $e_I \otimes \mathrm{id}_S$, $e_{12}$, $m_{12}$ (a mono) to $|\boldsymbol{I} \otimes A|$. The diagonal fill-in is then an arrow from $S$ to $S_{12}$, which is inverse to $\widehat{\mathbb{l}}$ by Fact 4.3.

6.6. *Right unit* $\widetilde{r}$.

This works exactly as for the left unit, see diagram on the right. The right triangle is the coherence condition (18). The desired right unit is given by $(\widehat{\mathbb{r}}, \boldsymbol{r}_A)$. The inverse of $\widehat{\mathbb{r}}$ is built as for $\widehat{\mathbb{l}}$.

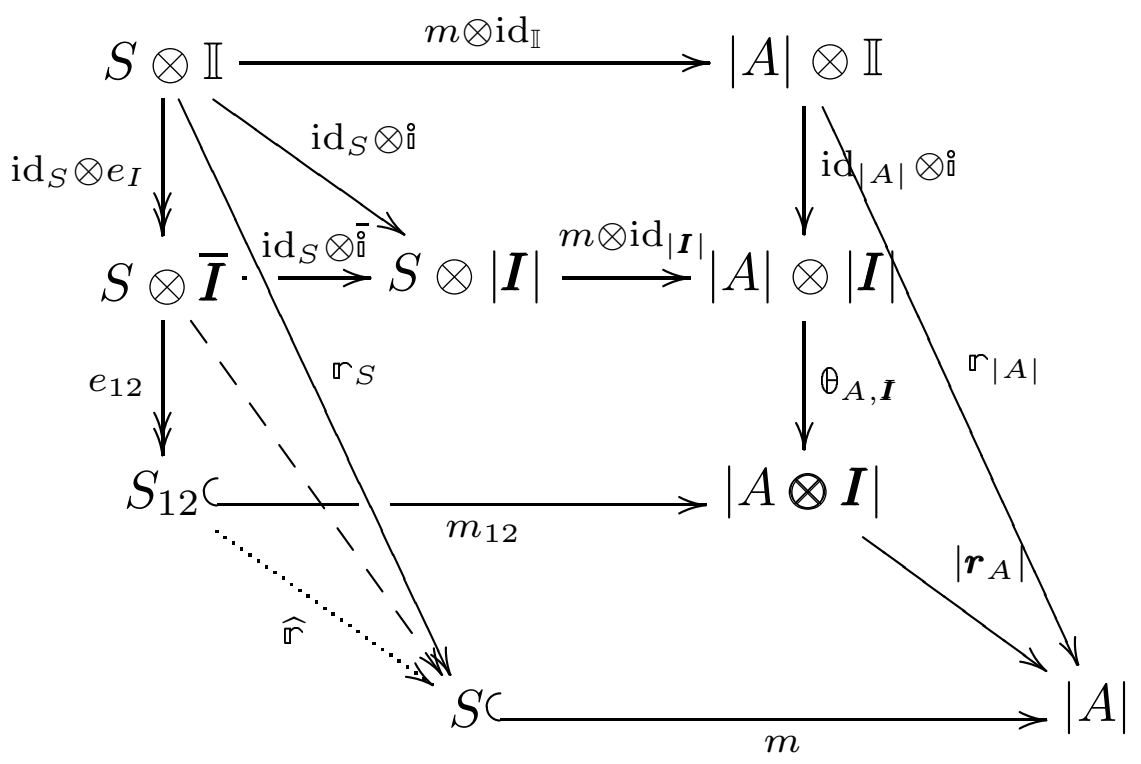

(24)

Finally, all required naturality, isomorphism, and coherence conditions hold by Fact 4.3.

We recap what we need to lift monoidal structure to the subscone:

---

**(i.a)** monoidal categories $(\boldsymbol{C}, \otimes, \boldsymbol{I}, \boldsymbol{\alpha}, \boldsymbol{\ell}, \boldsymbol{r})$ and $(\mathbb{C}, \otimes, \mathbb{I}, \mathbb{a}, \mathbb{l}, \mathbb{r})$, and a monoidal functor $|_| : \boldsymbol{C} \to \mathbb{C}$,

**(iii.a)** a monoidal mono factorization system on $\mathbb{C}$.

---

6.7. *Symmetric Monoidal Categories*

We now assume that we have got, and want to preserve *symmetric* monoidal structure. Recall that a symmetric monoidal category $(\mathbb{C}, \otimes, \mathbb{I}, \mathbb{a}, \mathbb{l}, \mathbb{r}, \mathbb{c})$ is a monoidal category $(\mathbb{C}, \otimes, \mathbb{I}, \mathbb{a}, \mathbb{l}, \mathbb{r})$, together with a *commutativity* natural transformation $\mathbb{c}_{A,B} : A \otimes B \to B \otimes A$ obeying the following coherence conditions.

The first coherence condition is $\mathbb{c}_{B,A} \circ \mathbb{c}_{A,B} = \mathrm{id}_{A \otimes B}$, which implies that $\mathbb{c}$ is actually a natural isomorphism. The others are:

$$\begin{array}{ccc} (A \otimes B) \otimes C & \xrightarrow{\mathbb{c}_{A,B} \otimes \mathrm{id}_C} & (B \otimes A) \otimes C \\ \downarrow{\scriptstyle \mathbb{a}_{A,B,C}} & & \downarrow{\scriptstyle \mathbb{a}_{B,A,C}} \\ A \otimes (B \otimes C) & & B \otimes (A \otimes C) \\ \downarrow{\scriptstyle \mathbb{c}_{A,B\otimes C}} & & \downarrow{\scriptstyle \mathrm{id}_B \otimes \mathbb{c}_{A,C}} \\ (B \otimes C) \otimes A & \xrightarrow[\mathbb{a}_{B,C,A}]{} & B \otimes (C \otimes A) \end{array} \tag{25}$$

$$\begin{array}{ccc} \mathbb{I} \otimes A & \xrightarrow{\mathbb{c}_{\mathbb{I},A}} & A \otimes \mathbb{I} \\ & {\scriptstyle \mathbb{l}_A} \searrow \quad \swarrow {\scriptstyle \mathbb{r}_A} & \\ & A & \end{array} \tag{26}$$

We now need $|_|$ to be a *symmetric* monoidal functor, that is, it should be monoidal and satisfy the extra coherence condition:

$$\begin{array}{ccc} |A_1| \otimes |A_2| & \xrightarrow{\mathbb{c}_{|A_1|,|A_2|}} & |A_2| \otimes |A_1| \\ \downarrow{\scriptstyle \mathbb{m}_{A_1,A_2}} & & \downarrow{\scriptstyle \mathbb{m}_{A_2,A_1}} \\ |A_1 \boldsymbol{\otimes} A_2| & \xrightarrow{|\boldsymbol{c}_{A_1,A_2}|} & |A_2 \boldsymbol{\otimes} A_1| \end{array} \tag{27}$$

6.8. *Commutativity* $\widetilde{\ell}$.

We lift the commutativity to the subscone, assuming $\boldsymbol{C}$ and $\mathbb{C}$ are symmetric monoidal, as follows. The back and front face are the definition of the two tensor products of $\langle S_1, m_1, A_1 \rangle$ and $\langle S_2, m_2, A_2 \rangle$, the top face is by naturality of commutativity, the right face is coherence (27). Finally $\widehat{\mathbb{c}}$ is given by a diagonal fill-in, and all expected diagrams commute by Fact 4.3.

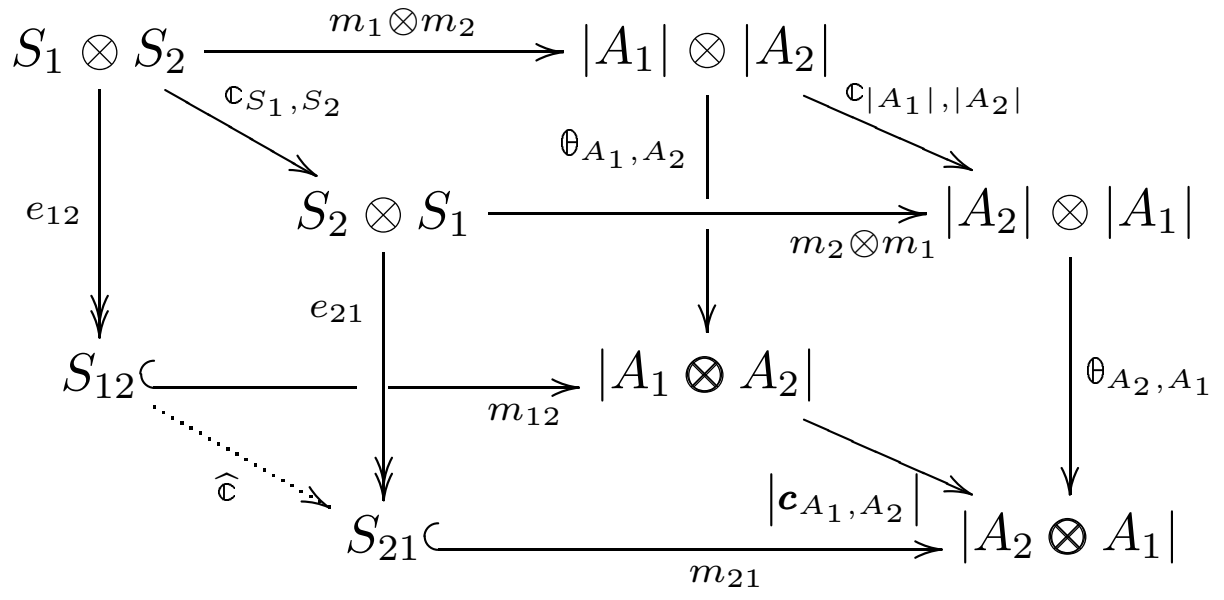

(28)

We recap what we need to lift *symmetric* monoidal structure to the subscone:

**(i.b)** symmetric monoidal categories $(\boldsymbol{C}, \otimes, \boldsymbol{I}, \boldsymbol{\alpha}, \boldsymbol{\ell}, \boldsymbol{r}, \boldsymbol{c})$
and $(\mathbb{C}, \otimes, \mathbb{I}, \mathbb{a}, \mathbb{l}, \mathbb{r}, \mathbb{c})$,
and a symmetric monoidal functor $|_| : \boldsymbol{C} \to \mathbb{C}$,
**(iii.a)** a monoidal mono factorization system on $\mathbb{C}$.

6.9. *Lifting Cartesian Products*

An important special case of symmetric monoidal structure is that given by finite products. This is given by one *terminal object* $\mathbf{1}$ such that, for every object $A$ in $\boldsymbol{C}$, there is a unique morphism $A \overset{!}{\longrightarrow} \mathbf{1}$, and by a *binary product* operation as explained in Section 5. For any two morphisms $f$ from $A$ to $A'$, $g$ from $B$ to $B'$, we write $f \times g$ for the morphism $\langle f \circ \pi_1, g \circ \pi_2 \rangle$ from $A \times B$ to $A' \times B'$.

It is well-known that binary product can be turned into a functor from $\boldsymbol{C} \times \boldsymbol{C}$ to $\boldsymbol{C}$, which is symmetric monoidal with unit 1, associativity $\langle \pi_1 \circ \pi_1, \langle \pi_2 \circ \pi_1, \pi_2 \rangle \rangle$, left unit $\pi_2$, right unit $\pi_1$, and commutativity $\langle \pi_2, \pi_1 \rangle$.

When $\boldsymbol{C}$ and $\mathbb{C}$ are both equipped with finite products, and $|_|$ is a functor from $\boldsymbol{C}$ to $\mathbb{C}$ that is monoidal with respect to these products, then the construction of Sections 6.1 and 6.7 yields a symmetric monoidal structure on $(\mathbb{C} \downarrow |_|)$ that we claim stems from a finite product structure on the subscone.

To this end, we assume that $|_|$ satisfies the coherence condition on the right, for $i \in \{1, 2\}$. We shall say that such a functor is *cartesian monoidal.* (Then $|_|$ is automatically symmetric monoidal.)

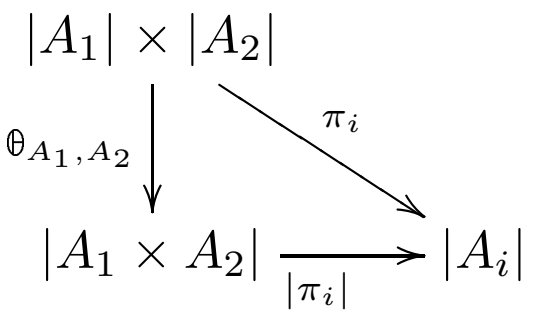

(29)

6.10. *Terminal object* $\widetilde{1}$.

Let $\widetilde{1}$ be the object $\langle \overline{I}, \bar{\mathbb{i}}, \mathbf{1} \rangle$ given by diagram (19). Specializing this diagram to the case at hand, this is given as the unique object up to iso making the following diagram commute:

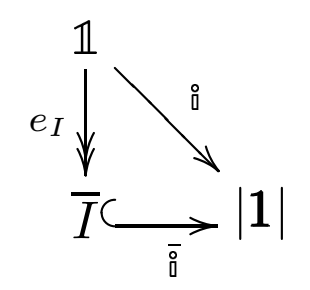

(30)

For any object $\langle S, m, A \rangle$ of the subscone, there is a unique arrow $\langle u, v \rangle$ from $\langle S, m, A \rangle$ to $\langle \overline{I}, \bar{\mathbb{i}}, \mathbf{1} \rangle$. Indeed, $v$ is the unique arrow ! from $A$ to $\mathbf{1}$, and $u$ is given by $e_I \circ !$; $u$ is also unique, by Fact 4.3.

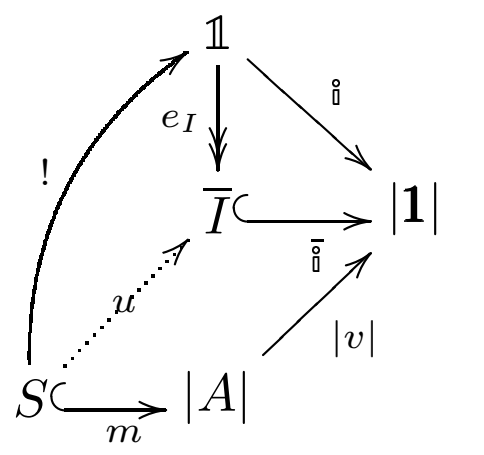

6.11. *Binary product* $\widetilde{\times}$.

Specializing the definition (20) of the lifted tensor product $\widetilde{\otimes}$ to the case at hand yields the object $\langle S_{12}, m_{12}, A_1 \otimes A_2\rangle = \langle S_1, m_1, A_1\rangle \widetilde{\times} \langle S_2, m_2, A_2\rangle$ defined by the diagram on the right.

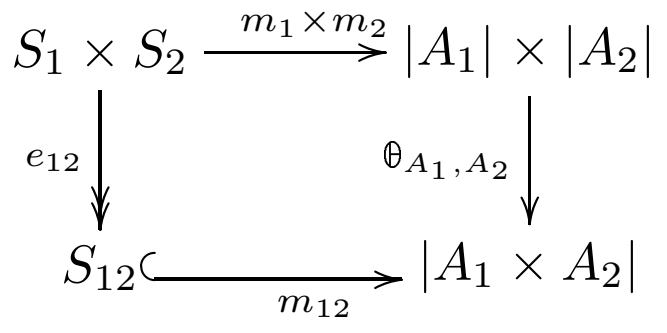

(31)

The $i$th projection $\langle \widetilde{\pi}_i, \pi_i\rangle$ is then given by the diagram on the right. The back square is a copy of (31), the right triangle is an instance of the coherence condition (29), while the top, slanted face is by standard properties of $\pi_i$. From two routes from $S_1 \times S_2$ to $|A_i|$, we get $\widetilde{\pi}_i$ by a diagonal fill-in.

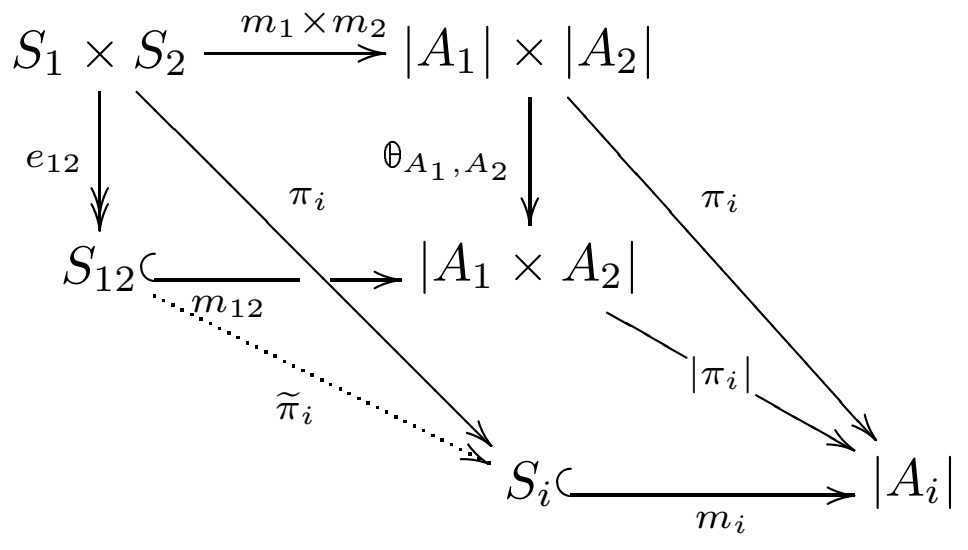

It remains to show that whenever we have two subscone morphisms $\langle \widetilde{f}_1, f_1\rangle$ from $\langle S, m, A\rangle$ to $\langle S_1, m_1, A_1\rangle$ and $\langle \widetilde{f}_2, f_2\rangle$ from $\langle S, m, A\rangle$ to $\langle S_2, m_2, A_2\rangle$, there is a unique morphism $\langle \widetilde{h}, h\rangle$ from $\langle S, m, A\rangle$ to the product $\langle S_{12}, m_{12}, A_1 \times A_2\rangle$ such that $\langle \widetilde{\pi}_i, \pi_i\rangle \circ \langle \widetilde{h}, h\rangle = \langle \widetilde{f}_i, f_i\rangle$ ($i \in \{1, 2\}$). Existence is assured: take $h = \langle f_1, f_2\rangle$, $\widetilde{h} = e_{12} \circ \langle \widetilde{f}_1, \widetilde{f}_2\rangle$, which satisfies the claim: this is an easy consequence of the diagram above. Uniqueness follows from the uniqueness of $h$ given by the definition of binary product in $\boldsymbol{C}$, and from Fact 4.3 guaranteeing the uniqueness of $\widetilde{h}$.

As is now usual, we recap what we need to lift products to the subscone:

---

**(i.c)** categories $\boldsymbol{C}$ and $\mathbb{C}$ with finite products,
and a cartesian monoidal functor $|_| : \boldsymbol{C} \to \mathbb{C}$,
**(iii.a)** a monoidal mono factorization system on $\mathbb{C}$.

---

## 7. Lifting Strong, Monoidal, and Commutative Monads to a Scone and a Subscone

Once we have got a monoidal structure on $\boldsymbol{C}$ and $\mathbb{C}$, we may consider *strong* monads $\boldsymbol{T}$ and $\mathbb{T}$ instead of just monads on each category. This is what we need to develop a theory of logical relations for Moggi's monadic $\lambda$-calculus. We shall also consider the more demanding cases of *monoidal* monads, and of *commutative* monads.

### 7.1. *Lifting Strong Monads*

That $\mathbb{T}$ is a strong monad means that a *strength* natural transformation $\mathbb{t}_{A,B} : A \otimes \mathbb{T}B \to \mathbb{T}(A \otimes B)$ is given such that the diagrams in Definition 3.2 in (Moggi, 1991) commute, that is:

$$\begin{array}{ccc} \mathbb{I}\otimes\mathbb{T}B & \xrightarrow{\mathbb{t}_{\mathbb{I},B}} & \mathbb{T}(\mathbb{I}\otimes B) \\ & \searrow^{\mathbb{l}_{\mathbb{T}B}} & \downarrow{\scriptstyle \mathbb{T}\mathbb{l}_B} \\ & & \mathbb{T}B \end{array} \tag{32}$$

$$\begin{array}{ccc} A\otimes B & \xrightarrow{\mathrm{id}_A\otimes\eta_B} & A\otimes\mathbb{T}B \\ & \searrow^{\eta_{A\otimes B}} & \downarrow{\scriptstyle \mathbb{t}_{A,B}} \\ & & \mathbb{T}(A\otimes B) \end{array} \tag{33}$$

$$\begin{array}{ccccc} (A\otimes B)\otimes\mathbb{T}C & & \xrightarrow{\quad\mathbb{t}_{A\otimes B,C}\quad} & & \mathbb{T}((A\otimes B)\otimes C) \\ \downarrow{\scriptstyle \alpha_{A,B,\mathbb{T}C}} & & & & \downarrow{\scriptstyle \mathbb{T}\alpha_{A,B,C}} \\ A\otimes(B\otimes\mathbb{T}C) & \xrightarrow[\mathrm{id}_A\otimes\mathbb{t}_{B,C}]{} & A\otimes\mathbb{T}(B\otimes C) & \xrightarrow[\mathbb{t}_{A,B\otimes C}]{} & \mathbb{T}(A\otimes(B\otimes C)) \end{array} \tag{34}$$

$$\begin{array}{ccccc} A\otimes\mathbb{T}^2B & \xrightarrow{\mathbb{t}_{A,\mathbb{T}B}} & \mathbb{T}(A\otimes\mathbb{T}B) & \xrightarrow{\mathbb{T}\mathbb{t}_{A,B}} & \mathbb{T}^2(A\otimes B) \\ \downarrow{\scriptstyle \mathrm{id}_A\otimes\mu_B} & & & & \downarrow{\scriptstyle \mu_{A\otimes B}} \\ A\otimes\mathbb{T}B & & \xrightarrow[\quad\mathbb{t}_{A,B}\quad]{} & & \mathbb{T}(A\otimes B) \end{array} \tag{35}$$

Formally, a *strong monad* is a four-tuple $(\mathbb{T},\eta,\mu,\mathbb{t})$ where $(\mathbb{T},\eta,\mu)$ is a monad and $\mathbb{t}$ is a strength making the above diagrams commute.

By lifting of $\boldsymbol{T}$ to $(\mathbb{C}\downarrow|_|)$ we now mean a strong monad, i.e. a monad $(\widetilde{T},\widetilde{\eta},\widetilde{\mu})$ together with a strength $\widetilde{t}_{X,Y} : X\widetilde{\otimes}\widetilde{T}Y \to \widetilde{T}(X\widetilde{\otimes}Y)$, such that diagram (2) commutes, equations (3) hold and

$$U\widetilde{t}_{X,Y} = \boldsymbol{t}_{UX,UY},$$

i.e., $U$ preserves strength.

To be able to give an appropriate lifting, we extend the monad morphism to a *strong monad* morphism, i.e., a monad morphism making the following additional diagram commute, which relates the strengths $\boldsymbol{t}_{A_1,A_2}$ and $\mathbb{t}_{|A_1|,|A_2|}$:

$$\begin{array}{ccc} |A_1|\otimes\mathbb{T}|A_2| & \xrightarrow{\mathbb{t}_{|A_1|,|A_2|}} & \mathbb{T}(|A_1|\otimes|A_2|) \\ \downarrow{\scriptstyle \mathrm{id}_{|A_1|}\otimes\sigma_{A_2}} & & \downarrow{\scriptstyle \mathbb{T}(\theta_{A_1,A_2})} \\ |A_1|\otimes|\boldsymbol{T}A_2| & & \mathbb{T}|A_1\boldsymbol{\otimes} A_2| \\ \downarrow{\scriptstyle \theta_{A_1,\boldsymbol{T}A_2}} & & \downarrow{\scriptstyle \sigma_{A_1\boldsymbol{\otimes} A_2}} \\ |A_1\boldsymbol{\otimes}\boldsymbol{T}A_2| & \xrightarrow[|\boldsymbol{t}_{A_1,A_2}|]{} & |\boldsymbol{T}(A_1\boldsymbol{\otimes} A_2)| \end{array} \tag{36}$$

Having lifted $\boldsymbol{T}$ to scones and subscones in previous sections, we only need to give a lifting of the strength $\boldsymbol{t}$. For scones this is straightforward—define $\widetilde{t}$ pointwise by

$$\widetilde{t}_{\langle S_1,m_1,A_1\rangle,\langle S_2,m_2,A_2\rangle} = \langle \mathbb{t}_{S_1,S_2}, \boldsymbol{t}_{A_1,A_2}\rangle.$$

Verifying that this is well-defined amounts to pasting together a naturality square for

$\mathbb{t}$ and a diagram (36):

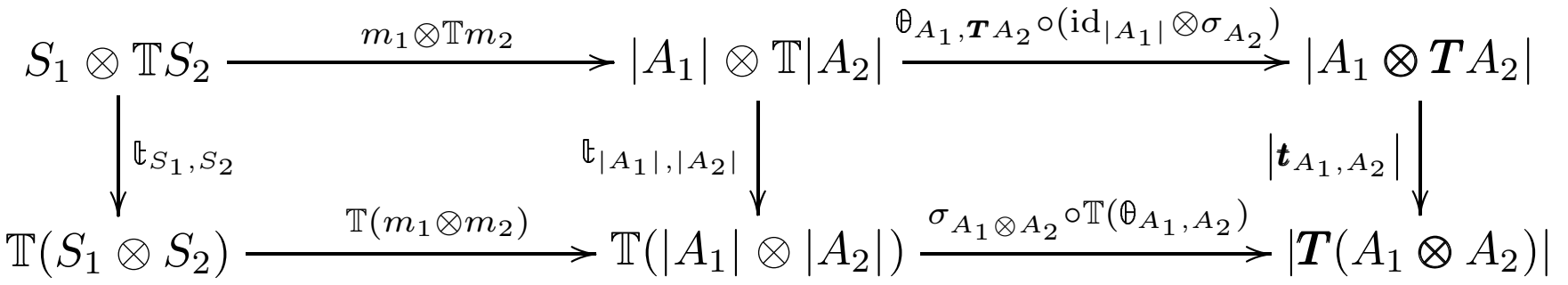

The upper side of this diagram is precisely $\langle S_1, m_1, A_1\rangle \widetilde{\otimes} \widetilde{T}\langle S_2, m_2, A_2\rangle$ in the scone while the lower side is $\widetilde{T}(\langle S_1, m_1, A_1\rangle \widetilde{\otimes} \langle S_2, m_2, A_2\rangle)$. (We let the interested reader define for herself the tensor product $\widetilde{\otimes}$ in the scone.) Checking naturality of $\widetilde{t}$ and strength laws is immediate since $\widetilde{t}$, $\widetilde{\alpha}$, $\widetilde{r}$, $\widetilde{\eta}$ and $\widetilde{\mu}$ are all defined pointwise.

Now we move to subscones. Call $\widetilde{T}$ the lifted monad defined in (7), (9), (10) and (11) in Section 4.

As in previous sections, $\widetilde{t}$ in subscones will differ from the case of scones only in its $\mathbb{C}$-component $\hat{\mathbb{t}}$, and this component will be induced as a unique diagonal guaranteed by diagram (6) in the diagram below.

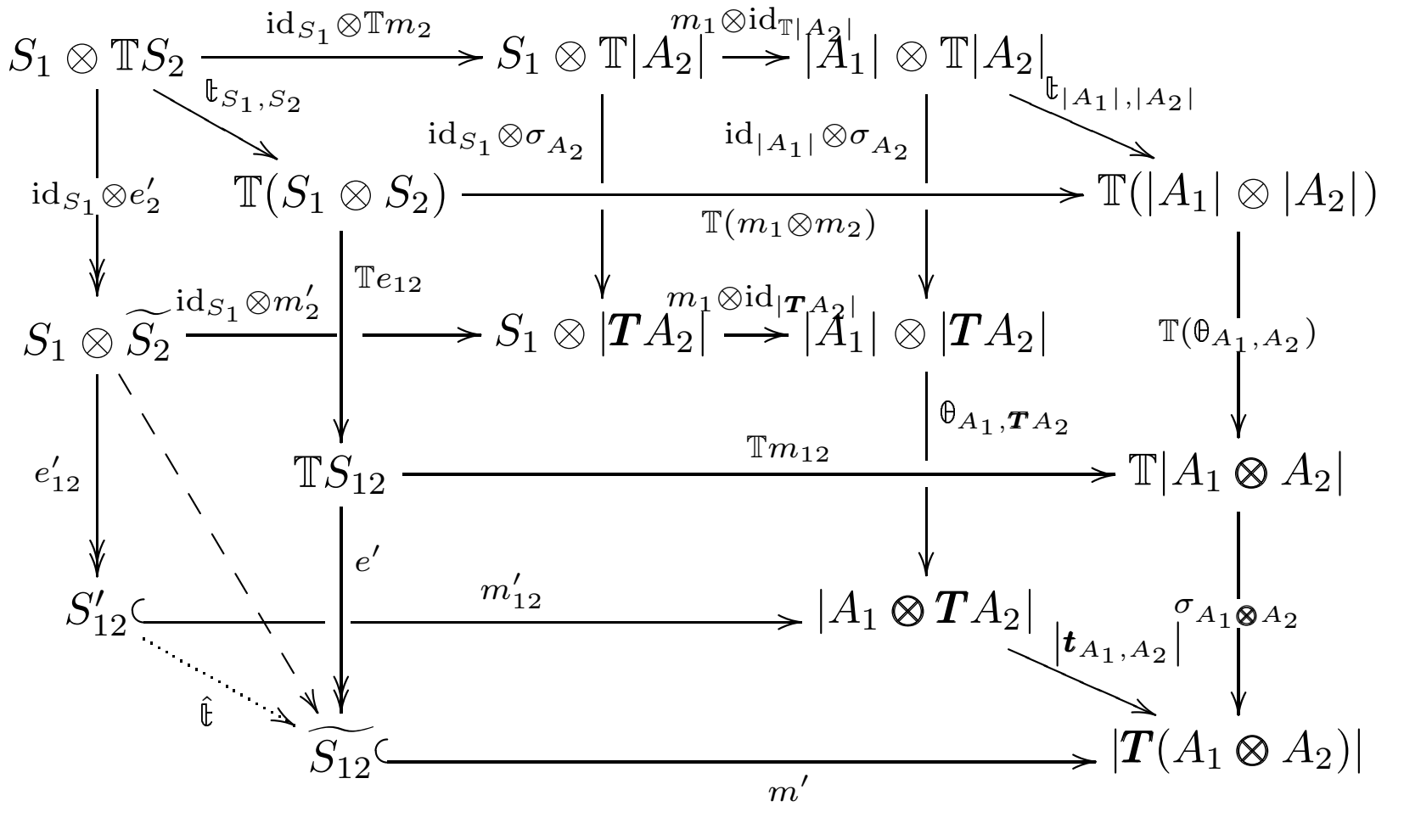

(37)

As ingredients of this diagram we have used:

— An instance 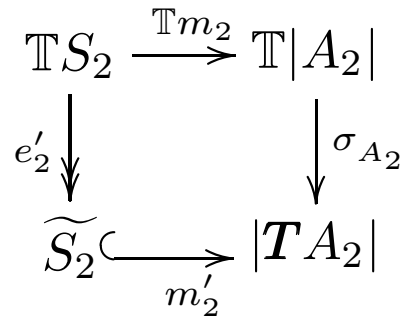 of Diagram (7) defining $\widetilde{T}\langle S_2, m_2, A_2\rangle$; this is tensored by $S_1$ on the left to get the upper left square of the back face. Notice that $\mathrm{id}_{S_1} \otimes e'_2$ is pseudoepi because our mono factorization system is monoidal.

— An instance 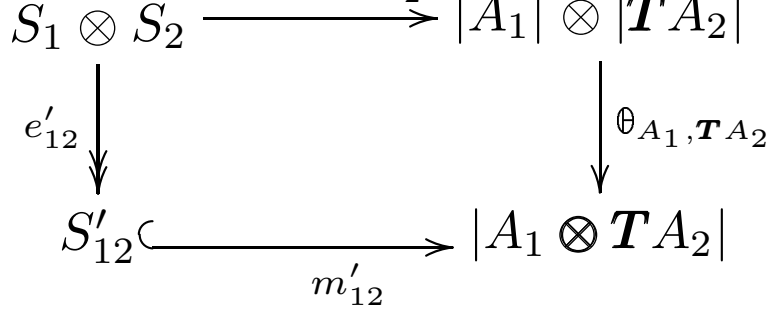 of Diagram (20) defining the tensor

product of $\langle S_1, m_1, A_1\rangle$ with $\widetilde{T}\langle S_2, m_2, A_2\rangle = \langle \widetilde{S_2}, m_2', \boldsymbol{T}A_2\rangle$; this is the lower square of the back face. (Note that the upper right square of the back face commutes trivially.)

— Another instance 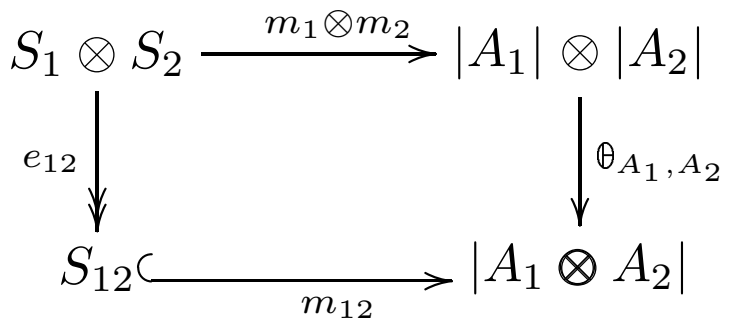

of Diagram (20) defining the tensor product $\langle S_{12}, m_{12}, A_1 \otimes A_2\rangle$ of $\langle S_1, m_1, A_1\rangle$ with $\langle S_2, m_2, A_2\rangle$; we apply $\mathbb{T}$ to this square to get the upper half of the front face.

— Another instance of Diagram (7) defining the application of $\widetilde{T}$ to the just mentioned tensor product $\langle S_{12}, m_{12}, A_1 \otimes A_2\rangle$: this is the lower half of the front face.

— A naturality square for $\mathbb{t}$ (top face), and

— An instance of Diagram (36), which defines the right face.

As usual, the dashed and the dotted arrows are given by diagonal fill-ins, therefore $\widetilde{t} = (\hat{\mathbb{t}}, \boldsymbol{t})$ is well-defined. Again, checking naturality of $\widetilde{t}$ and strength laws is immediate by Fact 4.3.

Here is the final set of ingredients for lifting a strong monad $(\boldsymbol{T}, \boldsymbol{\eta}, \boldsymbol{\mu}, \boldsymbol{t})$ on category $\boldsymbol{C}$ to $(\mathbb{C} \downarrow |_|)$:

**(i.a)** monoidal categories $(\boldsymbol{C}, \boldsymbol{\otimes}, \boldsymbol{I}, \boldsymbol{\alpha}, \boldsymbol{\ell}, \boldsymbol{r})$ and $(\mathbb{C}, \otimes, \mathbb{I}, \mathbb{a}, \mathbb{l}, \mathbb{r})$, and a monoidal functor $|_| : \boldsymbol{C} \to \mathbb{C}$,
**(ii.a)** a strong monad $(\mathbb{T}, \mathbb{\eta}, \mathbb{\mu}, \mathbb{t})$ on $\mathbb{C}$, related to $(\boldsymbol{T}, \boldsymbol{\eta}, \boldsymbol{\mu}, \boldsymbol{t})$ by a strong monad morphism $(|_|, \sigma)$ defined in (4) and (36),
**(iii.a)** a monoidal mono factorization system on $\mathbb{C}$.
**(iv)** $\mathbb{T}$ maps relevant pseudoepis to pseudoepis.

7.2. *Monoidal Monads*

Several strong monads are in fact monoidal—in fact all the monads of Section 10 are monoidal, except the continuation and continuation-like monads. While this notion is not needed in Moggi's account of computation (Moggi, 1991), this occurs naturally, and will be used in Section 8.4. A *monoidal* monad is a four-tuple $(\mathbb{T}, \mathbb{\eta}, \mathbb{\mu}, \mathbb{d})$, where $\mathbb{d}_{A,B} : \mathbb{T}A \otimes \mathbb{T}B \to \mathbb{T}(A \otimes B)$ is a *mediator* natural transformation, making the following diagrams commute:

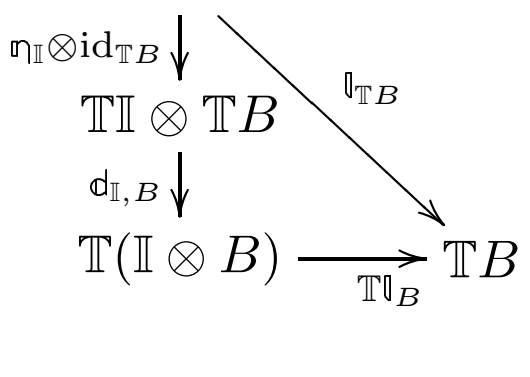

(38)

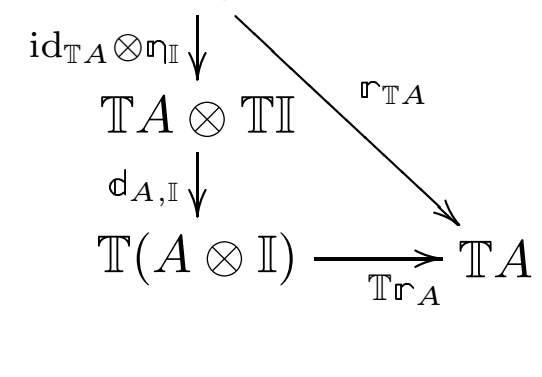

(39)

$$
\begin{array}{ccc}
A \otimes B & \xrightarrow{\eta_A \otimes \eta_B} & \mathbb{T}A \otimes \mathbb{T}B \\
& \searrow^{\eta_{A\otimes B}} & \downarrow{\scriptstyle \mathbb{d}_{A,B}} \\
& & \mathbb{T}(A \otimes B)
\end{array}
\tag{40}
$$

$$
\begin{array}{ccccc}
(\mathbb{T}A \otimes \mathbb{T}B) \otimes \mathbb{T}C & \xrightarrow{\mathbb{d}_{A,B} \otimes \mathrm{id}_{\mathbb{T}C}} & \mathbb{T}(A \otimes B) \otimes \mathbb{T}C & \xrightarrow{\mathbb{d}_{A\otimes B,C}} & \mathbb{T}((A \otimes B) \otimes C) \\
\downarrow{\scriptstyle \alpha_{\mathbb{T}A,\mathbb{T}B,\mathbb{T}C}} & & & & \downarrow{\scriptstyle \mathbb{T}\alpha_{A,B,C}} \\
\mathbb{T}A \otimes (\mathbb{T}B \otimes \mathbb{T}C) & \xrightarrow[\mathrm{id}_{\mathbb{T}A} \otimes \mathbb{d}_{B,C}]{} & \mathbb{T}A \otimes \mathbb{T}(B \otimes C) & \xrightarrow[\mathbb{d}_{A,B\otimes C}]{} & \mathbb{T}(A \otimes (B \otimes C))
\end{array}
\tag{41}
$$

$$
\begin{array}{ccccc}
\mathbb{T}^2 A \otimes \mathbb{T}^2 B & \xrightarrow{\mathbb{d}_{\mathbb{T}A,\mathbb{T}B}} & \mathbb{T}(\mathbb{T}A \otimes \mathbb{T}B) & \xrightarrow{\mathbb{T}\mathbb{d}_{A,B}} & \mathbb{T}^2(A \otimes B) \\
{\scriptstyle \mu_A \otimes \mu_B}\downarrow & & & & \downarrow{\scriptstyle \mu_{A\otimes B}} \\
\mathbb{T}A \otimes \mathbb{T}B & & \xrightarrow[\mathbb{d}_{A,B}]{} & & \mathbb{T}(A \otimes B)
\end{array}
\tag{42}
$$

Diagrams (38), (39), (41) state that $\mathbb{T}$ is a monoidal functor with mediating pair $(\mathbb{d}, \eta_{\mathbb{I}})$. Diagram (40) states that $\eta$ is a so-called monoidal natural transformation, while Diagram (42) states that $\mu$ is another monoidal natural transformation.

Given any monoidal monad $(\mathbb{T}, \eta, \mu, \mathbb{d})$ on $\mathbb{C}$, it is easy to check that $(\mathbb{T}, \eta, \mu, \mathbb{t})$ is a strong monad, where $\mathbb{t}_{A,B} = \mathbb{d}_{A,B} \circ (\eta_A \otimes \mathrm{id}_{\mathbb{T}B})$. Furthermore, $\mathbb{t}'_{A,B}$ defined as $\mathbb{d}_{A,B} \circ (\mathrm{id}_{\mathbb{T}A} \otimes \eta_B)$ is a *dual strength*, that is, a natural transformation $\mathbb{t}'_{A,B} : \mathbb{T}A \otimes B \to \mathbb{T}(A \otimes B)$ obeying the obvious duals of the strength laws (32), (33), (34), (35). (Formally, a dual strength is a strength on the dual monoidal category $(\mathbb{C}, \mathbb{I}, \otimes^{op}, \alpha^{-1}, \mathbb{r}, \mathbb{l})$, where $A \otimes^{op} B$ is defined as $B \otimes A$.)

Moreover, the strength $\mathbb{t}$ and the dual strength $\mathbb{t}'$ are compatible with the associativity, in the sense that the diagram below commutes.

$$
\begin{array}{ccccc}
(A \otimes \mathbb{T}B) \otimes C & \xrightarrow{\mathbb{t}_{A,B} \otimes \mathrm{id}_C} & \mathbb{T}(A \otimes B) \otimes C & \xrightarrow{\mathbb{t}'_{A\otimes B,C}} & \mathbb{T}((A \otimes B) \otimes C) \\
{\scriptstyle \alpha_{A,\mathbb{T}B,C}}\downarrow & & & & \downarrow{\scriptstyle \mathbb{T}\alpha_{A,B,C}} \\
A \otimes (\mathbb{T}B \otimes C) & \xrightarrow{\mathrm{id}_A \otimes \mathbb{t}'_{B,C}} & A \otimes \mathbb{T}(B \otimes C) & \xrightarrow{\mathbb{t}_{A,B\otimes C}} & \mathbb{T}(A \otimes (B \otimes C))
\end{array}
\tag{43}
$$

Finally, the strength and the dual strength commute, in the sense that the diagram on the right commutes. In fact, the common diagonal from $\mathbb{T}A \otimes \mathbb{T}B$ to $\mathbb{T}(A \otimes B)$ is just $\mathbb{d}_{A,B}$.

$$
\begin{array}{ccccc}
\mathbb{T}A \otimes \mathbb{T}B & & \xrightarrow{\mathbb{t}_{\mathbb{T}A,B}} & & \mathbb{T}(\mathbb{T}A \otimes B) \\
\downarrow & & & & \downarrow{\scriptstyle \mathbb{T}\mathbb{t}'_{A,B}} \\
{\scriptstyle \mathbb{t}'_{A,\mathbb{T}B}} & & & & \mathbb{T}^2(A \otimes B) \\
\downarrow & & & & \downarrow{\scriptstyle \mu_{A\otimes B}} \\
\mathbb{T}(A \otimes \mathbb{T}B) & \xrightarrow[\mathbb{T}\mathbb{t}_{A,B}]{} & \mathbb{T}^2(A \otimes B) & \xrightarrow[\mu_{A\otimes B}]{} & \mathbb{T}(A \otimes B)
\end{array}
\tag{44}
$$

In general, a monoidal monad can be defined equivalently as a monad with a strength and a dual strength that make the diagrams (43) and (44) commute. See Appendix A, in particular Appendix A.1 and Appendix A.2, for a proof.

It is natural to define a *monoidal monad morphism* from $(\boldsymbol{T}, \boldsymbol{\eta}, \boldsymbol{\mu}, \boldsymbol{d})$ to $(\mathbb{T}, \eta, \mu, \mathbb{d})$ as a monad morphism $\sigma$ from $(\boldsymbol{T}, \boldsymbol{\eta}, \boldsymbol{\mu})$ to $(\mathbb{T}, \mathbb{\eta}, \mathbb{\mu})$ making the following diagram commute:

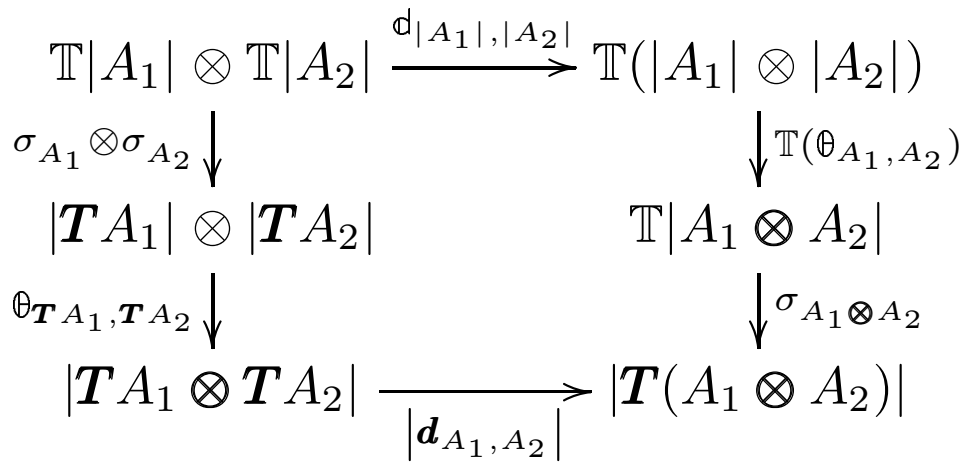

(45)

Every monoidal monad morphism $\sigma$ is also a strong monad morphism from $(\boldsymbol{T}, \boldsymbol{\eta}, \boldsymbol{\mu}, \boldsymbol{t})$ to $(\mathbb{T}, \mathbb{\eta}, \mathbb{\mu}, \mathbb{t})$, and also from $(\boldsymbol{T}, \boldsymbol{\eta}, \boldsymbol{\mu}, \boldsymbol{t}')$ to $(\mathbb{T}, \mathbb{\eta}, \mathbb{\mu}, \mathbb{t}')$, where $\boldsymbol{t}_{A,B} = \boldsymbol{d}_{A,B} \circ (\boldsymbol{\eta}_A \otimes \mathrm{id}_{\boldsymbol{T}B})$, $\mathbb{t}_{A,B} = \mathbb{d}_{A,B} \circ (\mathbb{\eta}_A \otimes \mathrm{id}_{\mathbb{T}B})$, $\boldsymbol{t}'_{A,B} = \boldsymbol{d}_{A,B} \circ (\mathrm{id}_{\boldsymbol{T}A} \otimes \boldsymbol{\eta}_B)$, $\mathbb{t}'_{A,B} = \mathbb{d}_{A,B} \circ (\mathrm{id}_{\boldsymbol{T}A} \otimes \mathbb{\eta}_B)$. In Appendix A.3, we show that the monoidal monad morphisms are exactly the natural transformations $\sigma$ that are both a strong monad morphism and a dual strong monad morphism.

One may think that lifting monoidal monads to scones and subscones is easy: get the strength and the dual strength from the mediator, and lift them as in Section 7.1. However, it is not immediately clear that what one would get would arise as strength and dual strengths of a mediator, in particular that (44) would commute. Let us therefore state the construction explicitly. This mimicks the construction of the lifted strength from Section 7.1. For scones, the lifted mediator is again defined pointwise by

$$\widetilde{d}_{\langle S_1, m_1, A_1\rangle, \langle S_2, m_2, A_2\rangle} = \langle \boldsymbol{d}_{S_1,S_2}, \mathbb{d}_{A_1,A_2}\rangle$$

For subscones, we mimick Diagram (37) in Diagram (46) below. We let the reader check all commutations.

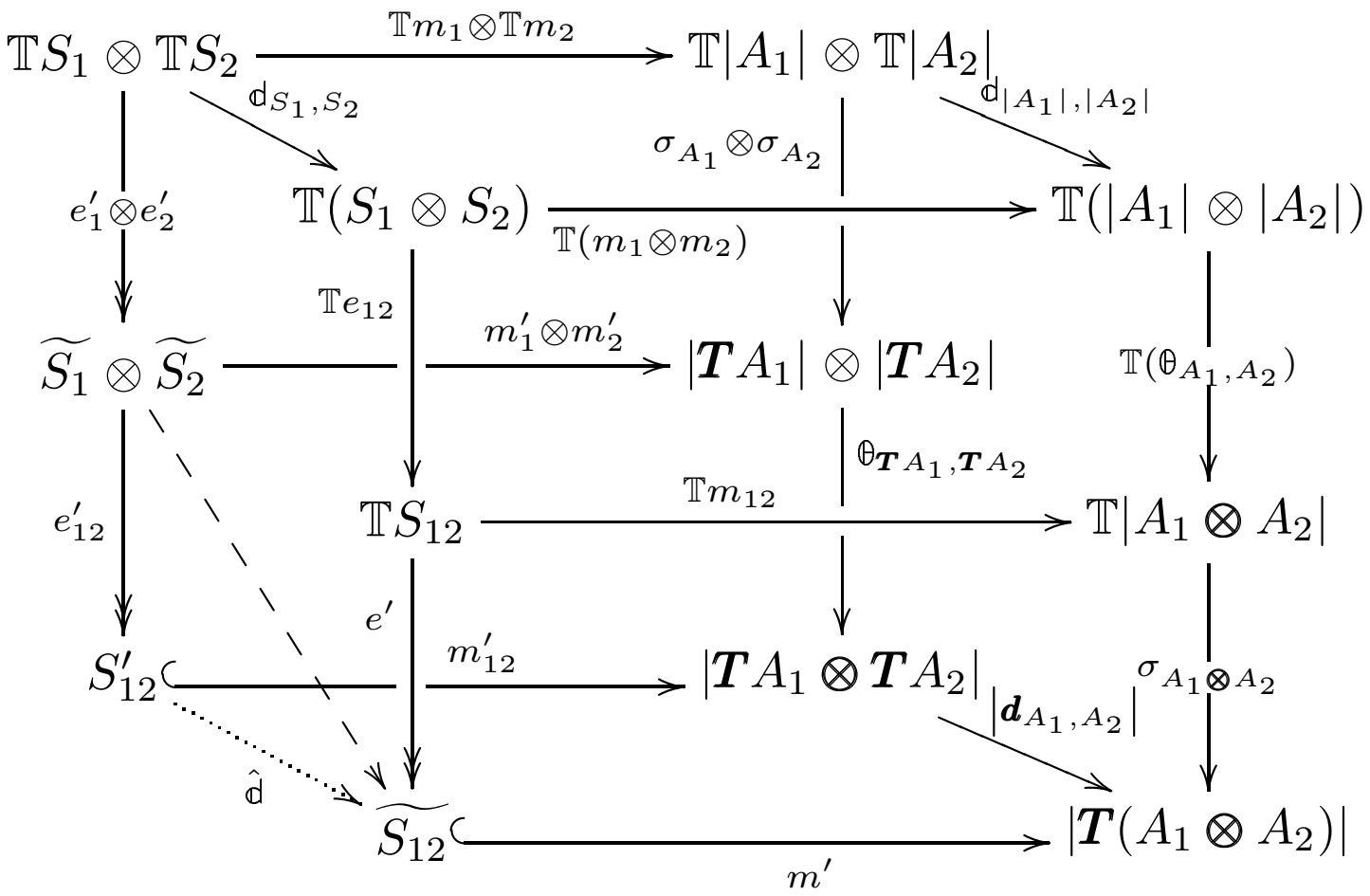

(46)

So, to lift a monoidal monad $(\boldsymbol{T}, \boldsymbol{\eta}, \boldsymbol{\mu}, \boldsymbol{d})$ on category $\boldsymbol{C}$ to $(\mathbb{C} \downarrow |_|)$, we require:

**(i.a)** monoidal categories $(\boldsymbol{C}, \otimes, \boldsymbol{I}, \boldsymbol{\alpha}, \boldsymbol{\ell}, \boldsymbol{r})$ and $(\mathbb{C}, \otimes, \mathbb{I}, \mathbb{\alpha}, \mathbb{l}, \mathbb{r})$,
and a monoidal functor $|_| : \boldsymbol{C} \to \mathbb{C}$,
**(ii.b)** a monoidal monad $(\mathbb{T}, \mathbb{\eta}, \mathbb{\mu}, \mathbb{d})$ on $\mathbb{C}$, related to
the monoidal monad $(\boldsymbol{T}, \boldsymbol{\eta}, \boldsymbol{\mu}, \boldsymbol{d})$ on $\boldsymbol{C}$ by
a monoidal monad morphism $(|_|, \sigma)$ defined in (4) and (45),
**(iii.a)** a monoidal mono factorization system on $\mathbb{C}$.
**(iv)** $\mathbb{T}$ maps relevant pseudoepis to pseudoepis.

7.3. *Commutative Monads*

In the case of *symmetric* monoidal categories $\boldsymbol{C}$ and $\mathbb{C}$, recall that if we also want to make the subscone a symmetric monoidal category, it suffices to replace **(i.a)** by **(i.b)**, which requires $|_|$ to be a *symmetric* monoidal functor. This case occurs notably if we want to lift a *commutative* monad to the subscone.

Recall that a strong monad $(\mathbb{T}, \mathbb{\eta}, \mathbb{\mu}, \mathbb{t})$ is commutative if and only if, letting $\mathbb{t}'_{A,B}$ be the dual strength $\mathbb{T}\mathbb{c}_{B,A} \circ \mathbb{t}_{B,A} \circ \mathbb{c}_{\mathbb{T}A,B}$, then Diagram (44) commutes.
Let $\mathbb{d}_{A,B}$ be the common diagonal $\mathbb{\mu}_{A\otimes B} \circ \mathbb{T}\mathbb{t}'_{A,B} \circ \mathbb{t}_{\mathbb{T}A,B} = \mathbb{\mu}_{A\otimes B} \circ \mathbb{T}\mathbb{t}_{A,B} \circ \mathbb{t}'_{A,\mathbb{T}B}$. We can check that $\mathbb{d}$ is then a mediator , whence every commutative monad is monoidal. In fact, a monoidal monad is commutative if and only if the following additional diagram commutes (see Appendix A.4).

$$\begin{array}{ccc} \mathbb{T}A \otimes \mathbb{T}B & \xrightarrow{\mathbb{d}_{A,B}} & \mathbb{T}(A \otimes B) \\ {\scriptstyle \mathbb{c}_{\mathbb{T}A,\mathbb{T}B}}\downarrow & & \downarrow{\scriptstyle \mathbb{T}\mathbb{c}_{A,B}} \\ \mathbb{T}B \otimes \mathbb{T}A & \xrightarrow[\mathbb{d}_{B,A}]{} & \mathbb{T}(B \otimes A) \end{array} \qquad (47)$$

For convenience, we shall now understand commutative monads as monoidal monads satisfying (47). The lifting of monoidal monads of Section 7.2 then yields a lifting of commutative monads, by Fact 4.3. Therefore, to lift a commutative monad $(\boldsymbol{T}, \boldsymbol{\eta}, \boldsymbol{\mu}, \boldsymbol{t})$ on category $\boldsymbol{C}$ to $(\mathbb{C} \downarrow |_|)$, we require:

**(i.b)** symmetric monoidal categories $(\boldsymbol{C}, \otimes, \boldsymbol{I}, \boldsymbol{\alpha}, \boldsymbol{\ell}, \boldsymbol{r}, \boldsymbol{c})$ and $(\mathbb{C}, \otimes, \mathbb{I}, \mathbb{\alpha}, \mathbb{l}, \mathbb{r}, \mathbb{c})$,
and a symmetric monoidal functor $|_| : \boldsymbol{C} \to \mathbb{C}$,
**(ii.c)** a commutative monad $(\mathbb{T}, \mathbb{\eta}, \mathbb{\mu}, \mathbb{d})$ on $\mathbb{C}$, related to
the commutative monad $(\boldsymbol{T}, \boldsymbol{\eta}, \boldsymbol{\mu}, \boldsymbol{d})$ on $\boldsymbol{C}$ by
a monoidal monad morphism $(|_|, \sigma)$ defined in (4) and (45),
**(iii.a)** a monoidal mono factorization system on $\mathbb{C}$.
**(iv)** $\mathbb{T}$ maps relevant pseudoepis to pseudoepis.

When $\mathbb{C}$ has all finite products and we consider the induced symmetric monoidal structure, we might require that the monoidal monad $(\mathbb{T}, \mathbb{\eta}, \mathbb{\mu}, \mathbb{d})$ is not just commutative but even *cartesian*, by which we mean that $\mathbb{T}$ is a cartesian monoidal functor with mediating pair $(\mathbb{d}, \mathbb{\eta}_{\mathbb{I}})$. (Recall that $\mathbb{T}$ is always a monoidal functor with this very mediating pair.) This means that $\mathbb{T}\pi_i \circ \mathbb{d}_{A,B} = \pi_i$, $i \in \{1, 2\}$. We just do not need this in our constructions; but it is often easier to prove that a monad is cartesian and infer that it is commutative than to prove that it is commutative directly: we shall see examples of cartesian monads in Section 10.

## 8. Building Monad Morphisms from Adjunctions

It is often the case that we have a (strong) monad on $\boldsymbol{C}$, and wish to build another one on $\mathbb{C}$ related to the latter by a monad morphism. The following results are then of some help.

Recall that, given two categories $\mathcal{C}$ and $\mathcal{D}$, a pair of functors $F : \mathcal{C} \to \mathcal{D}$ and $U : \mathcal{D} \to \mathcal{C}$ is an *adjunction* $F \dashv U$ if and only if there are natural transformations $\eta_{.} : . \to UF.$ (the *unit* of the adjunction) and $\epsilon_{.} : FU. \to .$ (the *counit*) such that $\epsilon_{F(A)} \circ F\eta_A = \mathrm{id}_{F(A)}$ and $U\epsilon_A \circ \eta_{U(A)} = \mathrm{id}_{U(A)}$. $F$ is said to be *left adjoint* to $U$, $U$ is *right adjoint* to $F$.

Then any adjunction $F \dashv U$ gives rise to a monad $(UF, \eta_{.}, U\epsilon_{F.})$ on $\mathcal{C}$. Conversely, there are two standard ways of retrieving an adjunction from a monad $(T, \eta, \mu)$ on $\mathcal{C}$, from Eilenberg-Moore algebras, or from the Kleisli category of the monad.

### 8.1. *Eilenberg-Moore algebras.*

A $T$*-algebra* is a morphism $T(A) \xrightarrow{s} A$, for some object $A$ of $\mathcal{C}$, satisfying the commutativity conditions:

$$\begin{array}{ccc} & & T(A) \\ & \overset{\eta_A}{\nearrow} & \downarrow s \\ A & \xrightarrow[\mathrm{id}_A]{} & A \end{array} \qquad \begin{array}{ccc} & & T^2(A) \\ & \overset{\mu_A}{\swarrow} & \downarrow T(s) \\ T(A) & & T(A) \\ \downarrow s & & \downarrow s \\ A & = & A \end{array} \tag{48}$$

$A$ is called the *support* of the algebra, $s$ its *structure map.* A morphism from $T(A) \xrightarrow{s} A$ to $T(B) \xrightarrow{u} B$ is a morphism $f : A \to B$ in $\mathbb{C}$ that commutes with structure maps, i.e., such that $f \circ s = u \circ T(f)$. $T$-algebras together with these morphisms forms a category $T$-**Alg**.

Then $F^T \dashv U^T$ is an adjunction, where $U^T : T\text{-}\mathbf{Alg} \to \mathcal{C}$ maps objects $T(A) \xrightarrow{s} A$ to $A$ and morphisms $f$ from $T(A) \xrightarrow{s} A$ to $T(B) \xrightarrow{u} B$ to the underlying morphism $f$ from $A$ to $B$ in $\mathcal{C}$; and where $F^T : \mathcal{C} \to T\text{-}\mathbf{Alg}$ maps the object $A$ to the $T$-algebra $T^2(A) \xrightarrow{\mu_A} T(A)$, and the morphisms $A \xrightarrow{f} B$ to $f$ seen as morphism from $F^T(A)$ to $F^T(B)$. The unit of the adjunction is $\eta$, while the counit $\epsilon$ is given on each $T$-algebra $T(A) \xrightarrow{s} A$ as the morphism $s$ itself, from $F^T U^T(\, T(A) \xrightarrow{s} A\,) = T^2(A) \xrightarrow{\mu_A} T(A)$ to $T(A) \xrightarrow{s} A$.

Moreover, the monad of this adjunction is the original monad $(T, \eta, \mu)$.

8.2. *Kleisli category.*

The objects of $\mathbf{Kleisli}(T)$ are the objects of $\mathcal{C}$, while the morphisms $A \xrightarrow{f} B$ of $\mathbf{Kleisli}(T)$ are the morphisms $A \xrightarrow{f} T(B)$ of $\mathcal{C}$. To avoid confusion, we write $\overline{f}$ the morphisms $f$ seen as a morphism in $\mathbf{Kleisli}(T)$. The identity $\overline{\mathrm{id}_A}$ on the object $A$ in $\mathbf{Kleisli}(T)$ is $\eta_A$, while composition $\overline{g} \circ \overline{f}$ is $\overline{\mu_C \circ T(g) \circ f}$, where $A \xrightarrow{f} T(B)$ and $B \xrightarrow{g} T(C)$ in $\mathcal{C}$.

Define $F_T : \mathcal{C} \to \mathbf{Kleisli}(T)$ as mapping the object $A$ to $A$, and the morphism $A \xrightarrow{f} B$ to the morphism from $A$ to $B$ in $\mathbf{Kleisli}(T)$ defined as $\eta_B \circ f$. Define $U_T$ : $\mathbf{Kleisli}(T) \to \mathcal{C}$ as mapping the object $A$ to $T(A)$, and the morphism $\overline{f}$ from $A$ to $B$ in $\mathbf{Kleisli}(T)$ to $\mu_B \circ T(f)$. Then $F_T \dashv U_T$ is an adjunction, whose unit is $\eta$, and whose counit $\epsilon$ is the identity morphism from $F_T U_T(A)$ to $T(A)$ in $\mathcal{C}$, seen as a morphism from $F_T U_T(A)$ to $A$ in $\mathbf{Kleisli}(T)$. The monad of $F_T \dashv U_T$ is again $(T, \eta, \mu)$.

8.3. *Monad Morphisms from Adjunctions*

**Proposition 8.1.** Let $(\boldsymbol{T}, \boldsymbol{\eta}, \boldsymbol{\mu})$ be a monad on a category $\boldsymbol{C}$, $|_| : \boldsymbol{C} \to \mathbb{C}$ be a functor with a left adjoint $\boldsymbol{D} : \mathbb{C} \to \boldsymbol{C}$. Let $\dot{\epsilon}_A : \boldsymbol{D}|A| \to A$ be the counit of the adjunction, $\dot{\eta}_E : E \to |\boldsymbol{D}(E)|$ be the unit of the adjunction.

Define $\mathbb{T} = |_| \circ \boldsymbol{T} \circ \boldsymbol{D} = |\boldsymbol{TD}|$, $\mathbb{\eta}_E = \left|\boldsymbol{\eta}_{\boldsymbol{D}(E)}\right| \circ \dot{\eta}_E$, $\mathbb{\mu}_E = \left|\boldsymbol{\mu}_{\boldsymbol{D}(E)} \circ \boldsymbol{T}\dot{\epsilon}_{\boldsymbol{TD}(E)}\right|$. Finally, let $\sigma_A = |\boldsymbol{T}\dot{\epsilon}_A| : \mathbb{T}|A| \to |\boldsymbol{T}A|$. Then $(\mathbb{T}, \mathbb{\eta}, \mathbb{\mu})$ is a monad on $\mathbb{C}$ and $(|_|, \sigma)$ is a monad morphism from $\boldsymbol{T}$ to $\mathbb{T}$.

*Proof.* Let $F \dashv U$ be any adjunction generating the monad, i.e., such that $UF = \boldsymbol{T}$, whose unit is $\boldsymbol{\eta}$, and whose counit $\epsilon$ is such that $\boldsymbol{\mu} = U\epsilon_F$. We may choose, e.g., $F^T \dashv U^T$ or $F_T \dashv U_T$. Compose the adjunction $\boldsymbol{D} \dashv |_|$ with $F \dashv U$, yielding an adjunction $F\boldsymbol{D} \dashv |U|$. The unit of this adjunction (on object $E$) is $\left|\boldsymbol{\eta}_{\boldsymbol{D}(E)}\right| \circ \dot{\eta}_E$, its counit (on object $A$) is $\epsilon_A \circ F(\dot{\epsilon}_{UA})$.

The monad of this adjunction is $(|UF\boldsymbol{D}|, \left|\boldsymbol{\eta}_{\boldsymbol{D}(.)}\right| \circ \dot{\eta}_., \left|U(\epsilon_{F\boldsymbol{D}(.)} \circ F(\dot{\epsilon}_{UF\boldsymbol{D}(.)}))\right|)$. But the monad of $F \dashv U$ is $(\boldsymbol{T}, \boldsymbol{\eta}, \boldsymbol{\mu})$, so $UF = \boldsymbol{T}$ and $\boldsymbol{\mu} = U\epsilon_F$. It follows that the monad of $F\boldsymbol{D} \dashv |U|$ is $(|\boldsymbol{TD}|, \left|\boldsymbol{\eta}_{\boldsymbol{D}(.)}\right| \circ \dot{\eta}_., \left|\boldsymbol{\mu}_{\boldsymbol{D}(.)} \circ \boldsymbol{T}(\dot{\epsilon}_{\boldsymbol{TD}(.)})\right|)$. This is $(\mathbb{T}, \mathbb{\eta}, \mathbb{\mu})$, which is therefore a monad.

It remains to show that $\sigma = |\boldsymbol{T}\dot{\epsilon}|$ is a monad morphism from $\boldsymbol{T}$ to $\mathbb{T}$. This is checked using the following diagrams. In the left diagram, the top triangle is one of the adjunction laws, the bottom square is by naturality of $|\boldsymbol{\eta}|$; so $\sigma_A$ (bottom arrow) composed with $\mathbb{\eta}_{|A|}$ (leftmost vertical path) equals $|\boldsymbol{\eta}_A|$ (rightmost path from $|A|$ to $|\boldsymbol{T}A|$).

In the right diagram, the top square is by naturality of $|\boldsymbol{T}\dot{\epsilon}|$, the bottom square is by naturality of $|\boldsymbol{\mu}|$, so $\sigma_A$ (bottom arrow) composed with $\mathbb{\mu}_{|A|}$ (leftmost vertical path) equals $|\boldsymbol{\mu}_A| \circ \sigma_{\boldsymbol{T}A} \circ \mathbb{T}\sigma_A$ (the other path from top left to bottom right). ☐

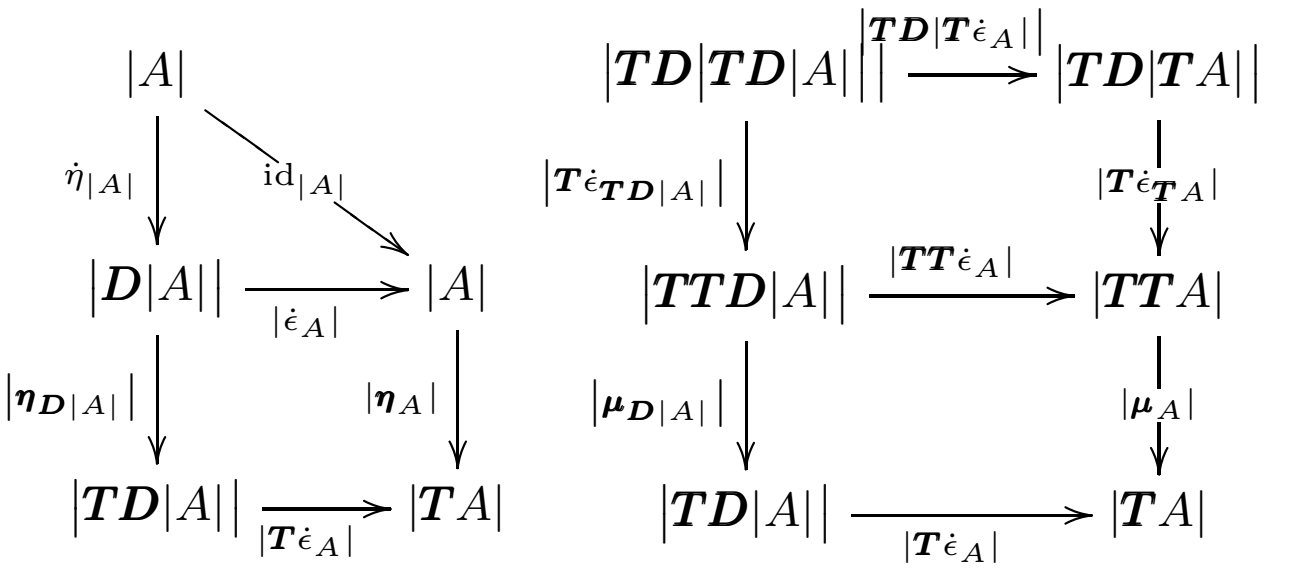

Note that we could have checked the required diagrams directly; the proof would be longer than going through adjunctions, as we did.

### 8.4. *Monoidal, and Strong Monad Morphisms from Monoidal Adjunctions*

We first reproduce the argument of Proposition 8.1 in the monoidal case. While monads correspond to adjunctions in well-defined ways, only *monoidal* monads can be linked to so-called *monoidal* adjunctions. This is the reason why we deal with monoidal monads first.

Let $(\mathcal{C}, I^{\mathcal{C}}, \otimes^{\mathcal{C}}, \alpha^{\mathcal{C}}, \ell^{\mathcal{C}}, r^{\mathcal{C}})$ and $(\mathcal{D}, I^{\mathcal{D}}, \otimes^{\mathcal{D}}, \alpha^{\mathcal{D}}, \ell^{\mathcal{D}}, r^{\mathcal{D}})$ be two monoidal categories. Let $F \dashv U$ be an adjunction, where $F : \mathcal{C} \to \mathcal{D}$, $U : \mathcal{D} \to \mathcal{C}$, with unit $\eta$, and counit $\epsilon$. This is a *monoidal adjunction* if and only if $F$ and $U$ are monoidal functors (with respective mediating pairs $(\theta^F, \iota^F)$ and $(\theta^U, \iota^U)$), and the unit $\eta$ and the counit $\epsilon$ are *monoidal natural transformations*, by which we mean that the following diagrams commute:

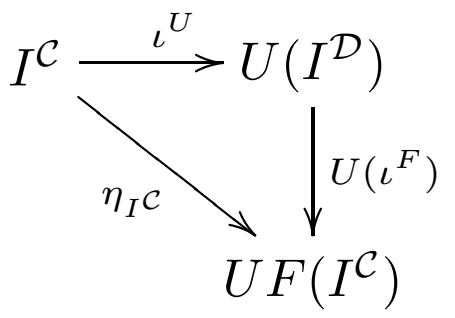

(49)

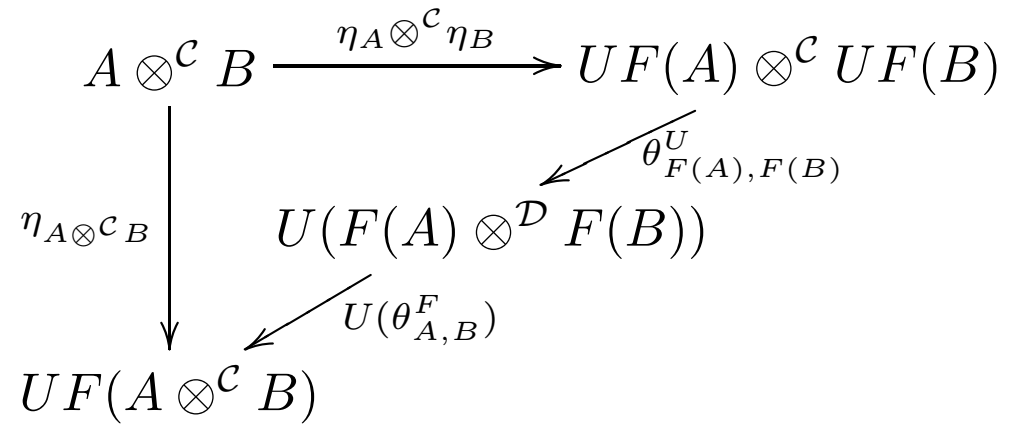

(50)

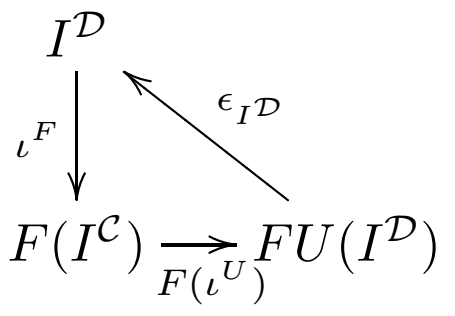

(51)

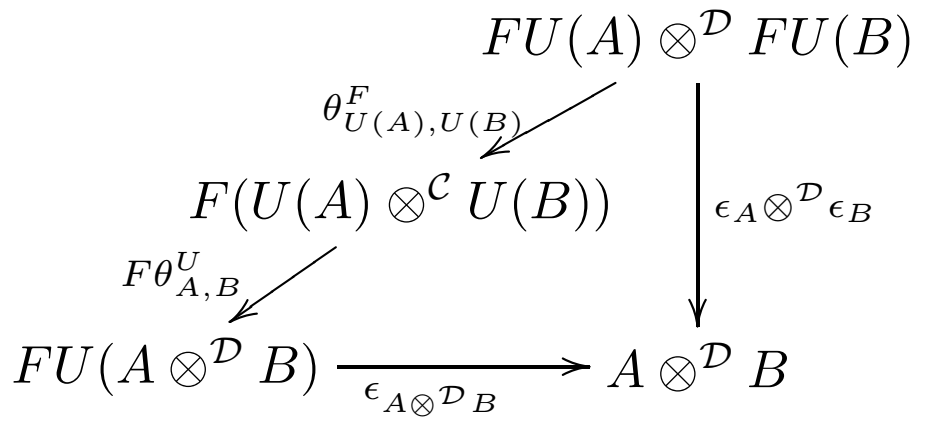

(52)

The value of monoidal adjunctions is their relation with monoidal monads (Section 7.2). Recall that a monoidal monad on $\mathcal{C}$ is a tuple $(T, \eta, \mu, d)$, where $(T, \eta, \mu)$ is a monad on $\mathcal{C}$ and $d_{A,B}$ is a natural transformation from $TA \otimes^{\mathcal{C}} TB$ to $T(A \otimes^{\mathcal{C}} B)$ such that the following diagrams commute:

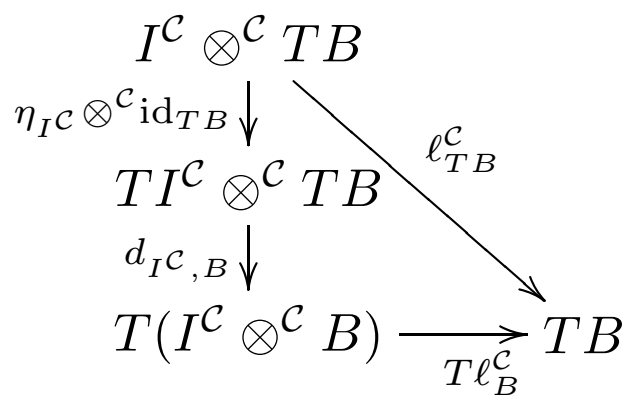

(38)

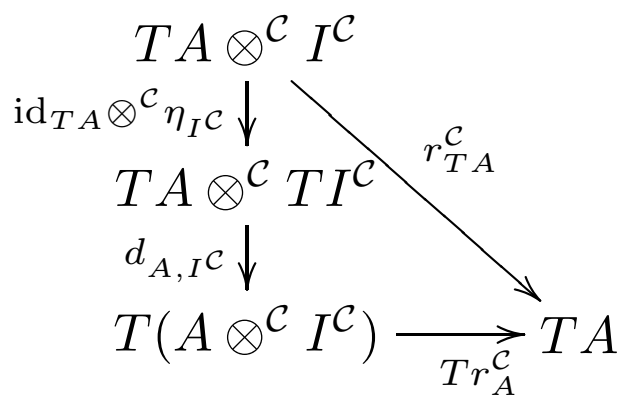

(39)

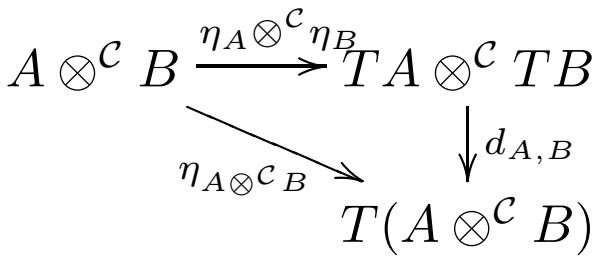

(40)

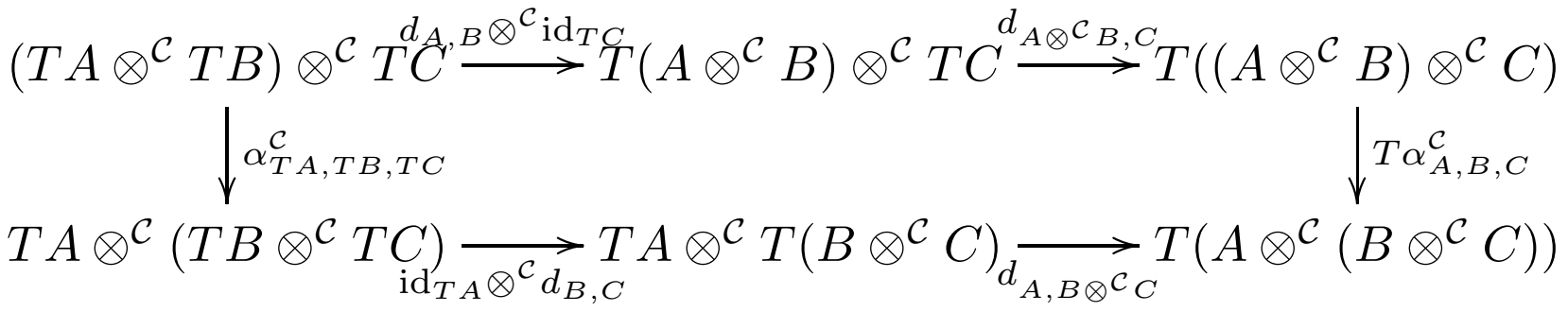

(41)

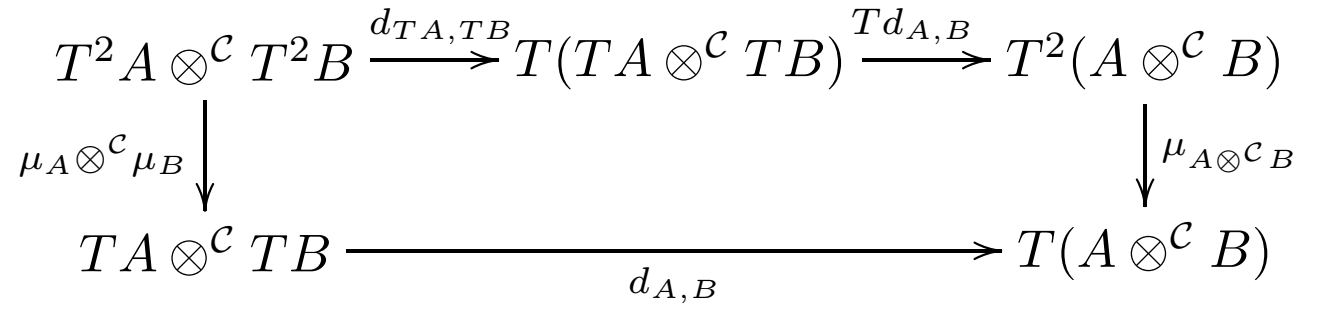

(42)

The following lemmas show respectively that every monoidal adjunction gives rise to a monoidal monad, that every monoidal monad yields a monoidal adjunction between the base category and the Kleisli category of the monad, and that monoidal adjunctions compose to yield monoidal adjunctions. Except for the first, the arguments are tedious computations, and therefore relegated to appendices[†].

**Lemma 8.2.** Let $F \dashv U$ be a monoidal adjunction, with unit $\eta$ and counit $\epsilon$, where $F : \mathcal{C} \to \mathcal{D}$, $U : \mathcal{D} \to \mathcal{C}$. Let $T$ be $UF$, $\mu_A$ be $U\epsilon_{F(A)}$, and $d_{A,B}$ be $U\theta^F_{A,B} \circ \theta^U_{F(A),F(B)}$. Then $(T, \eta, \mu, d)$ is a monoidal monad on $\mathcal{C}$.

*Proof.* $(T, \eta, \mu)$ is a monad by Proposition 8.1. We check the mediator laws (38), (39), (40), (41), (42) for $d$.

Diagram (38) is obtained by considering the following diagram. The top left triangle is a copy of (49), tensored by $UF(B)$ on the right. The square next to it on its right is a naturality square for $\theta^U$. The next trapezoid on the right (the top right trapezoid) is $U$ applied to a coherence square (17) for $\theta^F$, $\ell^{\mathcal{C}}$ and $\ell^{\mathcal{D}}$. The bottom face, atop the curved arrow $\ell^{\mathcal{C}}_{UF(B)}$, is another instance of a coherence square (17) for $\theta^F$, $\ell^{\mathcal{C}}$ and $\ell^{\mathcal{D}}$.

[†] These results are folklore, and were confirmed in discussions with Paul-André Melliès, François Lamarche, and Albert Burroni notably. However, we have been unable to find references on this. We have often been directed to the pioneering paper (Street, 1972), but could not find the expected results therein.

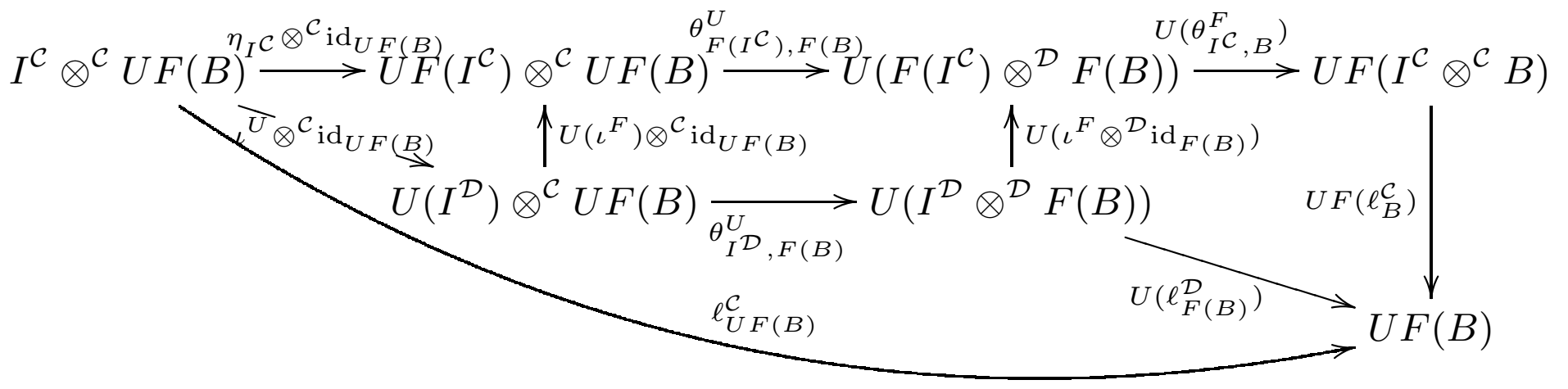

Now the topmost composition of arrows is $t_{I^{\mathcal{C}},B}$, the bottommost arrow from $I^{\mathcal{C}} \otimes^{\mathcal{C}} UF(B)$ to $UF(B)$ is $\ell^{\mathcal{C}}_{TB}$, and the rightmost vertical arrow is $T\ell^{\mathcal{C}}_B$.

Diagram (39). This is checked by similar arguments, replacing the coherence square (17) by (18).

Diagram (40). This is the diagram on the right, an instance of Diagram (50), stating that $\eta$ is a monoidal natural transformation. We recognize $d_{A,B}$ as the rightmost composition of vertical arrows, hence the desired Diagram (33).

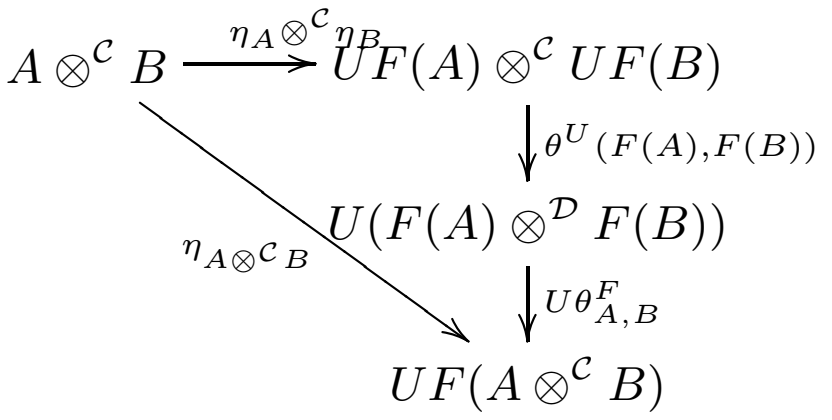

Diagram (41). For space reasons, we flip the diagram so that arrows involving strengths are vertical, and arrows involving associativities are horizontal. Also, we drop most subscripts, which are inferrable from context.

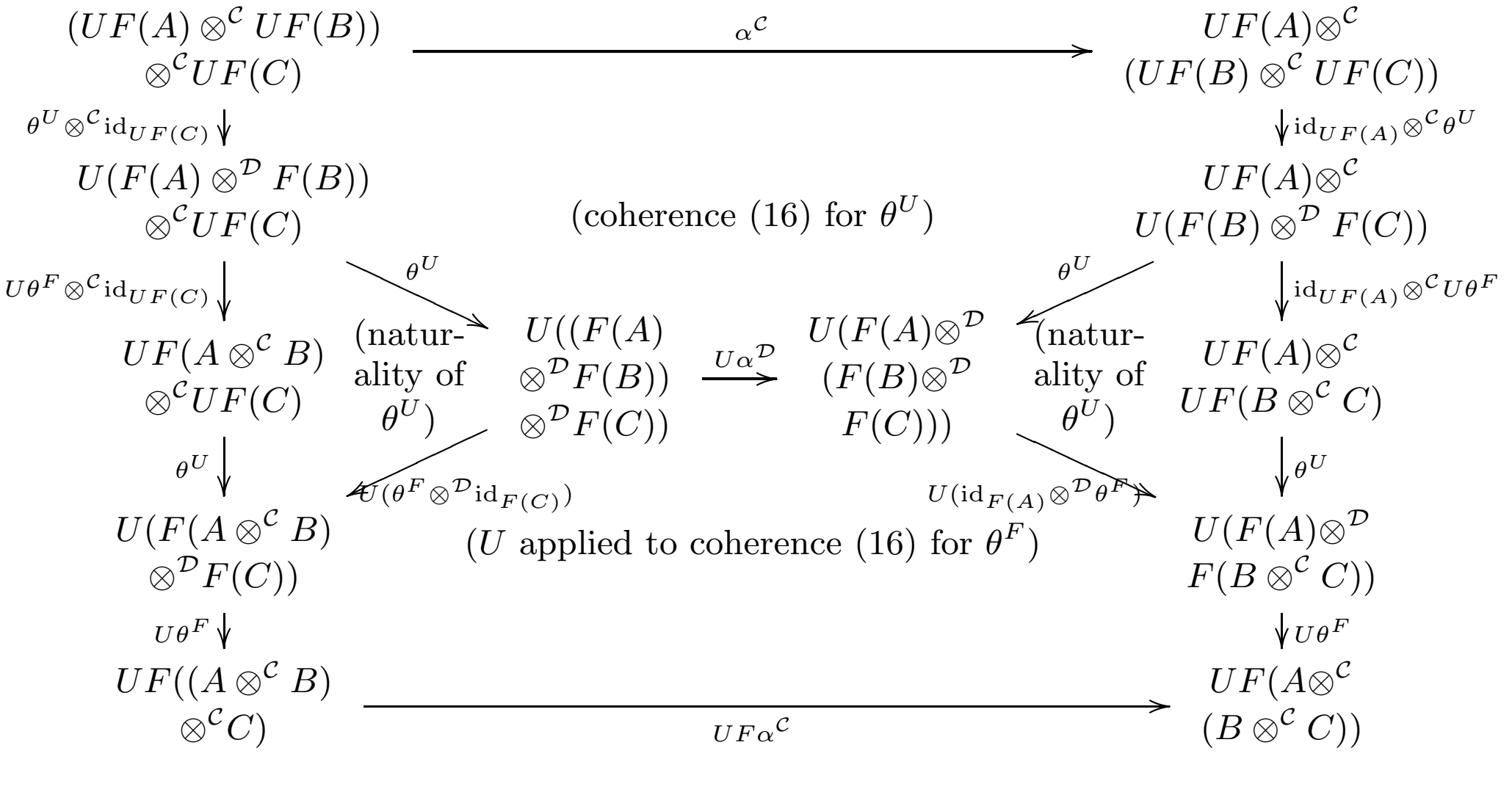

The vertical arrows on the left compose to form $d_{A \otimes^{\mathcal{C}} B,C} \circ (d_{A,B} \otimes^{\mathcal{C}} \mathrm{id}_{UF(C)})$, while the vertical arrows on the right compose to form $d_{A \otimes^{\mathcal{C}} B,C} \circ (\mathrm{id}_{UF(A)} \otimes^{\mathcal{C}} d_{B,C})$, whence the result.

Diagram (42). Similarly, we flip the diagram so that vertical arrows become horizontal and conversely:

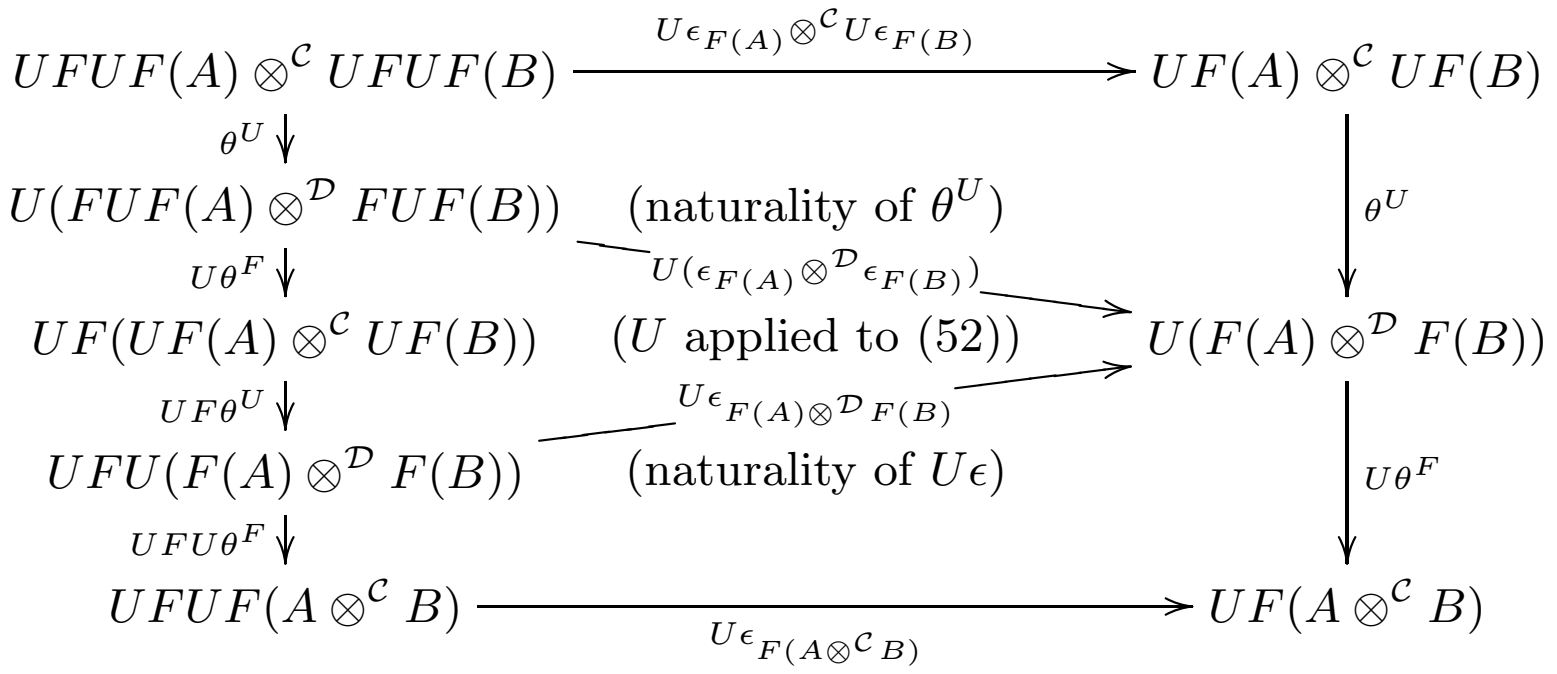

We recognize $Td_{A,B} \circ d_{TA,TB}$ as the leftmost composition of vertical arrows, and the rightmost vertical composition is $d_{A,B}$. Also, the top horizontal arrow is $\mu_A \otimes^{\mathcal{C}} \mu_B$, while the bottom arrow is $\mu_{A \otimes^{\mathcal{C}} B}$. ☐

**Lemma 8.3.** Let $(\mathcal{C}, I^{\mathcal{C}}, \otimes^{\mathcal{C}}, \alpha^{\mathcal{C}}, \ell^{\mathcal{C}}, r^{\mathcal{C}})$ be a monoidal category, and let $(T, \eta, \mu, d)$ be a monoidal monad on $\mathcal{C}$.

Let $\mathcal{D}$ be the Kleisli category of $T$, $I^{\mathcal{D}} = I^{\mathcal{C}}$, $\otimes^{\mathcal{D}}$ be defined on objects by $A \otimes^{\mathcal{D}} B = A \otimes^{\mathcal{C}} B$ and on morphisms by letting $f \otimes^{\mathcal{D}} g$ (in $\mathcal{D}$) be the morphism $d \circ (f \otimes^{\mathcal{C}} g)$ in $\mathcal{C}$; let $\alpha^{\mathcal{D}} = \eta \circ \alpha^{\mathcal{C}}$, $\ell^{\mathcal{D}} = \eta \circ \ell^{\mathcal{C}}$, $r^{\mathcal{D}} = \eta \circ r^{\mathcal{C}}$. Then $(\mathcal{D}, I^{\mathcal{D}}, \otimes^{\mathcal{D}}, \alpha^{\mathcal{D}}, \ell^{\mathcal{D}}, r^{\mathcal{D}})$, is a monoidal category.

Moreover, $F_T \dashv U_T$ is a monoidal adjunction. The mediating pairs of $F_T$ and $U_T$ are $(\theta^{F_T}, \iota^{F_T})$ and $(\theta^{U_T}, \iota^{U_T})$ respectively, where $\theta^{F_T}_{A,B} : F_T(A) \otimes^{\mathcal{D}} F_T(B) \to F_T(A \otimes^{\mathcal{C}} B)$ (in **Kleisli**$(T)$) is the morphism $\eta_{A \otimes^{\mathcal{C}} B}$ in $\mathcal{C}$, $\iota^{F_T} : I^{\mathcal{D}} \to F_T(I^{\mathcal{C}})$ (in **Kleisli**$(T)$) is $\eta_{I^{\mathcal{C}}}$, $\theta^{U_T}_{A,B} : U_T(A) \otimes^{\mathcal{C}} U_T(B) \to U_T(A \otimes^{\mathcal{D}} B)$ (in $\mathcal{C}$) is $d_{A,B}$, and $\iota^{U_T} : I^{\mathcal{C}} \to U_T(I^{\mathcal{D}})$ (in $\mathcal{C}$) is $\eta_{I^{\mathcal{C}}}$.

Finally, $F_T \dashv U_T$ generates the monoidal monad, in the sense that $U_T F_T = T$, $\eta$ is the unit of the adjunction and of $T$, $\mu_A = U_T \epsilon_{F(A)}$ where $\epsilon$ is the counit of the adjunction, and $d_{A,B} = U_T \theta^{F_T}_{A,B} \circ \theta^{U_T}_{F_T(A), F_T(B)}$.

*Proof.* Tedious. See Appendix B. ☐

**Lemma 8.4.** Let $\mathbb{C} \underset{|_|}{\overset{\boldsymbol{D}}{\rightleftarrows}} \boldsymbol{C} \underset{U}{\overset{F}{\rightleftarrows}} \mathcal{D}$ be a diagram of functors. Assume that these functors are monoidal; let $(\mathbb{\theta}, \mathbb{i})$ be the mediating pair of $|_|$, $(\boldsymbol{\theta}, \boldsymbol{i})$ that of $\boldsymbol{D}$, $(\theta^U, \iota^U)$ that of $U$, $(\theta^F, \iota^F)$ that of $F$.

Then $F\boldsymbol{D}$ and $|U|$ are monoidal functors, with respective mediating pairs $(F\boldsymbol{\theta} \circ \theta^F, F\boldsymbol{i} \circ \iota^F)$ and $(|\theta^U| \circ \mathbb{\theta}, |\iota^U| \circ \mathbb{i})$.

Furthermore, if $\boldsymbol{D} \dashv |_|$ and $F \dashv U$ are monoidal adjunctions, then $F\boldsymbol{D} \dashv |U|$ is a monoidal adjunction, too.

*Proof.* Straightforward. See Appendix C. ☐

The following proposition is then both similar and proved similarly to Proposition 8.1.

**Proposition 8.5.** Let $(\boldsymbol{C}, \otimes, \boldsymbol{I}, \boldsymbol{\alpha}, \boldsymbol{\ell}, \boldsymbol{r})$ and $(\mathbb{C}, \otimes, \mathbb{I}, \mathbb{a}, \mathbb{l}, \mathbb{r})$ be monoidal categories.

Let $(\boldsymbol{T}, \boldsymbol{\eta}, \boldsymbol{\mu}, \boldsymbol{d})$ be a monoidal monad on $\boldsymbol{C}$, $|_| : \boldsymbol{C} \to \mathbb{C}$ and $\boldsymbol{D} : \mathbb{C} \to \boldsymbol{C}$ be monoidal functors, yielding a monoidal adjunction $\boldsymbol{D} \dashv |_|$. Let $\dot{\epsilon}_A : \boldsymbol{D}|A| \to A$ be the counit of the adjunction, $\dot{\eta}_E : E \to |\boldsymbol{D}(E)|$ be the unit of the adjunction. Let $(\mathbb{\theta}, \mathbb{i})$ be the mediating pair of $|_|$, $(\boldsymbol{\theta}, \boldsymbol{i})$ be the mediating pair of $\boldsymbol{D}$.

Define $\mathbb{T} = |_| \circ \boldsymbol{T} \circ \boldsymbol{D} = |\boldsymbol{TD}|$, $\mathbb{\eta}_E = \left|\boldsymbol{\eta}_{\boldsymbol{D}(E)})\right| \circ \dot{\eta}_E$, $\mathbb{\mu}_E = \left|\boldsymbol{\mu}_{\boldsymbol{D}(E)} \circ \boldsymbol{T}\dot{\epsilon}_{\boldsymbol{TD}(E)}\right|$, $\mathbb{d}_{E,F} = |\boldsymbol{T\theta}_{E,F} \circ \boldsymbol{d}_{\boldsymbol{D}E,\boldsymbol{D}F}| \circ \mathbb{\theta}_{\boldsymbol{TD}E,\boldsymbol{TD}F}$. Finally, let $\sigma_A = |\boldsymbol{T}\dot{\epsilon}_A| : \mathbb{T}|A| \to |\boldsymbol{T}A|$. Then $(\mathbb{T}, \mathbb{\eta}, \mathbb{\mu}, \mathbb{d})$ is a monoidal monad on $\mathbb{C}$ and $(|_|, \sigma)$ is a monoidal monad morphism from $\boldsymbol{T}$ to $\mathbb{T}$.

*Proof.* Let $F = F_{\boldsymbol{T}}$, $U = U_{\boldsymbol{T}}$. By Lemma 8.3, $F \dashv U$ is a monoidal adjunction which generates the monoidal monad $(\boldsymbol{T}, \boldsymbol{\eta}, \boldsymbol{\mu}, \boldsymbol{d})$. Compose the monoidal adjunction $\boldsymbol{D} \dashv |_|$ with the monoidal adjunction $F \dashv U$, yielding the adjunction $F\boldsymbol{D} \dashv |U|$. This is also a monoidal adjunction by Lemma 8.4.

By Lemma 8.3, this monoidal adjunction generates a monoidal monad, and this is $(\mathbb{T}, \mathbb{\eta}, \mathbb{\mu}, \mathbb{d})$ as stated in the Proposition. Indeed, all cases except the mediator have been dealt with in Proposition 8.1, and the mediator is by definition $\left|U(F\boldsymbol{\theta} \circ \theta^F)\right| \circ \left|\theta^U\right| \circ \mathbb{\theta} = |UF\boldsymbol{\theta}| \circ \left|U\theta^F \circ \theta^U\right| \circ \mathbb{\theta} = |\boldsymbol{T\theta}| \circ |\boldsymbol{d}| \circ \mathbb{\theta}$.

It remains to check the monoidal monad morphism Diagram (45). This is given by the following diagram:

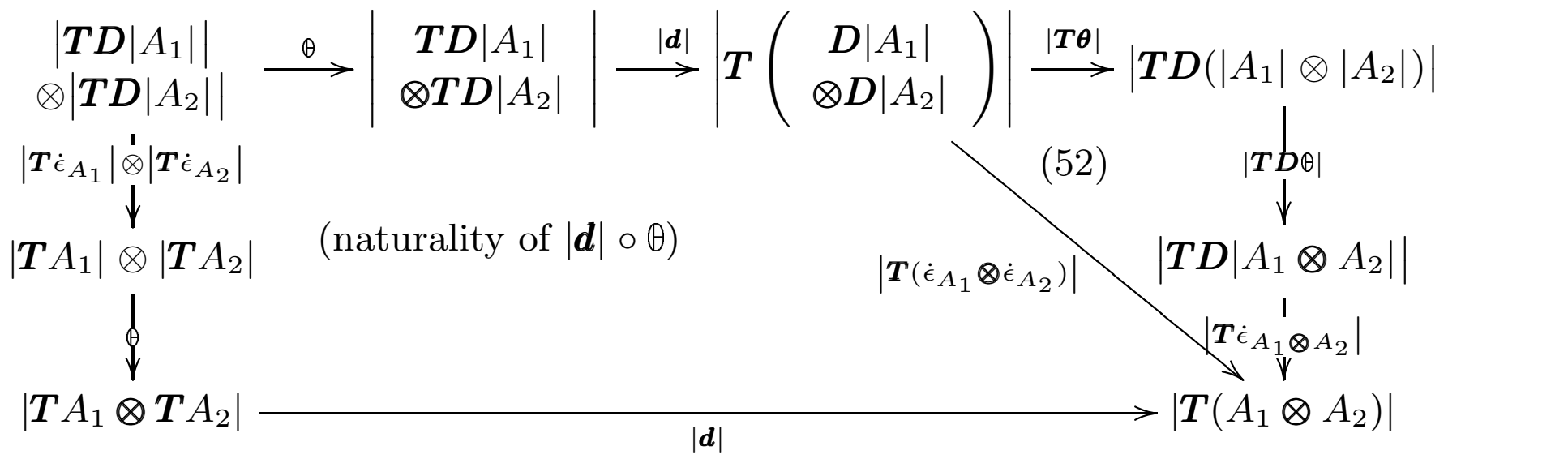

□

We can in fact prove something similar with just strong monads. Unfortunately, it seems that we cannot use the nice trick of going through some adjunction generating the strong monad. The proof therefore goes through extremely tedious diagram checking.

**Proposition 8.6.** Let $(\boldsymbol{C}, \otimes, \boldsymbol{I}, \boldsymbol{\alpha}, \boldsymbol{\ell}, \boldsymbol{r})$ and $(\mathbb{C}, \otimes, \mathbb{I}, \mathbb{a}, \mathbb{l}, \mathbb{r})$ be monoidal categories.

Let $|_|$ be a monoidal functor from $\boldsymbol{C}$ to $\mathbb{C}$, with mediating pair $(\mathbb{\theta}, \mathbb{i})$, $\boldsymbol{D}$ be a monoidal functor from $\mathbb{C}$ to $\boldsymbol{C}$, with mediating pair $(\boldsymbol{\theta}, \boldsymbol{i})$, and assume that $\boldsymbol{D} \dashv |_|$ is a monoidal adjunction.

Define $\mathbb{T} = |_| \circ \boldsymbol{T} \circ \boldsymbol{D} = |\boldsymbol{TD}|$, $\mathbb{\eta}_E = \left|\boldsymbol{\eta}_{\boldsymbol{D}(E)})\right| \circ \dot{\eta}_E$, $\mathbb{\mu}_E = \left|\boldsymbol{\mu}_{\boldsymbol{D}(E)} \circ \boldsymbol{T}\dot{\epsilon}_{\boldsymbol{TD}(E)}\right|$, $\sigma_A = |\boldsymbol{T}\dot{\epsilon}_A| : \mathbb{T}|A| \to |\boldsymbol{T}A|$.

Define also $\mathbb{t}_{E,F}$ as the composite $|\boldsymbol{TD\theta}_{E,F} \circ \boldsymbol{t}_{\boldsymbol{D}E,\boldsymbol{D}F}| \circ \mathbb{\theta}_{\boldsymbol{D}E,\boldsymbol{TD}F} \circ (\dot{\eta}_E \otimes \mathrm{id}_{\mathbb{T}F})$.

Then $(\mathbb{T}, \mathbb{\eta}, \mathbb{\mu}, \mathbb{t})$ is a strong monad on $\mathbb{C}$ and $(|_|, \sigma)$ is a strong monad morphism from $\boldsymbol{T}$ to $\mathbb{T}$.

*Proof.* Because of Proposition 8.1, we only have to check the strength laws (32), (33), (34), (35) for $\mathbb{t}$, and the strong monad morphism law (36). As we said, this is tedious, hence relegated to Appendix D. □

## 9. Lifting Closed Structures to the Subscone

If we are to lift the whole structure of a cartesian-closed category together with a strong monad on it, to the subscone, the only thing that remains is to lift exponential objects. As this is essentially the subject of (Mitchell and Scedrov, 1993) (together with the fact that subscones generalize logical relations), it would be legitimate to skip over this construction, knowing that it has been dealt with elsewhere.

However, we notice that the standard lifting construction of exponentials to the subscone requires $|_|$ to preserve products, at least up to natural isos. That is, it requires $|\mathbf{1}| \cong \mathbb{1}$, $|A \times B| \cong |A| \times |B|$. This is certainly the case for the functor $|_| = \boldsymbol{C}(\mathbf{1}, _)$, which is the standard choice in sconing constructions (Mitchell and Scedrov, 1993).

Until now, we have only assumed that $|_|$ was a monoidal (Section 6.1), resp. a symmetric monoidal (Section 6.7), resp. a cartesian monoidal (Section 6.9) functor. It would therefore be nice if we could dispense with the stringent requirement that $|_|$ preserved monoidal or cartesian structure exactly. This would also afford us some added generality.

It turns out that having a monoidal mono factorization system is all we need: exponentials lift to the subscone without any additional requirements compared to Section 6. This only requires a slight adjustment of the standard exponential lifting diagrams.

Recall that an *exponential*, or *internal hom object* (on the right), in a monoidal category $(\mathbb{C}, \otimes, \mathbb{I}, \mathbb{a}, \mathbb{l}, \mathbb{r})$, is an object $B^A$ together with a morphism $\mathsf{App}$ from $B^A \otimes A$ to $B$ and, for every morphism $u$ from $C \otimes A$ to $B$, a morphism $\Lambda(u)$ from $C$ to $B^A$ satisfying the two equations

$$\begin{array}{ccc} C \otimes A & \xrightarrow{\Lambda(u)\otimes \mathrm{id}_A} & B^A \otimes A \\ & \underset{u}{\searrow} & \downarrow \mathsf{App} \\ & & B \end{array} \tag{53}$$

for every morphism $u$ from $C \otimes A$ to $B$ ($\beta$*-equivalence*), and

$$\Lambda\left( C \otimes A \xrightarrow{v \otimes \mathrm{id}_A} B^A \otimes A \xrightarrow{\mathsf{App}} B \right) = v \tag{54}$$

for every morphism $v$ from $C$ to $B^A$ ($\eta$*-equivalence*). $\Lambda(u)$ is called the *currification* or the *abstraction* of $u$.

A more traditional definition is to require the existence of a *unique* morphism $\Lambda(u)$ as in Diagram (53). Uniqueness is indeed implied by (54): if there were two morphisms $v$ and $v'$ such that $u = \mathsf{App} \circ (v \otimes \mathrm{id}_A) = \mathsf{App} \circ (v' \otimes \mathrm{id}_A)$, then $v = \Lambda(\mathsf{App} \circ (v \otimes \mathrm{id}_A)) = \Lambda(u) = \Lambda(\mathsf{App} \circ (v' \otimes \mathrm{id}_A)) = v'$. Conversely, uniqueness of $\Lambda(u)$ implies Diagram (54): take $u = \mathsf{App} \circ (v \otimes \mathrm{id}_A)$ in Diagram (53).

Exponentials on the right are unique up to iso when they exist. A monoidal category is said to be *monoidal closed* (on the right) if and only if the exponential $B^A$ exists for all objects $A$ and $B$. Similarly, we call exponential on the left any object ${}^A B$ with a morphism $\mathsf{qqA}$ from $A \otimes {}^A B$ to $B$ such that, for every morphism $u$ from $A \otimes C$ to $B$, there is a unique morphism $(u)\Lambda$ from $C$ to ${}^A B$ such that $u = \mathsf{qqA} \circ (\mathrm{id}_A \otimes (u)\Lambda)$. In a *symmetric* monoidal category, it is equivalent to require the existence of exponentials on the right or on the left, and they coincide up to iso. A category with finite products that is also monoidal closed (for the monoidal structure induced by the product) is called *cartesian closed*.

Note that, in a monoidal closed category, there is a functor $_^A$ for each object $A$, which maps every object $B$ to $B^A$, and every morphism $B \xrightarrow{f} B'$ to $f^A = \Lambda(f \circ \mathsf{App})$, from $B^A$ to $B'^A$. In fact, there is a bifunctor from $\mathbb{C}^{op} \times \mathbb{C}$ to $\mathbb{C}$ mapping $A, B$ to $B^A$.

Moreover, the functor $_^A$ preserves monos: if $m$ is mono, then so is $m^A = \Lambda(m \circ \mathsf{App})$. It suffices to show that there is at most one morphism $f$ such that $\Lambda(m \circ \mathsf{App}) \circ f = h$ where $h$ is given. Indeed, $\mathsf{App} \circ (h \otimes \mathrm{id}) = \mathsf{App} \circ (\Lambda(m \circ \mathsf{App}) \otimes \mathrm{id}) \circ (f \otimes \mathrm{id}) = m \circ \mathsf{App} \circ (f \otimes \mathrm{id})$ by (53); if there were two such morphisms $f$ and $f'$, then $m \circ \mathsf{App} \circ (f \otimes \mathrm{id}) = m \circ \mathsf{App} \circ (f' \otimes \mathrm{id})$, so $\mathsf{App} \circ (f \otimes \mathrm{id}) = \mathsf{App} \circ (f' \otimes \mathrm{id})$ since $m$ is mono. Applying $\Lambda$ on both sides implies $f = f'$ by (54).

Once this is known, the standard way of lifting exponentials to the subscone is to require that $\boldsymbol{C}$ and $\mathbb{C}$ are cartesian closed, $\mathbb{C}$ has pullbacks, and that $|_|$ preserves finite products (exactly, or up to natural iso). This standard construction actually does not require cartesian closedness, and works equally well with monoidal closed categories, assuming $\mathbb{C}$ has pullbacks and $|_|$ preserves unit and tensor.

We recall this construction now. As $|_|$ preserves unit, the unit (terminal object in the cartesian closed case) $\widetilde{I}$ is $\langle \mathbb{I}, \mathrm{id}_{\mathbb{I}}, \boldsymbol{I} \rangle$ witness by the arrow $\mathbb{I} \xrightarrow{\mathrm{id}_{\mathbb{I}}} |\boldsymbol{I}| = \mathbb{I}$, tensor product (binary product in the cartesian closed case) is given by $\langle S_1, m_1, A_1 \rangle \widetilde{\otimes} \langle S_2, m_2, A_2 \rangle = \langle S_1 \otimes S_2, m_1 \otimes m_2, A_1 \boldsymbol{\otimes} A_2 \rangle$, and the exponential $\langle \widetilde{S}_2^1, \widetilde{m}_2^1, {A_2}^{A_1} \rangle = \langle S_2, m_2, A_2 \rangle^{\langle S_1, m_1, A_1 \rangle}$ is given by the square on the left below.

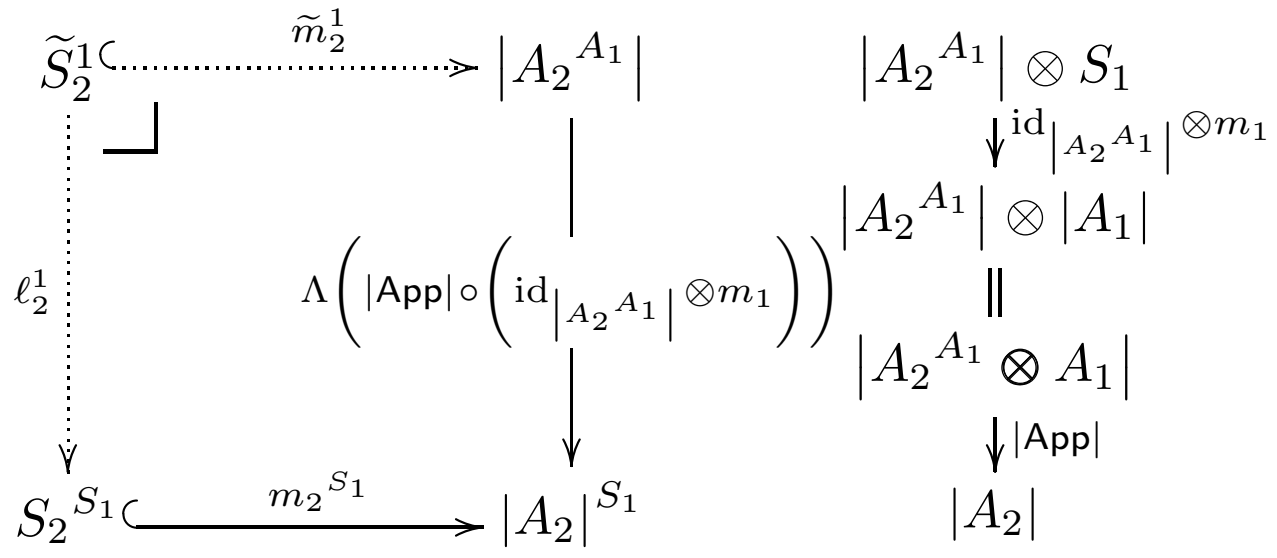

where the vertical morphism $\Lambda\left(|\mathsf{App}| \circ \left(\mathrm{id}_{|{A_2}^{A_1}|} \otimes m_1\right)\right)$ is $\Lambda$ applied to the composition of vertical morphisms on the right, the morphism ${m_2}^{S_1}$ is mono because $_^{S_1}$ preserves monos, and $\widetilde{m}_2^1$ is mono because pullbacks preserve monos. (We temporarily revert to the notation $\hookrightarrow$ to denote all monos.)

Application from $\langle \widetilde{S}_2^1, \widetilde{m}_2^1, {A_2}^{A_1}\rangle \widetilde{\otimes} \langle S_1, m_1, A_1\rangle$ in the subscone is given by

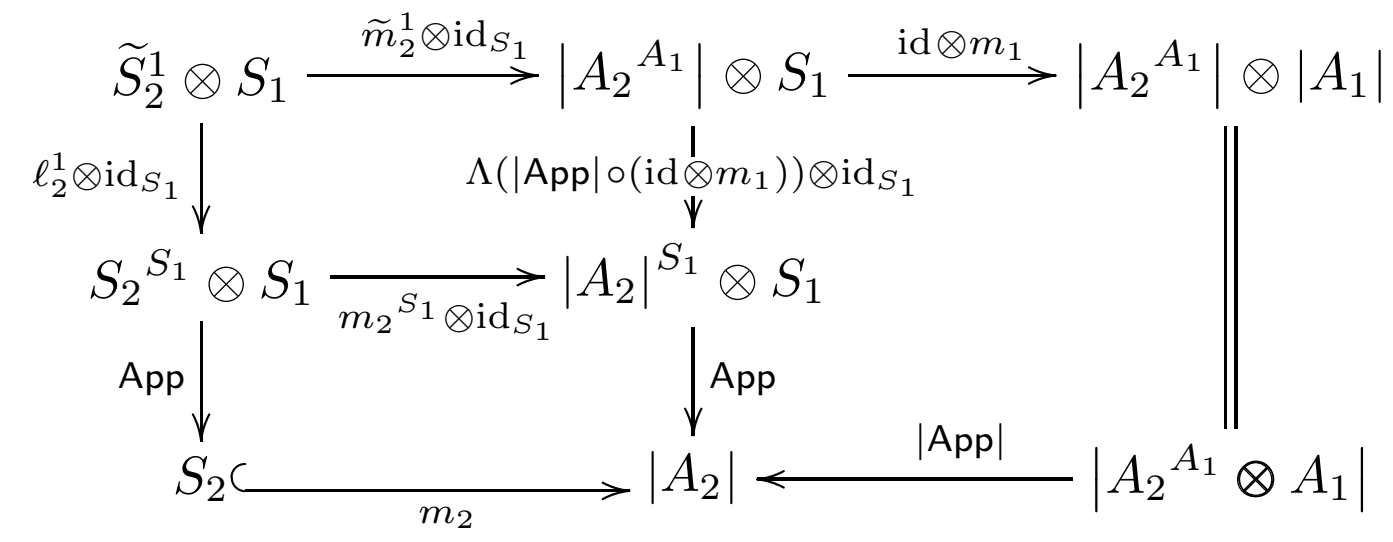

where the top left square is the definition of the exponential $\langle \widetilde{S}_2^1, \widetilde{m}_2^1, {A_2}^{A_1}\rangle$ tensor $S_1$ on the right. The bottom left square and the right square commute by (53). Then, application in the subscone is $\langle \mathsf{App} \circ (\ell_2^1 \otimes \mathrm{id}), |\mathsf{App}|\rangle$.

Abstraction of the morphism $\langle u, v\rangle$ in the subscone from $\langle S, m, A\rangle \widetilde{\otimes} \langle S_1, m_1, A_1\rangle$ to $\langle S_2, m_2, A_2\rangle$ is then the pair $\langle \tilde{u}, \Lambda(v)\rangle$ given by the diagram

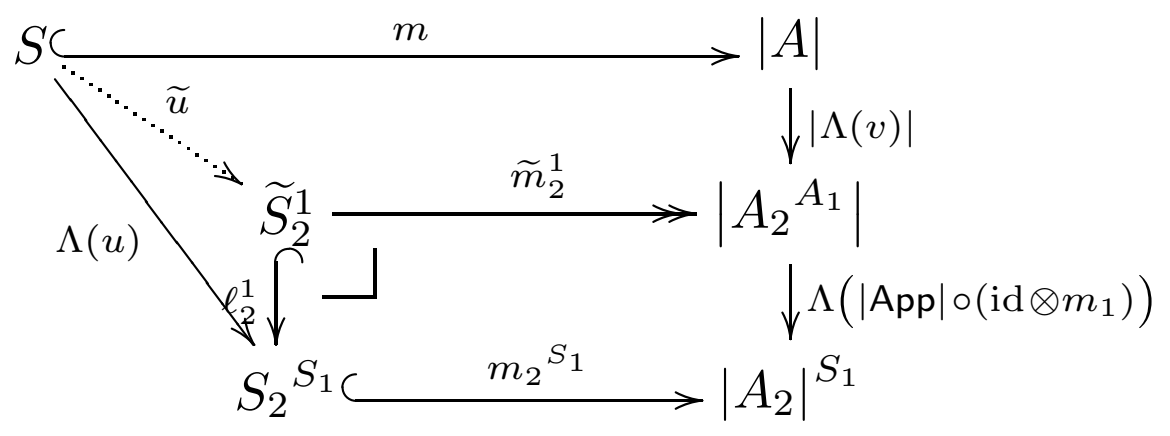

where $\widetilde{u}$ is given by the universal property of pullbacks. To this end, we must first check that the outer contour of the diagram commutes. We leave this to the reader.

The equations (53) and (54) then hold in the subscone, because they hold in $\boldsymbol{C}$, and using Fact 4.3.

In the case we are interested in here, $|_|$ does not preserve unit and tensor. Rather, we have required $|_|$ to be a monoidal functor, a strictly weaker notion. We have already seen in Section 6 that this was enough to lift monoidal structure to the subscone, using a monoidal mono factorization. We now realize that this is also enough to lift monoidal *closed* structure to the subscone. This requires only minor adjustments to the constructions above.

First, we require that the functor $_^A$ preserves *relevant* monos, for all $A$. While it always preserves monos, it is not clear that it should preserve relevant monos, hence the added assumption. (We return to our convention that $\hookrightarrow$ denotes relevant monos only.) We might also require that pullbacks preserve relevant monos, but this is not necessary, as we can use the mono factorization instead. Summing up, the exponential $\langle \widetilde{S}_2^1, \widetilde{m}_2^1, {A_2}^{A_1}\rangle = \langle S_2, m_2, A_2\rangle^{\langle S_1, m_1, A_1\rangle}$ is given by the diagram on the left below.

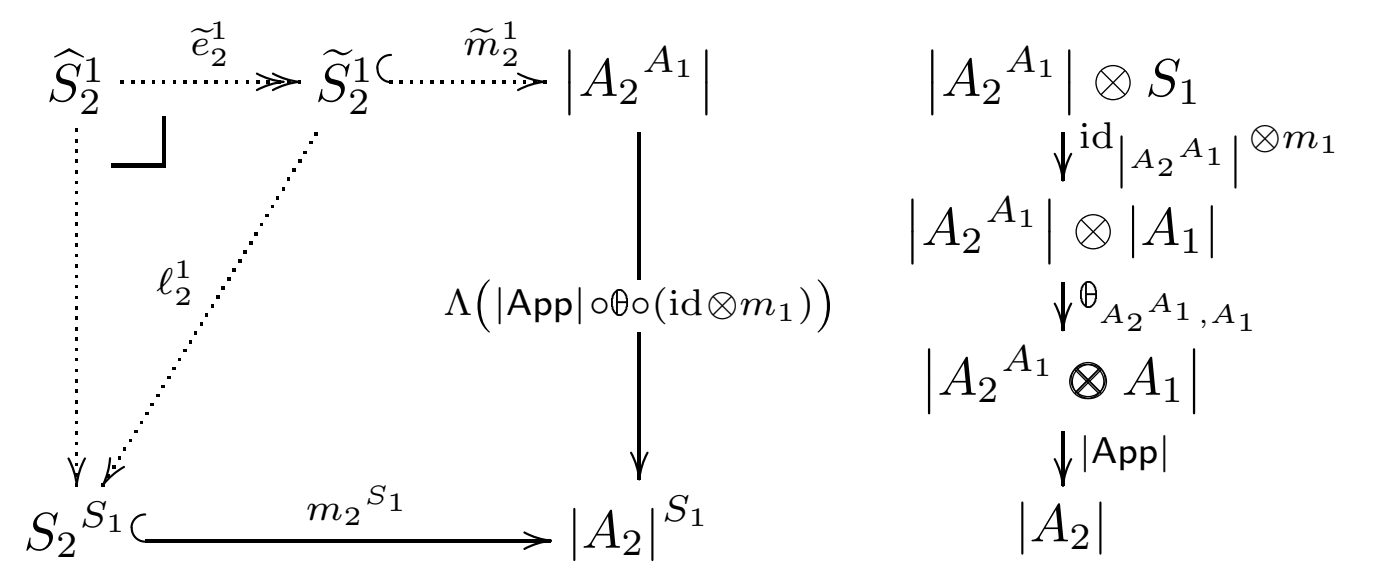

(55)

where the vertical morphism $\Lambda\left(|\mathsf{App}| \circ \theta \circ (\mathrm{id} \otimes m_1)\right)$ is $\Lambda$ applied to the composition of vertical morphisms on the right, the morphism ${m_2}^{S_1}$ is a relevant mono by our assumption that $_^{S_1}$ preserves relevant monos, the topmost horizontal composition of arrows (from $\widehat{S}_2^1$ to $|{A_2}^{A_1}|$) is given by pullback, and is factored as $\widetilde{e}_2^1$ followed by $\widetilde{m}_2^1$; and finally the morphism $\ell_2^1$ is given by a diagonal fill-in, from two paths from $\widehat{S}_2^1$ to $|A_2|^{S_1}$.

Application from $\langle \widetilde{S}_2^1, \widetilde{m}_2^1, {A_2}^{A_1}\rangle \widetilde{\otimes} \langle S_1, m_1, A_1\rangle$ is then given by $\langle \widehat{\mathsf{App}}, |\mathsf{App}|\rangle$ as given in the diagram below.

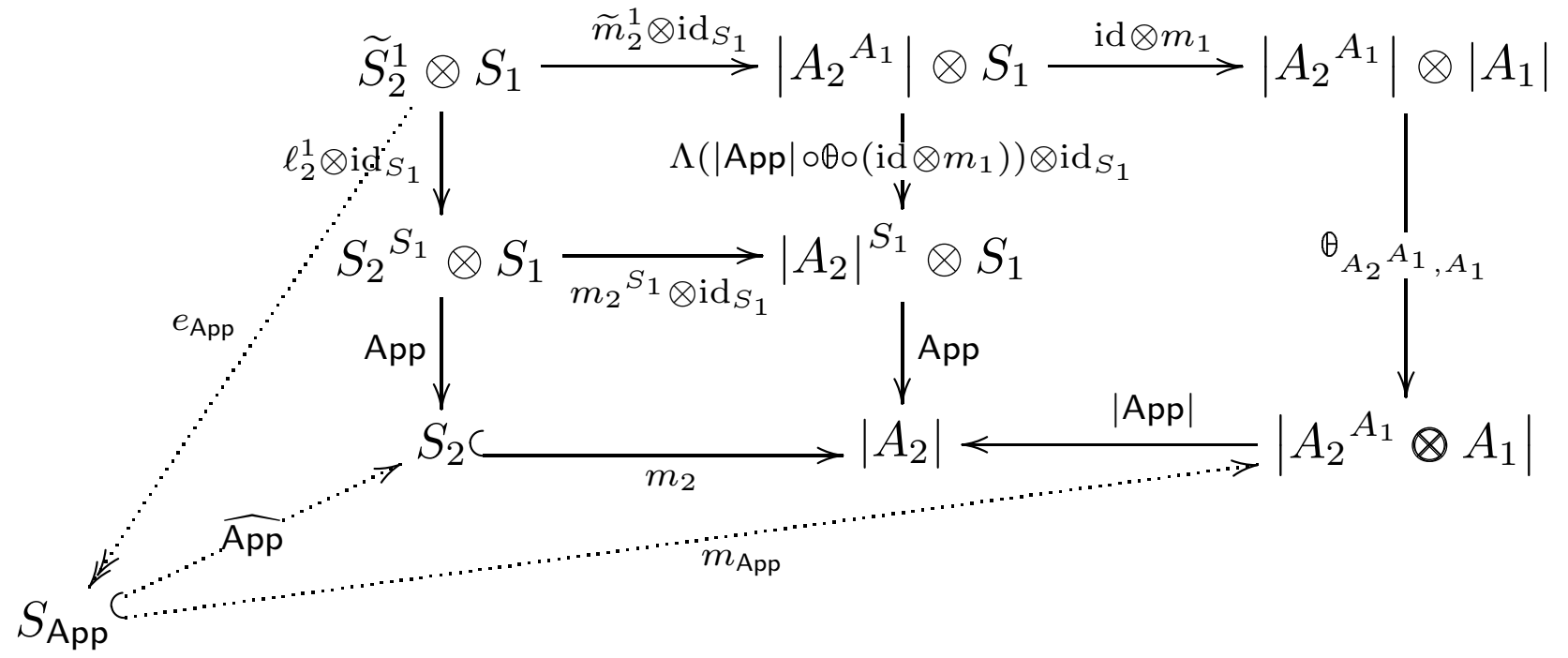

(56)

The top left square comes from the definition of the exponential $\langle \widetilde{S}_2^1, \widetilde{m}_2^1, {A_2}^{A_1}\rangle$, tensored by $S_1$ on the right. The bottom left square commutes because $\mathsf{App} \circ ({m_2}^{S_1} \otimes \mathrm{id}_{S_1}) = \mathsf{App} \circ (\Lambda(m_2 \circ \mathsf{App}) \otimes \mathrm{id}_{S_1}) = m_2 \circ \mathsf{App}$by (53). The right square commutes by (53) again. The outer contour, defined by $\widetilde{m}_2^1 \otimes m_1$ on the top, $\theta_{{A_2}^{A_1}, A_1}$ on the right, $e_{\mathsf{App}}$ on the left, $m_{\mathsf{App}}$ at the bottom, is the definition of the tensor product $\langle \widetilde{S}_2^1, \widetilde{m}_2^1, {A_2}^{A_1}\rangle \widetilde{\otimes} \langle S_1, m_1, A_1\rangle$ in the subscone. $\widehat{\mathsf{App}}$ is given by a diagonal fill-in, considering the two paths $e_{\mathsf{App}}$ followed by $|\mathsf{App}| \circ m_{\mathsf{App}}$ and $\mathsf{App} \circ (\ell_2^1 \otimes \mathrm{id}_{S_1})$ followed by $m_2$ from $\widetilde{S}_2^1 \otimes S_1$ to $|A_2|$.

Given any morphism $\langle u, v\rangle$ in the subscone from $\langle S, m, A\rangle \widetilde{\otimes} \langle S_1, m_1, A_1\rangle$ to $\langle S_2, m_2, A_2\rangle$, by definition the following diagram commutes. The top square is the definition of $\langle S, m, A\rangle \widetilde{\otimes} \langle S_1, m_1, A_1\rangle$.

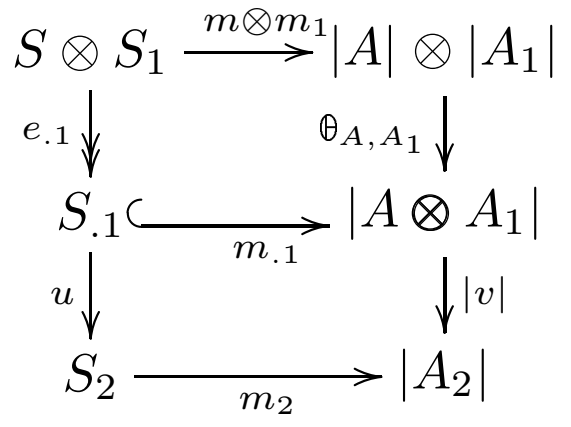

(57)

Abstraction of the morphism $\langle u, v\rangle$ in the subscone from $\langle S, m, A\rangle \widetilde{\otimes} \langle S_1, m_1, A_1\rangle$ to $\langle S_2, m_2, A_2\rangle$ is then the pair $\langle \widetilde{e}_2^1 \circ \tilde{u}, \Lambda(v)\rangle$ given by the diagram:

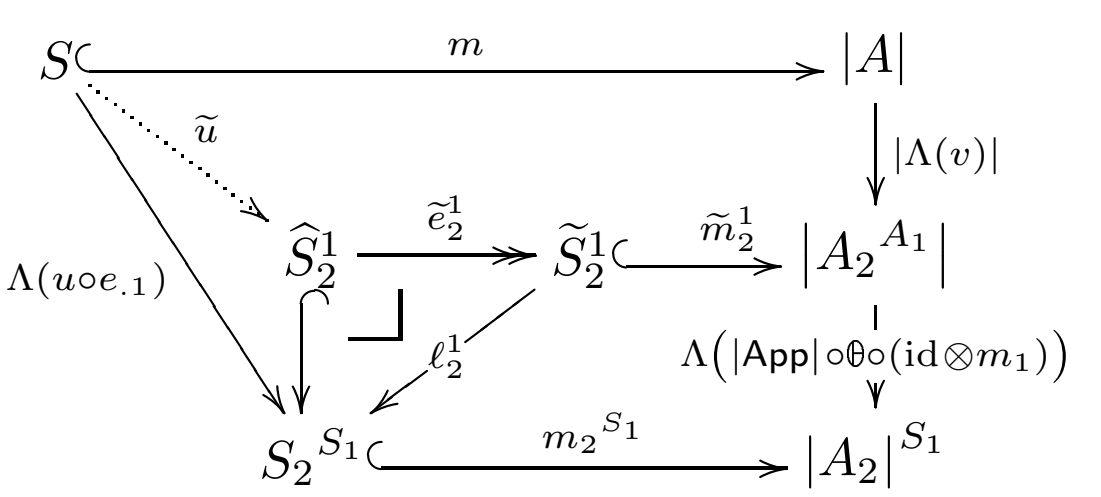

(58)

In this diagram, $\tilde{u}$ is given by the universal property of pullbacks, and this will be justified by the fact that the two outer paths from $S$ to $|A_2|^{S_1}$ are equal, which we have to check. First, we note the identity

$$\Lambda(s) \circ t = \Lambda(s \circ (t \otimes \mathrm{id}_C)) \tag{59}$$

whenever $t$ is a morphism from $A$ to $B$, and $s$ from $B \otimes C$ to $D$. Indeed, $\Lambda(s) \circ t = \Lambda(\mathsf{App} \circ ((\Lambda(s) \circ t) \otimes \mathrm{id}_C))$ (by (54)) $= \Lambda(\mathsf{App} \circ (\Lambda(s) \otimes \mathrm{id}_C) \circ (t \otimes \mathrm{id}_C)) = \Lambda(s \circ (t \otimes \mathrm{id}_C))$(by (53)). It follows that the lower path from $S$ to $|A_2|^{S_1}$ in Diagram 58 is

$$\begin{aligned}
{m_2}^{S_1} \circ \Lambda(u \circ e_{.1}) &= \Lambda(m_2 \circ \mathsf{App}) \circ \Lambda(u \circ e_{.1}) \\
&= \Lambda(m_2 \circ \mathsf{App} \circ (\Lambda(u \circ e_{.1}) \otimes \mathrm{id}_{S_1})) \qquad \text{(by (59))} \\
&= \Lambda(m_2 \circ u \circ e_{.1}) \qquad \text{(by (53))}
\end{aligned}$$

while the upper one is

$$\begin{aligned}
&\Lambda(|\mathsf{App}| \circ \theta \circ (\mathrm{id} \otimes m_1)) \circ |\Lambda(v)| \circ m \\
= \ &\Lambda(|\mathsf{App}| \circ \theta \circ (\mathrm{id} \otimes m_1) \circ ((|\Lambda(v)| \circ m) \otimes \mathrm{id}_{S_1})) \qquad \text{(by (59))} \\
= \ &\Lambda(|\mathsf{App}| \circ \theta \circ (|\Lambda(v)| \otimes \mathrm{id}_{|A_1|}) \circ (m \otimes m_1)) \\
= \ &\Lambda(|\mathsf{App}| \circ |\Lambda(v) \otimes \mathrm{id}_{A_1}| \circ \theta \circ (m \otimes m_1)) \qquad \text{(by naturality of } \theta) \\
= \ &\Lambda(|v| \circ \theta \circ (m \otimes m_1)) \qquad \text{(by (53))}
\end{aligned}$$

and these two quantities are equal by Diagram (57).

Finally, the equations (53) and (54) then hold in the subscone, because they hold in $\boldsymbol{C}$, and using Fact 4.3.

We sum up what we need to lift exponentials to the subscone, as usual. To lift exponentials on the right:

---

**(i.a.r)** monoidal closed (on the right) categories $(\boldsymbol{C}, \otimes, \boldsymbol{I}, \boldsymbol{\alpha}, \boldsymbol{\ell}, \boldsymbol{r})$ and $(\mathbb{C}, \otimes, \mathbb{I}, \alpha, \mathbb{l}, \mathbb{r})$, and a monoidal functor $|_| : \boldsymbol{C} \to \mathbb{C}$,

**(iii.a)** a monoidal mono factorization system on $\mathbb{C}$.

**(v)** $\mathbb{C}$ has pullbacks.

**(vi.r)** $_^A$ preserves relevant monos.

---

Exponentials are given by (55), application by (56), abstraction by (58).

The construction works equally well to lift exponentials on the left, reversing arguments to tensor products, so we require in this case:

**(i.a.l)** monoidal closed (on the left) categories $(\boldsymbol{C}, \otimes, \boldsymbol{I}, \boldsymbol{\alpha}, \boldsymbol{\ell}, \boldsymbol{r})$
and $(\mathbb{C}, \otimes, \mathbb{I}, \mathbb{\alpha}, \mathbb{l}, \mathbb{r})$ and a monoidal functor $|_| : \boldsymbol{C} \to \mathbb{C}$,
**(iii.a)** a monoidal mono factorization system on $\mathbb{C}$.
**(v)** $\mathbb{C}$ has pullbacks.
**(vi.l)** ${}^{A}_$ preserves relevant monos.

Clearly, in the symmetric monoidal closed case we require the following to get a symmetric monoidal closed subscone:

**(i.b)** symmetric monoidal closed categories $(\boldsymbol{C}, \otimes, \boldsymbol{I}, \boldsymbol{\alpha}, \boldsymbol{\ell}, \boldsymbol{r}, \boldsymbol{c})$
and $(\mathbb{C}, \otimes, \mathbb{I}, \mathbb{\alpha}, \mathbb{l}, \mathbb{r}, \mathbb{c})$, and a symmetric monoidal functor $|_| : \boldsymbol{C} \to \mathbb{C}$,
**(iii.a)** a monoidal mono factorization system on $\mathbb{C}$.
**(v)** $\mathbb{C}$ has pullbacks.
**(vi)** $_^{A}$ preserves relevant monos.

And in the cartesian closed case, we require:

**(i.c)** cartesian closed categories $\boldsymbol{C}$ and $\mathbb{C}$,
and a cartesian monoidal functor $|_| : \boldsymbol{C} \to \mathbb{C}$,
**(iii.a)** a monoidal mono factorization system on $\mathbb{C}$.
**(v)** $\mathbb{C}$ has pullbacks.
**(vi)** $_^{A}$ preserves relevant monos.

Putting all conditions together, we get a notion of subscone for categorical models of Moggi's meta-language, i.e., for cartesian closed categories with a strong monad, we require:

**(i.c)** cartesian closed categories $\boldsymbol{C}$ and $\mathbb{C}$,
and a cartesian monoidal functor $|_| : \boldsymbol{C} \to \mathbb{C}$,
**(ii.a)** a strong monad $(\mathbb{T}, \mathbb{\eta}, \mathbb{\mu}, \mathbb{t})$ on $\mathbb{C}$, related to $(\boldsymbol{T}, \boldsymbol{\eta}, \boldsymbol{\mu}, \boldsymbol{t})$ by
a strong monad morphism $(|_|, \sigma)$ defined in (4) and (36),
**(iii.a)** a monoidal mono factorization system on $\mathbb{C}$.
**(iv)** $\mathbb{T}$ maps relevant pseudoepis to pseudoepis.
**(v)** $\mathbb{C}$ has pullbacks.
**(vi)** $_^{A}$ preserves relevant monos.

## 10. Examples

At this point, we suspect the reader is relatively fed up with categorical diagrams and general abstract nonsense. It is therefore time to instantiate our constructions. We start with fairly easy cases in the category ***Set*** of sets in Section 10.1. We examine in more detail the non-determinism monad in Section 10.2, then the discrete probability monad in Section 10.3. A more demanding case is the name creation monad, which involves presheaves, and which we deal with in Section 10.4. We then engage in a series of examples demonstrating the difficulties of finding the right mono factorization system, and the need for varying categories, in particular for choosing $\boldsymbol{C}$ and $\mathbb{C}$ distinct, in situations of practical interest. Doing this, we shall see that our construction often produces definitions in the style of bisimulation, sometimes in the style of observational equivalence, and will in one case assume yet another style (existence of related refinements, see the end of Section 10.7). This will keep us busy from Section 10.5 to Section 10.8. These sections will unfortunately be rather mathematically demanding; it does not seem there was any escaping it.

### 10.1. *Lift, Exceptions, State, Non-Determinism, and Continuations in **Set***

As in Section 5, suppose $\mathbf{C}_1 = \mathbf{C}_2 = \mathbb{C} = \boldsymbol{Set}$, and $|_|_1$ and $|_|_2$ are identities. In particular, $|_|$ is the cartesian product functor from $\boldsymbol{Set}\times\boldsymbol{Set}$ to $\boldsymbol{Set}$. Below we summarize the action of $\widetilde{T}$ on a relation $S \hookrightarrow A_1\times A_2$, for different computational monads $\mathbb{T}$ of Moggi (Moggi, 1991). This is parameterized by a binary relation $R_E$ on exceptions in $E$ in the exception monad $A + E$, by a binary relation $R_{\mathrm{St}}$ on states in the state monad $(A\times\mathrm{St})^{\mathrm{St}}$, and by a binary relation $R_{\mathcal{R}}$ in the continuation monad $\mathcal{R}^{\mathcal{R}^A}$.

| Monad $\mathbb{T}$ | relation $\widetilde{S} \subseteq \mathbb{T}A_1\times\mathbb{T}A_2$ |
|---|---|
| $\mathbb{T}A = A_\bot = A\cup\{\bot\}$ | $\widetilde{S} = S \cup \{(\bot,\bot)\}$ |
| $\mathbb{T}A = A + E$ | $(v_1,v_2)\in\widetilde{S} \iff (v_1,v_2)\in S \vee (v_1,v_2)\in R_E$ |
| $\mathbb{T}A = (A\times\mathrm{St})^{\mathrm{St}}$ | $(f,g)\in\widetilde{S} \iff \forall s_1,s_2\in\mathrm{St}.$ <br> $(s_1,s_2)\in R_{\mathrm{St}} \Rightarrow (\pi_1(fs_1),\pi_1(gs_2))\in S \ \wedge\ (\pi_2(fs_1),\pi_2(gs_2))\in R_{\mathrm{St}}$ |
| $\mathbb{T}A = \mathbb{P}_{\mathrm{fin}}(A)$ | $(B_1,B_2)\in\widetilde{S} \iff$ <br> $\forall b_1\in B_1.\exists b_2\in B_2.(b_1,b_2)\in S\ \wedge$ <br> $\forall b_2\in B_2.\exists b_1\in B_1.(b_1,b_2)\in S$ |
| $\mathbb{T}A = \mathcal{R}^{\mathcal{R}^A}$ | $(\alpha_1,\alpha_2)\in\widetilde{S} \iff \big($ <br> $\forall k_1,k_2.(\forall a_1,a_2.(a_1,a_2)\in S \Rightarrow (k_1(a_1),k_2(a_2))\in R_{\mathcal{R}}) \Rightarrow$ <br> $(\alpha_1(k_1),\alpha_2(k_2))\in R_{\mathcal{R}}\big)$ |

We examine each case in more detail. We take surjections as pseudoepis, injections as relevant monos. This is the canonical choice for an epi-mono factorization system on ***Set***. Note that condition **(iv)**, that $\mathbb{T}$ maps relevant pseudoepis to pseudoepis, is always satisfied when $\mathbb{C} = \boldsymbol{Set}$. In fact, $\mathbb{T}$ preserves pseudoepis: every pseudoepi (surjective function) $e : A \to B$ has a section $m : B \to A$ (i.e., $e\circ m = \mathrm{id}_B$), by the Axiom of Choice. Then $\mathbb{T}e\circ\mathbb{T}m = \mathrm{id}_{\mathbb{T}B}$, showing that $\mathbb{T}e$ is surjective, hence pseudoepi.

10.1.1. *Lift monad $A_\bot$.* The monad morphism $\sigma$ from $(A\times B)_\bot$ to $A_\bot\times B_\bot$ maps $\bot$ to $(\bot,\bot)$, and every pair $(x,y)\in A\times B$ to itself. This is a commutative monad, hence monoidal, hence strong. The mediator $\mathrm{d}_{A,B}$ from $A_\bot\times B_\bot$ to $(A\times B)_\bot$ maps $(\bot,y)$ and $(x,\bot)$ to $\bot$, and $(x,y)$ where $x\in A$ and $y\in B$ to itself. Note that it is not a cartesian monad, because $(\pi_2)_\bot\circ\mathrm{d}_{A,B}$ maps $(\bot,y)$ to $\bot$, while $\pi_2(\bot,y) = y$. Nonetheless, all required conditions are satisfied to lift $\mathbb{T}$ to a commutative monad on the subscone.

10.1.2. *Exception monad $A + E$.* The monad morphism $\sigma$ from $(A\times B)+E$ to $(A+E)\times(B+E)$ maps the pair $(x,y)\in A\times B$ to itself, and the exception $e\in E$ to the pair $(e, e)$. The strength $\mathbb{t}_{A,B}$ from $A \times (B + E)$ to $(A \times B) + E$ maps $(x, y)$ with $x \in A$ and $y \in B$ to $(x, y)$, and $(x, e)$ with $x \in A$ and $e \in E$ to $e$. (This is *not* a commutative monad.) Therefore this lifts to a monoidal monad on the subscone. As shown in the table above, the lifted monad relates two values $v_1$ and $v_2$ if and only if both are non-exceptional values (in $A$) and are related by $S$, or both are exceptions related by some a priori relation $R_E$.

10.1.3. *State transformer monad* $(A \times \mathrm{St})^{\mathrm{St}}$. The monad morphism $\sigma$ from $((A \times B) \times \mathrm{St})^{\mathrm{St}}$ to $(A \times \mathrm{St})^{\mathrm{St}} \times (B \times \mathrm{St})^{\mathrm{St}}$ maps $f : \mathrm{St} \to (A \times B) \times \mathrm{St}$ to the pair of functions mapping $s \in \mathrm{St}$ to $(v_1, s')$ and to $(v_2, s')$ respectively, where $f(s) = ((v_1, v_2), s')$. The strength maps $(x, f) \in A \times (B \times \mathrm{St})^{\mathrm{St}}$ to the function mapping $s \in \mathrm{St}$ to $((x, y), s')$, where $(y, s') = f(s)$. (Again, this is not a commutative monad.)

It follows that our lifting constructions apply, yielding the lifted monad described in the table above. Note that $f : \mathrm{St} \to A_1 \times \mathrm{St}$, $g : \mathrm{St} \to A_2 \times \mathrm{St}$ can be read as the transition functions of deterministic transition systems, which go from a state $s$ to another state $s'$ and emit a value in $A_1$ (resp. in $A_2$). These transition functions are in relation by $\widetilde{S}$ if and only if, for any two states that are in relation via $R_{\mathrm{St}}$, the values emitted by firing the transitions by $f$ and $g$ are in relation by $S$, and the target states after the transition are in relation via $R_{\mathrm{St}}$ again. This states that $f$ and $g$ are in relation by $\widetilde{S}$ if and only if $R_{\mathrm{St}}$ is a *bisimulation* between states.

10.1.4. *Finite powerset (non-determinism) monad* $\mathbb{P}_{\mathrm{fin}}(A)$. The monad morphism from $\mathbb{P}_{\mathrm{fin}}(A \times B)$ to $\mathbb{P}_{\mathrm{fin}}(A) \times \mathbb{P}_{\mathrm{fin}}(B)$ maps every relation $R \subseteq A \times B$ to the pair consisting of its domain $\{x | \exists y \cdot (x, y) \in R\}$ and its codomain $\{y | \exists x \cdot (x, y) \in R\}$. The mediator $\mathbb{d}_{A,B}$ maps $X \subseteq A$ and $Y \subseteq B$ to the relation $X \times Y \subseteq A \times B$, and makes $\mathbb{T}$ a commutative monad.

Our construction in the case of the finite powerset monad $\mathbb{P}_{\mathrm{fin}}()$ in fact expands to: $(B_1, B_2) \in \widetilde{S}$ iff $B_1 = \{x | (x, y) \in R\}$ and $B_2 = \{y | (x, y) \in R\}$ for some $R \subseteq S$. (Recall that $\widetilde{T}$ maps relations $S$ to the direct image $\widetilde{S}$ of $\langle \mathbb{T}\pi_1, \mathbb{T}\pi_2 \rangle : TS \to \mathbb{T}A_1 \times \mathbb{T}A_2$, see the end of Section 5.) This is equivalent to the condition given in the table above, which is the more usual way of defining bisimulations.

Indeed, if $B_1 = \{x | (x, y) \in R\}$ and $B_2 = \{y | (x, y) \in R\}$ for some $R \subseteq S$ then for every $b_1 \in B_1$ by construction there is some $b_2 \in B_2$ such that $(b_1, b_2) \in R$, therefore $(b_1, b_2) \in S$ since $R \subseteq S$, and symmetrically for every $b_2 \in B_2$ there is some $b_1 \in B_1$ such that $(b_1, b_2) \in S$: $B_1$ and $B_2$ are bisimilar.

Conversely, if $B_1$ and $B_2$ are bisimilar (in the sense just given), then let $R$ be the restriction of $S$ to $B_1 \times B_2$. For every $b_1 \in B_1$, by bisimilarity there is some $b_2 \in B_2$ such that $(b_1, b_2) \in S$, so $(b_1, b_2) \in R$, therefore $b_1 \in \{x | (x, y) \in R\}$; so $B_1 \subseteq \{x | (x, y) \in R\}$. The reverse inclusion is obvious, so $B_1 = \{x | (x, y) \in R\}$. The other equality $B_2 = \{y | (x, y) \in R\}$ is by symmetry.

That logical relations on powersets define bisimulations was conjectured in (Lazić and Nowak, 2000) and, for pre-logical relations, in (Honsell and Sannella, 2002).

Note that there is nothing special with the *finite* powerset monad here. We might have taken the set of all subsets instead, or the set of all subsets of cardinality at least $\alpha$ and strictly less than $\beta$, where $\alpha < \beta$ are two cardinal numbers such that $\alpha \leq 1$, and every finite product of cardinals between $\alpha$ (inclusive) and $\beta$ (exclusive) is again so. The finite powerset monad is the case $\alpha = 0$, $\beta = \aleph_0$; the lift monad is the case $\alpha = 0$, $\beta = 2$.

Note also that although this monad is always commutative, it is a cartesian monad if and only if $\alpha \geq 1$. Indeed, $\mathbb{T}$ is a cartesian monad if and only if the domain of the relation $X \times Y$ is $X$, and its codomain is $Y$. This is wrong if $Y$ or $X$ is allowed to be empty, but holds as soon as $X$ and $Y$ are non-empty. In particular, the finite-and-non-empty powerset monad (for serial non-determinism—no state is final) is cartesian.

10.1.5. *The continuation monad* $\mathcal{R}^{\mathcal{R}^A}$. The monad morphism $\sigma$ from $\mathcal{R}^{\mathcal{R}^{A\times B}}$ to $\mathcal{R}^{\mathcal{R}^A} \times \mathcal{R}^{\mathcal{R}^B}$ maps $\alpha : \mathcal{R}^{A\times B} \to \mathcal{R}$ to the pair $(\alpha_1, \alpha_2)$, where $\alpha_1$ maps $k_1 \in \mathcal{R}^A$ to $\alpha(k_1 \circ \pi_1)$ and $\alpha_2$ maps $k_2 \in \mathcal{R}^B$ to $\alpha(k_2 \circ \pi_2)$. The strength $\mathbb{t}_{A,B}$ maps $(x, \alpha) \in A \times \mathcal{R}^{\mathcal{R}^B}$ to the function mapping $k \in \mathcal{R}^{A\times B}$ to $\alpha(\lambda y \in B \cdot k(x, y))$. This monad is not commutative.

Our construction yields the rather opaque condition in the table above, where $a_1$, $a_2$ are values, $k_1$, $k_2$ are continuations, and $\alpha_1$, $\alpha_2$ are programs, taking continuations to answers in $\mathcal{R}$. Intuitively, think of continuations as computation environments (a toplevel loop, a shell) that take the result of a program and print something (called an answer, in $\mathcal{R}$) on a computer terminal. To evaluate a program $\alpha$ in continuation (environment) $k$, apply $\alpha$ to $k$. For simplicity, assume that the relation $R_{\mathcal{R}}$ on answers is equality. The condition then states that two programs are related by $\widetilde{S}$ if and only if they give identical answers when evaluated in related continuations (environments), where two environments are related if and only if they print the same answer on values that are related by $S$, i.e., if and only if they do not make any difference between $S$-related values. This is a form of observational equivalence.

### 10.2. *Labelled transition systems and bisimulations*

The case $\mathbb{T}A = \mathbb{P}_{\text{fin}}(A)$ allows one to define labelled transition systems as elements of $(\mathbb{T}A)^{A\times L}$, with labels in $L$ and states in $A$, as functions mapping states $a$ and labels $\ell$ to the set of states $a'$ such that $a \stackrel{\ell}{\longrightarrow} a'$. Our monad lifting $\widetilde{S}$ in this case is parameterized by a binary relation on $R_L$ on labels and is defined by:

$$(f_1, f_2) \in \widetilde{S} \iff \forall a_1, a_2, \ell_1, \ell_2 \cdot (a_1, a_2) \in S \wedge (\ell_1, \ell_2) \in R_L \Rightarrow \left\{ \begin{array}{l} \forall b_1 \in f_1(a_1, \ell_1).\exists b_2 \in f_2(a_2, \ell_2).(b_1, b_2) \in S \wedge \\ \forall b_2 \in f_2(a_2, \ell_2).\exists b_1 \in f_1(a_1, \ell_1).(b_1, b_2) \in S \end{array} \right.$$

In case $R_L$ is the equality relation, the relation $\widetilde{S}$ relates $f_1$ and $f_2$ iff $S$ is a *strong bisimulation* between the labelled transition systems $f_1$ and $f_2$.

### 10.3. *Discrete Probabilities and Subprobabilities*

Defining $\boldsymbol{T}A$ to be space of all probability distributions over the space $A$ allows us to define probabilistic transition systems as objects of $(\boldsymbol{T}A)^A$. In principle, this should work just like ordinary transition systems (Section 10.2).

In fact, defining what the right spaces should be, and ensuring that the required monads exist and have the required properties, is much subtler. For one, as far as we know, the category of measurable spaces and measurable functions is not known to be cartesian closed. As we don't want to delve into measure theoretic considerations at this point, let us deal with a simpler case: the monad of *discrete* (sub-)probability distributions over ***Set***.

A discrete sub-probability distribution $\nu$ over some set $A$ is simply a function mapping each element $x \in A$ to some non-negative number $a_x \in \mathbb{R}^+$, so that $\sum_{x \in A} a_x \leq 1$. This sum, which is in general infinite, denotes the supremum of all partial sums $\sum_{x \in B} a_x$, when $B$ ranges over the finite subsets of $A$. It is an easy exercise to show that only countably many coefficients $a_x$ can be non-zero, then. A discrete probability distribution additionally satisfies $\sum_{x \in A} a_x = 1$. The Dirac distribution $\delta_{x_0}$ at $x_0$ maps $x_0$ to 1 and every $x \neq x_0$ to 0. Accordingly, we write $\sum_{x \in A} a_x \delta_x$ for the discrete distribution mapping $x$ to $a_x$. The measure of a subset $B$ of $A$ under $\nu = \sum_{x \in A} a_x \delta_x$ is $\nu(B) = \sum_{x \in B} a_x$.

Let $\boldsymbol{T}A$ be the set of all discrete (sub-)probability distributions over $A$, $\boldsymbol{\eta}_A$ map $x \in A$ to $\delta_x \in \boldsymbol{T}A$, and $\boldsymbol{\mu}_A$ be defined as mapping $\nu_2 = \sum_{\nu \in \boldsymbol{T}A} a_\nu \delta_\nu \in \boldsymbol{T}^2 A$ to the discrete distribution $\boldsymbol{\mu}_A(\nu_2)$ such that $\boldsymbol{\mu}_A(\nu_2)(B) = \sum_{\nu \in \boldsymbol{T}A} a_\nu \nu(B)$ for every $B \subseteq A$; in other words, $\boldsymbol{\mu}_A(\nu_2) = \sum_{x \in A} \left( \sum_{\nu \in \boldsymbol{T}A} a_\nu \nu(\{x\}) \right) \delta_x$. It is easy to check that $(\boldsymbol{T}, \boldsymbol{\eta}, \boldsymbol{\mu})$ is a monad on ***Set***. Moreover, it is strong: the strength $\boldsymbol{t}_{A,B}$ maps $x \in A$ and $\nu = \sum_{y \in B} b_y \delta_y \in \boldsymbol{T}B$ to $\sum_{y \in B} b_y \delta_{(x,y)}$.

Our construction in the case of the (sub-)probability distribution monad on ***Set*** yields the following. Again, $\boldsymbol{T} = \mathbb{T}$, and recall that $\widetilde{T}$ maps relations $S$ on $A_1 \times A_2$ to the direct image $\widetilde{S}$ of $\langle \mathbb{T}\pi_1, \mathbb{T}\pi_2 \rangle : \mathbb{T}S \to \mathbb{T}A_1 \times \mathbb{T}A_2$. In other words, given $\nu_1 \in \boldsymbol{T}A_1$, $\nu_2 \in \boldsymbol{T}A_2$, $(\nu_1, \nu_2) \in \widetilde{S}$ if and only if there is a discrete (sub-)probability distribution $\nu$ on $S$ such that $\nu_1(B_1) = \nu((B_1 \times A_2) \cap S)$ for every $B_1 \subseteq A_1$ and $\nu_2(B_2) = \nu((A_1 \times B_2) \cap S)$ for every $B_2 \subseteq A_2$.

Write $\nu_1$ as $\sum_{x_1 \in A_1} a_{x_1} \delta_{x_1}$, $\nu_2$ as $\sum_{x_2 \in A_2} b_{x_2} \delta_{x_2}$. Then $(\nu_1, \nu_2) \in \widetilde{S}$ if and only if there is a discrete (sub-)probability distribution $\nu = \sum_{\substack{x_1 \in A_1, x_2 \in A_2 \\ (x_1, x_2) \in S}} c_{x_1,x_2} \delta_{(x_1,x_2)}$ on $S$ such that $a_{x_1} = \sum_{x_2 \in A_2 / (x_1,x_2) \in S} c_{x_1,x_2}$ for all $x_1 \in A_1$ and $b_{x_2} = \sum_{x_1 \in A_1 / (x_1,x_2) \in S} c_{x_1,x_2}$ for all $x_2 \in A_2$. This may be pictured by requiring that we have a matrix of non-negative reals $c_{x_1,x_2}$ such that: the only non-zero entries $(x_1, x_2)$ of this matrix are such that $(x_1, x_2) \in S$; the sum of the $x_1$ row is $a_{x_1}$; the sum of the $x_2$ column is $b_{x_2}$. This is a well-known problem in probability theory: find $\nu$ on the product $A_1 \times A_2$ having given *marginals* $\nu_1$ and $\nu_2$ and whose *support* $\{(x_1, x_2) \in A_1 \times A_2 | \nu(\{(x_1, x_2)\}) \neq 0\}$ is included in $S$.

Let us play with the definition of $\widetilde{S}$ for a moment. Consider the disjoint sum $A_1 \oplus A_2$, and the smallest equivalence $\equiv_S$ relation containing $S$. It is easy to see that, for every $\equiv_S$-equivalence class $C$,

$$((C \cap A_1) \times A_2) \cap S = (A_1 \times (C \cap A_2)) \cap S \tag{60}$$

It follows that if there is a discrete distribution $\nu$ with support contained in $S$ and whose marginals are $\nu_1$ and $\nu_2$, then for any $\equiv_S$-equivalence class $C$, $\nu_1(C \cap A_1) = \nu_2(C \cap A_2)$. Indeed, $\nu_1(C \cap A_1) = \nu(((C \cap A_1) \times A_2) \cap S)$, $\nu_2(C \cap A_2) = \nu((A_1 \times (C \cap A_2)) \cap S)$; then use (60).

Conversely, if $\nu_1(C \cap A_1) = \nu_2(C \cap A_2)$ for every $\equiv_S$-equivalence class $C$, then whenever $x_1 \equiv_S x_2$, let $d_{x_1,x_2} = \nu_1(C \cap A_1) = \nu_2(C \cap A_2)$, where $C$ is the equivalence class of $x_1$ (equivalently, of $x_2$). Now define $\nu$ as $\sum_{\substack{(x_1,x_2) \in A_1 \times A_2 \\ x_1 \equiv_S x_2 \\ d_{x_1,x_2} \neq 0}} \frac{a_{x_1} b_{x_2}}{d_{x_1,x_2}} \delta_{(x_1,x_2)}$. Clearly the support of $\nu$ is contained in the relation $\equiv_S$, and $\nu_1$ and $\nu_2$ are the marginals of $\nu$. So $(\nu_1, \nu_2) \in \widetilde{\equiv_S}$.

Let $S$ be *saturated* if and only if, for any $x_1 \in A_1$, $x_2 \in A_2$, $x_1 \equiv_S x_2$ implies $(x_1, x_2) \in S$. We have just shown the following: for any saturated relation $S \subseteq A_1 \times A_2$, $(\nu_1, \nu_2) \in \widetilde{S}$ if and only if $\nu_1(C \cap A_1) = \nu_2(C \cap A_2)$ for every $\equiv_S$-equivalence class $C$. The restriction to saturated relations is not too demanding; indeed, we can always saturate any binary relation $S$ by considering the restriction of $\equiv_S$ to $A_1 \times A_2$ instead.

Analogously with Section 10.2, we may define a *probabilistic* labelled transition system as an element of $(\boldsymbol{T}A)^{A \times L}$. Then two such transition systems $f_1$ and $f_2$ are in relation if and only if:

$$\forall x_1 \in A_1, x_2 \in A_2, \ell_1, \ell_2 \in L.(x_1, x_2) \in S \quad \wedge \quad (\ell_1, \ell_2) \in R_L \Rightarrow (f_1(x_1, \ell_1), f_2(x_2, \ell_2)) \in \widetilde{S} \tag{61}$$

When $S$ is saturated, this condition is equivalent to: for every $x_1 \in A_1$, $x_2 \in A_2$, $\ell_1, \ell_2 \in L$, if $x_1 \equiv_S x_2$ and $(\ell_1, \ell_2) \in R_L$ then for every $\equiv_S$-equivalence class $C$, $f_1(x_1, \ell_1)(C \cap A_1) = f_2(x_2, \ell_2)(C \cap A_2)$. This is exactly Larsen and Skou's (Larsen and Skou, 1991) notion of probabilistic bisimulation, when $L_1 = L_2$, $R_L$ is the equality relation, and $A_1$ and $A_2$ are finite. To see this, write $p_\ell(x) = f_1(x, \ell)$ if $x \in A_1$, $p_\ell(x) = f_2(x, \ell)$ if $x \in A_2$, and rewrite the condition above as: if $x_1 \equiv_S x_2$ then for every $\equiv_S$-equivalence class $C$, $p_\ell(x_1)(C) = p_\ell(x_2)(C)$.

It is therefore fair to say that any saturated relation $S$ (and by extension, any relation at all) such that $f_1$ and $f_2$ are in relation at type $A \times L \to \boldsymbol{T}A$, as defined in (61), is again a probabilistic bisimulation.

A note on discreteness: the monad of discrete (sub-)probability distributions is enough to model coin flips $\sum_{x \in \{0,1\}} 1/2\delta_x$. One may argue that, in any finite computation of a $\lambda$-term augmented with coin flips, only finite linear combinations of Dirac measures (*simple* valuations) will ever crop up, so that discrete distributions are general enough to accomodate all practical forms of probabilistic choice. This would not be true in the case of infinite computations, where at least all least upper bounds of directed chains of simple distributions (*quasi-simple* valuations (Goubault-Larrecq, 2005)) would be needed. Notwithstanding the fact that we should incorporate fixpoints in the language and replace ***Set*** with some other category such as ***Cpo***, this would exceed the realm of discrete distributions anyway: for example, the Lebesgue measure on $[0, 1]$ is a quasi-simple valuation but is definitely not discrete.

### 10.4. *Logical relations for dynamic name creation*

Consider the categorical model of dynamic name creation defined in (Stark, 1996). Let $\mathcal{I}$ be the category of finite sets and injective functions, and $\boldsymbol{Set}^{\mathcal{I}}$ be the category of functors from $\mathcal{I}$ to ***Set*** and natural transformations (the category of *covariant presheaves* over $\mathcal{I}$).

For short, write $\boldsymbol{T}As$ for $\boldsymbol{T}(A)(s)$ and similarly for other notations. Let $+$ denote disjoint union in $\mathcal{I}$.[‡]

We define the strong monad $(\boldsymbol{T}, \boldsymbol{\eta}, \boldsymbol{\mu}, \boldsymbol{t})$ on $\boldsymbol{Set}^{\mathcal{I}}$ by:

- $\boldsymbol{T}A = \mathrm{colim}_{s'} A(_ + s') : \mathcal{I} \to \boldsymbol{Set}$.

  On objects, this is given by $\boldsymbol{T}As = \mathrm{colim}_{s'} A(s + s')$, i.e., $\boldsymbol{T}As$ is the set of all equivalence classes of pairs $(s', a)$ with $s' \in \mathcal{I}$ and $a \in A(s + s')$ modulo the smallest equivalence relation $\equiv$ such that $(s', a) \equiv (s'', A(\mathrm{id}_s + j)a)$ for every morphism $s' \xrightarrow{j} s''$ in $\mathcal{I}$ (intuitively, given a set of *names* $s$, elements of $\boldsymbol{T}As$ are formal expressions $(\nu s')a$ where all names in $s'$ are bound and every name free in $a$ is in $s + s'$—modulo the fact that $(\nu s', s'')a \equiv (\nu s')a$ for any additional set of new names $s''$ not free in $a$). We shall write $[s', a]$ the equivalence class of $(s', a)$.

  On morphisms $s_1 \xrightarrow{i} s_2$, $\boldsymbol{T}Ai$ maps $[s', a]$ to the equivalence class of $[s', A(i + \mathrm{id}_{s'})a]$.
- For any $f : A \to B$ in $\boldsymbol{Set}^{\mathcal{I}}$, $\boldsymbol{T}fs : \boldsymbol{T}As \to \boldsymbol{T}Bs$ is defined by $\boldsymbol{T}fs[s', a] = [s', f(s + s')a]$. This is compatible with $\equiv$ because $f$ is natural.
- $\boldsymbol{\eta}_A s : As \to \boldsymbol{T}As$ is defined by $\boldsymbol{\eta}_A sa = [\emptyset, a]$.
- $\boldsymbol{\mu}_A s : \boldsymbol{T}^2 As \to \boldsymbol{T}As$ is defined by $\boldsymbol{\mu}_A s[s', [s'', a]] = [s' + s'', a]$.
- $\boldsymbol{t}_{A,B} s : As \times \boldsymbol{T}Bs \to \boldsymbol{T}(A \times B)s$ is defined by $\boldsymbol{t}_{A,B} s(a, [s', b]) = [s', (A\mathsf{inl}ss'a, b)]$ where $\mathsf{inl}ss' : s \to s + s'$ is the canonical injection.

Furthermore, $\boldsymbol{T}$ is a commutative monad, whose mediator $\mathbb{d}_{A,B}s : \boldsymbol{T}As \times \boldsymbol{T}Bs \to \boldsymbol{T}(A \times B)s$ maps $([s', a], [s'', b])$ to $[s' + s'', (A(\mathrm{id}_s + \mathsf{inl}s's'')a, A(\mathrm{id}_s + \mathsf{inr}s''s')b)]$, where $\mathsf{inr}s''s' : s'' \to s' + s''$ is the other canonical injection. In fact, $\boldsymbol{T}$ is a cartesian monad. Indeed, $(\boldsymbol{T}\pi_1 \circ \mathbb{d}_{A,B})s$ maps $([s', a], [s'', b])$ to $[s' + s'', A(\mathrm{id}_s + \mathsf{inl}s's'')a] = [s', a]$, so $\boldsymbol{T}\pi_1 \circ \mathbb{d}_{A,B} = \pi_1$, and similarly $\boldsymbol{T}\pi_2 \circ \mathbb{d}_{A,B} = \pi_2$.

It is important to note how $\equiv$ works. The category $\mathcal{I}$ has pushouts: in particular, if $s_0 \xrightarrow{i_1} s_1$ and $s_0 \xrightarrow{i_2} s_2$ are two morphisms in $\mathcal{I}$, then there is a finite set $s_1 +_{s_0} s_2$ and two morphisms $s_1 \xrightarrow{j_1} s_1 +_{s_0} s_2$, $s_2 \xrightarrow{j_2} s_1 +_{s_0} s_2$ such that $j_1 \circ i_1 = j_2 \circ i_2$—take $s_1 +_{s_0} s_2$ to be the disjoint sum $s_1 + s_2$ modulo the equivalence relation relating $i_1(a_0) = i_2(a_0)$ for every $a_0 \in s_0$.

It follows that $(*)$ for every $a_1 \in A(s + s_1)$, $a_2 \in A(s + s_2)$, $(s_1, a_1) \equiv (s_2, a_2)$ if and only if there is a finite set $s_{12}$ and two arrows $s_1 \xrightarrow{j_1} s_{12}$ and $s_2 \xrightarrow{j_2} s_{12}$ such that $A(\mathrm{id}_{s_1} + j_1)a_1 = A(\mathrm{id}_{s_2} + j_2)a_2$.

We take $\boldsymbol{C}_1 = \boldsymbol{C}_2 = \mathbb{C} = \boldsymbol{Set}^{\mathcal{I}}$, hence objects in the subscone give rise to $\mathcal{I}$-indexed *Kripke logical relations*. Furthermore, $|_|_1 = |_|_2 = |_|$ is the identity functor and $\mathbb{T}$ is just $\boldsymbol{T}$. The category $\boldsymbol{Set}^{\mathcal{I}}$ has a mono factorization consisting of pointwise surjections and pointwise injections.

As in Section 5, the monad morphism $\sigma_{\langle A_1, A_2 \rangle} s : \mathbb{T}(A_1 \times A_2)s \to \boldsymbol{T}A_1 s \times \boldsymbol{T}A_2 s$ is equal to $\langle \mathbb{T}\pi_1, \mathbb{T}\pi_2 \rangle s$. That is to say

$$\sigma_{\langle A_1, A_2 \rangle} s[s', (a_1, a_2)] = ([s', a_1], [s', a_2])$$

where $s' \in \mathcal{I}$, $a_1 \in A_1(s + s')$, $a_2 \in A_2(s + s')$.

[‡] Note that $+$ is not a coproduct in $\mathcal{I}$. In fact, $\mathcal{I}$ does not have a coproduct. However $+$ is functorial in both components, associative, and has a neutral element, and this is all we need.

Again as in ***Set***, every functor $S$ from $\mathcal{I}$ to $A_1 \times A_2$ has a representation

$$\langle {\pi^S}_1, {\pi^S}_2 \rangle : S \hookrightarrow A_1 \times A_2$$

where each arrow $\langle {\pi^S}_1, {\pi^S}_2 \rangle s : Ss \hookrightarrow A_1 s \times A_2 s$ is an inclusion.

Hence, $S$ is a family of relations $Ss$ between $A_1 s$ and $A_2 s$, functorial in $s$ (for each $s' \xrightarrow{i} s''$, $Si$ is the appropriate restriction of $A_1 i \times A_2 i$). Recall from Section 5 that $\widetilde{S}$ is defined as the direct image of $\langle \mathbb{T}{\pi^S}_1, \mathbb{T}{\pi^S}_2 \rangle$. So, (†) $[s_1, a_1]\ \widetilde{S}s\ [s_2, a_2]$ if and only if for some $s' \in \mathcal{I}$, $a'_1 \in A_1 s'$, $a'_2 \in A_2 s'$, $(s_1, a_1) \equiv (s', a'_1)$, $(s_2, a_2) \equiv (s', a'_2)$ and $a'_1\ S(s + s')\ a'_2$.

Using $(*)$ above, $(s_1, a_1) \equiv (s', a'_1)$ means that there is a finite set $s'_1$ and two arrows $s_1 \xrightarrow{j_1} s'_1$, $s' \xrightarrow{j'_1} s'_1$ in $\mathcal{I}$ such that: (a) $A_1(\mathrm{id}_s + j_1)a_1 = A_1(\mathrm{id}_s + j'_1)a'_1$. Similarly, $(s_2, a_2) \equiv (s', a'_2)$ means there is a finite set $s'_2$ and two arrows $s_2 \xrightarrow{j_2} s'_2$, $s' \xrightarrow{j'_2} s'_2$ in $\mathcal{I}$ such that: (b) $A_2(\mathrm{id}_s + j_2)a_2 = A_2(\mathrm{id}_s + j'_2)a'_2$. Consider arrows $j'_1$ and $j'_2$, which both have $s'$ as source, and build their pushout $s_0 = s'_1 +_{s'} s'_2$, with arrows $s'_1 \xrightarrow{j''_1} s_0$, $s'_2 \xrightarrow{j''_2} s_0$. Let $j$ be $j''_1 \circ j'_1 = j''_2 \circ j'_2$. By applying $A_1(\mathrm{id}_s + j''_1)$ to both sides of (a), $A_1(\mathrm{id}_s + (j''_1 \circ j_1))a_1 = A_1(\mathrm{id}_s + j)a'_1$. By applying $A_2(\mathrm{id}_s + j''_2)$ to both sides of (b), $A_2(\mathrm{id}_s + (j''_2 \circ j_2))a_2 = A_2(\mathrm{id}_s + j)a'_2$. Since $a'_1\ S(s + s')\ a'_2$ and $S$ is functorial, $A_1(\mathrm{id}_s + j)a'_1\ S(s + s_0)\ A_2(\mathrm{id}_s + j)a'_2$, so $A_1(\mathrm{id}_s + (j''_1 \circ j_1))a_1\ S(s + s_0)\ A_2(\mathrm{id}_s + (j''_2 \circ j_2))a_2$.

So if $(s_1, a_1)\ \widetilde{S}s\ (s_2, a_2)$ then there are arrows $s_1 \xrightarrow{i_1} s_0$ and $s_2 \xrightarrow{i_2} s_0$, namely $i_1 = j''_1 \circ j_1$ and $i_2 = j''_2 \circ j_2$, such that $A_1(\mathrm{id}_s + i_1)a_1\ S(s + s_0)\ A_2(\mathrm{id}_s + i_2)a_2$.

Conversely, if the latter holds, then (†) above clearly holds for $s' = s_0$, $a'_1 = A_1(\mathrm{id}_s + i_1)a_1$ and $s'_2 = A_2(\mathrm{id}_s + i_2)a_2$.

$\widetilde{S}s \hookrightarrow \boldsymbol{T}A_1 s \times \boldsymbol{T}A_2 s$ is thus given by

$$\begin{array}{rcl} [s_1, a_1]\ \widetilde{S}s\ [s_2, a_2] & \iff & \exists s_0 \in \mathcal{I} \cdot \exists i_1 : s_1 \to s_0 \in \mathcal{I} \cdot \exists i_2 : s_2 \to s_0 \in \mathcal{I} \cdot \\ & & (A_1(\mathrm{id}_s + i_1)a_1)\ S(s + s_0)\ (A_2(\mathrm{id}_s + i_2)a_2) \end{array} \tag{62}$$

where $a_1 \in A_1(s + s_1)$ and $a_2 \in A_2(s + s_2)$.

From (62) we define a logical relation for Moggi's metalanguage, as suggested in Section 5, by induction on types $\tau$. Each relation $[\![\tau]\!]$ is a functor from $\mathcal{I}$ to ***Set*** $\times$ ***Set***, so that $[\![\tau]\!]\, s$ is a binary relation for each type $\tau$ and each finite set $s$. We have:

$$\begin{array}{rcl} (f_1, f_2) \in [\![\tau \to \tau']\!]\, s & \iff & \forall s', i : s \to s' \in \mathcal{I}, (a_1, a_2) \in [\![\tau]\!]\, s' \cdot \\ & & \quad (f_1 s'(i, a_1), f_2 s'(i, a_2)) \in [\![\tau']\!]\, s' \\ ((a_1, a'_1), (a_2, a'_2)) \in [\![\tau \times \tau']\!]\, s & \iff & (a_1, a_2) \in [\![\tau]\!]\, s \wedge (a'_1, a'_2) \in [\![\tau']\!]\, s \\ ([s_1, a_1], [s_2, a_2]) \in [\![\boldsymbol{T}\tau]\!]\, s & \iff & \exists s_0, i_1 : s_1 \to s_0, i_2 : s_2 \to s_0 \in \mathcal{I} \cdot \\ & & (A_1(\mathrm{id}_s + i_1)a_1, A_2(\mathrm{id}_s + i_2)a_2) \in [\![\tau]\!]\, (s + s_0) \end{array}$$

This is similar to the logical relations of (Pitts and Stark, 1993; Stark, 1998). While (62) is roughly similar to the notion of logical relation of (Pitts and Stark, 1993), this paper does not rest on Moggi's computational $\lambda$-calculus. On the other hand (Stark, 1998) does rest on the computational $\lambda$-calculus but does not define a suitable notion of logical relation.

Zhang and Nowak show in (Zhang and Nowak, 2003) that the logical relation above is in fact strictly weaker than Pitts and Stark's logical relation (Pitts and Stark, 1993) when restricted to the latter's nu-calculus; Zhang and Nowak also claim that by reinstantiating our construction of the subscone with $\boldsymbol{C}_1 = \boldsymbol{C}_2 = \boldsymbol{Set}^{\mathcal{I}}$ as above but $\mathbb{C} = \boldsymbol{Set}^{\mathcal{I}^{\rightarrow}}$, where $\mathcal{I}^{\rightarrow}$ is the comma category whose objects are the morphisms of $\mathcal{I}$, another Kripke logical relation is obtained that coincides with Pitts and Stark's on the nu-calculus up to first order. This is in fact wrong, but Zhang (Zhang, 2005) shows that this can be repaired by replacing $\mathcal{I}^{\rightarrow}$ by the category $\mathcal{PI}^{\rightarrow}$ whose objects are, again, morphisms of $\mathcal{I}$, but whose morphisms are pullback diagrams in $\mathcal{I}$. It extends it to the full monadic meta-language, and rests purely on semantic principles, while Pitts and Stark's definition of their logical relation relies on normalization properties of the nu-calculus.

### 10.5. *Fun With Categories I: Taking $\boldsymbol{C} \neq \boldsymbol{Set}$, the Example of $\boldsymbol{Cpo}$*

The constructions of this paper apparently allow for some considerable degree of freedom. We may choose distinct categories for $\boldsymbol{C}$ and $\mathbb{C}$, different monads $\boldsymbol{T}$ and $\mathbb{T}$, several monad morphisms $\sigma$, several mono factorization systems, and so on.

However, the examples we have given until now used none of these degrees of freedom. By taking $\boldsymbol{C} = \mathbb{C} = \boldsymbol{Set}$ notably, we have essentially narrowed down our possible choices to just one. So let us have fun, trying to vary various parameters, and see what happens. We start by keeping $\boldsymbol{C} = \mathbb{C}$, but with categories $\boldsymbol{C}$ other than $\boldsymbol{Set}$ or presheaf categories. We shall then look at several mono factorization systems.

The aim of this section is to show that, at least if we keep $\boldsymbol{C} = \mathbb{C}$, our freedom is considerably restricted.

Imagine $\boldsymbol{C}$ is $\boldsymbol{Cpo}$, the category of cpos and continuous maps. By *cpo* we mean any ordered set such that any directed family of elements has a least upper bound. See (Abramsky and Jung, 1994; Gierz et al., 1980; Gierz et al., 2003) for background information on cpos and related topics. We always assume cpos to be equipped with their Scott topology.

Imagine the monad we are interested in on $\boldsymbol{Cpo}$ is that of demonic non-determinism, whereby intuitively, non-deterministic choices are in the hand of a malevolent external adversary. The traditional monad $\boldsymbol{T}$ modeling this is the *Smyth powerdomain* construction: $\boldsymbol{T}A = \mathcal{Q}(A)$ consists of all compact non-empty upward-closed subsets of $A$, ordered by *reverse* inclusion $\supseteq$. One problem here is that $\mathcal{Q}(A)$ need not be a cpo. It is one if $A$, equipped with its Scott topology, is a *sober* topological space; for example, if $A$ is a continuous cpo. So we should restrict to some cartesian-closed category consisting of continuous cpos only. We know what these categories should be, by Jung's results (Jung, 1990), but this is a bit complicated for an illustration. (We shall return to these categories anyway in Section 10.7, out of necessity, but will avoid them as long as we can.)

Instead, take Heckmann's characterization of the upper power domain (Heckmann, 1990) (without the top element, which would be the empty compact in $\mathcal{Q}(A)$ above). Let $\mathbf{2} = \{0, 1\}$ be the *Sierpiński space*, i.e., the two-element cpo with $0 < 1$. Let $[A \to \mathbf{2}]$ be the set of Scott-continuous functions from $A$ to $\mathbf{2}$; this is canonically isomorphic to the space of Scott opens of $A$, through the map sending $f \in [A \to \mathbf{2}]$ to $f^{-1}\{1\}$. Define $\top$ as $\lambda x \cdot 1$ in $[A \to \mathbf{2}]$, $\perp$ as $\lambda x \cdot 0$; least upper bounds $\sqcup$ and greatest lower bounds $\sqcap$ are defined pointwise. Following Heckmann, $\mathcal{U}_*(A)$ is the set of continuous linear non-trivial second-order predicates, i.e., the set of Scott-continuous functionals $F$ from $[A \to \mathbf{2}]$ to $\mathbf{2}$ that preserve $\top$ and meets $\sqcap$ (linearity), and map $\perp$ to 0 (non-triviality). (See (Heckmann, 1990, Chapter 19) for the rationale behind this construction, and for related theorems. While our constructions will work for several other power constructions, we only take one as illustration.)

$\mathcal{U}_*$ is a monad $\boldsymbol{T}$ on $\boldsymbol{Cpo}$. Its action on morphisms $h : A \to B$ is given by $\boldsymbol{T}(h) = \lambda F \in \mathcal{U}_*(A), f \in [B \to \mathbf{2}] \cdot F(f \circ h)$. Its unit is given by $\boldsymbol{\eta}_A(x) = \lambda f \in [A \to \mathbf{2}] \cdot f(x)$ $(x \in A)$, and its multiplication is given by $\boldsymbol{\mu}_A(G) = \lambda f \in [A \to \mathbf{2}] \cdot G(\lambda F \in \mathcal{U}_*(A) \cdot F(f))$ $(G \in \mathcal{U}_*^2(A))$. The strength is given by $\boldsymbol{t}_{A,B} = \lambda(x, F) \in A \times \mathcal{U}_*(B) \cdot \lambda f \in [A \times B \to \mathbf{2}] \cdot F(\lambda y \in B \cdot f(x, y))$. Note that these are the same formulae as those defining the continuation monad.

It turns out that if $A$ is a continuous cpo, then $\mathcal{U}_*(A)$ and $\mathcal{Q}(A)$ are naturally isomorphic. They key result here is the Hofmann-Mislove Theorem (Abramsky and Jung, 1994, Corollary 7.2.10). The isomorphism maps any upward-closed compact subset $Q$ of $A$ to the continuous linear functional $\lambda f \in [A \to \mathbf{2}] \cdot Q \subseteq f^{-1}\{1\}$, and conversely, maps any continuous linear functional $F$ to the intersection of all opens $O$ of $A$ such that $F(\chi_O) = 1$. (The hard part of the Hofmann-Mislove theorem is to show that this is compact, as soon as $A$ is sober.) While this shows the connection between $\mathcal{U}_*$ and $\mathcal{Q}$, we shall not use this until Section 10.7.

To apply our construction of logical relations, we need a mono factorization system. In $\boldsymbol{Cpo}$, there are at least three:

— The *trivial mono factorization*. Here, relevant monos are isos, and pseudoepis are arbitrary morphisms. This exists in every category, and is completely uninteresting in every category: given any binary relation $S$ between $A_1$ and $A_2$, i.e., $S \overset{m}{\hookrightarrow} A_1 \times A_2$, then $S$ is in fact the whole of $A_1 \times A_2$ (up to iso).

— The *epi / regular mono factorization*. Recall that a mono is *regular* if and only if it is the equalizer of a parallel pair. Let us describe regular monos more concretely in $\boldsymbol{Cpo}$.

A regular mono $m : C \hookrightarrow B$ in $\boldsymbol{Cpo}$ is a Scott-closed subset of $B$, up to isomorphism. More precisely, $m$ is the composite of an iso from $C$ to some Scott-closed subset $m(C)$ of $B$, with the canonical inclusion of $m(C)$ in $B$.

An epi is any continuous function $e : A \twoheadrightarrow C$ such that $e(A)$ is *dense* in $C$, i.e., such that $cl(e(A)) = C$, where $cl$ denotes the (Scott) topological closure operator. Equivalently, such that $e^{-1}(U) \neq \emptyset$ for any non-empty Scott-open $U$ of $C$.

It is well-known that this yields a mono factorization system on $\boldsymbol{Cpo}$. We skip the details, which are not very interesting at this point.

— The *extremal epi / mono factorization*. Recall that the *extremal* pseudoepis are the morphisms $A \xrightarrow{e} B$ such that, whenever $e$ can be written as the composite $A \longrightarrow B' \overset{m}{\hookrightarrow} B$ where $m$ is mono, then $m$ is actually iso.

Many categories $\boldsymbol{C}$ have a mono factorization system where the pseudoepis are the extremal pseudoepis, as shown by Freyd and Kelly: by Proposition 2.3.4 of (Freyd and Kelly, 1972), this is the case if $\boldsymbol{C}$ has all finite limits and admits intersections of monos (i.e., limits of all classes of monos $( X_\alpha \overset{m_\alpha}{\hookrightarrow} B )_{\alpha \in \Omega}$); or if $\boldsymbol{C}$ has all finite colimits and admits colimits of all classes of extremal pseudoepis $( B \overset{e_\alpha}{\twoheadrightarrow} Y_\alpha )_{\alpha \in \Omega}$; or if $\boldsymbol{C}$ has all finite limits, all finite colimits, and composites of coequalizers are coequalizers. In all these cases, extremal pseudoepis are in fact epis; in the third case, they are regular, i.e., coequalizers of parallel pairs. By the way, the argument of Freyd and Kelly in fact shows that $\boldsymbol{C}$ has a mono factorization system as soon as $\boldsymbol{C}$ has pullbacks (not necessarily all limits) and admits non-empty intersections of monos: the factorization of any given morphism $A \xrightarrow{f} B$ is the intersection of the class of all monos $m_\alpha$ such that $f$ factors through $m_\alpha$, i.e., such that $f$ can be written $A \xrightarrow{e_\alpha} C_\alpha \overset{m_\alpha}{\hookrightarrow} B$. (In this case, it is unknown whether pseudoepis are epi.)
While monos in $\boldsymbol{Cpo}$ are simple to characterize—they are just the injective continuous maps—extremal epis have a much more complicated structure, which we therefore elude.

It turns out that we have been unable to show that the functor $\boldsymbol{T} = \mathcal{U}_*$ mapped relevant pseudoepis to pseudoepis, despite our efforts. We shall see in the next sections that this is not needed. All we have to do is make $\mathbb{C}$ slightly different from $\boldsymbol{C}$, and $\mathbb{T}$ slightly different from $\boldsymbol{T}$.

### 10.6. *Fun With Categories II: Making $\boldsymbol{C}$ and $\mathbb{C}$ Distinct, e.g., $\boldsymbol{C} = \boldsymbol{Cpo}$, $\mathbb{C} = \boldsymbol{Ord}$*

Consider $\mathbb{C} = \boldsymbol{Ord}$, the category of partial orders and monotonic maps. There is an obvious forgetful functor $|_|_0 : \boldsymbol{Cpo} \to \boldsymbol{Ord}$. In the following, if $A \subseteq B$, we write $\uparrow_B A$ the subset $\{x \in B | \exists y \in A \cdot y \leq x\}$.

In any partial order $A$, say that a subset is *finitary* if and only if it is of the form $\uparrow_A E$ with $E$ finite. In the following, we take $\mathbb{T}A$ as the space of non-empty finitary subsets of $A$, ordered by *reverse* inclusion $\supseteq$. Why should we take this? Well, we have seen above that the space $\boldsymbol{T}A = \mathcal{U}_*(A)$ could be identified with the space of compact non-empty upward-closed subsets of $A$, ordered by reverse inclusion. For our construction to work out, in particular to be able to find a suitable monad morphisms, $\mathbb{T}$ should in a sense mimic $\boldsymbol{T}$ in the category $\mathbb{C}$. By going from $\boldsymbol{C} = \boldsymbol{Cpo}$ to $\mathbb{C} = \boldsymbol{Ord}$, we keep the ordering but forget about the (Scott-) topology. It turns out that, if we just know about the ordering $\leq$, there are many possible topologies from which it may arise. Precisely, there are many topologies with $\leq$ as *specialization quasi-ordering* — the one defined so that $x$ is less than or equal to $y$ if and only if every open containing $x$ must also contain $y$. One is the Scott-topology. The finest is the Alexandroff topology, whose opens are all upward-closed subsets. In all these topologies, the only subsets that are guaranteed to be compact upward-closed are the finitary subsets. The latter then intuitively form a concept similar to the notion of compact upward-closed subsets. This should be checked formally, by verifying the axioms for monad morphisms, which we do right away.

Note in passing that there is a slightly distinct presentation of $\mathbb{T}A$, as set of finite non-empty *antichains* (sets $E$ where no two distinct elements are comparable), ordered in the *Smyth ordering* $\leq^\sharp$ defined by $E \leq^\sharp E'$ if and only if every element of $E'$ is greater than or equal to some element of $E$.

For every map $f : A \to B$, $\mathbb{T}f$ maps the finitary subset $\uparrow_A \{x_1, \ldots, x_n\}$ to $\uparrow_B \{f(x_1), \ldots, f(x_n)\}$. This defines a monad on ***Ord***, together with unit $\eta_A : A \to \mathbb{T}A$ defined by $\eta_A(x) = \uparrow_A \{x\}$, multiplication $\mu_A : \mathbb{T}^2 A \to \mathbb{T}A$ defined by $\mu_A(\uparrow_{\mathbb{T}A} \{\uparrow_A E_1, \ldots, \uparrow_A E_n\}) = \uparrow_A \bigcup_{i=1}^n E_i$. This is in fact a commutative monad, with mediator $d_{A,B} : \mathbb{T}A \times \mathbb{T}B \to \mathbb{T}(A \times B)$ defined by $d_{A,B}(\uparrow_A E, \uparrow_B F) = \uparrow_{A\times B} (E \times F)$. The corresponding strength maps $(x, \uparrow_B \{y_1, \ldots, y_m\})$ to $\uparrow_{A\times B} \{(x, y_1), \ldots, (x, y_m)\}$.

The required monad morphism ${\sigma_A}^0 : \mathbb{T}|A|_0 \to |\boldsymbol{T}A|_0$ maps $\uparrow_A E$ to the function $\lambda f \in [A \to \mathbf{2}] \cdot E \subseteq f^{-1}\{1\}$. (We identify truth with 1, so that $E \subseteq f^{-1}\{1\}$ denotes 1 if true, 0 if false.) The latter function is continuous, because $E$ is finite, so this indeed defines a map to $|\boldsymbol{T}A|_0$. This is clearly natural in $A$. Let us check the monad morphism laws (4). First, ${\sigma_A}^0 \circ \eta_{|A|_0}$ maps $x \in A$ to $\lambda f \in [A \to \mathbf{2}] \cdot \{x\} \subseteq f^{-1}\{1\} = \lambda f \in [A \to \mathbf{2}] \cdot f(x) = |\boldsymbol{\eta}_A|_0(x)$. Second, fix an element $\uparrow_{\mathbb{T}A} \{\uparrow_A E_1, \ldots, \uparrow_A E_n\}$ in $\mathbb{T}^2|A|_0$. This is mapped by $\mathbb{T}{\sigma_A}^0$ to $\uparrow_{|\boldsymbol{T}A|_0} \{\lambda f \cdot E_1 \subseteq f^{-1}\{1\}, \ldots, \lambda f \cdot E_n \subseteq f^{-1}\{1\}\}$; then by ${\sigma_{\boldsymbol{T}A}}^0$ to $\lambda F \in [\boldsymbol{T}A \to \mathbf{2}] \cdot \forall i, 1 \leq i \leq n \cdot (\lambda f \cdot E_i \subseteq f^{-1}\{1\}) \subseteq F^{-1}\{1\} = \lambda F \cdot \forall i, 1 \leq i \leq n \cdot F(\lambda f \cdot E_i \subseteq f^{-1}\{1\}) = 1$; then by $|\boldsymbol{\mu}_A|_0$ to $\lambda f \cdot \forall i, 1 \leq i \leq n \cdot E_i \subseteq f^{-1}\{1\}$. On the other hand, $\uparrow_{\mathbb{T}A} \{\uparrow_A E_1, \ldots, \uparrow_A E_n\}$ is mapped by $\mu_A$ to $\uparrow_A (E_1 \cup \ldots \cup E_n)$, then by ${\sigma_A}^0$ to $\lambda f \cdot E_1 \cup \ldots \cup E_n \subseteq f^{-1}\{1\}$, so the diagram on the right of (4) commutes.

We may also check that ${\sigma_A}^0$ is a strong monad morphism. Diagram (36) states that we may indeed compute the map sending $(x, \uparrow_{|A_2|_0} \{y_1, \ldots, y_m\}) \in |A_1|_0 \times \mathbb{T}|A_2|_0$ to $\lambda f \in [A_1 \times A_2 \to \mathbf{2}] \cdot \{(x, y_1), \ldots, (x, y_m)\} \subseteq f^{-1}\{1\}$ in two different ways.

Now we have at least three mono factorization systems on ***Ord***, which are much simpler as in ***Cpo***:

1 The trivial mono factorization system, which is, as usual, uninteresting.

2 The epi / regular mono factorization. Here the epis are just the surjective monotone maps, while the regular monos are embeddings, i.e., maps $m$ such that $x \leq y$ if and only if $m(x) \leq m(y)$.

It is not quite obvious that the epis are the surjective monotone maps. Here is a quick proof of the difficult implication. Assume $e$ epi from $A$ to $C$. For any two upward-closed subsets $U$, $V$ of $C$, the characteristic functions $\chi_U, \chi_V : C \to \mathbf{2}$ are monotone. If $\chi_U \circ e = \chi_V \circ e$ then $U = V$. In other words, if $e(A) \cap U = e(A) \cap V$ then $U = V$. Consider any $z \in C$, and build $U = \uparrow_C z$, $V = \uparrow_C (e(A) \cap U)$. It is easy to check that $V \subseteq U$, so $e(A) \cap V \subseteq e(A) \cap U$; the converse inclusion is obvious. So $U = V$. It follows that $z \in V$, i.e., there is an element $z' \leq z$ such that $z' \in e(A) \cap \uparrow_C z$, whence $z \in e(A)$.

The factorization of $A \xrightarrow{f} B$ goes through the middle object $C$ defined as the image $f(A)$ of $A$ by $f$, equipped with the ordering inherited from $B$.

This is a monoidal mono factorization (condition **(iii.a)**): the product of two surjective monotone maps $e_1 : A_1 \to C_1$ and $e_2 : A_2 \to C_2$ is clearly again surjective.

Condition **(vi)**, that $_^A$ preserves regular monos, is obvious from the remark that $_^A$ is a right-adjoint, so it preserves limits, in particular equalizers.

It remains to show that $\mathbb{T}$ preserves epis (condition **(iv)**). Fix an epi $e$ from $A$ to $B$ in ***Ord***. It is clear that every $\uparrow_B \{y_1, \ldots, y_n\} \in \mathbb{T}B$ is the image of $\uparrow_A \{x_1, \ldots, x_n\} \in \mathbb{T}A$ by $\mathbb{T}e$, where for each $i$, $x_i$ is any element such that $e(x_i) = y_i$. So $\mathbb{T}e$ is surjective, that is, epi.

3 The extremal epi / mono factorization. The monos are just the injective monotone maps, while the extremal epis are the quotient maps. A *quotient map* $A \xrightarrow{q} C$ is by definition the map sending $x \in A$ to its equivalence class in $C = A/\equiv$, for some equivalence relation $\equiv$ on $A$. (This is up to post-composition with an iso, naturally.) The ordering on $C$ is then given as the smallest transitive relation $\leq_C$ such that $x \leq y$ in $A$ implies $q(x) \leq_C q(y)$ in $C$. Equivalently, $q$ is a quotient map if and only if it is surjective and, for every subset $V$ of $C$, $V$ is upward-closed in $C$ if and only if $q^{-1}(V)$ is upward-closed in $A$. The factorization of $A \xrightarrow{f} B$ goes through the middle object $C$ defined as the image $f(A)$ of $A$ by $f$, equipped with the ordering $\leq_C$ defined as the smallest transitive relation such that $x \leq y$ in $A$ implies $f(x) \leq_C f(y)$ in $C$.
This is also a monoidal mono factorization (condition **(iii.a)**): if $q_1 : A_1 \to A_1/\equiv_1$ and $q_2 : A_2 \to A_2/\equiv_2$ are quotient maps, then $q_1 \times q_2$ is the quotient map from $A_1 \times A_2$ to $(A_1 \times A_2)/\equiv$, where $(x_1, x_2) \equiv (y_1, y_2)$ if and only if $x_1 \equiv_1 y_1$ and $x_2 \equiv y_2$.
$\mathbb{T}$ also preserves extremal epis (condition **(iv)**). Indeed, let $q : A \to C$ be an extremal epi, that is, a quotient map, for some equivalence relation $\equiv$ on $A$. In particular, $C = A/\equiv$. By construction, $q(x) \leq q(x')$ if and only if $x(\leq \cup \equiv)^* y$, where $(\leq \cup \equiv)^*$ is the reflexive transitive closure of the union of the two relations $\leq$ and $\equiv$. For any $y \in C$, let $x$ be any element such that $q(x) = y$. To show that $\mathbb{T}q$ is quotient, we must show that for any subset $\mathcal{U}$ of $\mathbb{T}C$, if $(\mathbb{T}q)^{-1}(\mathcal{U})$ is upward-closed then so is $\mathcal{U}$. Let $\uparrow_C \{y_1, \ldots, y_m\}$ be an element of $\mathbb{T}C$, and $\uparrow_C \{y'_1, \ldots, y'_n\}$ any larger element (wrt. $\supseteq$, i.e., a smaller subset). We may assume without loss of generality that $\{y_1, \ldots, y_m\}$ and $\{y'_1, \ldots, y'_n\}$ are antichains, and that $\{y_1, \ldots, y_m\} \leq^\sharp_C \{y'_1, \ldots, y'_n\}$: for each $j$, $1 \leq j \leq n$, there is $i$, $1 \leq i \leq m$, such that $y_i \leq_C y'_j$. Since $q$ is surjective, write $y_i = q(x_i)$, $y'_j = q(x'_j)$. In particular: $(*)$ for each $j$ there is $i$ such that $x_i(\leq \cup \equiv)^* x'_j$. Then $(\mathbb{T}q)^{-1}(\uparrow_C \{y_1, \ldots, y_m\})$ is the set of elements $x \in A$ such that $x_i(\leq \cup \equiv)^* x$ for some $i$, $1 \leq i \leq m$. Similarly, $(\mathbb{T}q)^{-1}(\uparrow_C \{y'_1, \ldots, y'_n\})$ is the set of elements $x \in A$ such that $x'_j(\leq \cup \equiv)^* x$ for some $j$, $1 \leq j \leq n$. By $(*)$ above, every such $x$ is also such that $x_i(\leq \cup \equiv)^* x$ for some $i$, $1 \leq i \leq m$. So $(\mathbb{T}q)^{-1}(\uparrow_C \{y_1, \ldots, y_m\})$ contains $(\mathbb{T}q)^{-1}(\uparrow_C \{y'_1, \ldots, y'_n\})$. Since $(\mathbb{T}q)^{-1}(\mathcal{U})$ is upward-closed (for the $\supseteq$ ordering), $(\mathbb{T}q)^{-1}(\uparrow_C \{y'_1, \ldots, y'_n\})$ must also be in $(\mathbb{T}q)^{-1}(\mathcal{U})$, that is, $\uparrow_C \{y'_1, \ldots, y'_n\}$ is in $\mathcal{U}$. So $\mathcal{U}$ is upward-closed, and $\mathbb{T}q$ is therefore quotient.
Explicitly, one can show that the equivalence relation $\equiv'$ defined by the quotient map $\mathbb{T}q$ on $\mathbb{T}A$, that is the relation defined by $\uparrow_A \{x_1, \ldots, x_m\} \equiv' \uparrow_A \{x'_1, \ldots, x'_n\}$ if and only if $\mathbb{T}q(\uparrow_A \{x_1, \ldots, x_m\}) = \mathbb{T}q(\uparrow_A \{x'_1, \ldots, x'_n\})$, can be defined equivalently by: for every $i$ there is $j$ such that $x_i(\leq \cup \equiv)^* x'_j$, and for every $j$ there is $i$ such that $x'_j(\leq \cup \equiv)^* x_i$.

This much gives us an instance of our construction for logical predicates, i.e., logical relations of arity one. As for the ***Set*** example, binary predicates are obtained by letting ***C*** be ***Cpo*** $\times$ ***Cpo***, keeping $\mathbb{C} =$ ***Ord***, defining $|A_1, A_2| = A_1 \times A_2$, ***T*** as $\mathbb{T} \times \mathbb{T}$, and the monad morphism $\sigma_{A_1,A_2}$ by $\sigma_{A_1,A_2}(\uparrow_{A_1 \times A_2} \{(x_1, y_1), \ldots, (x_m, y_m)\}) = (\lambda f_1 \in [A_1 \to

$\mathbf{2}] \cdot \{x_1, \ldots, x_m\} \subseteq f_1^{-1}\{1\}, \lambda f_2 \in [A_2 \to \mathbf{2}] \cdot \{y_1, \ldots, y_m\} \subseteq f_2^{-1}\{1\})$. Take for mediating pair $\mathbb{0}_{A_1,A_2,B_1,B_2} : |A_1, A_2| \otimes |B_1, B_2| \to |(A_1, A_2) \times (B_1, B_2)|$ defined as mapping $((x_1, x_2), (y_1, y_2))$ to $((x_1, y_1), (x_2, y_2))$, and the obvious morphism $\mathbb{i} : \mathbf{1} \to |\mathbf{1}|$. Then checking that $\sigma$ is a strong monad morphism presents no difficulty.

Let's now look at the form of the binary logical relations we get in this setting, depending on the mono factorization system we consider.

**Epi / regular mono factorization, binary case.** In this setting, a binary relation $S \overset{m}{\hookrightarrow} A_1 \times A_2$ between $A_1$ and $A_2$ is just a subset $S$ of $A_1 \times A_2$, up to iso. The ordering on $S$ is inherited from that on $A_1 \times A_2$.

Given such a subset $S \subseteq A_1 \times A_2$, with inclusion embedding $m$, our construction first builds $\sigma_{A_1,A_2} \circ \mathbb{T}m$: this maps any finitary subset $\uparrow_S \{(x_1, y_1), \ldots, (x_m, y_m)\} \in \mathbb{T}S$ (i.e., where $(x_i, y_i) \in S$ for every $i$) to $(\lambda f_1 \in [A_1 \to \mathbf{2}] \cdot \{x_1, \ldots, x_m\} \in f_1^{-1}\{1\}, \lambda f_2 \in [A_1 \to \mathbf{2}] \cdot \{y_1, \ldots, y_m\} \in f_2^{-1}\{1\})$. Then factor this arrow through an object $\widetilde{S}$, which embeds into $\boldsymbol{T}A_1 \times \boldsymbol{T}A_2$: $\widetilde{S}$ is the subset of all $(F_1, F_2) \in \boldsymbol{T}A_1 \times \boldsymbol{T}A_2$ such that there are finitely many pairs of related elements $(x_1, y_1) \in S, \ldots, (x_m, y_m) \in S$ having the property that $F_1(f_1) = 1$ if and only if $f_1$ maps every $x_i$ to 1, and $F_2(f_2) = 1$ if and only if $f_2$ maps every $y_i$ to 1.

**Extremal epi / mono factorization, binary case.** In this framework, and up to iso, a binary relation $S \overset{m}{\hookrightarrow} A_1 \times A_2$ between $A_1$ and $A_2$ is again a subset $S$ of $A_1 \times A_2$, *together* with an ordering $\preceq$ such that $(x_1, x_2) \preceq (y_1, y_2)$ implies $x_1 \leq y_1$ and $x_2 \leq y_2$. Call $(S, \preceq)$ a *sub-order* of $A_1 \times A_2$.

Given such a sub-order, it can be checked that $\widetilde{S}$ relates exactly the same functions as with the epi / regular mono factorization. Only the ordering on $\widetilde{S}$ changes: instead of being inherited from that on $\boldsymbol{T}A_1 \times \boldsymbol{T}A_2$, it is the ordering $\leq_C$ defined above, with $C = \widetilde{S}$. In other words, $\leq_{\widetilde{S}}$ is the smallest transitive relation such that $(\lambda f_1 \in [A_1 \to \mathbf{2}] \cdot \{x_1, \ldots, x_m\} \in f_1^{-1}\{1\}, \lambda f_2 \in [A_1 \to \mathbf{2}] \cdot \{y_1, \ldots, y_m\} \in f_2^{-1}\{1\}) \leq_{\widetilde{S}}$ $(\lambda f_1 \in [A_1 \to \mathbf{2}] \cdot \{x'_1, \ldots, x'_n\} \in f_1^{-1}\{1\}, \lambda f_2 \in [A_1 \to \mathbf{2}] \cdot \{y'_1, \ldots, y'_n\} \in f_2^{-1}\{1\})$ whenever $\uparrow_S\{(x_1, y_1), \ldots, (x_m, y_m)\} \supseteq \uparrow_S\{(x'_1, y'_1), \ldots, (x'_n, y'_n)\}$, that is whenever for every $j$, $1 \leq j \leq n$, there is an $i$, $1 \leq i \leq m$, such that $(x_i, y_i) \preceq (x'_j, y'_j)$.

We now make two remarks.

**Remark.** Comparing the effect of the two mono factorizations above, it is clear that they define exactly the same relations, i.e., the same sets underlying the space $\widetilde{S}$. The two constructions only differ in the orderings on $\widetilde{S}$. This is due to the fact that the set underlying the middle object $C$ in a factorization $A \overset{e}{\twoheadrightarrow} C \overset{m}{\hookrightarrow} B$ of a map $f : A \to B$ is the image of $A$ under $f$ in both cases. This need not be the case in all categories, see the epi / regular mono case in ***Cpo*** (Section 10.5).

**Remark.** The binary relation $\widetilde{S}$ that we got above, for whichever of the two mono factorizations we examined, is a bit disappointing. Indeed, it only relates functionals of the form $\lambda f \in [A \to \mathbf{2}] \cdot \{x_1, \ldots, x_m\} \in f^{-1}\{1\}$, where $A$ is $A_1$ or $A_2$; equivalently, of the form $\lambda f \in [A \to \mathbf{2}] \cdot \forall i \cdot f(x_i) = 1$. The latter only encodes *finite* non-deterministic choice, among $x_1, \ldots, x_m$, while elements of $\boldsymbol{T}A$ encode general non-deterministic choice, typically among the elements of a compact saturated subset, which is not necessarily finitary. $\widetilde{S}$ does not relate any element of $\boldsymbol{T}A$ to any other, unless it encodes finite non-determinism. The core of the problem is our choice of the category $\mathbb{C}$ as $\boldsymbol{Ord}$, which keeps all ordering information, but forgets about all continuity properties.

10.7. *Fun With Categories III: The Case $\boldsymbol{C} = \boldsymbol{FS}$, $\mathbb{C} = \boldsymbol{Top}$*

This can be repaired by choosing another category for $\mathbb{C}$, which will keep all necessary information related to continuity. The obvious choice is $\boldsymbol{Top}$, the category of topological spaces. One of the novelties here is that for one choice of mono factorization, the pseudoepis won't be surjective maps. Another one is that we shall see cases where $\mathbb{T}$ maps relevant pseudoepis to pseudoepis but does not preserve pseudoepis.

Before we delve into the subject proper, we must note that choosing $\mathbb{C} = \boldsymbol{Top}$ incurs several problems.

First, $\boldsymbol{Top}$ is not cartesian closed. This is not too serious. We can for example define a logical relation at arrow types $\tau_1 \to \tau_2$ that would be defined only when the denotation of $\tau_1$ is *exponentiable*. Recall that a topological space $X$ is exponentiable if and only if the exponential object $Y^X$ exists for every topological space $Y$. It is well-known that the exponentiable spaces are exactly the *core-compact* spaces (Escardó and Heckmann, 2002), i.e., those whose lattice of opens is continuous. In particular, every continuous cpo is locally compact, hence core-compact, when seen as a topological space. Another possibility is to replace $\boldsymbol{Top}$ by one of its cartesian-closed subcategories $\boldsymbol{Top}_{\mathcal{C}}$, of which there are many: Day's Theorem indeed states that, given any class $\mathcal{C}$ of topological spaces, the category $\boldsymbol{Top}_{\mathcal{C}}$ of so-called $\mathcal{C}$-generated spaces, with continuous maps as morphisms, is cartesian-closed as soon as $\mathcal{C}$ is a productive class. See (Escardó et al., 2004) for details. Examples of productive classes include continuous cpos (*domain generated spaces*), or compact spaces (*compactly generated spaces*). In all these cases, $\boldsymbol{Top}_{\mathcal{C}}$ is coreflective in $\boldsymbol{Top}$, meaning that there is a functor $\mathcal{C} : \boldsymbol{Top} \to \boldsymbol{Top}_{\mathcal{C}}$ that is right adjoint to the inclusion functor $\subseteq$ from $\boldsymbol{Top}_{\mathcal{C}}$ into $\boldsymbol{Top}$. This immediately implies that any mono factorization system on $\boldsymbol{Top}$ transports to one on $\boldsymbol{Top}_{\mathcal{C}}$: if $A \xrightarrow{f} B$ in $\boldsymbol{Top}_{\mathcal{C}}$ factors as $A \overset{e}{\twoheadrightarrow} C \overset{m}{\hookrightarrow} B$ in $\boldsymbol{Top}$, then it factors as $A \overset{e}{\twoheadrightarrow} \mathcal{C}(C) \overset{m}{\hookrightarrow} B$ in $\boldsymbol{Top}_{\mathcal{C}}$, and $m$ is clearly again a mono in $\boldsymbol{Top}_{\mathcal{C}}$. Since product $\times_{\mathcal{C}}$ in $\boldsymbol{Top}_{\mathcal{C}}$ is given by $A \times_{\mathcal{C}} B = \mathcal{C}(A \times B)$ (Escardó and Heckmann, 2002, Section 5), the adjunction $\subseteq \dashv \mathcal{C}$ is monoidal: $\mathcal{C}$ preserves product on the nose. It follows that we can transport a strong monad on $\boldsymbol{Top}$ to one on $\boldsymbol{Top}_{\mathcal{C}}$, gaining a cartesian-closed structure along the way.

We will refrain from doing so *explicitly*, since we wouldn't learn much. We shall therefore concentrate on the $\boldsymbol{Top}$ case. Note in passing that, taking $\mathcal{C}$ consisting of just the Sierpiński space $\mathbf{2}$, $\boldsymbol{Top}_{\mathcal{C}}$ will be the category of Alexandroff spaces (Escardó and Heckmann, 2002, Example (2)), which is equivalent to $\boldsymbol{Ord}$; in other words, taking $\mathcal{C} = \{\mathbf{2}\}$ gives back the construction of Section 10.6.

Second, the product of cpos in $\boldsymbol{Top}$ does not coincide with their product in $\boldsymbol{Cpo}$. Precisely, any cpo $A$ defines a topological space $|A|$, where $A$ is equipped with its Scott topology. But the product topology, i.e., that of $|A| \times |B|$, is in general strictly coarser than the Scott topology on the product, i.e., the topology on $|A \times B|$. The obvious choice of mediating morphism $\theta_{A,B} : |A| \times |B| \to |A \times B|$, namely the identity map, will therefore fail to be continuous in general. We repair this by considering $\boldsymbol{C}$ to be some cartesian-closed subcategory of continuous cpos, instead of the whole of $\boldsymbol{Cpo}$. Indeed, in this case it is well-known that the Scott topology and the product topologies agree: $|A| \times |B| = |A \times B|$.

There are many choices for $\boldsymbol{C}$, then. The first, historically, is the category of Scott domains, i.e., algebraic bounded-complete cpos. Other candidates are Jung's category $\boldsymbol{FS}$ of FS-domains, or that $\boldsymbol{cL}$ of continuous L-domains (Jung, 1990). We observe that $\boldsymbol{T}A$ is an FS-domain as soon as $A$ is, because $\boldsymbol{T}A$ is a closed subset of $[[A \to \mathbf{2}] \to \mathbf{2}]$, and closed subsets of FS-domains are again FS-domains. So $\boldsymbol{T}$ defines a monad on $\boldsymbol{FS}$. We shall therefore use $\boldsymbol{C} = \boldsymbol{FS}$ is the rest of this section. It won't be necessary to know exactly what FS-domains really are. The interested reader is refered to (Jung, 1990).

We take the corresponding monad $\mathbb{T}$ on $\mathbb{C} = \boldsymbol{Top}$ as follows. Let $\mathbb{T}A$ be the set of continuous linear non-trivial second-order predicates on $A$, just as $\boldsymbol{T}A$, but equipped with the weak topology instead of the Scott topology. The *weak topology* is the topology generated by the basis of opens $\Box O = \{F \in \mathbb{T}A | F(\chi_O) = 1\}$, where $\chi_O$ is the indicator function of the open $O$ of $A$. Note in passing that *any* $f \in [A \to \mathbf{2}]$ must be of the form $\chi_O$ for some open $O$ of $A$. Our choice of the weak topology is justified by the fact that $\mathbb{T}$ will preserve pseudoepis (condition **(iv)**), see below.

The weak topology is in general coarser than the Scott topology, which thwarts the existence of a monad morphism $\sigma_A : \mathbb{T}|A| \to |\boldsymbol{T}A|$. (The identity map does not fit, as it would fail to be continuous.) However, when $A$ is a continuous cpo, the two topologies coincide—another good reason to take $\boldsymbol{C}$ to consist solely of continuous cpos. The classical argument is as follows. Since $\boldsymbol{T}A$ is a continuous cpo, its Scott topology has a basis consisting of opens of the form $\Uparrow F$, where $F \in \boldsymbol{T}A$, $\Uparrow F = \{G \in \boldsymbol{T}A | F \ll G\}$, and $\ll$ is the usual way-below relation: $F \ll G$ if and only if every directed family whose supremum is above $G$ contains an element above $F$. By the Hoffman-Mislove theorem, we may equate linear functionals with upward-closed compacts, and then $Q \ll Q'$ if and only if the interior $\mathring{Q}$ of $Q$ contains $Q'$. Then $\Uparrow Q = \{Q' | Q' \subseteq \mathring{Q}\} = \Box \mathring{Q}$ is a weak open. It follows that, when $A$ is continuous, the identity map from $\mathbb{T}|A|$ to $|\boldsymbol{T}A|$ is continuous, and therefore defines a monad morphism from $\boldsymbol{T}$ to $\mathbb{T}$.

Now we have at least *four* mono factorization systems on $\boldsymbol{Top}$:

1 The trivial mono factorization system, which is, as usual, uninteresting.

2 The epi / regular mono factorization. This is close to the corresponding one in $\boldsymbol{Cpo}$ or in $\boldsymbol{Ord}$: regular monos are embeddings of closed subsets (i.e., $m : C \hookrightarrow B$ is a regular mono if and only if $m$ is a homeomorphism onto a closed subset of $B$), and epis are continuous maps with dense image. Note that $e : A \to C$ is epi if and only if $C$ equals the closure of $e(A)$ in $C$, or equivalently, $e^{-1}(V)$ is non-empty for every non-empty open $V$ of $C$.

3 The extremal epi / mono factorization. This is much simpler to describe as in $\boldsymbol{Cpo}$, and is similar to $\boldsymbol{Ord}$. The monos are just the injective continuous maps, while the extremal epis are the quotient maps. A *quotient map* $A \stackrel{q}{\twoheadrightarrow} C$ is by definition the map sending $x \in A$ to its equivalence class in $C = A/\equiv$, for some equivalence relation $\equiv$ on $A$. (Again, up to postcomposition with isos.) Equivalently, $q$ is a quotient map if and only if $q$ is surjective and, for every subset $V$ of $C$, $V$ is open in $C$ if and only if $q^{-1}(V)$ is open in $A$.

4 What we shall call the *intermediate* mono factorization: the pseudoepis are the continuous surjective maps, and the relevant monos are the embeddings of subspaces (i.e., $m : C \hookrightarrow B$ is an embedding, that is, a homeomorphism onto some subspace, not necessarily closed, of $B$).

All of these are candidates for our construction. As usual, the main property to check is that $\mathbb{T}$ should preserve pseudoepis, or at least map relevant pseudoepis to pseudoepis.

**Epi / regular mono factorization.** Let $e$ be a pseudoepi from $A$ to $C$, that is, an epi: for every non-empty open $V$ of $C$, $e^{-1}(V)$ is non-empty. It suffices to show that $\mathbb{T}e^{-1}(\Box O)$ is non-empty for any non-empty open $O$ of $C$. (This is the point in choosing to equip $\mathbb{T}C$ with the weak topology.) Now $\mathbb{T}e^{-1}(\Box O)$ is the set of all $F \in \mathbb{T}A$ such that $\lambda f \in [C \to \mathbf{2}] \cdot F(f \circ e)$ is in $\Box O$, i.e., such that $F(\chi_O \circ e) = 1$, or equivalently $F(\chi_{e^{-1}(O)}) = 1$. By assumption, $e^{-1}(O)$ is non-empty: let $x \in e^{-1}(O)$, then $F = \eta_A(x) = \lambda f \in [A \to \mathbf{2}] \cdot f(x)$ is in $\mathbb{T}e^{-1}(\Box O)$.

Condition **(iv)** obtains. We also have to check condition **(iii.a)**, that this mono factorization system is monoidal. But products of epis are always epi, so this is clear.

**Intermediate mono factorization.** Here pseudoepis are surjective continuous maps. This is one case where $\mathbb{T}$ does *not* preserve pseudoepis. Take $A = \mathbb{N}$ with the discrete topology, and $C = \mathbb{N}$ with the cofinite topology, whose opens are $\emptyset$ and all complements of finite sets of integers. Now consider $e : A \to C$ the identity map. Since the topology of $A$ is finer than that of $C$, $e$ is continuous; $e$ is also clearly surjective. We claim that $\mathbb{T}e$ is not surjective. It is an easy exercise to show that the elements of $\mathbb{T}A$ are all of the form $\lambda f \in [A \to \mathbf{2}] \cdot E \subseteq f^{-1}\{1\}$, where $E$ is some finite non-empty subset of $A$. (And this comes from free if we consider that $A$ is sober, from the correspondence between $\mathbb{T}A$ and $\mathcal{Q}(A)$, and the fact that the upward-closed compacts of $A$ are the finite subsets $E$ of $A$.) Now consider the functional $G \in \mathbb{T}C$ that maps $f \in [C \to \mathbf{2}]$ to 1 if $f$ is not identically 0, and maps the zero function to 0. $G$ is trivially continuous, and preserves $\top$. That $G$ preserves binary infima rests on the fact that the intersection of any two non-empty opens is again non-empty in $C$. Assume now that $G$ is the image by $\mathbb{T}e$ of some functional $\lambda f \in [A \to \mathbf{2}] \cdot E \subseteq f^{-1}\{1\}$, in other words, $G = \lambda g \in [C \to \mathbf{2}] \cdot E \subseteq (g \circ e)^{-1}\{1\}$. For any integer $n \in \mathbb{N}$, $G(\chi_{C \setminus \{n\}}) = 1$, so $E$ cannot contain $n$; i.e., $E$ is empty, a contradiction.

However, $\mathbb{T}$ maps *relevant* pseudoepis to pseudoepis. The crucial, albeit trivial, observation is:

**Fact 10.1.** If $\mathbb{T}$ preserves relevant monos and $\sigma_A$ is a relevant mono, then all relevant pseudoepis are iso. In particular, $\mathbb{T}$ maps relevant pseudoepis to isos, hence to pseudoepis.

Indeed, a relevant pseudoepi is the pseudoepi part of the factorization of $\sigma_A \circ \mathbb{T}m$, where $m$ is a relevant mono. Since by assumption $\sigma_A \circ \mathbb{T}m$ is a relevant mono, we conclude by uniqueness of factorizations up to iso.

In our case, $\sigma_A$ is the identity map between $\mathbb{T}|A|$ and $|\boldsymbol{T}A|$; remember these two coincide because the Scott topology agrees with the weak topology on continuous cpos. It remains to show that $\mathbb{T}m$ is a relevant mono as soon as $m : C \to B$ is. Remember that a relevant mono $m$ is a homeomorphism onto its image, i.e., an injective, continuous and open map. $\mathbb{T}m$ is continuous. It is injective, too: if $\mathbb{T}m(F) = \mathbb{T}m(F')$, then for every open $O'$ of $A$, $F(\chi_{0'} \circ m) = F'(\chi_{0'} \circ m)$. In particular, for every open $O$ of $C$, taking $O' = m(O)$ (since $m$ is an open map), $F(\chi_{m(O)} \circ m) = F'(\chi_{m(O)} \circ m)$. Now $\chi_{m(O)} \circ m = \chi_{m^{-1}(m(O))} = \chi_O$ since $m$ is injective, so $F(\chi_0) = F'(\chi_O)$: so $F = F'$. Finally, $\mathbb{T}m$ is open. Since $\Box O \cap \Box O' = \Box(O \cap O')$, the sets $\Box O$ form a basis of the weak topology, that is, every open subset of $\mathbb{T}C$ is a union $\bigcup_{i \in I} \Box O_i$. It is easy to check that $\mathbb{T}m(\bigcup_{i \in I} \Box O_i)$ equals $\bigcup_{i \in I} \Box m(O_i)$, which is clearly open. So $\mathbb{T}m$ is a relevant mono.

We have therefore established condition **(iv)**: $\mathbb{T}$ maps relevant pseudoepis to pseudoepis, although $\mathbb{T}$ does not map pseudoepis to pseudoepis.

To lift the cartesian structure to the subscone, we also have to check condition **(iii.a)**, that this mono factorization system is monoidal. This is again clear: the product of two surjective continuous maps is again surjective and continuous.

**Extremal epi / mono factorization.** Since $\mathbb{T}$ does not preserve surjective maps, it is unlikely that it will preserve extremal epis. Moreover, Fact 10.1 does not apply, since $\mathbb{T}$ does not preserve injective maps. Take indeed $m : C \to B$ be the identity from $C = \mathbb{N}$ with the discrete topology, to $B = \mathbb{N}$ with its Alexandroff topology. Then $\mathbb{T}m$ maps $\lambda f \cdot \{2\} \subseteq f^{-1}\{1\}$ and $\lambda f \cdot \{2, 3\} \subseteq f^{-1}\{1\}$ to the same element $\lambda g \cdot \{2\} \subseteq g^{-1}\{1\}$, because for any Alexandroff-continuous (i.e., monotone) function $g : B \to \mathbf{2}$, $g(2) = 1$ if and only if $g(2) = g(3) = 1$.

Let us now consider the form of binary logical relations in the above topological settings. Again, we look at each available mono factorization. We will not consider the extremal epi / mono factorization. We will in fact also avoid the intermediate mono factorization system. In the intermediate system, $\mathbb{T}$ does not preserve pseudoepis, and Fact 10.1 does not apply, because $\sigma_{A_1,A_2}$ will not be a relevant mono in general.

In the binary case, $|A_1, A_2| = A_1 \times A_2$ (there is no ambiguity in writing $\times$ here, as products of continuous cpos coincide with topological products), and for every FS-domains $A_1$ and $A_2$, $\sigma_{A_1,A_2}$ maps $F \in \mathbb{T}(A_1 \times A_2)$ to the pair of maps $\lambda f_1 \in [A_1 \to \mathbf{2}] \cdot F(f_1 \circ \pi_1)$ and $\lambda f_2 \in [A_2 \to \mathbf{2}] \cdot F(f_2 \circ \pi_2)$. Here $\pi_1$ and $\pi_2$ denote first and second projections, respectively.

**Epi / regular mono factorization, binary case.** A binary relation $S \overset{m}{\hookrightarrow} A_1 \times A_2$ between $A_1$ and $A_2$ is, up to iso, a closed subset $S$ of $A_1 \times A_2$. It is easy to check that our construction builds a lifted relation $\widetilde{S}$ between $\mathbb{T}A_1$ and $\mathbb{T}A_2$ defined by $(F_1, F_2) \in \widetilde{S}$ if and only if $(F_1, F_2)$ is in the closure of the set

$$\widetilde{S}_0 = \Big\{(G_1, G_2) \in \mathbb{T}A_1 \times \mathbb{T}A_2 | \exists G \in \mathbb{T}S \cdot \quad (63)$$

$$\forall f_1 \in [A_1 \to \mathbf{2}] \cdot G_1(f_1) = G(\lambda(x_1, x_2) \in S \cdot f_1(x_1))$$

$$\wedge \forall f_2 \in [A_2 \to \mathbf{2}] \cdot G_2(f_2) = G(\lambda(x_1, x_2) \in S \cdot f_2(x_2))\Big\}$$

It is easier to understand this through the Hofmann-Mislove isomorphism. Any element $G_i$ of $\mathbb{T}A_i$ can be written as $\lambda\chi_O \cdot Q_i \subseteq O$ for some non-empty compact upward-closed subset $Q_i$ of $A_i$, $i = 1, 2$. Similarly $G$ corresponds to some non-empty compact upward-closed subset $Q$ of $S$. The condition $\forall f_1 \in [A_1 \to \mathbf{2}] \cdot G_1(f_1) = G(\lambda(x_1, x_2) \in S \cdot f_1(x_1))$ is then equivalent to: for every open subset $O_1$ of $A_1$, $Q_1 \subseteq O_1$ if and only if $Q \subseteq (O_1 \times A_2) \cap S$. Since the upward-closed subsets of a topological space are exactly those that are the intersection of all opens containing them, this is equivalent to $Q \subseteq (Q_1 \times A_2) \cap S$. Summing this up,

$$\begin{aligned} \widetilde{S}_0 \;=\; & \{(\lambda\chi_O \cdot Q_1 \subseteq O, \lambda\chi_O \cdot Q_2 \subseteq O) \mid \\ & \quad \exists Q \text{ non-empty compact upward-closed subset of } S \cdot Q \subseteq (Q_1 \times Q_2) \cap S\} \end{aligned}$$

By Tychonoff's Theorem, $Q_1 \times Q_2$ is compact, so $(Q_1 \times Q_2) \cap S$ is compact since $S$ is closed. Since $Q_1 \times Q_2$ is upward-closed in $A_1 \times A_2$, $(Q_1 \times Q_2) \cap S$ is upward-closed in $S$, so in fact $\widetilde{S}_0$ consists of those pairs $(\lambda\chi_O \cdot Q_1 \subseteq O, \lambda\chi_O \cdot Q_2 \subseteq O)$ such that $(Q_1 \times Q_2) \cap S$ is non-empty. In other words, and up to the Hofmann-Mislove Theorem, $\widetilde{S}_0$ relates those compacts $Q_1$ and $Q_2$ such that $Q_1$ contains an element $x_1$ and $Q_2$ contains $x_2$ so that $(x_1, x_2)$ are related by $S$.

Remember, then, that $\widetilde{S}$ is the topological closure of $\widetilde{S}_0$. However, we claim that $\widetilde{S}_0$ is closed in the weak topology. To show this, it is enough to show that $\widetilde{S}_0$ is Scott-closed, since the weak topology and the Scott topology coincide on $\mathbb{T}A$ for any continuous cpo $A$. First, $\widetilde{S}_0$ is downward-closed (with respect to the product $\supseteq \times \supseteq$ of reversed inclusion). Second, if $(Q_{1i}, Q_{2i})$, $i \in I$, forms a directed family (wrt. $\supseteq$) of pairs of non-empty compact upward-closed subsets such that $(Q_{1i} \times Q_{2i}) \cap S \neq \emptyset$ for all $i$, we note that $((Q_{1i} \times Q_{2i}) \cap S)_{i \in I}$ forms a directed family of non-empty compact subsets of $A_1 \times A_2$, because $S$ is closed in $A_1 \times A_2$. Since $A_1 \times A_2$ is sober, its least upper bound (its intersection) is again non-empty and compact, see e.g., (Abramsky and Jung, 1994, Corollary 7.2.11). It follows that $\left(\bigcap_{i \in I} Q_{1i} \times \bigcap_{i \in I} Q_{2i}\right) \cap S = \bigcap_{i \in I} (Q_{1i} \times Q_{2i}) \cap S$, so the least upper bound $\bigcap_{i \in I} Q_{1i} \times \bigcap_{i \in I} Q_{2i}$ is again in $\widetilde{S}_0$. We conclude that $\widetilde{S}_0$ is Scott-closed, hence closed, so that $\widetilde{S} = \widetilde{S}_0$.

Note that $\widetilde{S}$ does *not have the form of a bisimulation relation*; rather, $Q_1$ and $Q_2$ are related by $\widetilde{S}$ if and only if $Q_1$ and $Q_2$ have related refinements, where a *refinement* $x$ of a compact $Q$ is an element $x \in Q$, i.e., if $\uparrow x$ is an element above $Q$ in $\mathcal{Q}(X)$.

Transposing this back to the realm of functionals, we see that $(G_1, G_2) \in \widetilde{S}$ if and only if there is a refinement $x_1$ of $G_1$, i.e., an element $x_1$ of $A_1$ such that $G_1 \leq \eta_{A_1}(x_1)$, and a refinement $x_2$ of $G_2$ such that $(x_1, x_2) \in S$.

It would have been tempting to guess that $\widetilde{S}$ would be characterized by the following relation $\widehat{\widehat{S}}$, in the style of observational equivalence. Define $\widehat{S}$ as the set of all $(f_1, f_2)$ such that whenever $(x_1, x_2) \in S$ then $f_1(x_1) = f_2(x_2)$; then $\widehat{\widehat{S}}$ is the set of $(G_1, G_2)$ such that whenever $(f_1, f_2) \in \widehat{S}$, $G_1(f_1) = G_2(f_2)$. Clearly, $(G_1, G_2) \in \widetilde{S}_0$ implies $(G_1, G_2) \in \widehat{\widehat{S}}$. However, the converse does not seem to hold.

10.8. *Fun With Categories IV: Metric Spaces*

Among the degrees of freedom afforded by our constructions, observe that we have the option to consider categories that are not cartesian-closed, but, say, just symmetric monoidal closed.

The categories of ultrametric spaces, and of complete ultrametric spaces, are cartesian-closed. It is even possible to equip these categories with a probability monad. The main difficulty is to show that the space of all probability measures on an ultrametric space can be equipped with a distance that makes it ultrametric. See (de Vink and Rutten, 1999, Lemma 4.3). Let us take the stance of working inside the larger category $\boldsymbol{Met}$ of metric spaces—for whatever practical reason. Our reason is to provide an example: we wish to show how our construction works on the symmetric monoidal closed, but not cartesian-closed, category $\boldsymbol{Met}$.

Precisely, we consider the category $\boldsymbol{Met}$ of metric spaces and non-expansive maps. We write $d_A$ for the distance on $A$, and allow distances to take the value $+\infty$. (This is as in (Lawvere, 1973), who considers additional relaxations on the notion of distance.) In other words, a distance $d_A$ on $A$ is any function from $A$ to $\mathbb{R}_+ \cup \{+\infty\}$, such that $d_A(x, y) = 0$ if and only if $x = y$, for every $x, y \in A$; $d_A(x, y) = d_A(y, x)$ for every $x, y \in A$; and $d_A(x, z) \leq d_A(x, y) + d_A(y, z)$ for every $x, y, z \in A$. A map $f$ is *non-expansive* from $A$ to $B$ if and only if $d_B(f(x), f(x')) \leq d_A(x, x')$ for every $x, x' \in A$.

Every metric space $A$ has a topology, generated by its open balls $\mathcal{B}(x, \epsilon) = \{y | d_A(x, y) < \epsilon\}$, for which it is Hausdorff. There is a cartesian product on $\boldsymbol{Met}$: $A \times B$ is the set of pairs $(x, y)$, $x \in A$, $y \in B$, with distance $d_{A \times B}$ defined by $d_{A\times B}((x, y), (x', y')) = \max(d_A(x, x'), d_B(y, y'))$. There is no corresponding notion of exponential.

On the other hand, $\boldsymbol{Met}$ is symmetric monoidal closed. The tensor product $A \otimes B$ is the set of pairs $(x, y)$, $x \in A$, $y \in B$, with distance $d_{A\otimes B}$ defined by $d_{A\otimes B}((x, y), (x', y')) = d_A(x, x') + d_B(y, y')$. The exponential $C^B$ is the set of non-expansive maps $f$ from $B$ to $C$, with distance $d_{C^B}(f, f') = \sup_{x\in B} d_C(f(x), f'(x))$. (This is well-defined because we allow distances to take the value $+\infty$.) The underlying topology of $A \otimes B$, as well as of $A \times B$, is the product topology of $A$ and $B$.

As far as monads are concerned, we shall look at the probability monad on $\boldsymbol{Met}$. I.e., we consider general, not just discrete probabilities. This will provide a non-trivial example of our construction. A *measure* on $A$ is by convention a measure on the Borel $\sigma$-algebra of its topology. A *probability* measure maps $A$ to 1. A natural choice for a probability monad $\mathcal{M}_1$ on $\boldsymbol{Met}$ is to let $\mathcal{M}_1(A)$ be the set of all probability measures on $A$, equipped with the *Hutchinson metric*:

$$d_{\boldsymbol{T}A}(\boldsymbol{\nu}, \boldsymbol{\xi}) = \sup_{g\in \boldsymbol{Met}(A,[0,1])} \left| \int_{x\in A} g(x) d\boldsymbol{\nu} - \int_{x\in A} g(x) d\boldsymbol{\xi} \right|$$

making $\mathcal{M}_1(A)$ a metric space. (Some call it the Kantorovitch metric; one may also call it the $L^1$ metric.) It can be checked that $\mathcal{M}_1$ gives rise to a monoidal monad. We skip the details.

Now take $\mathbb{C} = \boldsymbol{C} = \boldsymbol{Met}$. There is a natural epi-mono factorization on $\boldsymbol{Met}$: relevant monos are *isometric embeddings*, that is, maps $m : C \to B$ such that $d_B(m(x), m(y)) =$ $d_C(x, y)$ for all $x, y \in C$. Pseudoepis from $A$ to $C$ are *surjective* non-expansive maps from $A$ to $C$. It is fairly easy to see that this yields a monoidal mono factorization system on $\boldsymbol{Met}$.

It is then tempting to choose $\mathbb{T}$ and $\boldsymbol{T}$ to be $\mathcal{M}_1$, but this would expose some difficulties. Indeed, we know that $\mathcal{M}_1$ preserves pseudoepis (condition **(iv)**), that is, surjective non-expansive maps, in some cases but not in general. E.g., if $f$ is continuous surjective from $A$ to $B$, and both $A$ and $B$ are compact Hausdorff, then $\mathcal{M}_1(f)$ is surjective (Bourbaki, 1969, 2.4, Lemma 1); similarly if the underlying topological space of $A$ is an analytic space (Bourbaki, 1969, 2.4, Proposition 9).

Instead, we choose $\mathbb{T}$ as a continuation-like monad that is as close to $\mathcal{M}_1$ as possible. (Remember the $\mathcal{Q}$ and $\mathcal{U}_*$ monads in $\boldsymbol{Top}$.) Let $\mathcal{C}_b(A)$ be the vector space of all bounded non-expansive functions from the metric space $A$ to $\mathbb{R}$, with the sup norm: for all $g \in \mathcal{C}_b(A)$, $||g|| = \sup_{x \in A} |g(x)|$. Let $\mathcal{L}_A$ be the vector space of all continuous linear forms on $\mathcal{C}_b(A)$, that is, of all continuous linear functions $F : \mathcal{C}_b(A) \to \mathbb{R}$. Recall that a linear function $F$ from $\mathcal{C}_b(A)$ to $\mathbb{R}$ is continuous if and only if it is Lipschitz, i.e., if and only if $||F|| = \sup_{||g||=1} |F(g)|$ is a well-defined real (i.e., not $+\infty$). Then $||F||$ is called the *norm* of $F$. Let $\mathcal{L}^1_A$ be the subset of those $F$ in $\mathcal{L}_A$ of norm 1. Finally, let $\mathbb{T}A$ be the subset of those $F$ in $\mathcal{L}^1_A$ which are positive, i.e., such that $g \geq 0$ implies $F(g) \geq 0$. Both $\mathcal{L}^1_A$ and $\mathbb{T}A$ inherit the distance on $\mathcal{L}_A$ that is induced by norm. Equivalently, $\mathbb{T}A$ is a metric space whose distance is defined by $d_{\mathbb{T}A}(F, G) = \sup_{g \in \boldsymbol{Met}(A,[-1,1])} |F(g) - G(g)|$. $\mathbb{T}$ then defines an endofunctor on $\boldsymbol{Met}$ by letting $\mathbb{T}f$ map $F \in \mathbb{T}A$ to $\lambda g \in \mathcal{C}_b(B) \cdot F(g \circ f)$, for every non-expansive map $f$ from $A$ to $B$.

The space $\mathcal{M}_1(A)$ of all probability measures on $A$ embeds in $\mathbb{T}A$ by $\boldsymbol{\nu} \mapsto \lambda g \in \mathcal{C}_b(A) \cdot \int_{x \in A} g(x) d\boldsymbol{\nu}$. In fact, the distance on probability measures on $A$ inherited from the metric structure on $\mathbb{T}A$ is a slight variant of the Hutchinson metric introduced above. Moreover, if $A$ is compact, then every element of $\mathbb{T}A$ arises from a probability measure in this way, by the Riesz Representation Theorem. This makes $\mathbb{T}A$ a metric space that is arguably close enough to a space of probability measures.

The point of this definition of $\mathbb{T}$ is that $\mathbb{T}$ preserves pseudoepis in $\boldsymbol{Met}$. The proof is similar to that of (Bourbaki, 1969, 2.4, Lemma 1).

**Lemma 10.2.** If $f$ is any surjective non-expansive map from $A$ to $B$, then $\mathbb{T}f$ is surjective from $\mathbb{T}A$ to $\mathbb{T}B$.

*Proof.* Let $\lambda_f$ be the function mapping $g \in \mathcal{C}_b(B)$ to $g \circ f$. Note that $\lambda_f(g)$ is bounded and non-expansive, hence in $\mathcal{C}_b(A)$. Also, $||\lambda_f(g)|| = \sup_{x \in A} |g(f(x))| = \sup_{y \in B} |g(y)|$ (since $f$ is surjective) $= ||g||$, so that $\lambda_f$ is an isometric embedding of $\mathcal{C}_b(B)$ into $\mathcal{C}_b(A)$.

Let $H$ be the range of $\lambda_f$ in $\mathcal{C}_b(A)$. $H$ is a linear subspace of $\mathcal{C}_b(A)$, which is by construction isometrically isomorphic to $\mathcal{C}_b(B)$. In particular, $\lambda_f^{-1}$ is a continuous linear map from $H$ to $\mathcal{C}_b(B)$.

Given $G \in \mathbb{T}B$, $G \circ \lambda_f^{-1}$ is therefore a continuous linear form on $H$. By the Hahn-Banach Theorem, $G \circ \lambda_f^{-1}$ can be extended to a continuous linear form $F$ on $\mathcal{C}_b(A)$, with the same norm, i.e., $||F|| = ||G \circ \lambda_f^{-1}||$. The latter means that $||F|| = \sup_{h \in \mathcal{C}_b(A), ||h||=1} |F(h)| = \sup_{h \in H, ||h||=1} |G(\lambda_f^{-1}(h))| = \sup_{g \in \mathcal{C}_b(B), ||g \circ f||=1} |G(g)| = ||G|| = 1$.

As far as positivity is concerned, we first note that $||G|| = G(\lambda y \in B \cdot 1)$. Indeed, first, $||\lambda y \in B \cdot 1|| = 1$, so $||G|| \geq G(\lambda y \in B \cdot 1)$. To show the converse inequality, note first that for every $g \in \mathcal{C}_b(B)$, we may write $g$ as the difference $g_+ - g_-$ of two positive functions, so $|G(g)| = |G(g_+) - G(g_-)| \leq \max(G(g_+), G(g_-))$ (since $G(g_+), G(g_-) \geq 0$); if $||g|| = 1$, then $g_+(y) \leq 1$ and $g_-(y) \leq 1$ for all $y \in B$, so $|G(g)| \leq G(\lambda y \in B \cdot 1)$; taking the sup over all $g$, $||G|| \leq G(\lambda y \in B \cdot 1)$.

Then $F(\lambda x \in A \cdot 1) = F(\lambda_f(\lambda y \in B \cdot 1)) = G(\lambda y \in B \cdot 1)$. So $F(\lambda x \in A \cdot 1) = ||G|| = ||F||$. If $F$ was not positive, there would be some $g \in \mathcal{C}_b(A)$, $g \geq 0$, with $F(g) < 0$. We may assume without loss of generality that $g \leq 1$, so $\lambda x \in A \cdot 1 - g(x)$ is of norm at most 1. But then $||F|| \geq F(\lambda x \in A \cdot 1 - g(x)) = F(\lambda x \in A \cdot 1) - F(g) > F(\lambda x \in A \cdot 1)$, a contradiction. So $F$ is positive.

Finally, $\mathbb{T}f(F) = \lambda g \in \mathcal{C}_b(B) \cdot F(g \circ f) = \lambda g \in \mathcal{C}_b(B) \cdot G(\eta(g \circ f)) = \lambda g \in \mathcal{C}_b(B) \cdot G(g) = G$. Since $G$ is arbitrary, $\mathbb{T}f$ is surjective. □

$\mathbb{T}$ can be used to form a monoidal monad on ***Met***. Define $\eta_A$ as the function mapping $x \in A$ to $\lambda g \in \mathcal{C}_b(A) \cdot g(x)$. This is natural in $A$, and non-expansive in $x$. Define $\mu_A$ as mapping $F \in \mathbb{T}^2 A$ to $\lambda g \in \mathcal{C}_b(\mathbb{T}A) \cdot F(\lambda G \in \mathbb{T}A \cdot G(g))$. It requires a bit more work to show that $\mu_A$ is well-defined and non-expansive in $F$; it is however clear that this is natural in $A$. Furthermore, the monad laws are satisfied, so that $(\mathbb{T}, \eta, \mu)$ is a monad on ***Met***. The strength $t_{A,B}$ maps $(x, G) \in A \otimes \mathbb{T}B$ to $\lambda h \in \mathcal{C}_b(A \otimes B) \cdot G(\lambda y \in B \cdot h(x, y))$.

To define a notion of binary metric logical relations, we take $\boldsymbol{C} = \boldsymbol{Met} \times \boldsymbol{Met}$, define a monad $\boldsymbol{T}$ on $\boldsymbol{C}$ pointwise (i.e., $\boldsymbol{T}(A_1, A_2) = (\mathbb{T}A_1, \mathbb{T}A_2)$). It remains to define a monad morphism from $\boldsymbol{T}$ to $\mathbb{T}$.

Define $|A_1, A_2|$ as the cartesian product $A_1 \times A_2$, much as in ***Set***. Note that the only difference between cartesian product and tensor product is that their distances differ: $d_{A_1 \times A_2}((x, y), (x', y'))$ is the max of the distances $d_{A_1}(x, x')$ and $d_{A_2}(y, y')$, while $d_{A_1 \otimes A_2}((x, y), (x', y'))$ is the sum of the same distances.

Similarly as in ***Set***, define the monad morphism $\sigma_{A_1,A_2}$ as $\langle \mathbb{T}\pi_1, \mathbb{T}\pi_2 \rangle$. That is, for every $F \in \mathbb{T}(A_1 \times A_2)$, $\sigma_{A_1,A_2}(F) = (\lambda g_1 \in \mathcal{C}_b(A_1) \cdot F(\lambda(x, y) \cdot g_1(x)), \lambda g_2 \in \mathcal{C}_b(A_2) \cdot F(\lambda(x, y) \cdot g_2(y)))$. We let the reader check all required properties. Then $(|_|, \sigma)$ is a monoidal monad morphism. It is also easy to check that ***Met*** has pullbacks: the pullback of $f : A \to C$ and $g : B \to C$ is the subspace of $A \times B$ consisting of all $(x, y)$ such that $f(x) = g(y)$; and $_^A$ preserves relevant monos: $_^A$ maps $f : B \to C$ to $\lambda g \in B^A \cdot f \circ g$, which is an isometric embedding as soon as $f$ is.

This yields the following notion of binary metric logical relation $[\![\tau]\!]$ between metric spaces $[\![\tau]\!]_1$ and $[\![\tau]\!]_2$ indexed by types in a linear version of Moggi's meta-language. The types are:

$$\tau ::= b|1|\tau \otimes \tau|\tau \multimap \tau|\boldsymbol{T}(\tau)$$

where $b$ ranges over base types, $\multimap$ is linear implication, and $\boldsymbol{T}()$ is a syntactic, strong monad. It is easy to craft a calculus based on this type algebra. The linear $\lambda$-terms are

given by:

$$s, t, u, v, \ldots \quad ::= \quad x \mid * \mid \texttt{let}\, * = s\, \texttt{in}\, t \mid s \otimes t \mid \texttt{let}\, x \otimes y = s\, \texttt{in}\, t \mid st \mid \lambda x \cdot s$$
$$\mid \texttt{val}\, s \mid \texttt{let}\, \texttt{val}\, x = s\, \texttt{in}\, t$$

The typing rules are those of Moggi's calculus, except for its linear flavor. Contexts $\Gamma$ are multisets of bindings $x : \tau$, with pairwise distinct variable parts $x$, and comma denotes multiset union.

$$\frac{}{x : A \vdash x : A}\,(Var)$$

$$\frac{\Gamma \vdash s : 1 \quad \Delta, x : 1 \vdash t : \tau}{\Gamma, \Delta \vdash \texttt{let}\, * = s\, \texttt{in}\, t : \tau}\,(1\mathcal{E}) \qquad \frac{}{\vdash * : 1}\,(1)$$

$$\frac{\Gamma \vdash s : \tau \otimes \tau' \quad \Delta, x : \tau, y : \tau' \vdash t : \tau''}{\Gamma, \Delta \vdash \texttt{let}\, x \otimes y = s\, \texttt{in}\, t : \tau''}\,(\otimes\mathcal{E}) \qquad \frac{\Gamma \vdash s : \tau \quad \Delta \vdash t : \tau'}{\Gamma, \Delta \vdash s \otimes t : \tau \otimes \tau'}\,(Pair)$$

$$\frac{\Gamma \vdash s : \tau \multimap \tau' \quad \Delta \vdash t : \tau}{\Gamma, \Delta \vdash st : \tau'}\,(App) \qquad \frac{\Gamma, x : \tau \vdash s : \tau'}{\Gamma \vdash \lambda x \cdot s : \tau \multimap \tau'}\,(Abs)$$

$$\frac{\Gamma \vdash s : \boldsymbol{T}(\tau) \quad \Delta, x : \tau \vdash t : \boldsymbol{T}(\tau')}{\Gamma, \Delta \vdash \texttt{let}\, \texttt{val}\, x = s\, \texttt{in}\, t : \boldsymbol{T}(\tau')}\,(Let) \qquad \frac{\Gamma \vdash s : \tau}{\Gamma \vdash \texttt{val}\, s : \boldsymbol{T}(\tau)}\,(Val)$$

Conversion rules are given by judgments $\Gamma \vdash s \to t : \tau$ (which we write $s \to t$ when $\Gamma$ and $\tau$ are clear or irrelevant), of which the most important are:

$$\begin{array}{rl}
\texttt{let}\, * = *\, \texttt{in}\, s \to s & \vdash s \to * : 1 \\
\texttt{let}\, x \otimes y = s \otimes t\, \texttt{in}\, u \to u[x := s, y := t] & \texttt{let}\, x \otimes y = s\, \texttt{in}\, x \otimes y \to s \\
(\lambda x \cdot s)t \to s[x := t] & \lambda x \cdot sx \to s \quad (x \text{ not free in } s) \\
\texttt{let}\, \texttt{val}\, x = \texttt{val}\, s\, \texttt{in}\, t \to t[x := s] & \texttt{let}\, \texttt{val}\, x = s\, \texttt{in}\, \texttt{val}\, x \to s \\
\texttt{let}\, \texttt{val}\, x = \texttt{let}\, \texttt{val}\, y = s\, \texttt{in}\, t\, \texttt{in}\, u \quad \to & \texttt{let}\, \texttt{val}\, y = s\, \texttt{in}\, \texttt{let}\, \texttt{val}\, x = t\, \texttt{in}\, u
\end{array}$$

There are other rules, notably the infamous commutative conversion rules—which originate in (Prawitz, 1965)—and which we won't state. The interpretation in symmetric monoidal closed categories follows the same lines as those of the ordinary $\lambda$-calculus with a monadic type, replacing cartesian products by tensor products. Our constructions then yield a notion of subscone for this kind of calculus and models, requiring conditions **(i.b)**, **(iii.a)**, **(iv)**, **(v)**, and **(vi)**.

Note that $[\![\tau \otimes \tau']\!]_i = [\![\tau]\!]_i \otimes [\![\tau']\!]_i$, $[\![\tau \multimap \tau']\!]_i = {[\![\tau']\!]_i}^{[\![\tau]\!]_i}$, $[\![\boldsymbol{T}(\tau)]\!]_i = \mathbb{T}\,[\![\tau]\!]_i$. Then the definition of the subscone specializes to:

$$\begin{array}{rcl}
(f_1, f_2) \in [\![\tau \multimap \tau']\!] & \iff & \forall a_1 \in [\![\tau]\!]_1, a_2 \in [\![\tau]\!]_2 \cdot \\
& & \quad (a_1, a_2) \in [\![\tau]\!] \Rightarrow (f_1(a_1), f_2(a_2)) \in [\![\tau']\!] \\
((a_1, a_1'), (a_2, a_2')) \in [\![\tau \otimes \tau']\!] & \iff & (a_1, a_2) \in [\![\tau]\!] \wedge (a_1', a_2') \in [\![\tau']\!]
\end{array}$$

Finally, $(F_1, F_2) \in [\![\boldsymbol{T}(\tau)]\!]$ if and only if there is $F \in \mathbb{T}\,[\![\tau]\!]$ such that

$$\begin{array}{rcl}
F_1(g_1) & = & F(\lambda(x, y) \in [\![\tau]\!] \cdot g_1(x)) \text{ for every } g_1 \in \mathcal{C}_b([\![\tau]\!]_1) \\
F_2(g_2) & = & F(\lambda(x, y) \in [\![\tau]\!] \cdot g_2(y)) \text{ for every } g_2 \in \mathcal{C}_b([\![\tau]\!]_2)
\end{array}$$

(Up to isomorphism, relevant monos are inclusions, so we may consider that $[\![\tau]\!] \subseteq [\![\tau]\!]_1 \times [\![\tau]\!]_2$.) This condition implies that, if $(F_1, F_2) \in [\![\boldsymbol{T}(\tau)]\!]$, then for every $g_1 \in \mathcal{C}_b([\![\tau]\!]_1)$ and $g_2 \in \mathcal{C}_b([\![\tau]\!]_2)$ such that $g_1(a_1) = g_2(a_2)$ for every $(a_1, a_2) \in [\![\tau]\!]$, then $F_1(g_1) = F_2(g_2)$. It turns out that this is equivalent to it, just as in the case of the continuation monad: $(F_1, F_2) \in [\![\boldsymbol{T}(\tau)]\!]$ if and only if, for every $g_1 \in \mathcal{C}_b([\![\tau]\!]_1)$ and $g_2 \in \mathcal{C}_b([\![\tau]\!]_2)$ such that $g_1(a_1) = g_2(a_2)$ for every $(a_1, a_2) \in [\![\tau]\!]$, then $F_1(g_1) = F_2(g_2)$.

The difficult if direction is proved as follows. Assume that $F_1(g_1) = F_2(g_2)$ for every $g_1 \in \mathcal{C}_b([\![\tau]\!]_1)$ and $g_2 \in \mathcal{C}_b([\![\tau]\!]_2)$ such that $g_1(a_1) = g_2(a_2)$ for every $(a_1, a_2) \in [\![\tau]\!]$. Let $S$ be $[\![\tau]\!]$, and $H$ be the subspace of $\mathcal{C}_b(S)$ of those functions of the form $\lambda(x, y) \in S \cdot g_1(x) + g_2(y)$, where $g_1$ ranges over $\mathcal{C}_b([\![\tau]\!]_1)$, $g_2$ ranges over $\mathcal{C}_b([\![\tau]\!]_2)$. Let $F_0(g)$ be defined as $F_1(g_1)+F_2(g_2)$ for every $g = \lambda(x, y) \in S \cdot g_1(x)+g_2(y)$ in $H$. This is independent of the choice of $g_1$ and $g_2$, since if $g$ can also be written as $\lambda(x, y) \in S \cdot g'_1(x) + g'_2(y)$, then by construction $g_1(a_1) - g'_1(a_1) = g'_2(a_2) - g_2(a_2)$ for every $(a_1, a_2) \in S = [\![\tau]\!]$, so $F_1(g_1 - g'_1) = F_2(g'_2 - g_2)$, which implies $F_1(g_1)+F_2(g_2) = F_1(g'_1)+F_2(g'_2)$. It is easy to see that $F_0$ is a continuous linear form, so it extends to a continuous linear form $F$ with the same norm on the whole of $\mathcal{C}_b(S)$, by the Hahn-Banach Theorem. By the same argument as in Lemma 10.2, $F$ is a positive linear form. Clearly $F_1(g_1) = F(\lambda(x, y) \in [\![\tau]\!] \cdot g_1(x))$ for every $g_1 \in \mathcal{C}_b([\![\tau]\!]_1)$, and $F_2(g_2) = F(\lambda(x, y) \in [\![\tau]\!] \cdot g_2(y))$ for every $g_2 \in \mathcal{C}_b([\![\tau]\!]_2)$, whence the claim.

In other words, the logical relation at $\boldsymbol{T}(\tau)$ types is defined by a formula in the style of observational equivalence.

It is reasonable to represent the set of probabilistic transition systems on a metric space $A$ as the metric space $\mathbb{T}A^A$ of all non-expansive maps from the set of states $A$ to the set of probability measures on $A$. The subscone construction provides a condition when two such probabilistic transition systems are related. Given any relation $S$ on $A$ (defining a metric subspace of $A \times A$), define the relation $\widetilde{S}$ on $\mathbb{T}A^A$ by: for any non-expansive maps $f_1, f_2$ from $A$ to $\mathbb{T}A$, $(f_1, f_2) \in \widetilde{S}$ if and only if, for every $(a_1, a_2) \in S$, there is $F \in \mathbb{T}S$ such that

$$\begin{aligned} f_1(a_1)(g_1) &= F(\lambda(x, y) \in S \cdot g_1(x)) \text{ for every } g_1 \in \mathcal{C}_b(A) \\ f_2(a_2)(g_2) &= F(\lambda(x, y) \in S \cdot g_2(y)) \text{ for every } g_2 \in \mathcal{C}_b(A) \end{aligned}$$

or, equivalently, for every $g_1, g_2 \in \mathcal{C}_b(A)$ such that $g_1(x) = g_2(y)$ for every $(x, y) \in S$, then $f_1(a_1)(g_1) = f_2(a_2)(g_2)$. In case $A$ is compact, by the Riesz Representation Theorem, we may replace positive continuous linear functionals on $\mathcal{C}_b(A)$ by probability measures. Rephrasing the above, we get: for any non-expansive maps $f_1, f_2$ from $A$ to the space of probability measures on $A$, $(f_1, f_2) \in \widetilde{S}$ if and only if, for every $(a_1, a_2) \in S$, there is a probability measure $\nu$ on $S$ such that

$$\begin{aligned} f_1(a_1)(X_1) &= \nu((X_1 \times A) \cap S) \text{ for every measurable subset } X_1 \text{ of } A \\ f_2(a_2)(X_2) &= \nu((A \times X_2) \cap S) \text{ for every measurable subset } X_2 \text{ of } A \end{aligned}$$

We retrieve a notion of probabilistic bisimulation: $S$ is a *bisimulation* on the space $A$ of states between the transition systems $f_1$ and $f_2$ if and only if $\widetilde{S}$ relates $f_1$ and $f_2$.

In passing, we have just shown that the observational-equivalence-style definition of $\widetilde{S}$

coincides with the bisimulation-style definition, using the Riesz Representation Theorem. This can be seen as a form of a completeness theorem.

The notion above is formally analogous to the notion of probabilistic bisimulation of (Desharnais et al., 2002); i.e., the formulas we use and theirs for defining bisimulations is the same. This is also formally analogous to metric bisimulations as introduced by de Vink and Rutten (de Vink and Rutten, 1999) for ultrametric spaces, and extended by Worrell (Worrell, 2000) to generalized metric spaces. A difference is that the latter authors use a coalgebraic approach; in particular, their construction requires a $\mathbb{T}$ functor that preserves isometric embeddings (our relevant monos), while we require it to preserve pseudoepis (surjective non-expansive maps).

## 11. Related work.

One succesful approach to categorical generalization of logical relations is the setting of (Hermida, 1993) based on fibrations. We do not assume in this paper the forgetful functor to be a fibration, similarly to other authors (e.g. (Plotkin et al., 2000)). Even if we lose thereby the equivalence to logic, this is justified, since we do not focus on logical rules corresponding to logical relations in this paper.

We have already said that there is no unique notion of monad lifting. One of the simplest is the lifting, proposed in (Crole and Pitts, 1992), $\widetilde{T}$ of $\boldsymbol{T}$, which maps the object $\langle S, f, A\rangle$ of the scone to $\langle S, |\boldsymbol{\eta}_A| \circ f, \boldsymbol{T}A\rangle$. This is a special case of our notion of lifting on the scone $(\mathbb{C} \downarrow |_|)$ (Section 3), taking $\mathbb{T}$ the identity monad and $\sigma_A = |\boldsymbol{\eta}_A|$. Turi (Turi, 1996) considers lifting monads to the category of coalgebras of a given endofunctor. This is a special case of our framework, when $\boldsymbol{C} = \mathbb{C}$ (and $\boldsymbol{T} = \mathbb{T}$) and moreover only objects of the form $S \xrightarrow{f} |S|$ are taken into consideration, and only morphisms of the form $\langle g, g\rangle$. This defines the category of $|_|$-coalgebras as a proper subcategory of scones. Turi uses a simpler version of our monad morphisms, namely distributivity law of a monad over an endofunctor; monad morphisms involve two monads and a functor between distinct categories. Turi's distributivity laws are similar enough to monad morphisms that we had called the latter *distributivity laws* for monads in the conference version of this paper. This also influenced (Goubault-Larrecq and Goubault, 2003), where comonad morphisms are used, but are called distributivity laws for comonads. (This relies on pullbacks, whereas we use mono factorization systems.) Calling these monad morphisms distributivities turned out to be a bad choice, as distributivity laws denote a close but different concept, due to Jon Beck (Beck, 1969); also, while distributivity laws tend to be rare, monad morphisms abound.

In (Power and Watanabe, 2002) different possible ways to combine a monad and a comonad were studied in a systematic way. In particular, the authors used a notion of a distributivity of a monad over a comonad (and dual distributivity of the comonad over the monad). This is a stronger notion than monad morphisms, as it requires commutativity of two copies of diagrams (4), one for the monad and another one for the comonad. On the other hand, both the monad and the comonad live in a single category, unlike in our case.

We also note that neither Pitts nor Turi deal with subscones.

(Kinoshita and Power, 1999) consider lifting a monad on ***Set*** to the category of relations in ***Set***. This is a particular case of our setting. However, the main topic of this paper is a more complex situation of lax logical relations and data-refinement for the computational lambda calculus. A subsequent paper (Power and Tanaka, 2000) develops further the ideas of (Kinoshita and Power, 1999). It addresses in particular lax logical relations for the computational lambda calculus and for the linear lambda calculus.

In the same way that we lift a monad to relations, Rutten (Rutten, 1998) defines an extension of an endofunctor in ***Set*** to a category of relations. The latter has relations as *morphisms* between sets. An endofunctor extends to relations iff it preserves weak pullbacks(which in particular implies preserving monos), and if so, the extension is unique. (This is actually a special case of a more general fact, proved in (Carboni et al., 1990) for regular categories.) The approach taken by Rutten is different from ours, where relations are *objects* rather than morphisms. Hence, Rutten imposes a different functoriality condition: the action of a lifted endofunctor on a composition of two relations must coincide with a composition of actions of the lifted endofunctor on these two relations. This amounts to closedness under composition of relations yielded by the lifted endofunctor.

(Pitts, 1991) proposes the Evaluation Logic for reasoning about programs with computational effects. It is based on the computational lambda-calculus (Moggi, 1991). Semantics of Pitts' Evaluation Logic is studied further in (Moggi, 1995).

(Pitts, 1996) considers litfing of certain constructions on domains to relations on these domains. He focuses on categories on domains and does not consider (strong) monads. Instead of category of relations he assumes a relational structure, relaxing a notion of mapping (morphism) of relations.

## 12. Conclusions

The main contribution of this paper is a natural extension of logical relations able to deal with monadic types. We illustrate its naturality and its practical value by demonstrating that various notions of bisimulations and a non-trivial notion of logical relation for dynamic name creation are instances of our construction. Besides, our construction provides a natural integration between notions of simulations for transition systems (possibly probabilistic), higher-order computation (the import of the $\lambda$-calculus), and limited forms of side-effects (e.g., dynamic names), yielding streamlined criteria for observational equivalence of those combined systems.

## Acknowledgments

We are grateful to Masahito Hasegawa, John Power, Ian Stark, Paul-André Melliès, François Lamarche, Vincent Danos, and Albert Burroni, and the anonymous referees for their useful comments.

## Appendix A. Monoidal Monads, Commutative Monads

A.1. *From Mediators to Strengths and Dual Strengths*

Let $(\mathbb{T}, \eta, \mu, \mathbb{d})$ be a monoidal monad. Let $\mathbb{t}_{A,B} = \mathbb{d}_{A,B} \circ (\eta_A \otimes \mathrm{id}_{\mathbb{T}B})$, $\mathbb{t}'_{A,B} = \mathbb{d}_{A,B} \circ (\mathrm{id}_{\mathbb{T}A} \otimes \eta_B)$. We show that $\mathbb{t}$ is a strength, $\mathbb{t}'$ a dual strength, and that Diagrams (43) and (44) commute.

Diagram (32). This is exactly Diagram (38).

Diagram (33). This is exactly Diagram (40).

Diagram (34). This is shown by the diagram below. The topmost row is $\mathbb{t}_{A\otimes B,C}$, while the bottommost composition of morphisms from $\mathbb{T}((A \otimes B) \otimes C)$ to $\mathbb{T}(A \otimes (B \otimes C))$ is $\mathrm{id}_A \otimes \mathbb{t}_{B,C}$ followed by $\mathbb{t}_{A,B\otimes C}$.

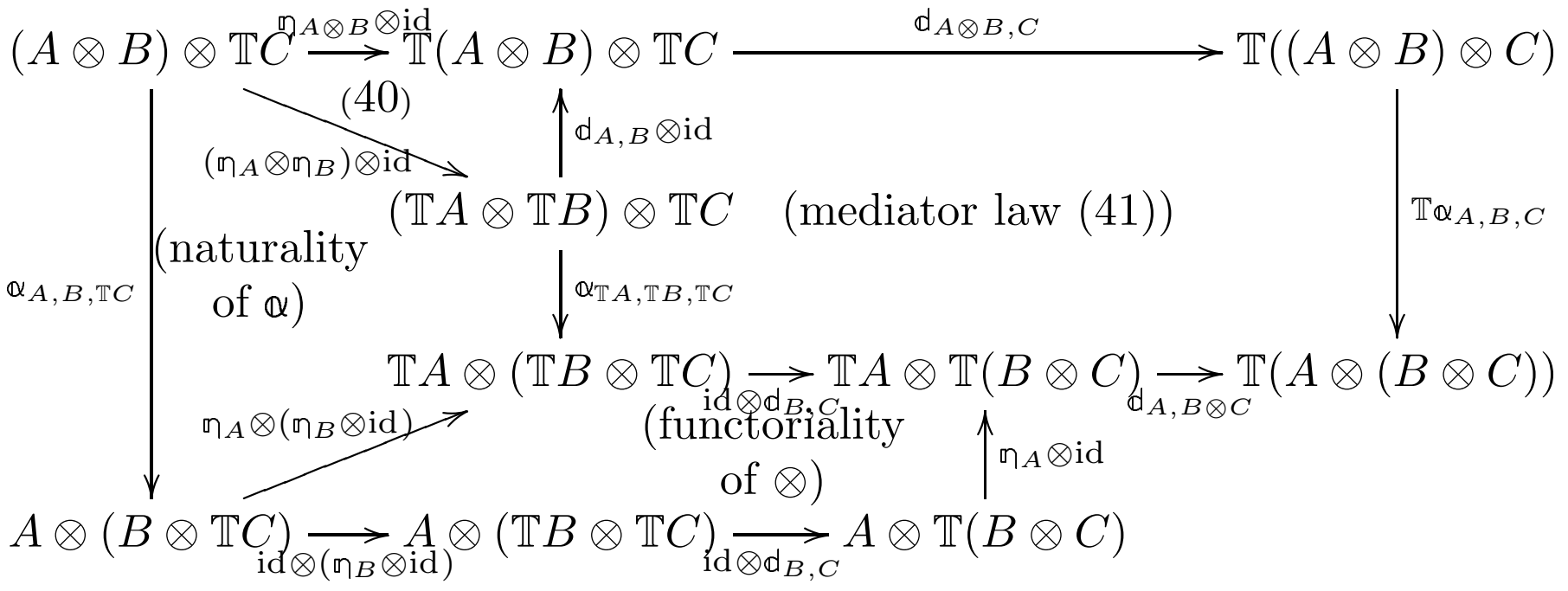

Diagram (35).

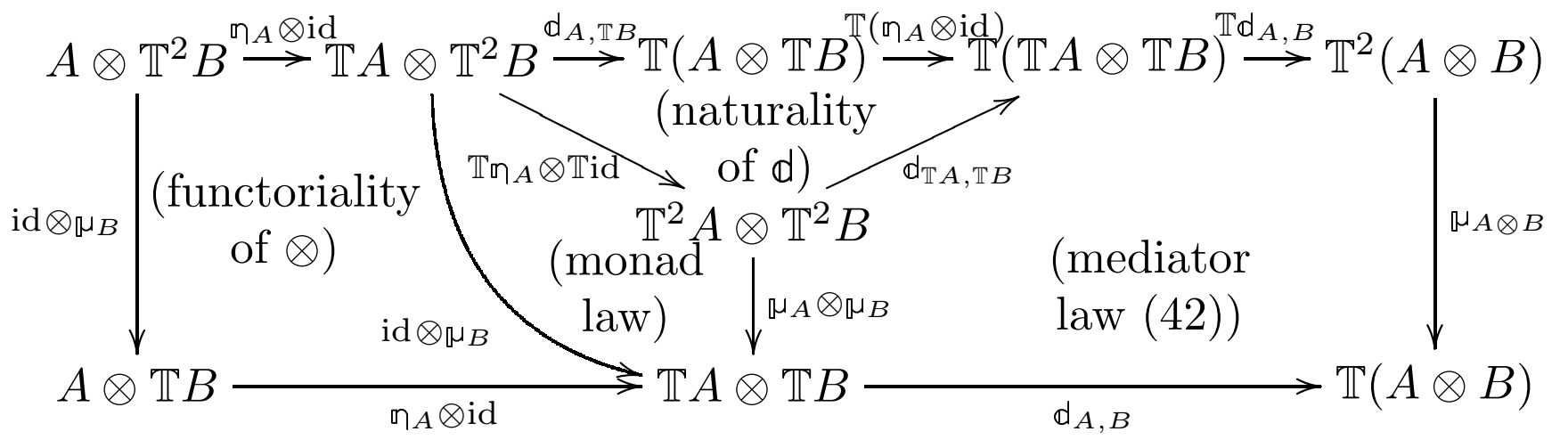

So $\theta$ is a strength. That $\theta'$ is a dual strength is proved similarly. We use Diagram (39) instead of Diagram (38) to prove the dual form of Diagram (32).

Diagram (43). In the diagram below, the top row is $\mathbb{t}_{A,B} \otimes \mathrm{id}$ followed by $\mathbb{t}'_{A\otimes B,C}$, while the bottom row is $\mathrm{id} \otimes \mathbb{t}'_{B,C}$ followed by $\mathbb{t}_{A,B\otimes C}$.

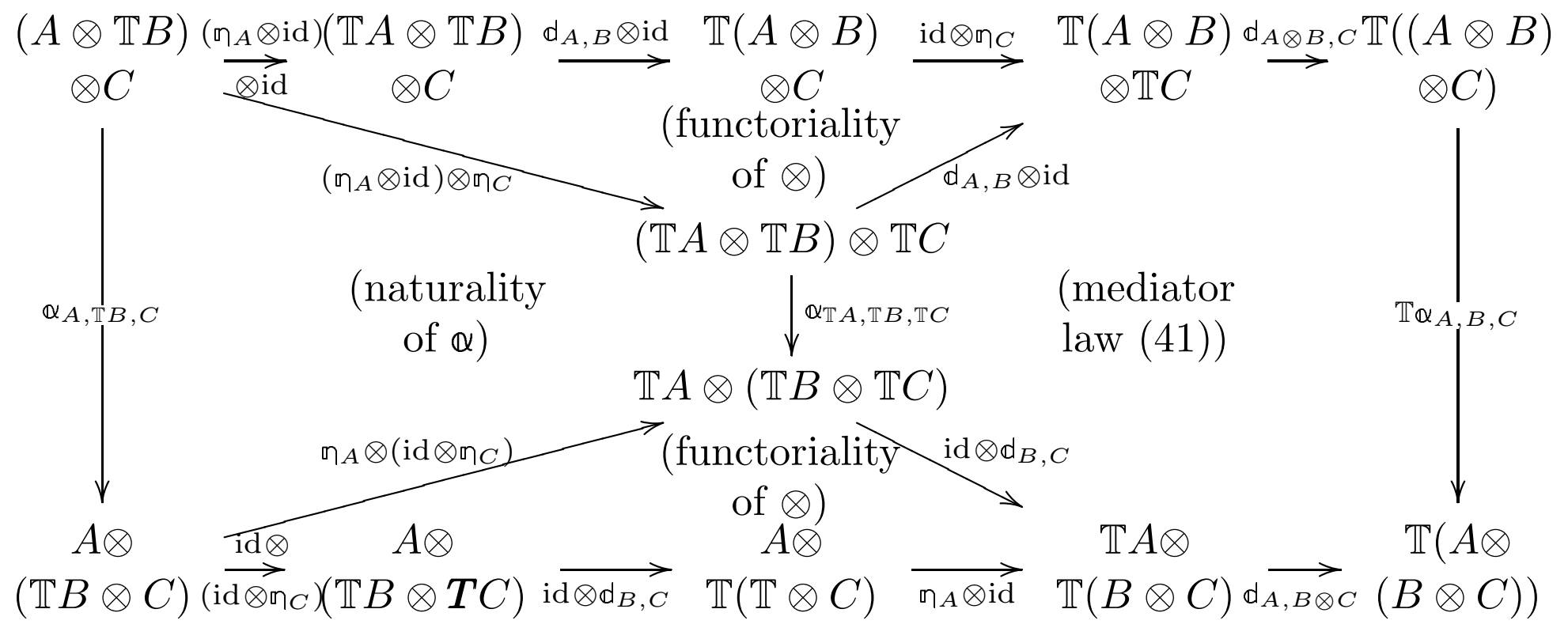

Diagram (44). This is the diagram below, where the top row is $\mathbb{t}_{\mathbb{T}A,B}$, the leftmost vertical arrows form $\mathbb{t}'_{A,\mathbb{T}B}$, the rightmost ones form $\mathbb{T}\mathbb{t}'_{A,B}$ followed by $\mu_{A\otimes B}$, and the bottom row is $\mathbb{T}\mathbb{t}_{A,B}$ followed by $\mu_{A\otimes B}$. Note that, as announced, the common diagonal from the top left corner to the bottom right corner is $\mathbb{d}_{A,B}$.

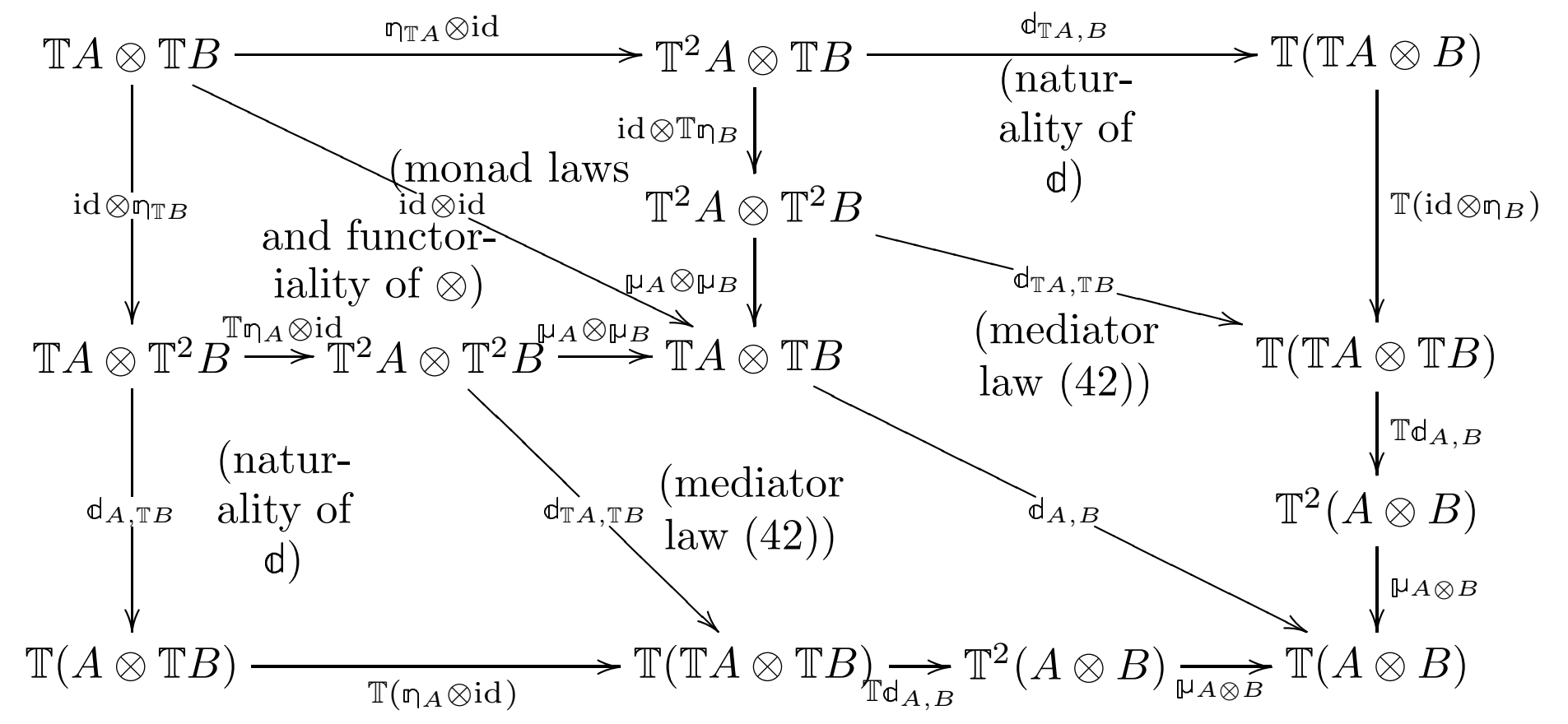

A.2. *From Strengths and Dual Strengths to Mediators*

Let $(\mathbb{T}, \eta, \mu)$ be a monad such that $\mathbb{t}$ is a strength and $\mathbb{t}'$ is a dual strength making Diagrams (43) and (44) commute. Let $\mathbb{d}$ be the common diagonal of Diagram (44), i.e., $\mathbb{d}_{A,B} = \mu_{A\otimes B} \circ \mathbb{T}\mathbb{t}'_{A,B} \circ \mathbb{t}_{\mathbb{T}A,B} = \mu_{A\otimes B} \circ \mathbb{T}\mathbb{t}_{A,B} \circ \mathbb{t}'_{A,\mathbb{T}B}$. We show that $\mathbb{d}$ is a mediator, which means that we check Diagrams (38), (39), (40), (41), and (42). By convention, call (32'), (33'), (34'), (35') the dual diagrams satisfied by the dual strength $\mathbb{t}'$.

First, notice that $\mathbb{t}_{A,B} = \mathbb{d}_{A,B} \circ (\eta_A \otimes \mathrm{id}_{\mathbb{T}B})$. Indeed, using the definition $\mathbb{d}_{A,B} = \mu_{A\otimes B} \circ \mathbb{T}\mathbb{t}'_{A,B} \circ \mathbb{t}_{\mathbb{T}A,B}$, $\mathbb{d}_{A,B} \circ (\eta_A \otimes \mathrm{id}_{\mathbb{T}B})$ is the top row of the following diagram, while $\mathbb{t}_{A,B}$ is the bottom composition of arrows from $A \otimes \mathbb{T}B$ to $\mathbb{T}(A \otimes B)$.

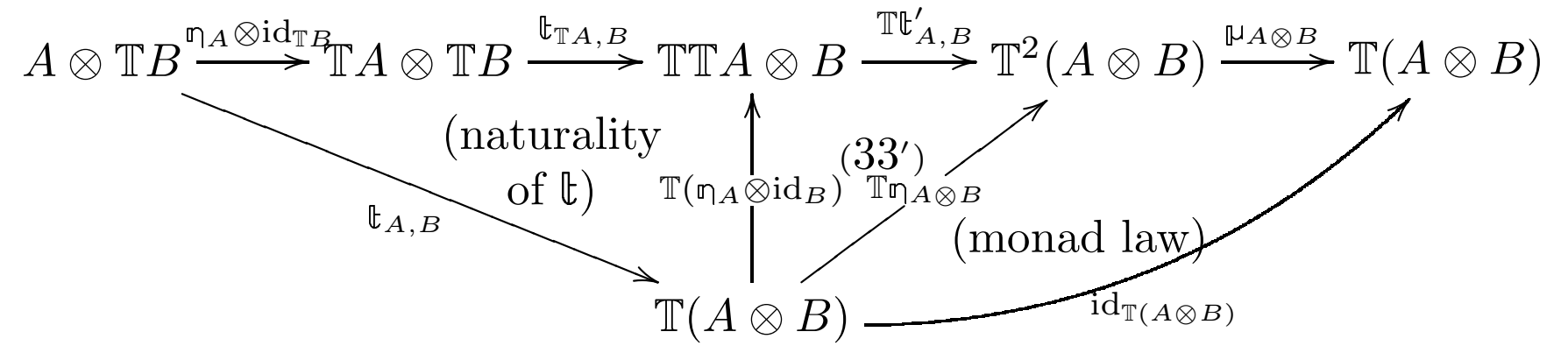

Diagram (38) is then just Diagram (32).

Diagram (39). Similarly, $\mathbb{t}'_{A,B} = \mathbb{d}_{A,B} \circ (\mathrm{id}_{\mathbb{T}A} \otimes \eta_B)$ (here we use the other characterization of $\mathbb{d}_{A,B}$ as $\mu_{A\otimes B} \circ \mathbb{T}\mathbb{t}_{A,B} \circ \mathbb{t}'_{A,\mathbb{T}B}$ and (33)), so Diagram (39) is just Diagram (32').

Diagram (40). Immediate using (33) and $\mathbb{t}_{A,B} = \mathbb{d}_{A,B} \circ (\eta_A \otimes \mathrm{id}_{\mathbb{T}B})$: $\mathbb{d}_{A,B} \circ (\eta_A \otimes \eta_B) = \mathbb{t}_{A,B} \circ (\eta_A \circ \mathrm{id}_B) = \eta_{A\otimes B}$.

Diagram (41). This one is harder, and is proved by considering the diagram below. The top row is $\mathbb{d}_{A,B} \otimes \mathrm{id}_{\mathbb{T}C}$, the rightmost vertical arrows form $\mathbb{d}_{A\otimes B,C}$ followed by $\mathbb{T}\alpha_{A,B,C}$, while the leftmost ones form $\alpha_{\mathbb{T}A,\mathbb{T}B,\mathbb{T}C}$ followed by $\mathrm{id}_{\mathbb{T}A} \otimes \mathbb{d}_{B,C}$, and the bottommost row is $\mathbb{d}_{A,B\otimes C}$.

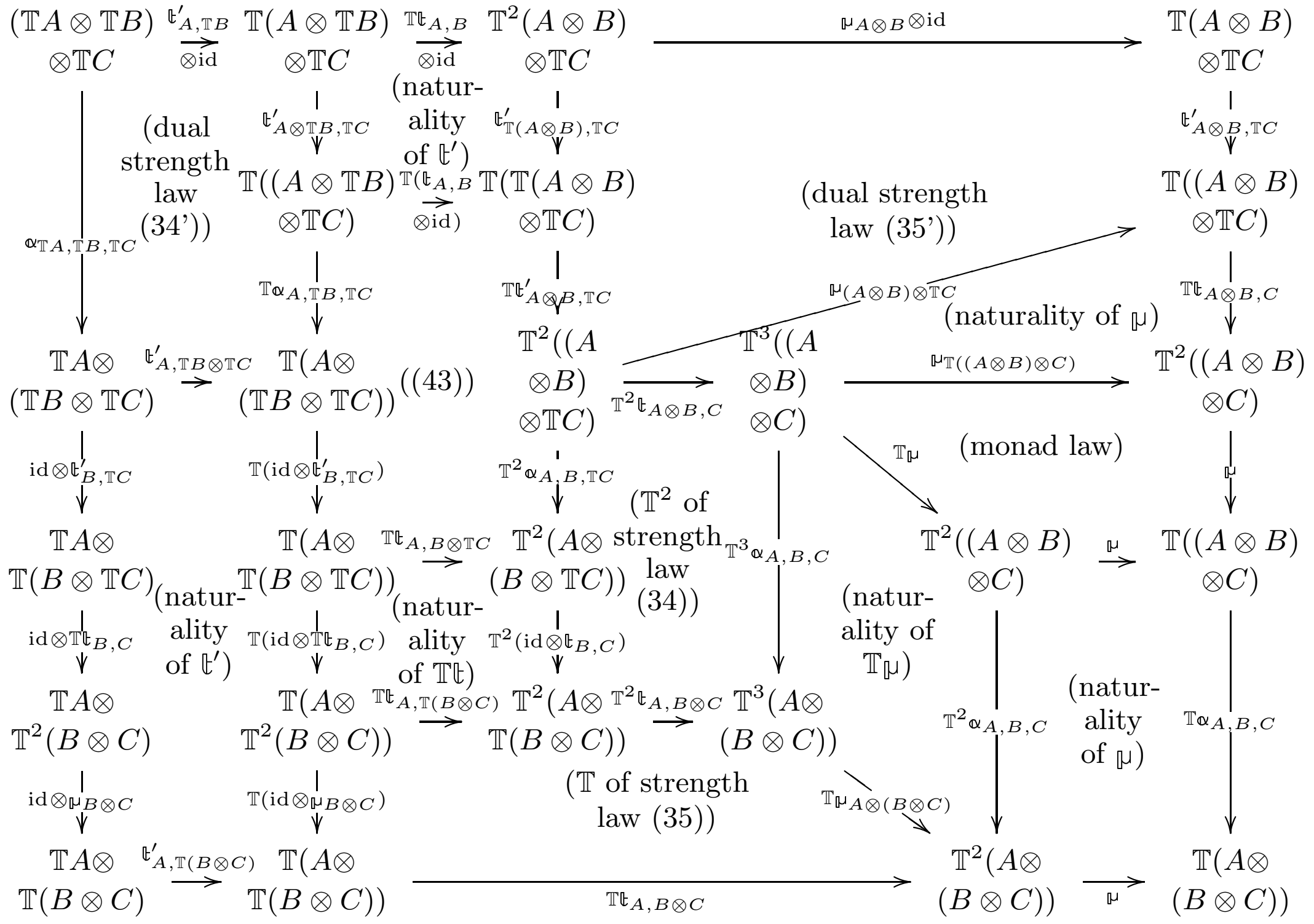

Diagram (42). First, we notice that

$$\mu_{A\otimes B}\circ\mathbb{T}\mathbb{d}_{A,B}\circ\mathbb{t}'_{\mathbb{T}A,\mathbb{T}B} \quad=\quad \mathbb{d}_{A,B}\circ(\mu_A\otimes\mathrm{id}_{\mathbb{T}B}) \tag{64}$$

This is shown using the diagram on the right. The left-hand side is the path from the top left corner to the bottom right corner on the right that goes down then right, and the right-hand side is the path going right then down.

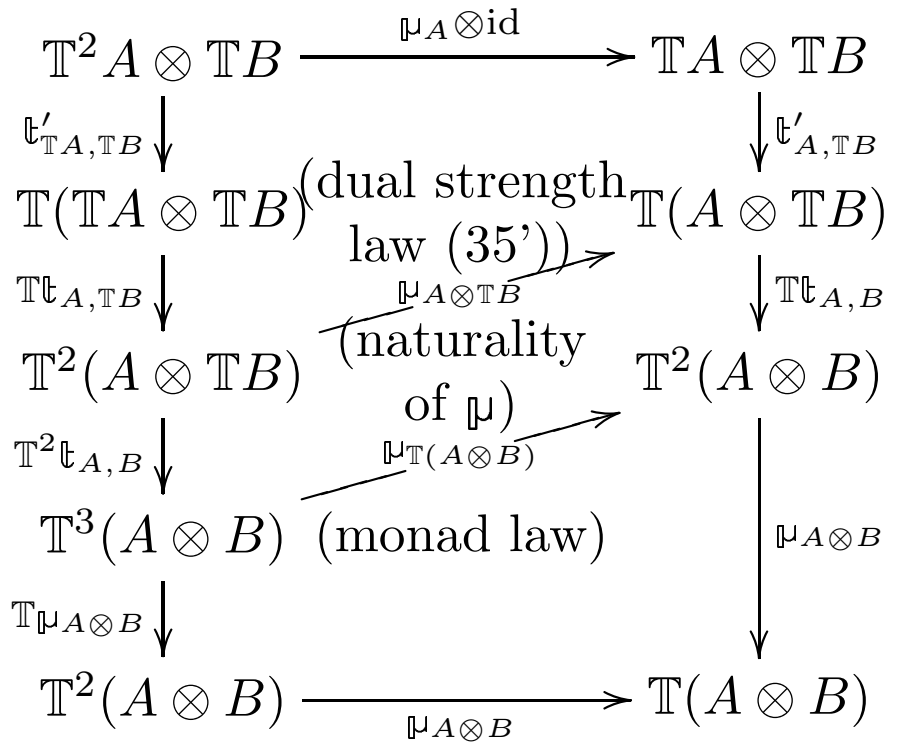

By symmetry, the following also holds:

$$\mu_{A\otimes B}\circ\mathbb{T}\mathbb{d}_{A,B}\circ\mathbb{t}_{\mathbb{T}A,\mathbb{T}B} \quad=\quad \mathbb{d}_{A,B}\circ(\mathrm{id}_{\mathbb{T}A}\otimes\mu_B) \tag{65}$$

Diagram (42) then follows. The top row below is $\mathbb{d}_{\mathbb{T}A,\mathbb{T}B}$, while the leftmost vertical arrows form $\mathbb{\mu}_A \otimes \mathbb{\mu}_B$.

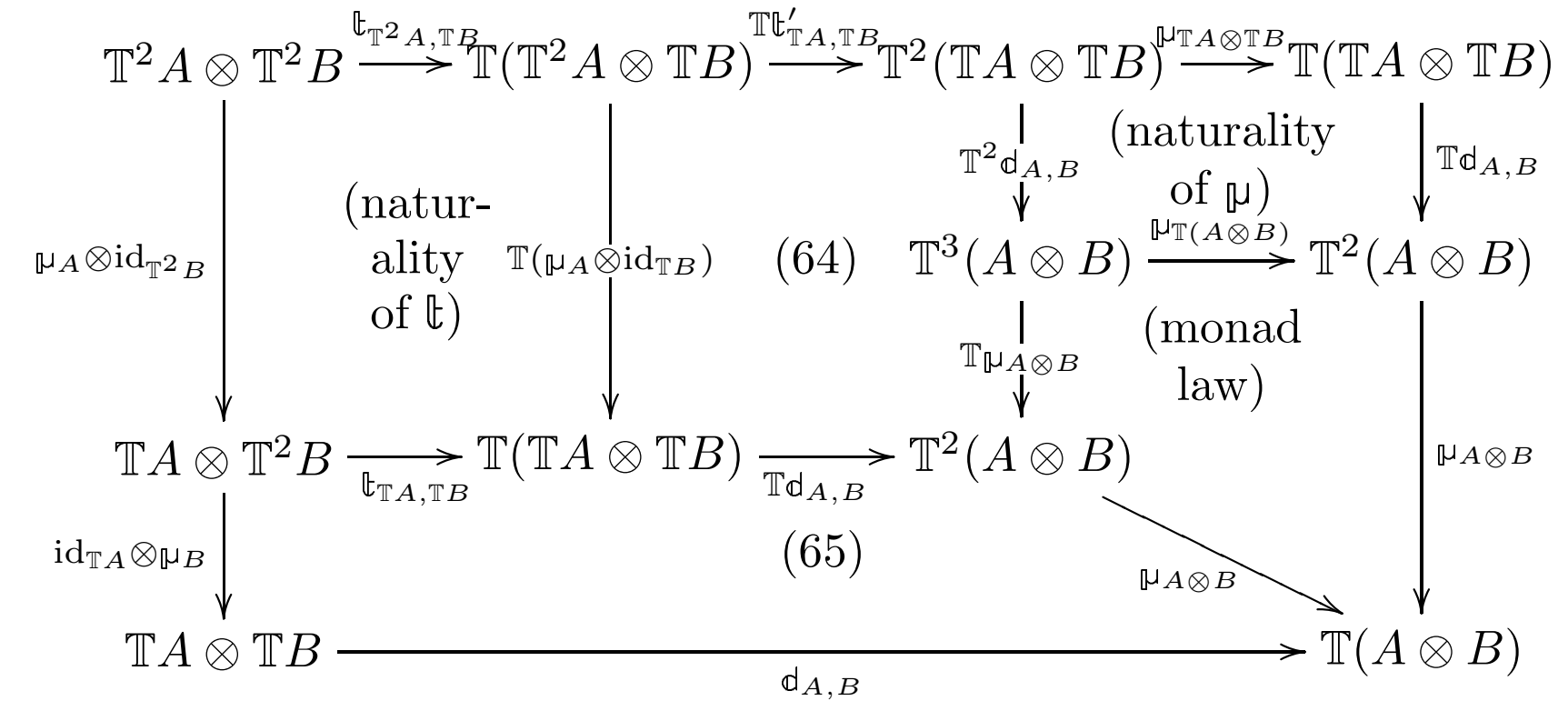

A.3. *Monoidal Monad Morphisms*

We first show that every monoidal monad morphism from $(\boldsymbol{T}, \boldsymbol{\eta}, \boldsymbol{\mu}, \boldsymbol{d})$ to $(\mathbb{T}, \mathbb{\eta}, \mathbb{\mu}, \mathbb{d})$ is both a strong monad morphism from $(\boldsymbol{T}, \boldsymbol{\eta}, \boldsymbol{\mu}, \boldsymbol{t})$ to $(\mathbb{T}, \mathbb{\eta}, \mathbb{\mu}, \mathbb{t})$ and (by symmetry) a dual strong monad morphism from $(\boldsymbol{T}, \boldsymbol{\eta}, \boldsymbol{\mu}, \boldsymbol{t}')$ to $(\mathbb{T}, \mathbb{\eta}, \mathbb{\mu}, \mathbb{t}')$, where $\mathbb{t}_{A,B} = \mathbb{d}_{A,B} \circ (\mathbb{\eta}_A \otimes \mathrm{id}_{\mathbb{T}B})$, $\mathbb{t}'_{A,B} = \mathbb{d}_{A,B} \circ (\mathrm{id}_{\mathbb{T}A} \otimes \mathbb{\eta}_B)$. This is by the following diagram.

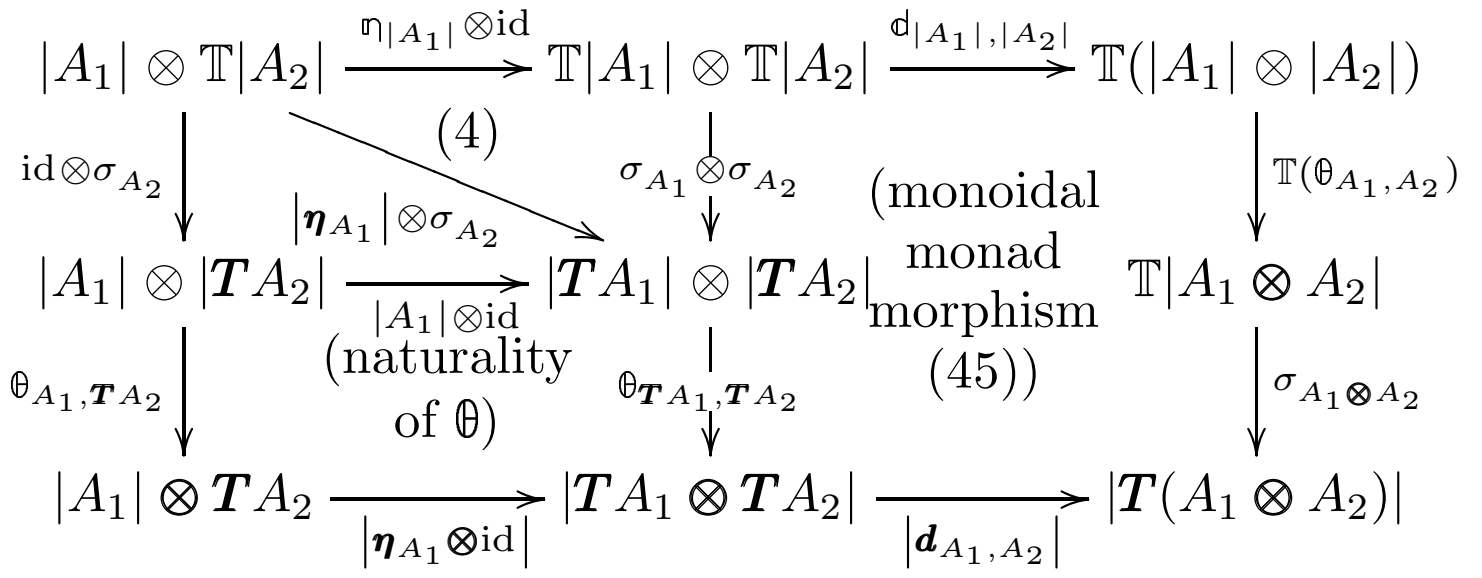

Conversely, let $\sigma$ be both a strong monad morphism from $(\boldsymbol{T}, \boldsymbol{\eta}, \boldsymbol{\mu}, \boldsymbol{t})$ to $(\mathbb{T}, \mathbb{\eta}, \mathbb{\mu}, \mathbb{t})$ and a dual strong monad morphism from $(\boldsymbol{T}, \boldsymbol{\eta}, \boldsymbol{\mu}, \boldsymbol{t}')$ to $(\mathbb{T}, \mathbb{\eta}, \mathbb{\mu}, \mathbb{t}')$. Let $\mathbb{d}$ be the common diagonal of Diagram (44), i.e., $\mathbb{d}_{A,B} = \mathbb{\mu}_{A\otimes B} \circ \mathbb{T}\mathbb{t}'_{A,B} \circ \mathbb{t}_{\mathbb{T}A,B} = \mathbb{\mu}_{A\otimes B} \circ \mathbb{T}\mathbb{t}_{A,B} \circ \mathbb{t}'_{A,\mathbb{T}B}$. Define $\boldsymbol{d}$ similarly.

In the diagram below, the top row is $\mathbb{d}_{|A_1|,\mathbb{T}|A_1|}$, while the leftmost vertical arrows from $\sigma_{A_1} \otimes \sigma_{A_2}$ followed by $\theta_{\boldsymbol{T}A_1,\boldsymbol{T}A_2}$. The bottom row is $|\boldsymbol{d}_{A_1,A_2}|$ followed by $|\boldsymbol{\mu}_{A_1 \otimes A_2}|$.

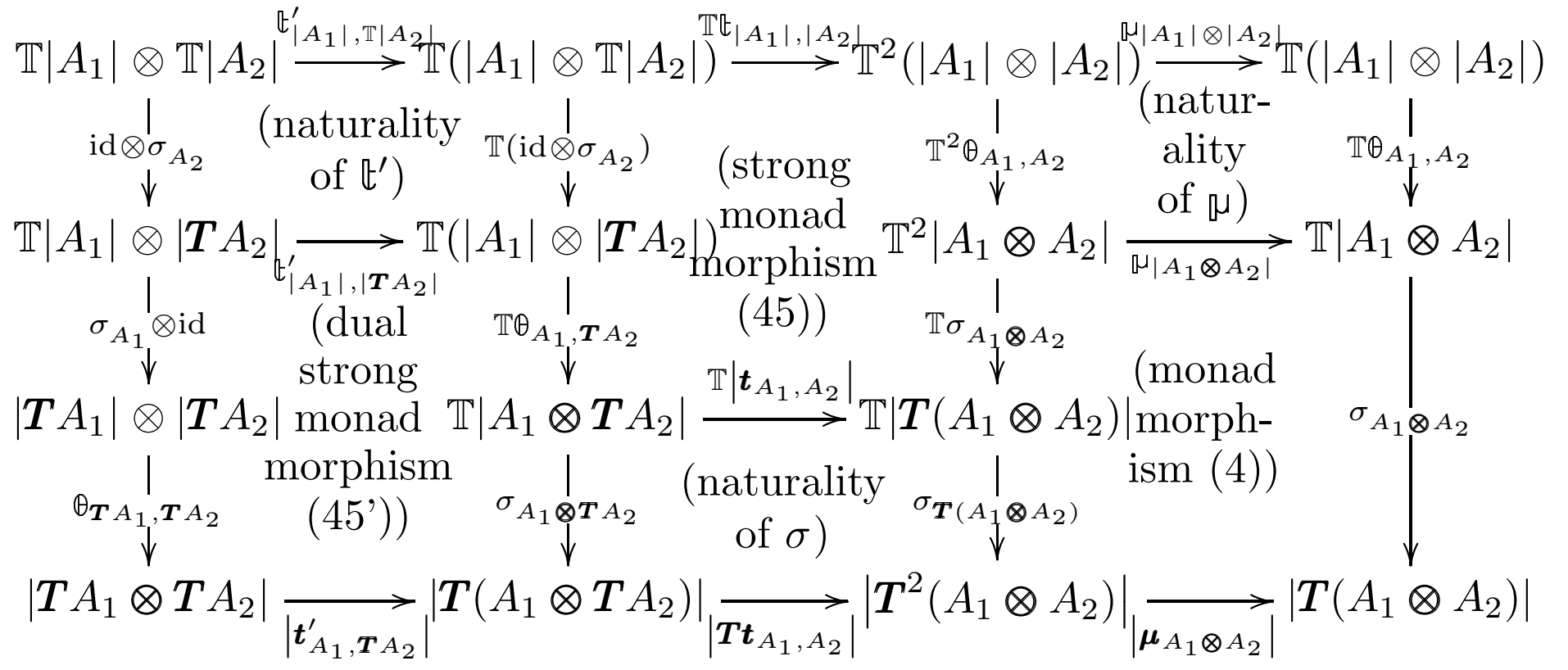

A.4. *Commutative Monads are Monoidal*

Let $(\mathbb{T}, \eta, \mu, \theta)$ be a strong monad on a symmetrical monoidal category. Let $\mathbb{t}'_{A,B}$ be the dual strength $\mathbb{T}\mathbb{c}_{B,A} \circ \mathbb{t}_{B,A} \circ \mathbb{c}_{\mathbb{T}A,B}$. That $\mathbb{T}$ is a commutative monad means that Diagram (44) commutes. To show that $\mathbb{d}_{A,B} = \mu_{A\otimes B} \circ \mathbb{T}\mathbb{t}'_{A,B} \circ \mathbb{t}_{\mathbb{T}A,B} = \mu_{A\otimes B} \circ \mathbb{T}\mathbb{t}_{A,B} \circ \mathbb{t}'_{A,\mathbb{T}B}$ is a mediator, it then suffices to show that Diagram (43) commutes. In the diagram below, the top row is $\mathbb{t}_{A,B} \otimes \mathrm{id}_C$ followed by $\mathbb{t}'_{A\otimes B,C}$, the bottom row is $\mathrm{id}_A \otimes \mathbb{t}'_{B,C}$ followed by $\mathbb{t}_{A,B\otimes C}$. The two triangles involving $\mathbb{c}$ on the left side are instances of the identity $\mathbb{c} \circ \mathbb{c} = \mathrm{id}$.

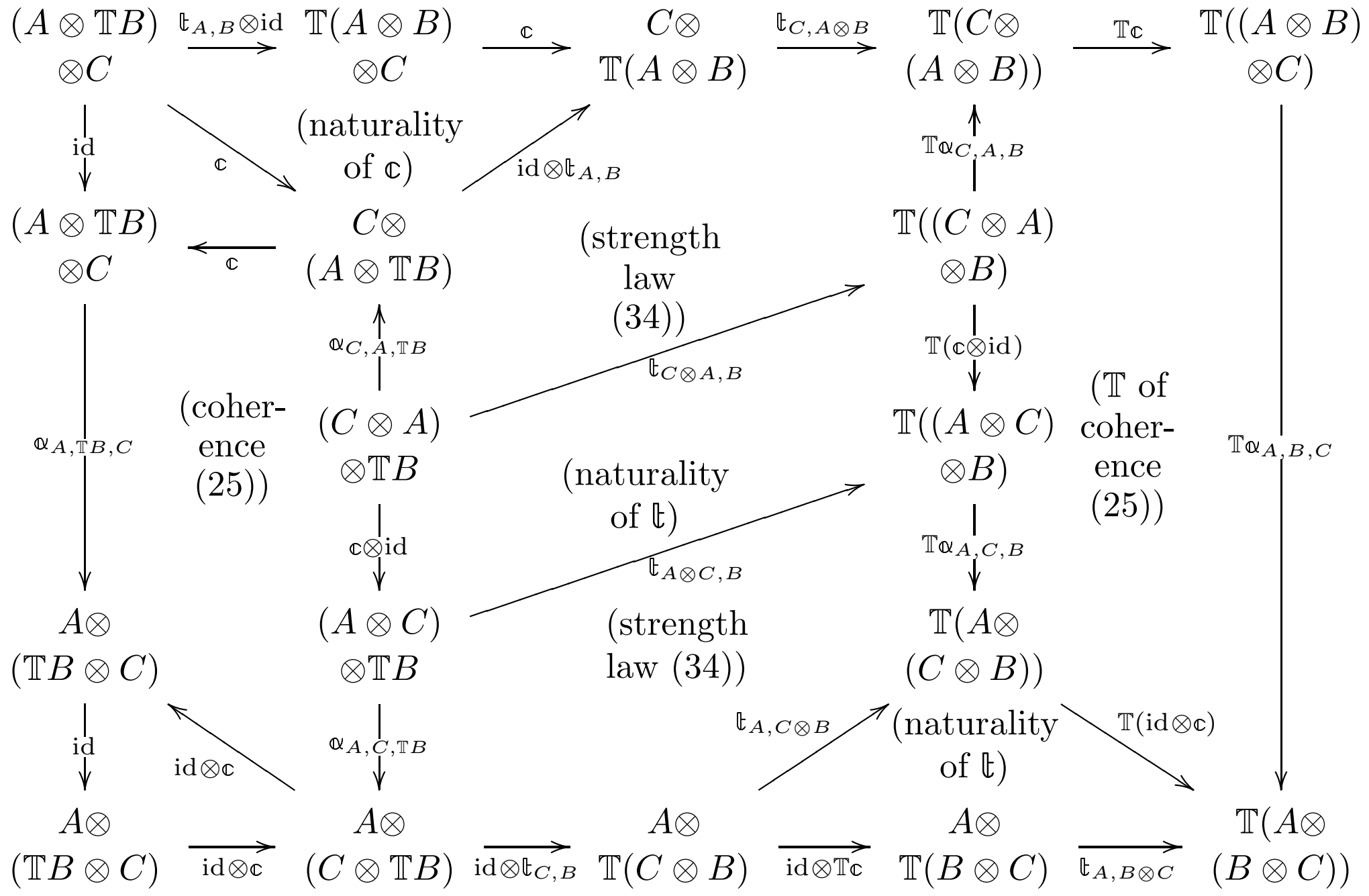

We now observe that Diagram (47) commutes:

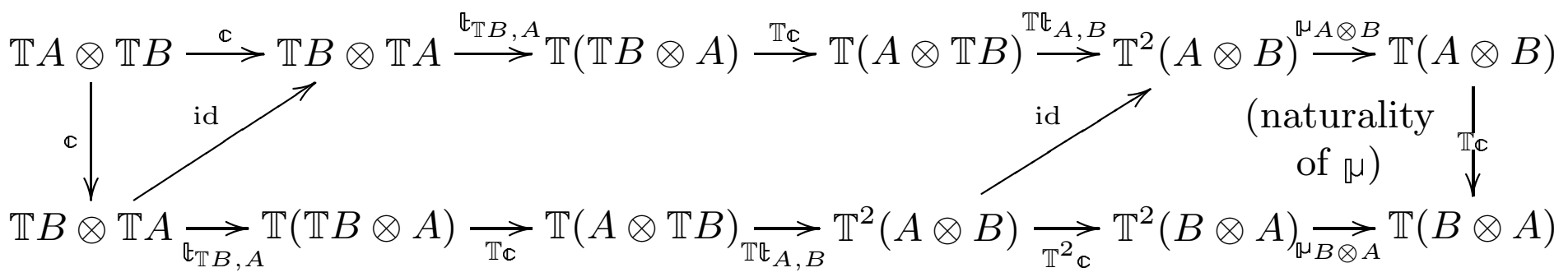

Indeed, the leftmost triangle is by $\mathbb{c}\circ\mathbb{c} = \mathrm{id}$, the inner parallelogram obviously commutes, and the top and bottom rows are the two ways of writing $\mathbb{d}_{A,B}$.

Conversely, given any monoidal monad making Diagram (47) commute, we claim that the derived strength $\mathbb{t}$ and dual strength $\mathbb{t}'$ are related by $\mathbb{t}'_{A,B} = \mathbb{T}\mathbb{c}_{B,A} \circ \mathbb{t}_{B,A} \circ \mathbb{c}_{\mathbb{T}A,B}$. Equivalently, since $\mathbb{c} \circ \mathbb{c} = \mathrm{id}$, we check that $\mathbb{t}'_{A,B} \circ \mathbb{c}_{B,\mathbb{T}A} = \mathbb{T}\mathbb{c}_{B,A} \circ \mathbb{t}_{B,A}$. This is obvious:

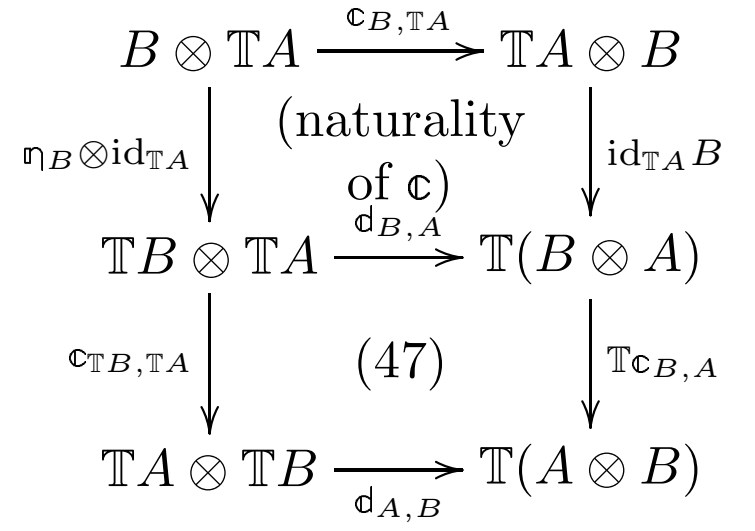

## Appendix B. Proof of Lemma 8.3

We first check that $(\mathcal{D}, I^{\mathcal{D}}, \otimes^{\mathcal{D}}, \alpha^{\mathcal{D}}, \ell^{\mathcal{D}}, r^{\mathcal{D}})$ is a monoidal category. Write $\bar{\circ}$ the notion of composition in $\mathcal{D} = \mathbf{Kleisli}(T)$, i.e., $g \bar{\circ} f = \mu \circ Tg \circ f$.

— Functoriality of $\otimes^{\mathcal{D}}$. First, $\otimes^{\mathcal{D}}$ applied to the identity on $(A, B)$ is $\mathrm{id}_A \otimes^{\mathcal{D}} \mathrm{id}_B = \mathrm{id}_{F_T(A)} \otimes^{\mathcal{D}} \mathrm{id}_{F_T(B)} = F_T(\mathrm{id}_A) \otimes^{\mathcal{D}} F_T(\mathrm{id}_B) = F_T(\mathrm{id}_A \otimes^{\mathcal{C}} \mathrm{id}_B) = F_T(\mathrm{id}_{A \otimes^{\mathcal{C}} B}) = \mathrm{id}_{F_T(A \otimes^{\mathcal{C}} B)} = \mathrm{id}_{A \otimes^{\mathcal{D}} B}$.
Second, $\otimes^{\mathcal{D}}$ preserves composition. Indeed, let $f, f', g, g'$ be morphisms from $A$ to $B$, from $A'$ to $B'$, from $B$ to $C$ and from $B'$ to $C'$ respectively in $\mathcal{D}$. Then the composition of $g \otimes^{\mathcal{D}} g'$ with $f \otimes^{\mathcal{D}} f'$ in $\mathcal{D}$ is the morphism $\mu_{C \otimes^{\mathcal{C}} C'} \circ T(d_{C,C'} \circ (g \otimes^{\mathcal{C}} g')) \circ d_{B,B'} \circ (f \otimes^{\mathcal{C}} f')$ in $\mathcal{C}$. The following diagram then commutes:

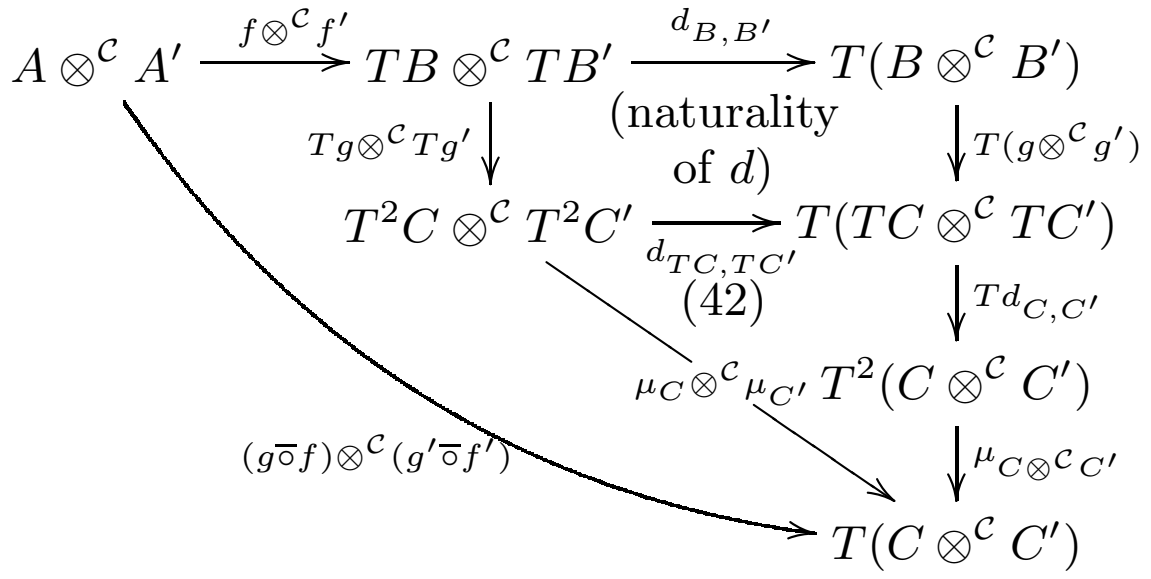

The top right route from $A \otimes^{\mathcal{C}} A'$ to $T(C \otimes^{\mathcal{C}} C')$ is $(g \otimes^{\mathcal{D}} g') \bar{\circ} (f \otimes^{\mathcal{D}} f')$, while the straight line diagonal is $(g \bar{\circ} f) \otimes^{\mathcal{D}} (g' \bar{\circ} f')$.

— Naturality of $\alpha^{\mathcal{D}}$. Let $f$, $g$, $h$ be morphisms from $A$ to $A'$, $B$ to $B'$ and $C$ to $C'$ respectively in $\mathcal{D}$. Then $\alpha^{\mathcal{D}}$ composed in $\mathcal{D}$ with $(f \otimes^{\mathcal{D}} g) \otimes^{\mathcal{D}} h$ is $\mu \circ T(\eta \circ \alpha^{\mathcal{C}}) \circ d \circ ((d \circ (f \otimes^{\mathcal{C}} g)) \otimes^{\mathcal{C}} h) = T\alpha^{\mathcal{C}} \circ d \circ (d \otimes^{\mathcal{C}} \mathrm{id}) \circ ((f \otimes^{\mathcal{C}} g) \otimes^{\mathcal{C}} h)$ (the route going all the way down then right from the top left corner to the bottom right corner below), while $f \otimes^{\mathcal{D}} (g \otimes^{\mathcal{D}} h)$ composed with $\alpha^{\mathcal{D}}$ in $\mathcal{D}$ is $\mu \circ T(d \circ (f \otimes^{\mathcal{C}} (d \circ (g \otimes^{\mathcal{C}} h)))) \circ \eta \circ \alpha^{\mathcal{C}} = \mu \circ Td \circ T(\mathrm{id} \otimes^{\mathcal{C}} d) \circ T(f \otimes^{\mathcal{C}} (g \otimes^{\mathcal{C}} h)) \circ \eta \circ \alpha^{\mathcal{C}}$ (the other route, going right then down).

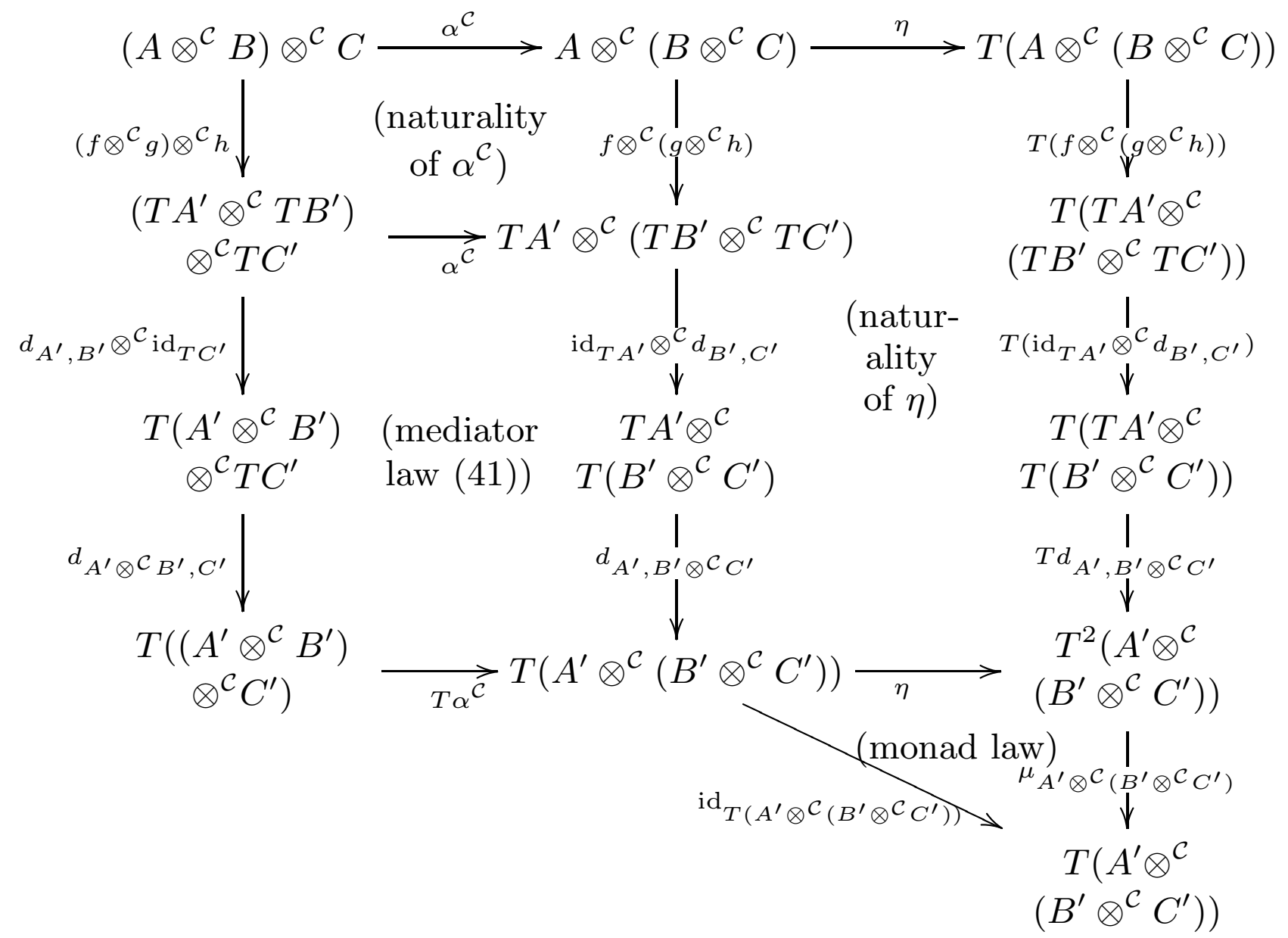

— Naturality of $\ell^{\mathcal{D}}$. Let $f$ be any morphism from $B$ to $B'$ in $\mathcal{D}$. Then the composition of $f$ with $\ell^{\mathcal{D}}$ in $\mathcal{D}$ is $\mu \circ Tf \circ \eta \circ \ell^{\mathcal{C}}$ (route going right then down), while the composition of $\ell^{\mathcal{D}}$ with $(\mathrm{id}^{\mathcal{D}} \otimes^{\mathcal{D}} f)$ is $\mu \circ T(\eta \circ \ell^{\mathcal{C}}) \circ d \circ (\eta \otimes^{\mathcal{C}} f)$ (dented route going down then right).

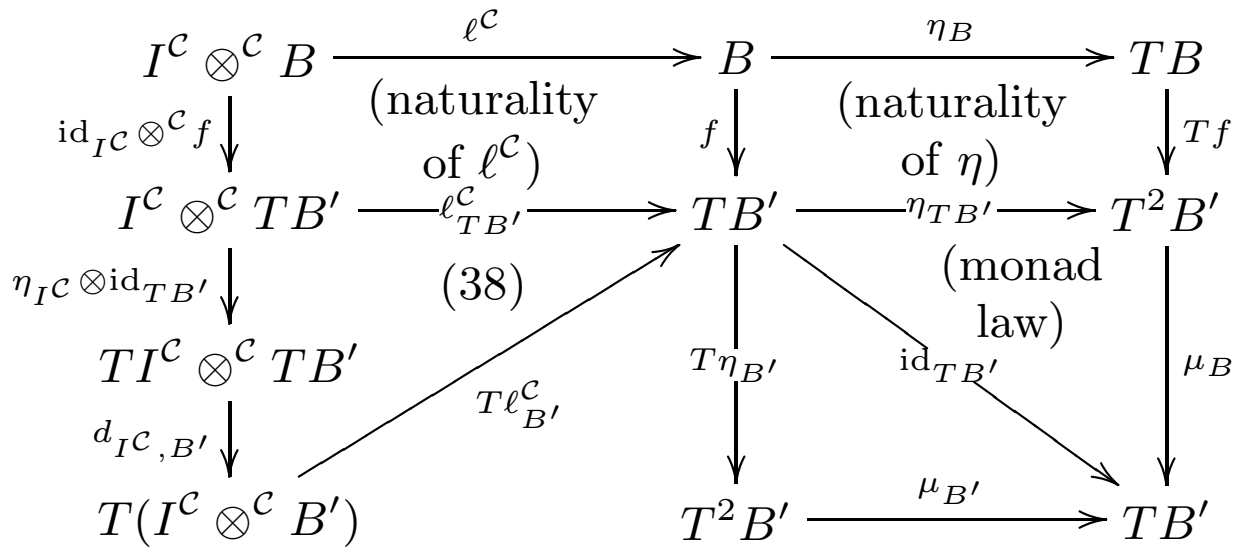

— Naturality of $r^{\mathcal{D}}$ is obtained similarly, using (39) instead of (38).

— $\alpha^{\mathcal{D}}$, $\ell^{\mathcal{D}}$, $r^{\mathcal{D}}$ are isomorphisms: indeed $\alpha^{\mathcal{D}} = F_T(\alpha^{\mathcal{C}})$, $\alpha^{\mathcal{C}}$ is an isomorphism, and every functor preserves isomorphisms. Similarly for $\ell^{\mathcal{D}}$ and $r^{\mathcal{D}}$. Since they are natural, they are natural isomorphisms.

— Pentagon identity (14). We have to check that (dropping subscripts) $\alpha^{\mathcal{D}} \bar{\circ} \alpha^{\mathcal{D}} = (\mathrm{id}^{\mathcal{D}} \otimes^{\mathcal{D}} \alpha^{\mathcal{D}}) \bar{\circ} \alpha^{\mathcal{D}} \bar{\circ} (\alpha^{\mathcal{D}} \otimes^{\mathcal{D}} \mathrm{id}^{\mathcal{D}})$. Note that the left-hand side is $\mu \circ T(\eta \circ \alpha^{\mathcal{C}}) \circ \eta \circ \alpha^{\mathcal{C}} = T\alpha^{\mathcal{C}} \circ \eta \circ \alpha^{\mathcal{C}} = \eta \circ \alpha^{\mathcal{C}} \circ \alpha^{\mathcal{C}}$ (by naturality of $\eta$, $Tf \circ \eta = \eta \circ f$ for any $f$). Now notice that by (40), $d \circ (\eta \otimes^{\mathcal{C}} \eta) = \eta$. Also, $\mu \circ T\eta = \mathrm{id}$. So the right-hand side is:

$$\begin{aligned}
& \mu \circ T(d \circ (\eta \otimes^{\mathcal{C}} (\eta \circ \alpha^{\mathcal{C}}))) \circ [\alpha^{\mathcal{D}} \bar{\circ} (\alpha^{\mathcal{D}} \otimes^{\mathcal{D}} \mathrm{id}^{\mathcal{D}})] \\
= \;& \mu \circ T(d \circ (\eta \otimes^{\mathcal{C}} \eta)) \circ T(\mathrm{id} \otimes^{\mathcal{C}} \alpha^{\mathcal{C}}) \circ [\alpha^{\mathcal{D}} \bar{\circ} (\alpha^{\mathcal{D}} \otimes^{\mathcal{D}} \mathrm{id}^{\mathcal{D}})] \\
= \;& \mu \circ T\eta \circ T(\mathrm{id} \otimes^{\mathcal{C}} \alpha^{\mathcal{C}}) \circ [\alpha^{\mathcal{D}} \bar{\circ} (\alpha^{\mathcal{D}} \otimes^{\mathcal{D}} \mathrm{id}^{\mathcal{D}})] \qquad \text{(by (40))} \\
= \;& T(\mathrm{id} \otimes^{\mathcal{C}} \alpha^{\mathcal{C}}) \circ [\alpha^{\mathcal{D}} \bar{\circ} (\alpha^{\mathcal{D}} \otimes^{\mathcal{D}} \mathrm{id}^{\mathcal{D}})] \\
= \;& T(\mathrm{id} \otimes^{\mathcal{C}} \alpha^{\mathcal{C}}) \circ \mu \circ T(\eta \circ \alpha^{\mathcal{C}}) \circ (\alpha^{\mathcal{D}} \otimes^{\mathcal{D}} \mathrm{id}^{\mathcal{D}}) \\
= \;& T(\mathrm{id} \otimes^{\mathcal{C}} \alpha^{\mathcal{C}}) \circ T\alpha^{\mathcal{C}} \circ (\alpha^{\mathcal{D}} \otimes^{\mathcal{D}} \mathrm{id}^{\mathcal{D}}) \\
= \;& T(\mathrm{id} \otimes^{\mathcal{C}} \alpha^{\mathcal{C}}) \circ T\alpha^{\mathcal{C}} \circ d \circ ((\eta \circ \alpha^{\mathcal{C}}) \otimes^{\mathcal{C}} \eta) \\
= \;& T(\mathrm{id} \otimes^{\mathcal{C}} \alpha^{\mathcal{C}}) \circ T\alpha^{\mathcal{C}} \circ d \circ (\eta \otimes^{\mathcal{C}} \eta) \circ (\alpha^{\mathcal{C}} \otimes^{\mathcal{C}} \mathrm{id}) \\
= \;& T(\mathrm{id} \otimes^{\mathcal{C}} \alpha^{\mathcal{C}}) \circ T\alpha^{\mathcal{C}} \circ \eta \circ (\alpha^{\mathcal{C}} \otimes^{\mathcal{C}} \mathrm{id}) \qquad \text{(by (40))} \\
= \;& \eta \circ (\mathrm{id} \otimes^{\mathcal{C}} \alpha^{\mathcal{C}}) \circ \alpha^{\mathcal{C}} \circ (\alpha^{\mathcal{C}} \otimes^{\mathcal{C}} \mathrm{id}) \qquad \text{(by naturality of } \eta)
\end{aligned}$$

and we are done, since by the pentagon identity (14) for $\alpha^{\mathcal{C}}$, $\alpha^{\mathcal{C}} \circ \alpha^{\mathcal{C}} = (\mathrm{id} \otimes^{\mathcal{C}} \alpha^{\mathcal{C}}) \circ \alpha^{\mathcal{C}} \circ (\alpha^{\mathcal{C}} \otimes^{\mathcal{C}} \mathrm{id})$.

— Triangle identity (15). We must check that $(\mathrm{id}^{\mathcal{D}} \otimes^{\mathcal{D}} \ell^{\mathcal{D}}) \bar{\circ} \alpha^{\mathcal{D}} = r^{\mathcal{D}}$. The left-hand side is $\mu \circ T(d \circ (\eta \otimes^{\mathcal{C}} (\eta \circ \ell^{\mathcal{C}})) \circ \eta \circ \alpha^{\mathcal{C}} = \mu \circ T(d \circ (\eta \otimes^{\mathcal{C}} \eta) \circ (\mathrm{id} \otimes^{\mathcal{C}} \ell^{\mathcal{C}})) \circ \eta \circ \alpha^{\mathcal{C}} = \mu \circ T(\eta \circ (\mathrm{id} \otimes^{\mathcal{C}} \ell^{\mathcal{C}})) \circ \eta \circ \alpha^{\mathcal{C}}$ (by (40)) $= T(\mathrm{id} \otimes^{\mathcal{C}} \ell^{\mathcal{C}}) \circ \eta \circ \alpha^{\mathcal{C}}$ (by the monad laws) $= \eta \circ (\mathrm{id} \otimes^{\mathcal{C}} \ell^{\mathcal{C}}) \circ \alpha^{\mathcal{C}}$ (by naturality of $\eta$) $= \eta \circ r^{\mathcal{C}}$ (by the triangle identity (15) in $\mathcal{C}$) $= r^{\mathcal{D}}$, which is the right-hand side.

We must now check that $F_T \dashv U_T$ is a monoidal adjunction. We already know that it is an adjunction.

— $(\theta^{F_T}, \iota^{F_T})$ is mediating for $F$. Indeed, both $\theta^{F_T}_{A,B}$ and $\iota^{F_T}$ are just identity morphisms in $\mathcal{D}$, from which the coherence conditions (16), (17) and (18) are clear. The first reduces to showing that $F_T(\alpha^{\mathcal{C}}) = \alpha^{\mathcal{D}}$, the second to $F_T(\ell^{\mathcal{C}}) = \ell^{\mathcal{D}}$, the third to $F_T(r^{\mathcal{C}}) = r^{\mathcal{D}}$, which hold by construction.

— $(\theta^{U_T}, \iota^{U_T})$ is mediating for $U_T$. Indeed, the first coherence condition (16) is just Diagram (41), since $U_T \alpha^{\mathcal{D}} = U_T F_T \alpha^{\mathcal{C}} = T\alpha^{\mathcal{C}}$. The second coherence condition (17) is exactly Diagram (38), and the third, (18), is exactly Diagram (39).

— The unit $\eta$ and the counit $\epsilon$ are monoidal natural transformations. (Recall that $\epsilon_A$ is the identity morphism from $TA$ to $TA$ in $\mathcal{C}$, seen as a morphism from $TA$ to $A$ in $\mathcal{D}$.) Indeed, Diagram (49) (in $\mathcal{C}$) means that $\eta_{I^{\mathcal{C}}} = U_T(\iota^{F_T}) \circ \iota^{U_T} = U_T(\mathrm{id}) \circ \eta_{I^{\mathcal{C}}}$, which is obvious. Diagram (50) states that $U_T(\theta^{F_T}_{A,B}) \circ \theta^{U_T} \circ (\eta_A \otimes^{\mathcal{C}} \eta_B) = \eta_{A \otimes^{\mathcal{C}} B}$, i.e., $d_{A,B} \circ (\eta_A \otimes^{\mathcal{C}} \eta_B) = \eta_{A \otimes^{\mathcal{C}} B}$, which is just Diagram (40). Turning to diagrams in $\mathcal{D}$, Diagram (51) means that $\epsilon_{I^{\mathcal{D}}} \bar{\circ} F_T(\iota^{U_T}) \bar{\circ} \iota^{F_T}$ is the identity in $\mathcal{D}$; seen as a morphism in $\mathcal{C}$, this is

$$\begin{aligned}
\mu \circ T(\mathrm{id}) \circ [F_T(\iota^{U_T}) \bar{\circ} \iota^{F_T}] &= \mu \circ [F_T(\iota^{U_T}) \bar{\circ} \iota^{F_T}] \\
&= \mu \circ \mu \circ T(F_T(\iota^{U_T})) \circ \iota^{F_T} \\
&= \mu \circ \mu \circ T(\eta_{TI^{\mathcal{C}}} \circ \eta_{I^{\mathcal{C}}}) \circ \eta_{I^{\mathcal{C}}} = \eta_{I^{\mathcal{C}}}
\end{aligned}$$

(by the monad laws), and the latter is just the identity on $I^{\mathcal{D}}$ in $\mathcal{D}$, as required. Finally, Diagram (52) means that $\epsilon_{A \otimes^{\mathcal{D}} B} \bar{\circ} F_T \theta^{U_T} \bar{\circ} \theta^{F_T} = \epsilon_A \otimes^{\mathcal{D}} \epsilon_B$. Seen as a morphism in $\mathcal{C}$,

the left-hand side is

$$\begin{aligned}
& \mu_{A\otimes^{\mathcal{C}}B} \circ T\epsilon_{A\otimes^{\mathcal{D}}B} \circ [F_T\theta^{U_T}\bar{\circ}\theta^{F_T}] \\
= \; & \mu_{A\otimes^{\mathcal{C}}B} \circ T\mathrm{id} \circ [F_T\theta^{U_T}\bar{\circ}\theta^{F_T}] \\
= \; & \mu_{A\otimes^{\mathcal{C}}B} \circ [F_T\theta^{U_T}\bar{\circ}\theta^{F_T}] \\
= \; & \mu_{A\otimes^{\mathcal{C}}B} \circ \mu_{T(A\otimes^{\mathcal{C}}B)} \circ T(F_T\theta^{U_T}) \circ \theta^{F_T} \\
= \; & \mu_{A\otimes^{\mathcal{C}}B} \circ \mu_{T(A\otimes^{\mathcal{C}}B)} \circ T(\eta_{T(A\otimes^{\mathcal{C}}B)} \circ d_{A,B}) \circ \eta_{A\otimes^{\mathcal{C}}B}
\end{aligned}$$

which we recognize as the long trip in the diagram below from the top left corner that goes down, right, and up to the top right corner.

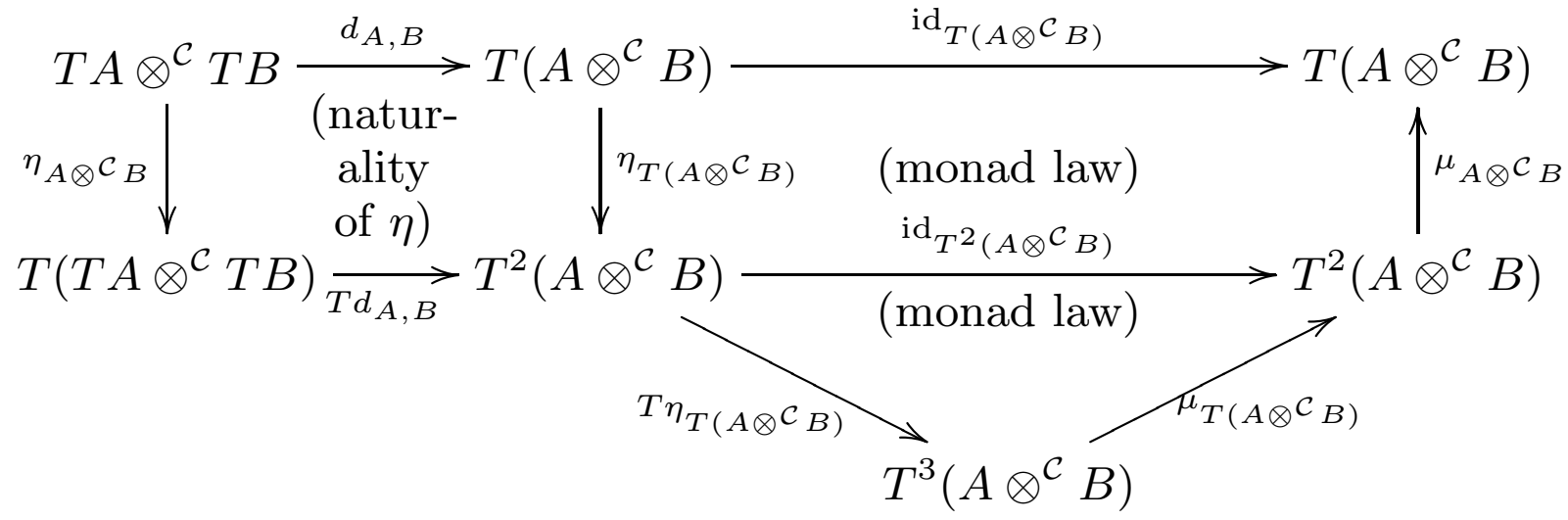

Then the composition of arrows from the top left corner to the top right corner is $d_{A,B}$. This is exactly the desired morphism $\epsilon_A \otimes^{\mathcal{D}} \epsilon_D$ of $\mathcal{D}$, which, seen as a morphism of $\mathcal{C}$, is $d_{A,B} \circ (\mathrm{id}_{TA} \otimes^{\mathcal{C}} \mathrm{id}_{TB})$ by definition.

We check finally that $F_T \dashv U_T$ generates the monoidal monad. The only thing to check is $d_{A,B} = U_T\theta^{F_T}_{A,B} \circ \theta^{U_T}_{F_T(A),F_T(B)}$. This is clear, since the right-hand side is $U_T(\mathrm{id}_{A\otimes^{\mathcal{D}}B}) \circ d_{A,B} = d_{A,B}$. $\square$

**Appendix C. Composition of Monoidal Adjunctions**

We first show that $|U|$ is a monoidal functor, with mediating pair $(\left|\theta^U\right| \circ \mathbb{\theta}, \left|\iota^U\right| \circ \mathbb{i})$.

We must check all three coherence conditions (16), (17), (18).

Diagram (16).

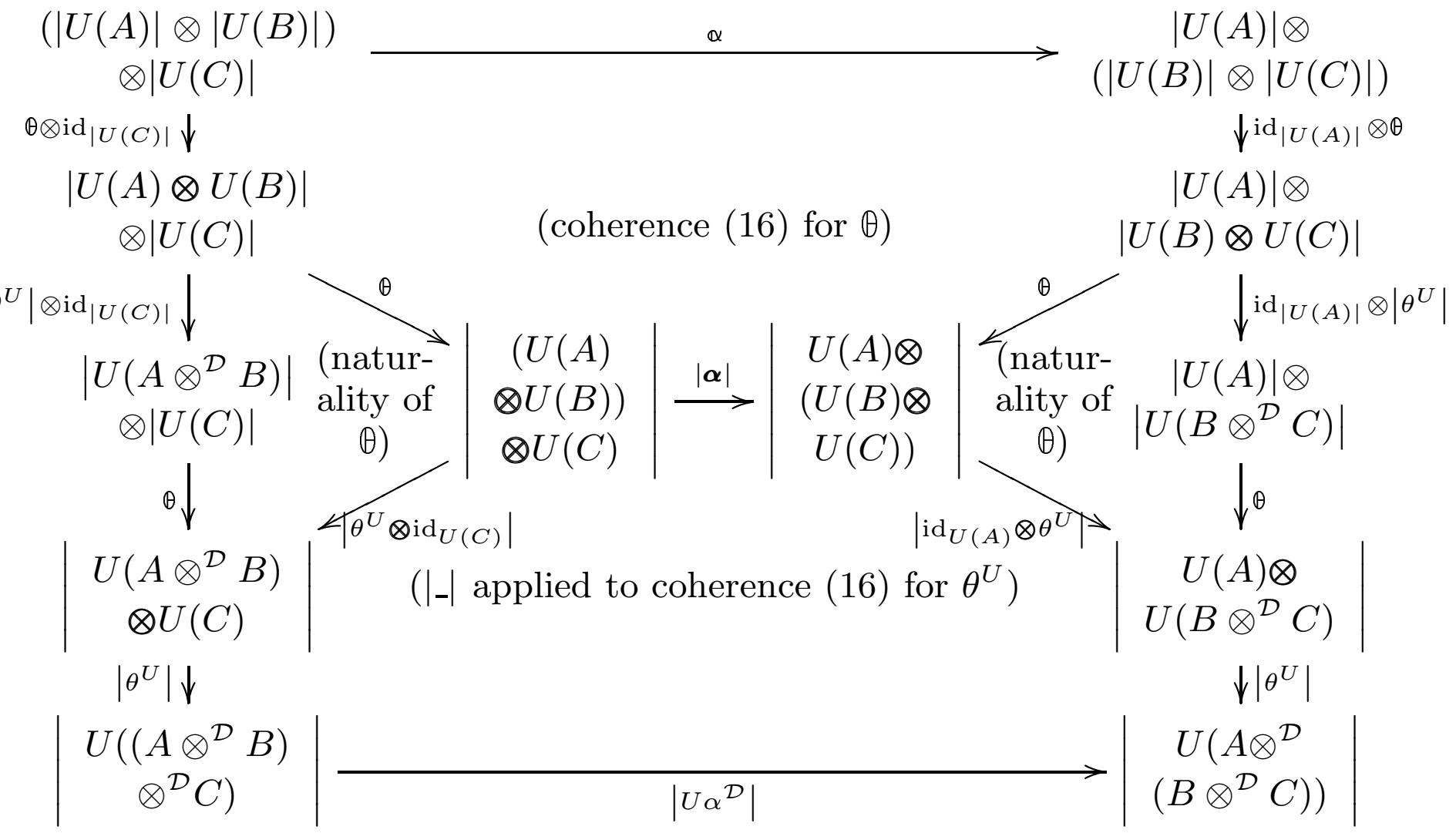

Diagram (17).

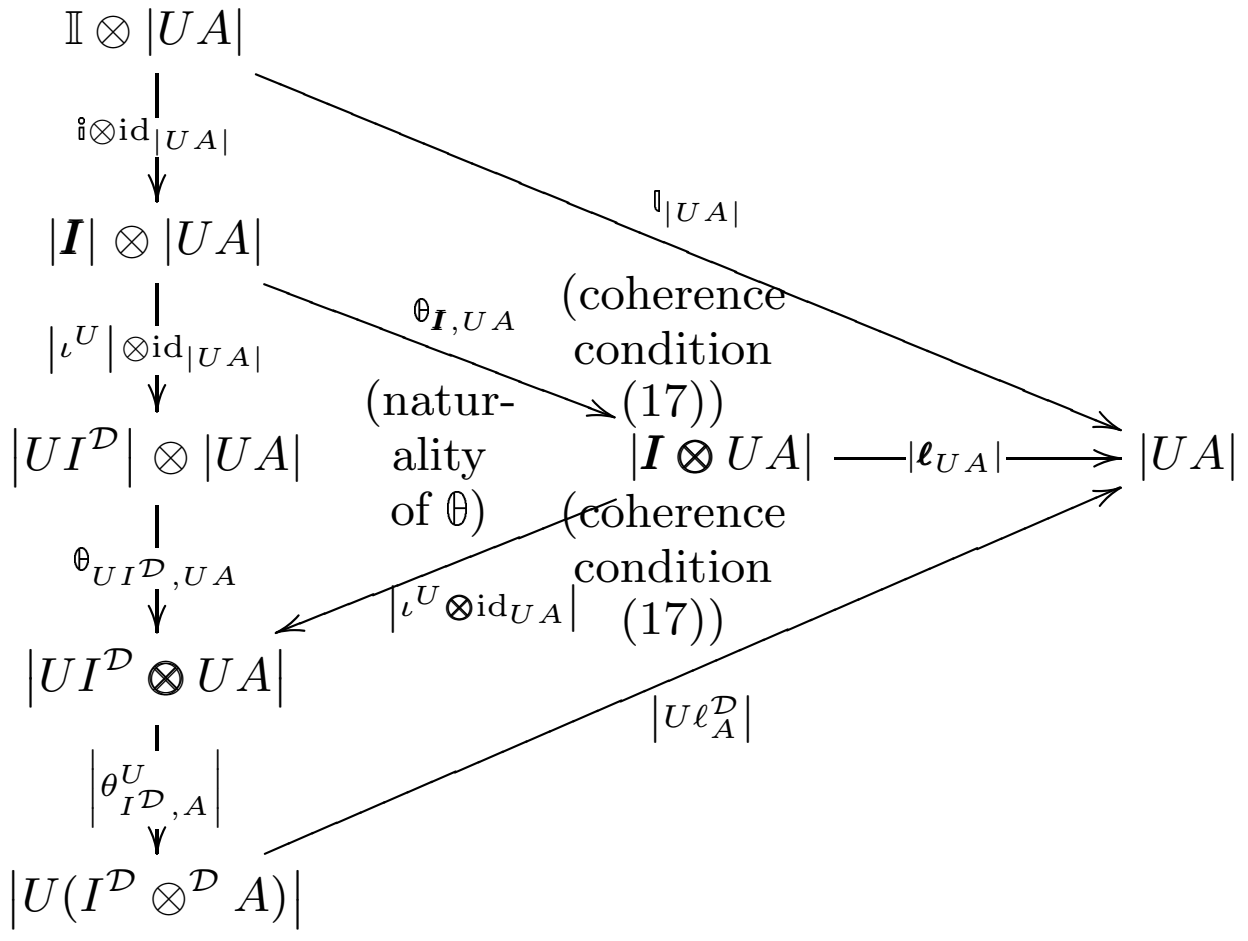

Diagram (18). This is checked by a diagram similar to the latter.

By the same reasoning, $F\boldsymbol{D}$ is also a monoidal functor, with mediating pair $(F\boldsymbol{\theta} \circ \theta^F, F\boldsymbol{i} \circ \iota^F)$.

We now check the monoidal adjunction conditions (49), (50), (51), (52). Let $\dot{\epsilon}$ the counit, and $\dot{\eta}$ the unit of $\boldsymbol{D} \dashv |_|$, and $\boldsymbol{\epsilon}$ the counit, and $\boldsymbol{\eta}$ the unit of $F \dashv U$.

Diagram (49).

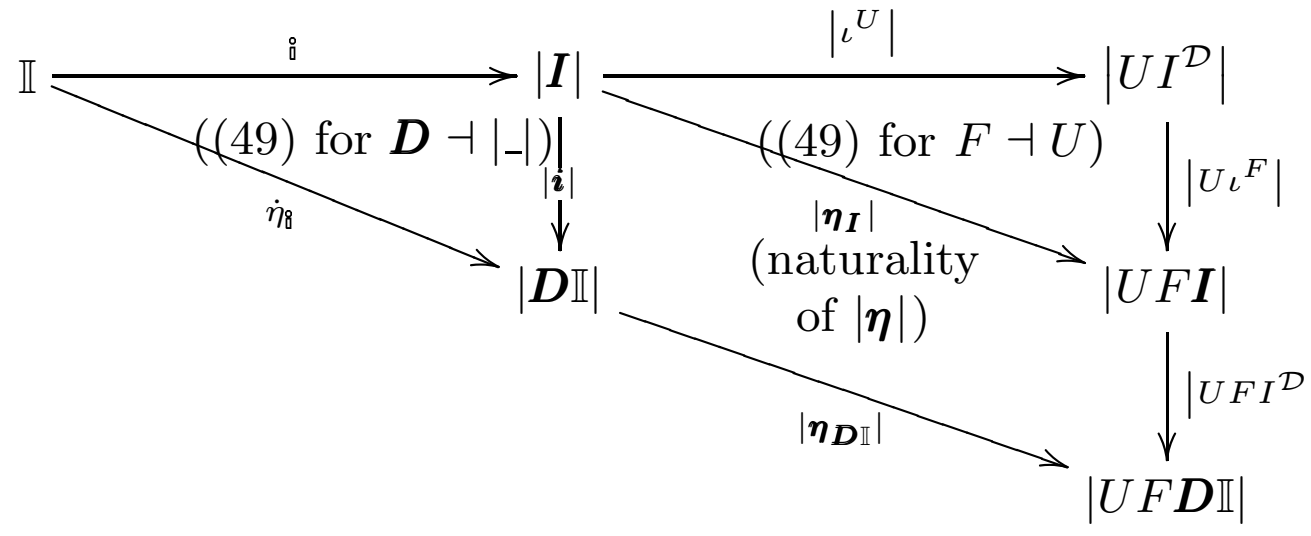

Diagram (50).

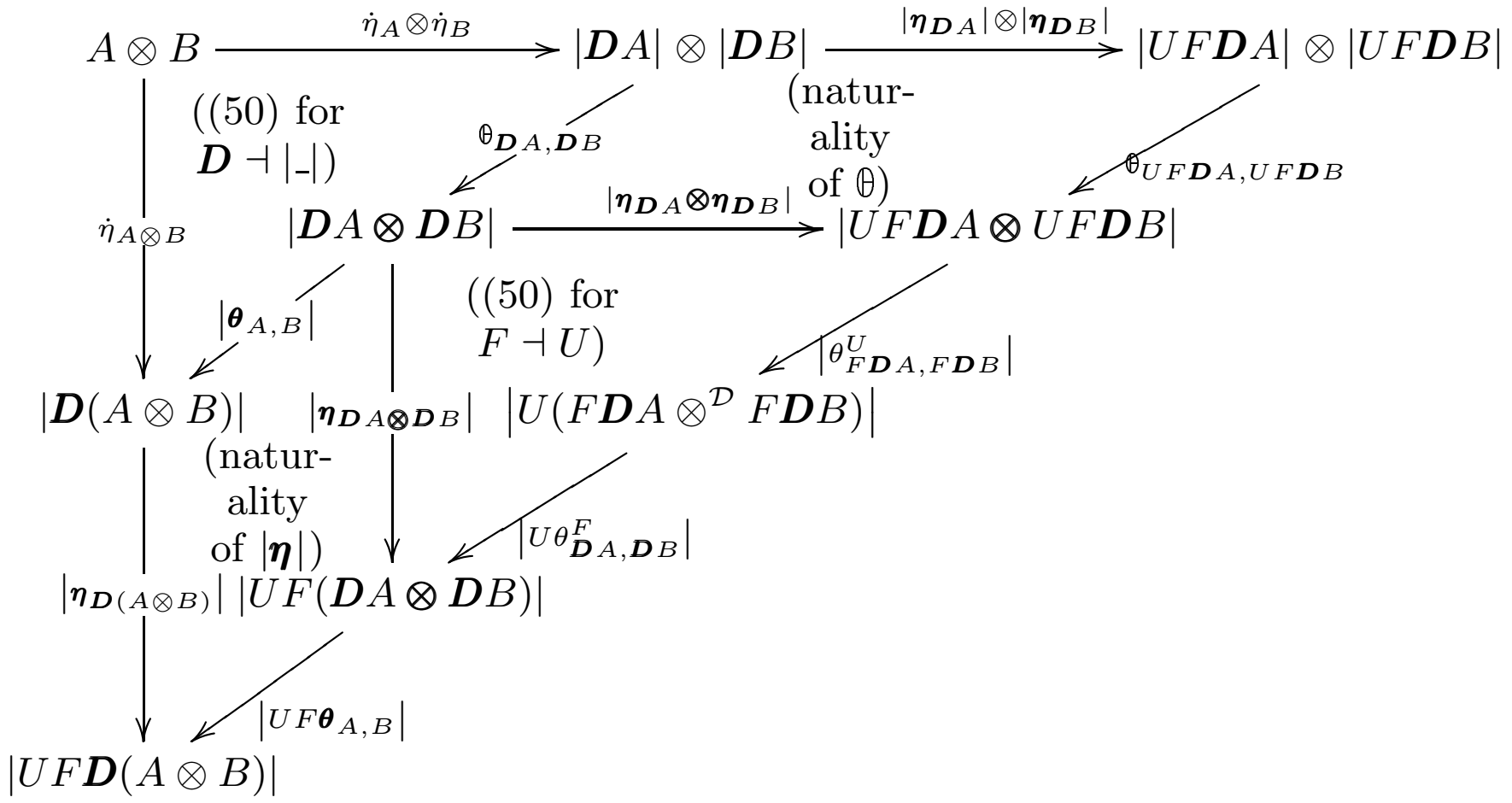

Diagrams (51) and (52) are dealt with similar, and give rise to similar verifications.

## Appendix D. Proof of Proposition 8.6

Diagram (32) is obtained by considering the following diagram. We drop subscripts on natural transformations so as to save space; they can be inferred from context.

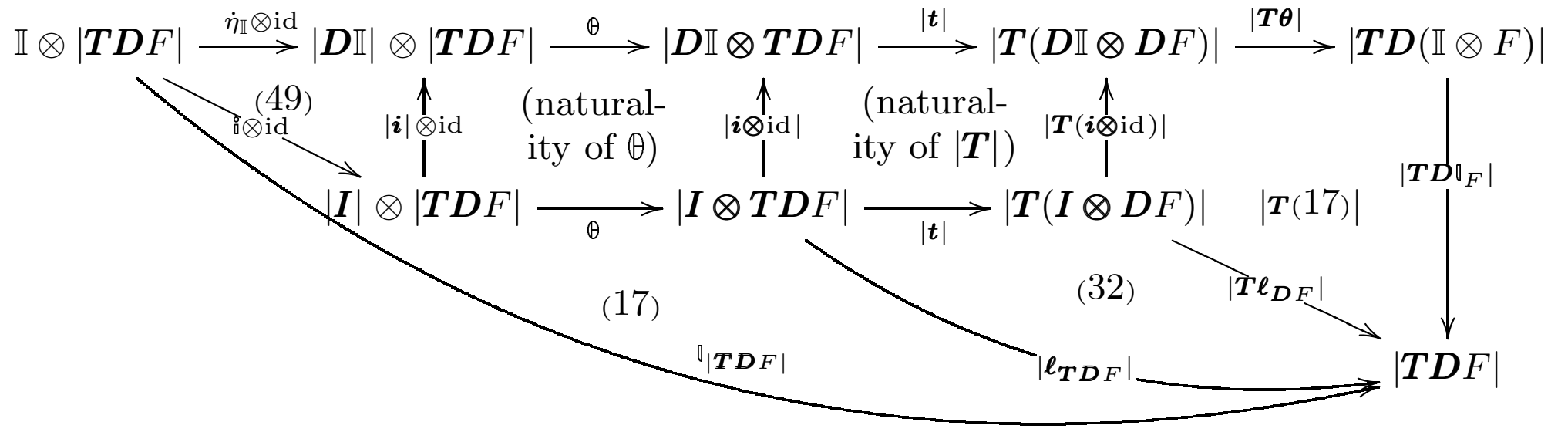

We have also decorated the interior of each face with the justification why its edges commute. For example, the top left triangle is a copy of (49), one of the diagrams stating that the adjunction is monoidal. The bottom face, whose bottom edge is a curved arrow, commutes by the coherence square (17) for $\mathbb{\theta}$. The rightmost parallelogram commutes by $|\boldsymbol{T}|$ applied to another coherence square (17), this time for $\boldsymbol{\theta}$. Now the topmost composition of arrows is $\mathbb{t}_{\mathbb{I},F}$, the bottommost arrow from $\mathbb{I} \otimes |\boldsymbol{TD}F|$ to $|\boldsymbol{TD}F|$ is $\mathbb{l}_{\mathbb{T}F}$, and the rightmost vertical arrow is $\mathbb{T}\mathbb{l}_F$.

Diagram (33). This is the diagram below. We recognize $\mathbb{t}_{E,F}$ as the rightmost composition of vertical arrows, $\mathrm{id}_E \otimes \mathbb{\eta}_F$ as the topmost composition of horizontal arrows, and $\mathbb{\eta}_{E\otimes F}$ is the diagonal from $E \otimes F$ to $|\boldsymbol{TD}(E \otimes F)|$.

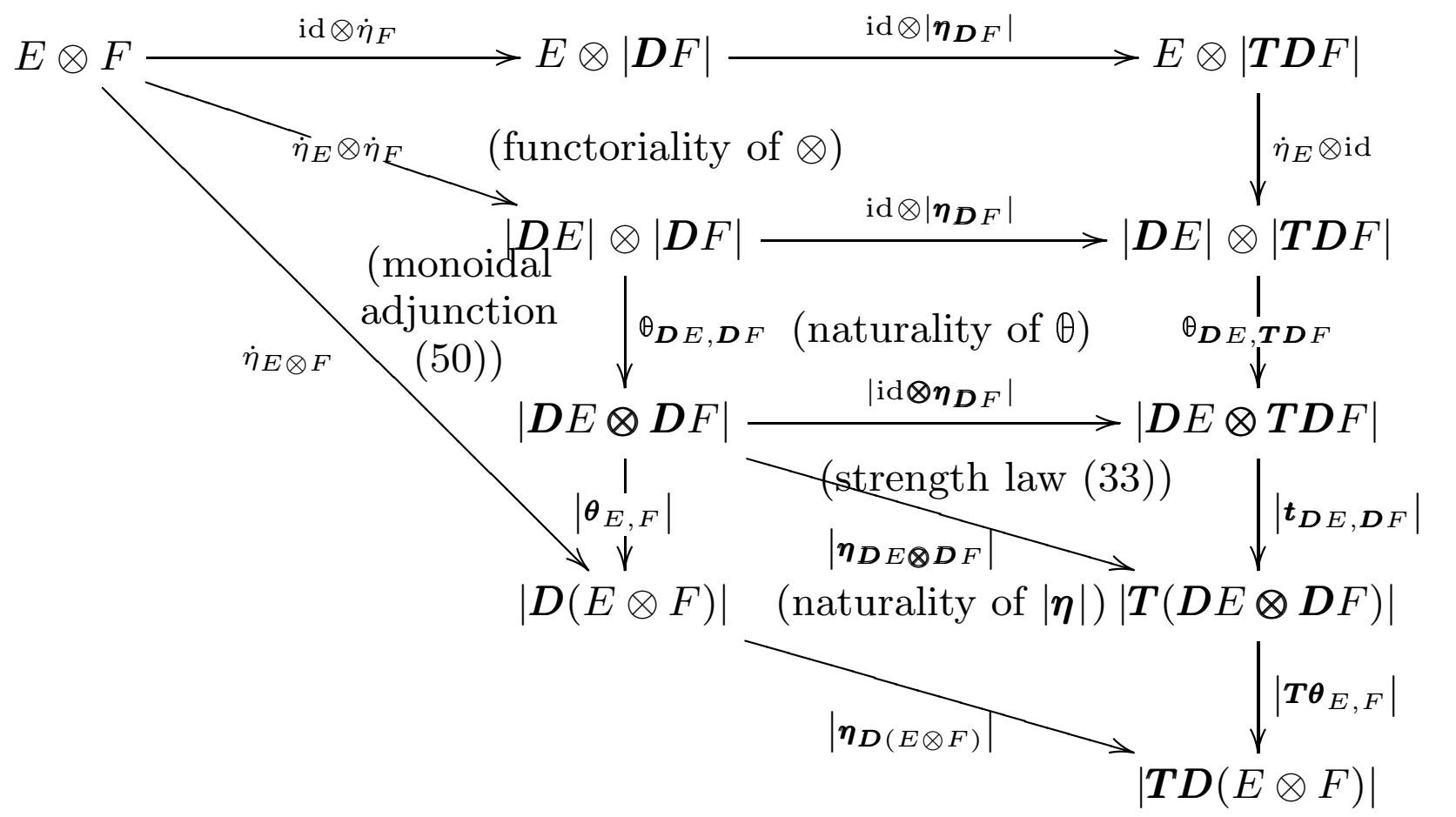

Diagram (34). For reasons of space, we flip the diagram so that arrows involving strengths are vertical, and arrows involving associativities are horizontal. Again, we drop most subscripts from natural transformations, which are inferrable from context.

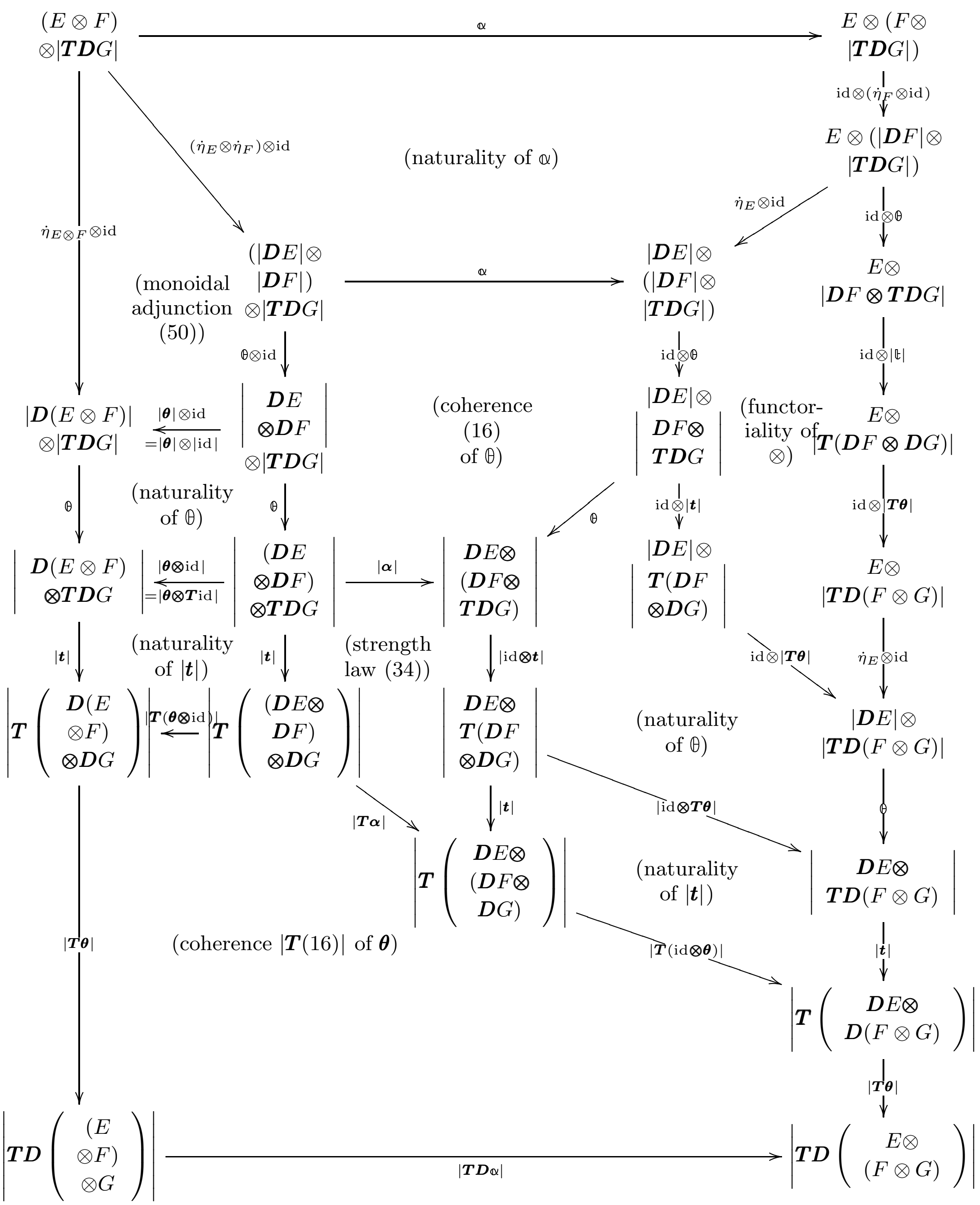

The vertical arrows on the left compose to form $t_{A\otimes^{\mathcal{C}} B,C}$, while the vertical arrows on the right compose to form $\mathrm{id} \otimes^{\mathcal{C}} t_{B,C}$ followed by $t_{A,B\otimes^{\mathcal{C}} C}$, whence the result.

Diagram (35). This is the diagram below. Again, the diagram is flipped sideways, and certain subscripts have been dropped. The topmost arrow from $E \otimes |\boldsymbol{TD}|\boldsymbol{TD}F||$ to $E \otimes |\boldsymbol{TD}F|$ is $\mathrm{id}_E \otimes \mathbb{\mu}_F$ by definition, while the bottommost arrow from $|\boldsymbol{TD}|\boldsymbol{TD}(E \otimes F)||$ to $|\boldsymbol{TD}(E \otimes F)|$ is $\boldsymbol{\mu}_{E\otimes F}$. The rightmost composition of vertical arrows, from $E \otimes |\boldsymbol{TD}F|$ to $|\boldsymbol{TD}(E \otimes F)|$ is $\boldsymbol{\mu}_{E\otimes F}$, is $\mathbb{t}_{E,F}$, and the lefmost one is $\mathbb{t}_{E,\mathbb{T}F}$ followed by $\mathbb{T}\mathbb{t}_{E,F}$. (Recall that $\mathbb{T} = |\boldsymbol{TD}|$.)

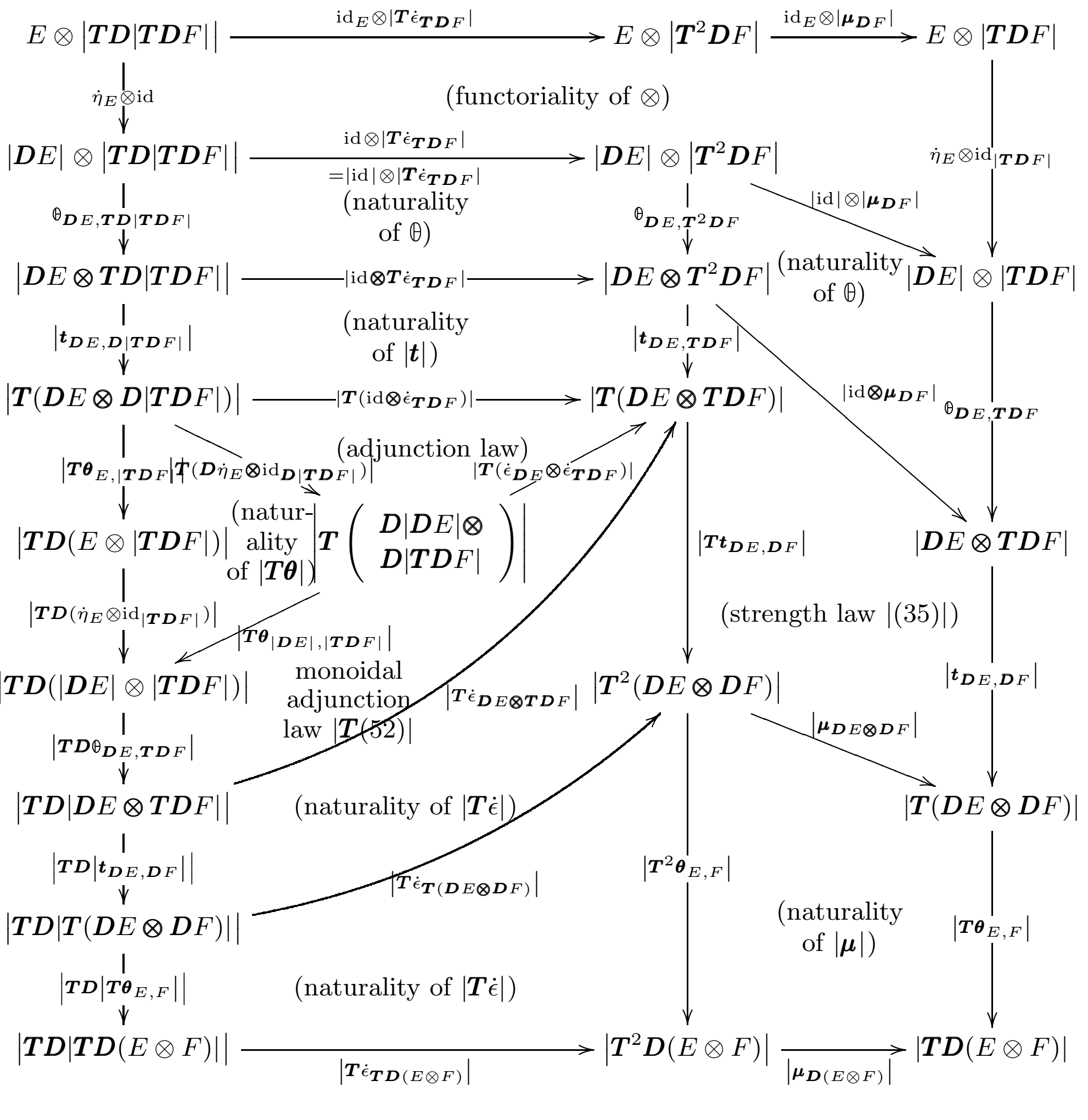

Diagram (36). Again, we flip the diagram sideways. We recognize $|\boldsymbol{t}_{A_1,A_2}|$ as the rightmost vertical arrow, $\mathbb{t}_{|A_1|,|A_2|}$ as the leftmost composition of vertical arrows, $\mathrm{id}_{|A_1|} \otimes \sigma_{A_2}$ followed by $\mathbb{\theta}_{A_1,\boldsymbol{T}A_2}$ on the top row, and $\mathbb{T}(\mathbb{\theta}_{A_1,A_2})$ followed by $\sigma_{A_1 \otimes A_2}$ on the bottom row.

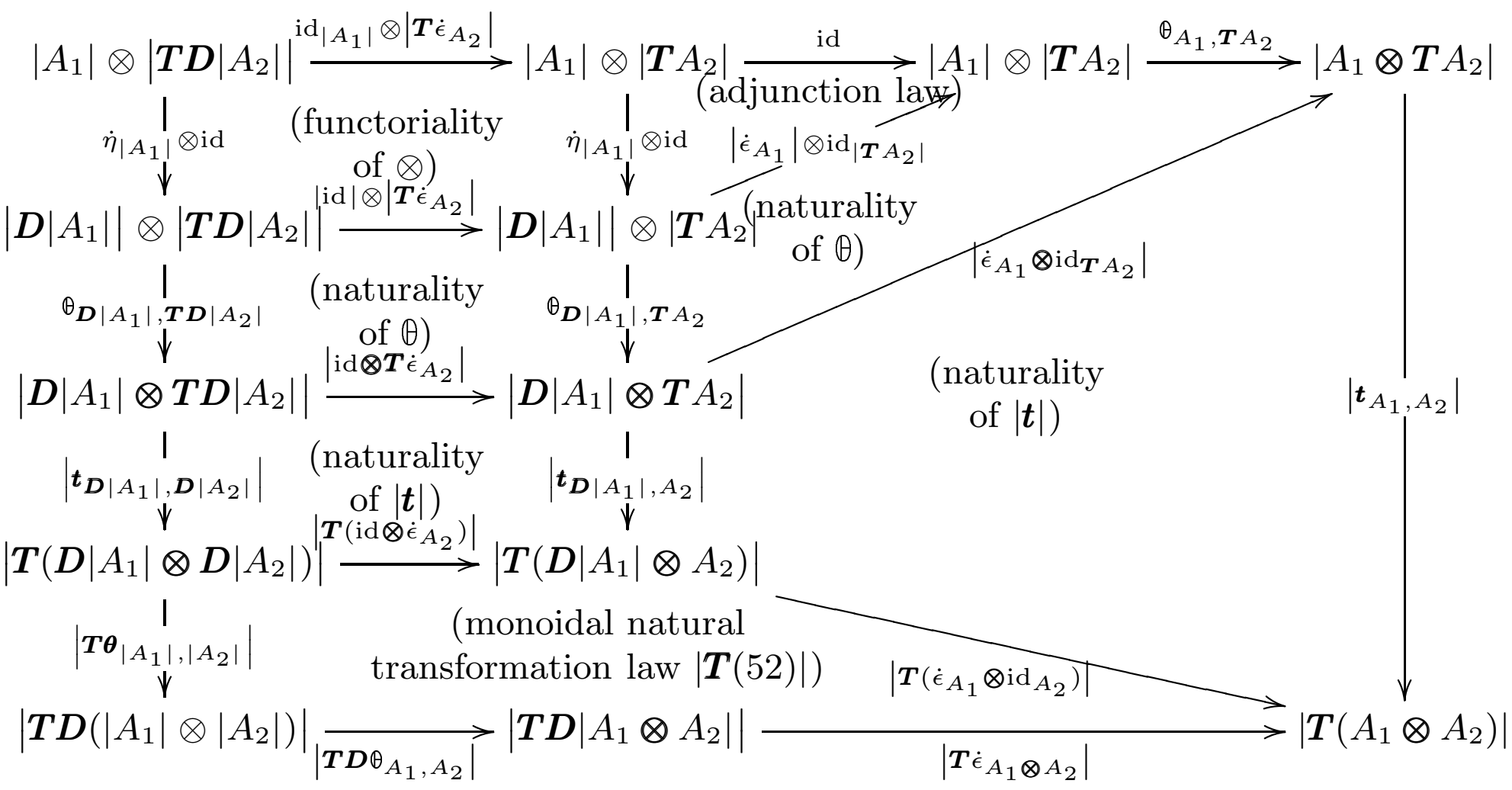

(functoriality of ⊗)
(adjunction law)
(naturality of θ)
(naturality of θ)
(naturality of |t|)
(naturality of |t|)
(monoidal natural transformation law |T(52)|)

□